\documentclass[10pt]{article}
\usepackage{psfig,graphicx,epsfig,amsmath,amssymb}

\bibstyle{apsrev.bib}

\def\oppropto{\mathop{\propto}} 
\def\opsimeq{\mathop{\simeq}}

\def\opsim{\mathop{\sim}} 

\def\fig#1#2{\includegraphics[height=#1]{#2}}

\newcommand{\ms}{\noalign{\vspace{3pt plus2pt minus1pt}}}
\newcommand{\hs}{\noalign{\vspace{12pt plus2pt minus2pt}}}

\input epsf.sty

\catcode`\@=11
\def\numberbysection{\@addtoreset{equation}{section}
\def\theequation{\thesection.\arabic{equation}}}
 \numberbysection           

\begin{document}

\title{Strong disorder RG approach of random systems}

\author{Ferenc Igl\'oi \\
 Research Institute for Solid State Physics and Optics, \\
 H-1525 Budapest,
P.O.Box 49, Hungary \\
Institute for Theoretical Physics, \\
Szeged University
H-6720 Szeged, Hungary  \\
\\
C\'ecile Monthus \\
 Service de Physique Th\'eorique, \\
  Unit\'e de recherche associ\'ee au CNRS, \\
  DSM/CEA Saclay, 91191 Gif-sur-Yvette, France}

\maketitle

\begin{abstract}
  There is a large variety of quantum and classical systems in which
  the quenched disorder plays a dominant r\^ole over quantum, thermal,
  or stochastic fluctuations : these systems display strong spatial
  heterogeneities, and many averaged observables are actually governed
  by rare regions.  A unifying approach to treat the dynamical and/or
  static singularities of these systems has emerged recently,
  following the pioneering RG idea by Ma and Dasgupta and the detailed
  analysis by Fisher who showed that the Ma-Dasgupta RG rules yield
  asymptotic exact results if the broadness of the disorder grows
  indefinitely at large scales.  Here we report these new developments
  by starting with an introduction of the main ingredients of the
  strong disorder RG method.  We describe the basic properties of
  infinite disorder fixed points, which are realized at critical
  points, and of strong disorder fixed points, which control the
  singular behaviors in the Griffiths-phases.  We then review in
  detail applications of the RG method to various disordered models,
  either (i) quantum models, such as random spin chains, ladders and
  higher dimensional spin systems, or (ii) classical models, such as
  diffusion in a random potential , equilibrium at low temperature and
  coarsening dynamics of classical random spin chains, trap models,
  delocalization transition of a random polymer from an interface,
  driven lattice gases and reaction diffusion models in the presence
  of quenched disorder.  For several one-dimensional systems, the
  Ma-Dasgupta RG rules yields very detailed analytical results,
  whereas for other, mainly higher dimensional problems, the RG rules
  have to be implemented numerically.  If available, the strong
  disorder RG results are compared with another, exact or numerical
  calculations.

\end{abstract}


\newpage

\tableofcontents

\newpage


\section*{ Organization of the review}
\addcontentsline{toc}{section}{Organization of the review}

This work aims to review the recent developments obtained via strong
disorder renormalization group methods in the theory of disordered
systems. The strong disorder RG method, which has been introduced by
Ma and Dasgupta in 1979, has become a very efficient method with a
clear physical status only much later with the works of Daniel Fisher,
who showed that the RG method becomes asymptotically exact if the
distribution of disorder broadens without limits at large scales, and
who computed several exact critical exponents and scaling functions in
random quantum spin chains.  Following Fisher's results, intensive
research started, first in random quantum models and then in classical
disordered systems.  In all concerned fields the asymptotically exact
strong disorder RG results have provided new insights into the
physical mechanisms and helped to interpret the existing numerical
findings. It is likely that the strong disorder RG methods will give
new impulses in the future to the numerical and theoretical research
of disordered systems.

Early developments of the strong disorder RG method for random quantum
spin chains have been summarized in a conference proceedings
\cite{danielreview} and briefly described in \cite{bhatt,husephysrep}. However, a
comprehensive review with applications of different fields of research
is still lacking and we aim to fill this gap by this work.

The review is organized into three parts : 

(i) { \bf The first part is a general introduction to the essential
  concepts of the strong disorder RG method :}

In sections \ref{secmotiv} and \ref{secprinciples},
we explain the interest of defining disorder-dependent
 real space RG procedures, and
why strong disorder RG rules are an ideal tool to study systems in which disorder
dominates at large scales over quantum, thermal, or stochastic fluctuations.

Then in section \ref{chapRGrules},
we present the simplest specific examples of strong disorder RG
rules, the associated Infinite and Strong disorder fixed points, and
how exact non-trivial critical exponents emerge in this framework.
This section \ref{chapRGrules} being somewhat more technical, it can be skip at
the first reading, although it is essential to understand the 
meaning and the properties of strong disorder RG flows.

(ii)  {\bf The second part is devoted to the detailed study of various quantum models}

(ii)  { \bf The third part is devoted to the detailed study of various classical models}

The sections of Parts II and III devoted to specific models have
been written to be as independent as possible.  As a consequence,
after the general introduction contained in Part I, the reader
interested in a specific model may jump directly to the corresponding
section.  For readers who wish to learn strong disorder RG methods in
details in order to be able to use them, we recommend to read the
first sections of each part, devoted respectively to the random
transverse-field Ising chain in Sec.\ref{chapRTFIC} and to the Sinai
walk in Sec.\ref{chapsinai}, because these two sections are the more
detailed ones.

(iv) { \bf The Appendices} contain more technical aspects or more
specialized discussions.  In particular, Appendix A deals with the
scaling concept in disorder systems, making a comprehensive
presentation of scaling at different types of singular points
(conventional random fixed point, infinite disorder fixed point,
Griffiths singularities, large spin fixed point). The various types of
scaling are especially important to find the appropriate scaling
analysis of numerical data.

(iv) {\bf  The References}  

Before the references to original articles, we have added a short list
of books and reviews that give a more general background on topics or
models that we discuss only from the point of view of strong disorder
RG.

\part{ BASIC IDEAS OF STRONG DISORDER RG}

\section{ Motivations for disorder-dependent RG procedures}

\label{secmotiv}

\subsection*{Various descriptions of disordered systems}

The presence of disorder in a system can give rise to completely new
phenomena, like for instance the Anderson
localization\cite{andersonloc} in condensed matter physics, or the
aging behaviors in statistical
physics\cite{leticialeshouches,crisantiritort}. The study of
disordered systems has thus generated various specific approaches
since fifty years, for reviews
see\cite{binderyoung,replica,luckbook,crisantibook,sgandrf,kawashima}
. The first example is the Dyson-Schmidt method \cite{dyson,schmidt},
based on the notion of invariant measure : it yields exact results for
the one dimensional systems that can be described by infinite products
of random transfer matrices \cite{luckbook, crisantibook}. To
understand the interest and the specificity of strong disorder
renormalizations, which constitute the subject of this review, it is
convenient to classify the various ways of dealing with disordered the
systems into two main categories:

$\bullet$ there are on one hand {\bf approaches which start by
averaging over the disorder}, because their aim is to compute
self-averaging observables, such as for instance the
free energy if one is interested into the thermodynamics. 
There exist a certain number of specific prescriptions to carry
out this average over the disorder, such as the replica method \cite{replica}, the supersymmetric method \cite{susy}, and the dynamical method
\cite{dynamicreview} (for a parallel presentation of the three
methods, see \cite{kurchan3methods}). After this average on the disorder,
there are no more spatial heterogeneities, but the
homogeneous system that has to be studied contains
in counterpart new effective interactions.

$\bullet$ there are on the other hand {\bf approaches which try to
describe spatial heterogeneities of the disorder}, like certain famous
arguments and various real-space renormalizations : 
we will now discuss both in some more details, since these
approaches belong to the same `family' of the strong disorder RG.

\subsection*{From scaling arguments on disorder fluctuations...}

\label{argumentslocaux}

Among the arguments which have played a great role in the
understanding of disordered systems, one may quote

 (a) {\bf the Lifshitz argument} \cite{lifshitz, lifbook},
which allows to predict the essential singularities of the
density of states near spectrum  edges.
The idea consists in identifying
the disorder
configurations that support states in this energy region, and 
in estimating the probabilities of
these favorable configurations. 

(b) {\bf the Griffiths phases}, in which rare ordered regions 
induce essential singularities for the statics \cite{griffiths}
as well as very slow relaxation behaviors for the dynamics  \cite{randeira,
braygriffithsdyna}.

(c) {\bf the Harris criterion} \cite{harris} on the relevance
of weak disorder around a pure critical point. It consists
in estimating the
fluctuations of the critical temperature induced by
the disorder spatial fluctuations.

(d) {\bf the theorem of Chayes {\it  et al.}} \cite{chayes}, which shows that
the space fluctuations of the disorder imply the bound $\nu \geq
2/d$ for the critical exponent $\nu$ for disordered systems in
dimension $d$.

(e) {\bf the Imry-Ma argument} \cite{imryma}, which allows to predict
the presence of domain walls at zero temperature in random field
systems, by balancing the local energy fluctuations from the random
fields with the energy cost of domain walls.

(f) {\bf the theorem by Aizenman and Wehr} \cite{aizenmanwehr} about the rounding
of first-order phase transitions by quenched disorder. In $2d$, arbitrarily
weak (continuous) disorder softens the phase transition to second order.

In fact, these various arguments only involve two different
statistical properties. Indeed, the arguments (a) and (b),
which are very close \cite{theo89, luckbook}, are both based on the notion of 
rare events: in any infinite configuration of
disorder, there exist arbitrarily large ordered domains with
exponentially small probabilities. On the other hand, the arguments
(c), (d), (e) and (f) all refer to the average or typical behavior in $\sqrt
N$ for the sum of a large number $N$ of (independent) random
variables.

These simple probabilistic arguments are well founded and almost
`` irrefutable". To the best of our knowledge, the only argument which
has given rise to a controversy \cite{imryrf} is the Imry-Ma argument
which was in disagreement with the ``dimensional reduction"
predicted by the field theory approaches, either at all the orders in perturbation theory \cite{aharony, villain88} or in the
supersymmetric formalism \cite{parisisourlas}. The rigorous studies
\cite{imbrie, bricmont} finally gave reason to... the
Imry-Ma argument! This example shows that the simple heuristic 
arguments may have some non-trivial
physical content, which is not always easy to obtain by
more sophisticated methods. 

In conclusion, these probabilistic arguments allow to
understand clearly the physics, because they identify the
local fluctuations of the disorder which are responsible for such or
such phenomenon. On the other hand, it is often difficult to go beyond
the qualitative ideas to do more precise computations! To make some progress
 while remaining in the same spirit, the most natural idea
is of course to consider real space renormalizations.

\subsection*{... to real space renormalizations on the disorder}

The choice to work in real space to define a renormalization
procedure, which already presents a great interest in pure
systems \cite{niemeijer, burkhardt}, becomes the unique choice
in the presence of disorder if one wishes to describe space
heterogeneities.

\paragraph*{Block Renormalizations}

The block renormalizations, based on decimation or on
Migdal-Kadanoff ideas seem the most frequent procedures for 
disordered systems. The Migdal-Kadanoff procedures 
\cite{migdal, kadanoff} indeed constitute simple approximations to
carry out block renormalizations on regular lattices. They
also represent exact renormalizations on certain hierarchical lattices
\cite{kaufman}. Among the various systems that have been studied
in this way, one may quote for instance
the Potts model \cite{kinzel}, the diluted ferromagnet
\cite{jaya}, and especially spin glasses, which gave rise to a
large number of works : their aim was either to determine phase diagrams
\cite{youngsgrgmk, braymoore84, brayfeng}, or to study various properties of the
spin-glass phase \cite{thill, moore, sasaki}, especially the
chaotic character \cite{braymoore87} of the RG flow trajectories 
\cite{mckay, banavarbray, muriel, thill} which is a great novelty
with respect to pure systems. Also random-field systems are studied by
this type of renormalization group method\cite{nattermann}.

In addition, various disordered systems have been studied via
renormalization procedures on hierarchical networks, in particular the 
Potts model \cite{derridagardner}, directed polymers in random media
\cite{derridagriffiths, cookderrida, dasilveirajpb}, or wetting
on disordered substrates \cite{derridawetting, tangchate}.

Let us also mention the block renormalization 
for the trap model \cite{machta}, for
random quantum spin chains \cite{hirsch} and for the random
contact process\cite{hooyberghs}. In this renormalization approach,
the lattice, and thus the disorder, is treated in a homogeneous
fashion and characterized by a Gaussian distribution the variance of
which is renormalized during the transformation. For strongly disordered
system, which are spatially heterogeneous the block RG yields only a rather
approximative picture of the critical behavior.

\paragraph*{Functional Renormalization for interfaces in
random media}

For models of interfaces in random media, there exist
a field theoretical functional RG method \cite{danielFRG}
 which studies the flow of the
 disorder correlator. We refer to the review \cite{wiese}
for the description of the various recent developments, and to the
references \cite{balents, pldwiese} for comparisons with the replica method.

\paragraph*{Renormalization for disordered XY models}

The introduction of disorder in two-dimensional XY models, in which there exist Kosterlitz-Thouless transitions in the pure case, leads to 
a Coulomb gas RG describing the flow for the
probability distribution for fugacities \cite{carpentier1} to
take into account the influence of spatial heterogeneities on the topological
defects.
This approach has also been used to study the glass transition
of a particle in a random potential presenting logarithmic 
correlations \cite{carpentier2}, a problem that has otherwise
been studied by various methods \cite{mudry95,mudry96,mudry97}.  

\paragraph*{Phenomenological RG for spin glasses}

The `` droplet theory " \cite{droplets} for spin glasses
in finite dimensions, originates from a phenomenological
renormalization introduced by Mc Millan \cite{mcmillan} and developed
by Bray and Moore \cite{braymooreheidelberg}. In the formulation of
Bray and Moore \cite{braymooreheidelberg}, the essential idea is
that the probability distribution $P_L(J)$ of the effective couplings $J$
at scale $L$ converges towards a fixed form, with a width
$J(L)=J L^{y}$ which depends on the scale $L$.
The exponent $y$ and the limiting distribution, which are
calculable exactly in $d=1$, have been studied numerically in $d=2$ 
($y<0$)
and in $d=3$ ($y>0$ strong coupling) \cite{braymooreheidelberg}. This idea
of a strong coupling fixed point (or zero temperature fixed point), described
by a scaling form for the probability distribution of
a disorder variable, corresponds in fact exactly to the description of strong disorder renormalizations, whenever they can be applied.

\paragraph*{Ma-Dasgupta Renormalization for quantum spin chains}

The renormalization procedure introduced by Ma-Dasgupta-Hu
in 1979 \cite{madasgupta} to study the quantum spin chain $S=1/2$
with random antiferromagnetic couplings has for
essential property to renormalize space
 in an inhomogeneous way in order to adapt
better to the local disorder fluctuations.
Indeed, usual RG methods treat space in a
homogeneous way, by replacing for instance each block of spins of
a given size by one super-spin. If this homogeneous character
 is natural for pure systems, one may however question its legitimacy
in the presence of disorder which breaks the translation
invariance. Ma, Dasgupta and Hu have thus defined a 
renormalization procedure based on the energy, and not on the size of a spatial
cell. The procedure consists in eliminating in an iterative way the
degrees of freedom of higher energy, to obtain in the end an effective
theory at low energy (this idea is thus somewhat reminiscent of
the RG defined for the Kondo problem \cite{wilson1}): this low energy effective
theory for the spin chain is nowadays called 
 ``the random singlet phase".

This renormalization procedure has actually remained
not well known and not well understood
during many years... until the works of Daniel Fisher
in 1994-1995 \cite{danielrtfic, danielantiferro} which gave 
to this method :

\label{introdaniel}

{\bf (i) a well defined theoretical status, via the notion of ``
infinite disorder fixed points"}

Whereas the method was first considered as an
approximate procedure without control, Daniel Fisher showed that the 
renormalization flow takes a scaling form which converge towards an
``infinite disorder fixed point " (this means that the disorder
increases indefinitely at large scale), which made 
the method asymptotically exact \cite{danielrtfic, danielantiferro}. 
In addition,
 the application of the method to Random Transverse Field Ising Chain allowed him to explicitly show the exactness of the
obtained results via a direct comparison with some observables
that had been rigorously computed for the McCoy and Wu model \cite{mccoywu,
shankar}. (This disordered McCoy-Wu model is a two-dimensional
Ising model with columnar disorder,
which is equivalent to RTFIC.)

{\bf (ii) remarkable possibilities of explicit
computations}

For the RTFIC \cite{danielrtfic}, the
Ma-Dasgupta procedure allows to obtain a lot of new results
with respect to the rigorous methods \cite{mccoywu, shankar, theohenri},
in particular the exact critical exponent $\beta=(3-\sqrt 5)/2$ for
the spontaneous magnetization. More surprisingly, Daniel Fisher even
computed observables that are unknown for the corresponding
pure model, i.e. for the pure
two-dimensional Ising model! (for instance the explicit
scaling function describing the magnetization as a function of the external
field in the critical region).

These works of Daniel Fisher have thus generated a great interest for these
methods in the field of the quantum spins, as we will describe in the following. We will also show that the strong disorder renormalizations
 are in fact not limited to the quantum disordered spin systems, but
are also an ideal tool to study also a large class of disordered
systems in statistical physics.


\section{Principles of strong disorder RG}

\label{secprinciples}

In this section, we present the
essential physical ideas common to the various strong disorder RG
 that have been proposed in the different physical contexts.

In all models considered in this review, the various
 strong disorder RG procedures are based on the same idea: at large scale,
the disorder dominates with respect to the thermal or quantum
fluctuations. In particular, the strong disorder renormalizations
are intrinsically specific to disordered systems and
cannot even be defined for the pure systems which do not present space
heterogeities.

\subsection{The essential idea : Dominance of the disorder over thermal or quantum fluctuations}
\label{dominance}
In some sense, the pure systems, which are
controlled by the thermal or quantum fluctuations are
characterized by a large `` degeneracy", since all
sites are equivalent, whereas the presence of disorder breaks completely this
degeneracy. To be more specific, let us discuss some examples of this idea in the various fields.

\subsubsection*{Example for the Ground state of a quantum
system}

In the pure antiferromagnetic quantum chain of spin $S=1/2$, 
the ground state can be qualitatively seen as an appropriate linear
combination of all the states that correspond to a possible way
of pairing the spins two by two to form singulets. 

On the contrary,
in the presence of disorder, the Ma-Dasgupta renormalization procedure
 associates to each realization of the disorder 
a ground
state which corresponds to only one way of pairing the spins two by two in
singulets. Thus, in a fixed disordered sample, a given spin is
completely correlated with only one another spin of the chain, which
may be at a rather large distance, but is almost
not correlated with the other spins, even its immediate neighbors on
the chain.

 \subsubsection*{Example for the equilibrium of a classical spin chain}

 In the pure Ising ferromagnetic chain, a domain wall 
can be found with equal probabilities on all the bonds because they all are
equivalent. The usual energy/entropy argument between
the cost energy $2J$ of a domain wall and the entropy $S
\sim K \ln L$ associated with the arbitrary position of the domain wall in
a finite system of length $L$, allows to
understand the absence of long range order at
any finite temperature and the behavior of the typical size
 $L_T \sim e^{2J/T}$ of domains.

On the contrary, the presence of random fields completely destroys this
equivalence between all the sites. The Imry-Ma argument 
\cite{imryma}, which replaces the previous energy/entropy argument
is an energy/energy argument : the comparison between the
domain wall energy cost $2J$, and the typical energy
$\sqrt{\sigma L}$
which can be gained by taking advantage of a favorable fluctuation of the sum
$\sum_{i=1}^l h_i$ of the random fields on a domain of size $L$
leads to the absence of long range order even at zero
temperature, with a typical domain length 
$L_{IM} \sim J^2/\sigma$. We will see later that the strong disorder RG
 allows to construct the
positions of the Imry-Ma domain walls in each given sample.

 \subsubsection*{Examples for random walks in random media}

For random walks in random media, it is useful to introduce
two different characteristic lengths

$\bullet$ the first important length $x(t)$ represents the typical distance traveled
during time $t$. This scales characterizes the anomalous diffusion properties.
For instance, in the following, we will encounter both logarithmic
$x(t) \sim (\ln {t})^2$ and algebraic $x(t) \sim t^{a} $ behaviors.

$\bullet$ the second important length $y(t)=x_1(t)-x_2(t)$ 
represents the distance between two independent particles diffusing in the same
disordered sample from the same initial condition.
In the pure case, this length scale of course coincides with the first one
$y(t) \sim \sqrt{t} \sim x(t)$, but this is not the case anymore
in disordered samples. The notion of localization, which is the crucial
property to apply a strong disorder RG analysis, concerns the behavior
of this relative distance $y(t)$ : if the variable $y$ remains finite with probability $p=1$ in the limit of infinite time, the localization is `total',
and can be associated with an infinite disorder fixed point; if the variable $y$ remains finite with probability $0<p<1$ in the limit of infinite time, the localization is `partial', and can be associated with a finite disorder fixed point.
In this review, we will encounter both cases.

For a random walk in a Brownian potential (the so called Sinai walk), there
are sample-dependent areas that concentrate almost all the
probability weight, and we will see later how to characterize them by a
renormalization procedure. In particular, the thermal
fluctuations are completely sub-dominant : the distance between two
independent particles (i.e. with two independent thermal histories)
which diffuse in the same disordered sample, remains a finite random variable in the limit of infinite time : this phenomenon is known as the Golosov localization \cite{golosovlocali}.
 
In the unidimensional trap model characterized
by a broad distribution of trapping times $p(\tau>) \sim
1/\tau^{1+\mu}$ with $0 <\mu< 1$, the diffusion front in a given sample
is concentrated on a finite number of important traps.
Again, we will see later how to study the statistical
properties of these traps via a strong disorder RG.
In contrast, in the phase $\mu>1$ where the averaged trapping time is
finite, there is no localization anymore in the limit of infinite time, and 
the strong disorder approach looses its meaning.

 \subsection{Notions of Infinite and Strong disorder fixed points}

If one is interested into the behavior of a disordered
system at large scale, there are a priori three possibilities for
the evolution of the effective disorder compared to the thermal
fluctuations. Indeed, when the scale increases, this effective
disorder can either become

(i) smaller and smaller without bound : the system is then controlled by a pure
fixed point.

(ii) larger and larger without bound : the system is then controlled by an infinite disorder
fixed point.

(iii) or it may converge towards a finite level: the system is
then controlled by a finite disorder fixed point.

In certain models, any initial disorder, even very small, drives the system
towards an infinite disorder fixed point (ii) at large scale : in particular, this is the case for
the random antiferromagnetic quantum chain random $S  =1/2$, for the Sinai model, etc....

A finite disorder fixed point  (iii) is usually characterized
 by a parameter which varies continuously, like the
parameter $\mu$ in the trap model. The fixed point can often be considered
as a strong disorder fixed point in a certain region of the parameters. 
For instance, in the trap model
 or in the Sinai model in external field,
the dynamics is controlled by a strong disorder fixed point in the phase $0<\mu<1$ which presents a partial localization of the thermal packet:
in this phase, there is a finite probability that two thermal trajectories in the same sample remain at a relative finite distance in the limit of infinite time. 

 In conclusion, the strong disorder renormalization methods concern: 

 $\bullet$ the infinite disorder fixed points  (ii).  

$\bullet$ the finite disorder fixed points (iii) that present strong disorder
properties, such as Griffiths phases in quantum models
or localization phenomena for random walks in random media.

\subsection{ How to know if the disorder dominates at large scale?} 

\subsubsection*{ Via a priori theoretical arguments ?} 

The relative importance of thermal fluctuations with respect to disorder 
fluctuations at large scale can not be seen directly on the microscopic model.
It is actually not very well-known for the majority of disordered systems. 
 Even for the random field systems, for which there exists 
at zero temperature an Imry-Ma argument discussed above, there does not exist, to our knowledge, any general argument which would include the thermal fluctuations of the domain walls and which would estimate the the entropy of decomposing a sample into Imry-Ma domains.

  \subsubsection*{ Via numerical studies?} 

In the numerical studies which have a priori direct informations on various thermal configurations for a fixed realization of the disorder, 
it is actually quite rare to find this information sample by sample, because the published results are in general devoted to the various disorder averaged observables. What is of course a pity from the point of view of the strong disorder approaches, for which the essential information is precisely the importance of the thermal fluctuations in a fixed sample. Indeed, the averages over the samples always present the ``risk" to be dominated by rare events and to give a false idea of typical behaviors. 
 For instance, in the Sinai model, the thermal width averaged over the samples, which is a very natural observable in numerical
simulations to characterize the spreading of the thermal packet, diverges at large time. This result could be interpreted at first sight as an absence of
localization asymptotically, whereas in fact, the distance between two independent particles in the same sample remains a finite random variable at infinite time, which corresponds to a very strong localization in law.

 \subsubsection*{ Assumption of strong disorder and its check} 

As a consequence, the reasoning in strong disorder approaches 
 is usually as follows:  one starts by assuming that the disorder dominates at large scale, and one checks at the end the consistency of this assumption. 
  More precisely, the strong disorder renormalization procedures 
 contain their own test of validity : if the probability distributions have a width which grows indefinitely by the RG flow, the obtained results 
are asymptotically exact, whereas if the width of the probability distributions
converge towards a finite value, the obtained results are only approximate,
and the order of magnitude of the approximation can be characterized.

\subsection{ Essential features of infinite disorder fixed points}
\label{features}
In several random systems the explicit construction of the strong disorder RG is not evident, in another cases
the form of singularities at the critical point may depend on the strength of disorder.
In these cases combined numerical and analytical
studies could help to identify the type of the random fixed point. An infinite disorder fixed point is
characterized by the following properties.

\begin{itemize}

\item Strong dynamical anisotropy

The typical length-scale is related to the logarithm of the typical time-scale, thus the dynamical exponent
is formally infinity.

\item Ultra-slow dynamics

The dynamical processes, such as autocorrelations for random magnets or diffusion of the random random walk
takes place in a very slow, logarithmic time-scale.

\item Broad distribution of physical quantities

Distribution of a physical quantity, say the order-parameter, $m$, is logarithmically broad. Generally, in a
finite system of length, $L$, the appropriate scaling combination is $L^{\psi} \ln m$. As a consequence
typical and average values are different and generally involve different type of singularities.

\item Dominant effect of rare regions

The average value of a physical quantity is generally dominated by the rare events (or rare regions of a
large sample). In a rare event the physical quantity has a value of $O(1)$ and to obtain the critical
singularities it is generally enough to determine the fraction of rare events. Therefore calculations at
infinite disorder fixed points are often comparatively easier, than at a conventional random fixed point.

\end{itemize}

In conclusion, the 
strong disorder RG approaches have a meaning
whenever the disorder heterogeneities determine the dominant state of the system, whereas the thermal or quantum fluctuations only provide subleading corrections.  Their goal is then to build the dominant state of the system for each realization of the disorder.  To implement this program, it is now necessary to specify the computation methods to carry out the renormalization on the disorder.

\section{ General features of Ma-Dasgupta RG rules}

\label{chapRGrules}

Whereas renormalizations in pure systems 
usually involve a finite number of coupling constants, 
 renormalizations in disordered systems involve probability distributions, i.e. functions which belong to an infinite dimensional space.
The analysis of the RG flow and of the fixed points obviously become
much harder.  This difficulty usually leads to completely numerical studies, or,
at the analytical level, to additional approximations which consist in projecting on finite spaces, i.e. one chooses a certain analytical form for distributions containing a few parameters whose evolutions are studied.  In this section, we will see that the strong disorder RG rules generate, in a certain number of 
favorable cases, RG flows that are simple enough to be analyzed completely.
Moreover, the fixed point distributions have usually an interesting probabilistic interpretation. 

 \subsection{ What is a strong disorder RG rule  ?}

The various strong disorder RG which have been  developed in different physical contexts, are characterized by the two essential properties :

 $\bullet$ the renormalization concerns the extreme value of a random variable.  This extreme value which evolves by renormalization and constitutes the scale of renormalization:  it is the `` cut-off"of the renormalized distribution. 

 $\bullet$ the renormalization is local in space:  at each stage, it is only the immediate neighbors of the extreme random variable which is concerned by the RG procedure.

Let us now present the two basic examples, 
one for a quantum spin chain, one for a dynamical classical model,
that actually leads to the same RG rules for appropriate variables.

\subsubsection*{ Quantum Example : RG rule for the Random Transverse field Ising Chain
(RTFIC)}
\label{RG_RTFIC}

In the RTFIC model, there are random couplings $J_i>0$
and random transverse fields $h_i>0$, that alternate along the chain.
The physics of the model will be discussed in details
in Section \ref{chapRTFIC}.
Here our goal is simply to describe the strong disorder RG rules
\cite{danielrtfic} :
at each step, the maximum
$ \Omega=max \{ J_i,h_j \}$ is chosen.
If it is a random transverse field $h_2=\Omega$, it is eliminated with its two
neighbor couplings $(J_2,J_3)$
to produce the new effective coupling  
\begin{eqnarray}
J'=\frac{J_2 J_3}{ \Omega}
\label{rulertfic1} 
\end{eqnarray}
If the maximum is a coupling $J_2=\Omega$, 
 it is eliminated with its two
neighbor $(h_1,h_2)$
to produce the new effective transverse field    
\begin{eqnarray}
h'=\frac{h_1 h_2}{ \Omega} 
\label{rulertfic2} 
\end{eqnarray}
So the new effective couplings and transverse fields that are introduced
are statistically independent from all other
disorder variables remaining in the chain.
This property is essential to write closed RG equations
for the probability distributions $P_{\Omega}(J)$ et $P_{\Omega}(h)$
of the couplings and fields at RG scale $\Omega$.

\subsubsection*{ Classical Example : RG rule for the Sinai model}

\begin{figure}
\centerline{\includegraphics[height=8cm]{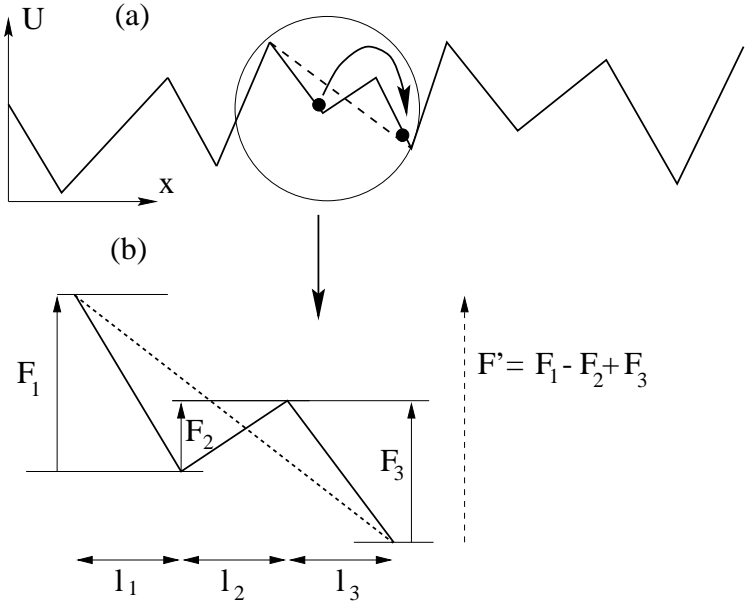}} 
\caption{\it  RG rules for a 1D random potential.} 
\label{figrulerg}
\end{figure}

In the Sinai model, the disorder consists in random forces $f_i$.
The strong disorder RG approach (that will be described in details in Section
\ref{chapsinai}) yields that
 the important degrees of freedom are the extrema of the associated
potential $U(i)=\sum_{j=0}^i f_j$ at large scale. 
 To define the extrema of the initial potential, it is thus necessary to start by grouping all the consecutive forces $f_i$ of the same sign together :  
the landscape then consists in alternating 
descending $(F_i^+, l_i^+)$ and ascending $(F_i^-, l_i^-)$ bonds.  
Barriers $F$ and lengths $l$  are now positive random variables which respectively represent the potential differences and the distances between two consecutive
local extrema of the initial model.  The RG rule of the landscape is then the following one \cite{rgsinaishort,rgsinailong}: 
 one chooses the smallest barrier $\Gamma=min \{ F_i^+, F_i^- \} $. If the smallest barrier is a descending bond $F_2^+=\Gamma$, one eliminates it with its two neighboring ascending bonds $F_1^-$ and $F_2^-$ to form a new ascending bond 
 \begin{eqnarray} 
(F^-)' = F_1^- + F_2^- -\Gamma 
\label{rulesinai1}
\end{eqnarray} 
If the smallest barrier is an ascending bond $F_1^-=\Gamma$, one eliminates it 
with its two neighboring descending bonds $F_1^+$ and $F_2^+$ to form a new descending bond
\begin{eqnarray} 
(F^+)' =F_1^+ + F_2^+ -\Gamma 
\label{rulesinai2}
\end{eqnarray} 
These rules are completely equivalent to the decimation rules  
of the RTFIC (\ref{rulertfic1},\ref{rulertfic2})
by a simple logarithmic
transformation of the variables
\begin{eqnarray} 
F^+_i &&=- \ln J_i \\
 F^-_i &&=- \ln h_i \\
\Gamma && =- \ln \Omega
\end{eqnarray}


  If one wishes to keep in addition
the information on the initial distances in physical space, one writes the RG rule for the length during the decimation process:  the length of the new renormalized bond is simply equal to the sum of the three eliminated bonds
 \begin{eqnarray} 
l =l_1+l_2+l_3 
\label{rulelonguor} 
\end{eqnarray} 
This new length is statistically independent 
of the other lengths remaining in the system, but is 
correlated with the barrier $F$ existing on the same bond. 
 Thus, one writes a closed system of two flow equations for the two joint
laws $P_{\Gamma}^{\pm}(F,l)$. 

 \subsection{ Iteration and convergence towards a fixed point} 
 
 Once the RG rules have been defined, one has to study the flow upon iteration of the RG rules.
 This can be done either analytically for a large class of one-dimensional systems,
 or numerically for more complex systems, in particular models in dimension $d>1$.
 In all cases, one is looking for a convergence towards a fixed point, which is generally located
 at $\Omega^*=0$, in appropriate rescaled variables. During renormalization, the complementary variables, such as
 couplings and transverse-fields for the RTFIC or descending and ascending bonds for the Sinai model,
 can be decimated symmetrically, which corresponds to a critical point, or asymmetrically, which
 happens in the Griffiths phases. We will now describe the fixed points corresponding to the RG rules
 described above as examples.

 \subsection{ The basic Infinite Disorder Fixed Point ( Critical point )}

The symmetric case $\overline{f_i}=0$ of the Sinai model
corresponds to the quantum critical point $ \overline{\ln J}=\overline{ \ln h}$ of the RTFIC model.  At large scale, the descending and ascending bonds become statistically equivalent, and the only relevant parameter 
at large scale is the variance $\overline{f_i^2}=2 \sigma$ of
the initial model, as in the Central Limit theorem.  The fixed point is then characterized by a joint distribution $P^*(\eta,  \lambda)$  of the scaling variables $\eta = \frac{F-\Gamma}{\Gamma}$ for the barriers and $\lambda=\frac{ \sigma l}{\Gamma^2}$ for the lengths.  This joined law 
 is defined by its Laplace transform with respect to the length $\lambda$ \cite{danielrtfic} 
\begin{eqnarray}
\int_0^{+\infty} d\lambda e^{-p \lambda} {  P}^*(\eta, \lambda) = \theta(\eta>0) \frac{ { \sqrt p}} { \sinh { \sqrt p}} e^{ - \eta { \sqrt p} \coth { \sqrt p}} 
\label{ptfixeetalambda} 
\end{eqnarray}
In particular, the distribution of the barriers alone is simple exponential  
\begin{eqnarray}
{  P}^*(\eta) = \theta(\eta>0) e^{ - \eta} 
\label{ptfixeeta} 
\end{eqnarray}
whereas the distribution of the lengths alone takes the form of an infinite series of exponentials 
\begin{eqnarray}
{  P}^*(\lambda) 
= LT^{-1}_{p \to \lambda} 
\left( \frac{ 1}{ \cosh {\sqrt p}} \right) && = \sum_{n=-\infty}^{+\infty} \pi (-1)^n \left(n+\frac{1}{2} \right) e^{ - \pi^2 \left(n+\frac{1}{2} \right)^2 \lambda} 
\\ && = \frac{1}{{\sqrt \pi} \lambda^{3/2}} \sum_{m=-\infty}^{+\infty} (-1)^m \left(m+\frac{1}{2} \right) e^{ - \left(m+\frac{1}{2} \right)^2 \frac{1}{\lambda}} \nonumber 
\label{ptfixelambda} 
\end{eqnarray}
(the two series correspond to each other via a Poisson inversion formula).  The convergence towards the fixed point solution (\ref{ptfixeetalambda}) is of order $1/\Gamma$ for random walk models \cite{danielrtfic}. For the 1D Brownian motion which already represents the universal scaling limit of random walks, the fixed point (\ref{ptfixeetalambda}) is an exact result at any scale $\Gamma$, as shown by a direct calculation via constrained path integrals 
\cite{rgtoy}.

 \subsection{ The basic Strong Disorder Fixed Point ( Griffiths phases)}
\label{bSDFP}

 In the biased case $\overline{f_i}=f_0>0$, the 
RG equations lead to a family of solutions depending on one parameter 
denoted by $ \delta$ \cite{danielrtfic}:  the two distributions 
for ascending and descending bonds have for Laplace transforms
 \begin{eqnarray}
\int_0^{+\infty} dl e^{-p L} P_{\Gamma}^{*\pm}(F, l) = \theta(F > \Gamma) \frac{ { \sqrt { p+\delta^2}} e^{\mp \delta \Gamma}} { \sinh { \sqrt { p+\delta^2}}} e^{ - (F-\Gamma) \left[ { \sqrt { p+\delta^2}} \coth { \sqrt { p+\delta^2}} \mp \delta \right ]} 
\label{ptfixeetalambdadelta} 
\end{eqnarray}
In particular, the distributions of barriers alone have the following exponential forms 
\begin{eqnarray}
P_{\Gamma}^{*+}(F) && = \theta(F > \Gamma) \frac{ 2 \delta} { e^{2 \delta \Gamma} -1} e^{ - (F-\Gamma) \frac{ 2 \delta} { e^{2  \Gamma} -1}} \opsimeq_{\Gamma \to \infty} \frac{ 2 \delta} { e^{2 \delta \Gamma}} e^{ - (F-\Gamma) \frac{ 2 \delta} { e^{2 \delta \Gamma}}} \\ P_{\Gamma}^{*-}(F) && = \theta(F > \Gamma) \frac{ 2 \delta} { 1-e^{-2 \delta \Gamma}} e^{ - (F-\Gamma) \frac{ 2 \delta} { 1-e^{-2 \delta \Gamma}}} \opsimeq_{\Gamma \to \infty} \theta(F > \Gamma) 2 \delta e^{ - (F-\Gamma) 2 \delta} 
\nonumber
\label{soluppdelta} 
\end{eqnarray}
From the point of view of the renormalized landscape, the meaning of the parameter $2 \delta$ is thus clear:  the probability distribution of large barriers 
against the bias $f_0$ remains stable at large scale (apart from the presence of the cut-off $\Gamma$) and $(2 \delta)$ is the coefficient of the exponential asymptotic decay of this distribution.  It is now necessary to specify the meaning of this parameter $\delta$ for the microscopic model.  Close to the critical point  corresponding to $ \delta=0$, the parameter $\delta$ can be developed at the first order in the bias \cite{danielrtfic} 
\begin{eqnarray}\delta = \frac{ f_0}{\sigma} +O(f_0^2) 
\label{smalldelta}
\end{eqnarray}
However, if one deviates from the immediate vicinity of the critical point, the correct non-perturbative definition of the parameter $ \delta$ in term of the initial distribution $Q(f)$ of the variables $(f_i)$ of the microscopic model is that $\delta$ represents the solution of the equation \cite{rgsinailong}
 \begin{eqnarray}\overline{ e^{ - 2 \delta f_i}} \equiv \int_{-\infty}^{+\infty} df Q(f) e^{ - 2 \delta f} = 1 
\label{bonnedefdelta} 
\end{eqnarray}
Of course, this definition exactly corresponds, with the change of notation $2 \delta=\mu/T$, with the definition of the dimensionless parameter $\mu$ representing the exponent of the anomalous diffusion $x \sim t^{\mu}$ of the biased Sinai model \cite{kestenetal, derridapomeau, jpbreview}. 
 The development into the two first cumulants yields again the simple expression (\ref{smalldelta}), which is exact at all the orders only for an initial Gaussian distribution.  Again, if one directly considers the 1D biased Brownian motion, the solutions (\ref{ptfixeetalambdadelta}) are exact results on any scale $\Gamma$, as shown by a direct calculation by constrained path integrals 
\cite{rgtoy}.

  Whereas the solution (\ref{ptfixeetalambda}) of the symmetric
 case $\delta=0$ is an infinite disorder fixed point, the solution (\ref{soluppdelta}) of the asymmetric case $\delta > 0$ is a finite disorder fixed point, because the distribution of large barriers against the bias has a
finite width  $\frac{1}{2 \delta}$ asymptotically. The strong disorder RG thus gives exact results about
position dependent quantities only in the limit $\delta \to 0$. The dynamical quantities, such as the
autocorrelation function, however is expected to be exact.  We will discuss in Section
\ref{chapsinaibiais} how to generalize this RG procedure when the parameter $\delta$ is small, but non-zero.

\subsection{Auxiliary variables and critical exponents}
\label{auxiliary}

In strong disorder RG, `auxiliary variables' are variables associated with the bonds which will evolve according to rules in parallel
with the RG rule of the main variable.  The first important example is 
the length $l$  that we have already met (\ref{rulelonguor}).  It is an auxiliary variable because it is not the length that determines the renormalization, but the associated barrier $F$:  indeed at each stage, one does not choose the minimal length, but the minimal barrier $F$.  Another important auxiliary variable for the RTFIC is the magnetization $\mu$ of the spin clusters :  this variable exists only in association with the random fields $h_i$ and evolves
with the rule \cite{danielrtfic}
\begin{eqnarray}
 \mu =\mu_1+\mu_3 
\label{rulertficm} 
\end{eqnarray}
More generally, in the various models discussed in this review, one is led to study various auxiliary variables which evolve according to the general rule \begin{eqnarray}
\mu =a \mu_1+b \mu_2+c \mu_3 
\label{rulegeneauxi} 
\end{eqnarray}
where $(a, b, c)$ are constants.  We have already seen how the joint 
probability distribution of barriers and lengths characterizes the statics of the renormalized landscape at a given scale (\ref{ptfixeetalambda}):  in particular, in the renormalized landscape at scale $\Gamma$, in which there are only barriers $F>\Gamma$, the lengths have for scaling $l \sim \Gamma^2$, which is the usual Brownian scaling as it should. However, apart from this very special case $a=b=c=1$, the auxiliary variables (\ref{rulegeneauxi}) lead to non-trivial exponents  $\mu \sim \Gamma^{\phi}$ who reflect the `dynamical' properties of the RG 
procedure over all preceding scales $\Gamma'<\Gamma$, and which contain much more information than the landscape at the scale $\Gamma$ alone.  For instance, for the RTFIC, the very simple rule (\ref{rulertficm}) corresponding to $a=c=1$ and $b=0$ already leads for the magnetization $\mu \sim \Gamma^{\phi}$ to the irrational exponent equal to the golden mean \cite{danielrtfic}
\begin{eqnarray}
\phi(a=c=1, b=0)=\frac{1+\sqrt 5}{2}
\label{golden} 
\end{eqnarray}
This example already shows that a simple one-dimensional random walk contains non-trivial exponents if one is interested into `subtle' properties   
involving the dynamics of the renormalized landscape.

More generally, for auxiliary variables of type (\ref{rulegeneauxi}), one obtains that, when the condition $a+c=2$ is satisfied (this happens rather often in practice for the most natural observables), the associated exponent satisfies a quadratic equation and reads
\begin{eqnarray}
\phi(a+c=2, b)=\frac{1+\sqrt { 5+4 b}}{2}
\label{aux}
\end{eqnarray}
This exponent generalizes the exponent $\phi(a=1,b=1,c=1)=2$ of the length (\ref{rulelonguor} ).
Otherwise, when $a+c \neq 2$, the exponent $\phi(a, b, c)$ satisfies a more complicated equation involving the confluent hypergeometric function $U(A, B, z)$ \cite{rgsinailong,rgreadiff}.

As a final remark, let us mention that 
the strong disorder RG procedures have actually a close relationship with some 
pure growth models, which were introduced in a completely independent way, 
as we explain in Appendix \ref{growthmodels}.

\part{ RG STUDY OF QUANTUM MODELS}

\section{ Random Transverse Field Ising Chain}

\label{chapRTFIC}

The RTFIC is the paradigmatic example of random quantum spin chains for which the most complete
analytic results are known. The early results by McCoy and Wu\cite{mccoywu} and
others\cite{shankar} have been greatly extended by Fisher using the strong disorder RG method\cite{danielrtfic},
which has inspired subsequent investigations\cite{senthil,motrunich,rsrgdyna,igloi02}. These results,
in particular at the critical point are of importance for other disordered problems, as well,
for which the RG trajectories flow into the infinite disorder fixed point of the RTFIC. Examples are
 the random quantum Potts, clock and Ashkin-Teller models
in Sec.\ref{PCAT}, the asymmetric simple exclusion process  with particle-wise disorder in Sec.\ref{ASEP2},
and the random contact process in Sec.\ref{RCP}.

\subsection{ Model}

\label{RTFIC_2}

The RTFIC is defined by the Hamiltonian:
\begin{equation}
H_I=-\sum_i J_i \sigma_i^x \sigma_{i+1}^x-\sum_i h_i \sigma_i^z\;.
\label{hamilton_I}
\end{equation}
Here the $\sigma_i^x$, $\sigma_i^z$ are Pauli matrices at site $i$ and
the $J_i$ exchange couplings and the $h_i$ transverse-fields are
random variables with distributions $\pi(J)$ and $\rho(h)$,
respectively. The Hamiltonian in eq(\ref{hamilton_I}) is closely related
to the transfer matrix of a classical two-dimensional layered Ising
model, as described in details in Sec.\ref{M_W}.

\subsubsection{Free-fermion representation}

\label{free_fer}

The usual way of study the RTFIC is to map the Hamiltonian in
eq(\ref{hamilton_I}) through a Jordan-Wigner transformation
and a following canonical transformation \cite{lieb61} into a free
fermion model:
\begin{equation}
H=\sum_{q=1}^L \epsilon_q \left(\eta_q^+\eta_q- \frac{1}{2}\right)\;,
\label{fermion}
\end{equation}
where $\eta_q^+$ and $\eta_q$ are fermion creation and annihilation
operators, respectively. Here we consider a finite chain of length $L$ with free or fixed boundary conditions, i.e. with $J_L=0$, when the fermion energies $\epsilon_q$ are
obtained via the solution of an eigenvalue problem\cite{igloiturban96}, which necessitates
the diagonalization of a $2L \times 2L$ tridiagonal matrix {\bf T} with
non-vanishing matrix-elements $T_{2i-1,2i}=T_{2i,2i-1}=h_i$,
$i=1,2,\dots,L$ and $T_{2i,2i+1}= T_{2i+1,2i}=J_i$, $i=1,2,\dots,L-1$, thus

\begin{equation}
{\bf T} = \left(
\begin{matrix}
 0  & h_1 &     &       &       &       &     \cr
h_1 &  0  & J_1 &       &       &       &     \cr
    & J_1 &  0  & h_2   &       &       &     \cr
    &     & h_2 &  0    &\ddots &       &     \cr
    &     &      &\ddots&\ddots &J_{L-1}&     \cr
    &     &      &      &J_{L-1}&   0   & h_L \cr
    &     &      &      &       &  h_L  &  0  \cr
\end{matrix}
\right)\quad,\qquad
\label{trid}
\end{equation}

One is confined to the $\epsilon_q \ge 0$ part of the
spectrum. In the free-fermion representation one can also calculate the
(local) magnetization and different types of correlation and autocorrelation
functions\cite{lieb61,igloi98}. Here we just quote two simple results, with the
help of them we shall illustrate features of scaling of the RTFIC.

\subsubsection{Surface magnetization, phase diagram and critical exponents}

\label{m_s}

Fixing the end-spin at site $i=L$, thus taking the transverse field, $h_L=0$, the magnetization at the
other surface, at $i=1$, can be expressed by the simple exact formula\cite{peschel84,igloi98}:
\begin{equation}
m_s=\left[1+\sum_{i=1}^{L-1} \prod_{j=1}^l
\left( h_j \over J_j \right)^2\right]^{-1/2}\;.
\label{peschel}
\end{equation}
The surface magnetization thus involves a so-called Kesten random variable,
whose properties are discussed in more details in Appendix \ref{appkesten}.

Analyzing Eq.(\ref{peschel}) in a disordered system in the thermodynamic limit we can define a quantum control-parameter,
\begin{equation}
\delta={[\ln h]_{av}-[\ln J]_{av} \over \rm{var}[\ln h]+\rm{var}[\ln J]}\;,
\label{delta_I}
\end{equation}
so that the system is ferromagnetic, the average surface magnetization is non-zero for $\delta<0$, and it is
paramagnetic with vanishing average surface magnetization for $\delta>0$.

At the critical point, $\delta=0$, in a large, finite sample 
the surface magnetization is determined by the largest term in the sum of Eq.(\ref{peschel}),
which typically grows as $\exp(AL^{1/2})$, which follows from the central limit theorem. Consequently
the typical surface magnetization is exponentially small in $L^{1/2}$
\begin{equation}
m_s^{typ}(L) \sim \exp( -{\rm cst} L^{1/2}),
\label{ms_typ}
\end{equation}
the appropriate scaling combination is $(\ln m_s)/L^{1/2}$ and the distribution of $m_s$ is
logarithmically broad. For analytical results about the distribution function of the surface
magnetization see\cite{dharyoung,microcano}.

As argued in\cite{igloi98} there are rare realizations, in which the sum of random variables:
$\varepsilon_j=\ln \frac{h_j}{J_j}$ have surviving walk character and therefore $m_s^{rare}=O(1)$. The
average of $m_s$ at the critical point is dominated by these rare realizations and its value follows
from the fraction of surviving walks, $P_{surv}(L) \sim L^{-1/2}$:
\begin{equation}
[m_s]_{\rm av}(L) \sim L^{-x_m^s},\quad x_m^s=1/2.
\label{ms_av}
\end{equation}
Here $x_m^s$ is the anomalous dimension of the surface magnetization.

Repeating the analyses\cite{igloi98} in the ferromagnetic phase close to the critical point, we obtain:
\begin{equation}
[m_s]_{\rm av}(\delta) \sim -\delta^{\beta_s},\quad \beta_s=1,
\label{ms_beta}
\end{equation}
and from the finite-size dependence of $[m_s]_{\rm av}(L,\delta)$ we obtain for the average correlation length,
\begin{equation}
\xi \sim |\delta|^{-\nu},\quad \nu=2,
\label{xi_av}
\end{equation}
so that the scaling relation: $\beta_s=x_m^s \nu$ is satisfied. Finally, typical correlations have a faster
decay:
\begin{equation}
\xi_{\rm typ} \sim |\delta|^{-\nu_{\rm typ}},\quad \nu_{\rm typ}=1.
\label{xi_typ}
\end{equation}
The critical exponents of the RTFIC are summarized in table \ref{TAB1d}, in which we have included the bulk
magnetization exponent, $\beta$, and scaling dimension, $x_m$, together with the exponent, $\psi$, defined
in Eq.(\ref{scales}). Note, that the scaling relations, $\beta=\nu x_m$ and $\beta_s=\nu x_m^s$ are
satisfied.

\begin{table}
\caption{Critical exponents of the RTFIC}

\label{TAB1d}
\centerline{
\begin{tabular}{|c|c|c|c|c|c|c|}
\noalign{\bigskip}\hline\noalign{\smallskip}
$\beta$ & $\beta_s$ & $x_m$ & $x_m^s$ & $\nu$ & $\nu_{typ}$ & $\psi$   \\
\noalign{\smallskip}\hline\noalign{\smallskip}
$(3-\sqrt{5})/2$ & $1$ & $(3-\sqrt{5})/4$ & $1/2$ & $2$ & $1$ & $1/2$ \\
\noalign{\smallskip}\hline
\end{tabular}
}
\end{table}

\subsubsection{Low-energy excitations and dynamics}

The lowest energy gap of the RTFIC, $\epsilon_1(L)$ in Eq.(\ref{trid}), can be estimated for free boundary
conditions, provided $\epsilon_1(L)$ goes to zero faster, than $1/L$. It is given by\cite{igloi97a}:
\begin{equation}
\epsilon_1(L) \sim m_s \overline{m_s}
\prod_{i=1}^{L-1} {h_i\over J_i}\;.
\label{lambda1}
\end{equation}
Here $m_s$ and $\overline{m_s}$ denote the finite-size surface
magnetizations at both ends of the chain, as defined in
eq(\ref{peschel}) (for $\overline{m_s}$ simply replace $h_j/J_j$ by
$h_{L-j}/J_{L-j}$ in this eq.).

{\it Critical point} 

Here $\epsilon_1(L)$ typically behaves as the surface magnetization, consequently:
\begin{equation}
\epsilon(\delta=0,L) \sim \exp(- {\rm const}\cdot L^{1/2})\;.
\label{epscrit}
\end{equation}
Thus time-scale, $t_r \sim\epsilon^{-1}$, and length-scale, $L \sim \xi$, are related as:
\begin{equation}
\ln t_r \sim \xi^{\psi},\quad \psi=1/2\\;,
\label{scales}
\end{equation}
which corresponds to extreme anisotropic scaling with an anisotropy, or dynamical exponent $z=\infty$.
We remark that in conventional anisotropic systems the usual relation is: $t_r \sim \xi^z$, with $z < \infty$.

{\it Paramagnetic phase}

Here, for ${\rm max}\{J\} > {\rm min}\{h\}$ there are rare regions of
size, $l_{rare} \sim \ln L$, in which the local couplings are stronger than the local transverse fields, which
prefers (local) ferromagnetic ordering.
The corresponding energy gap is exponentially small in $l_{rare}$, $\epsilon \sim \exp(-A l_{rare})$, thus
between time- and length-scale the relation is:
\begin{equation}
t_r \sim \xi^z\;.
\label{scales_gr}
\end{equation}
This is the usual scaling relation in the Griffiths-phase of the system\cite{griffiths}. 
In Eq.(\ref{scales_gr}) the dynamical exponent, $z=z(\delta)$, depends on the distance of the critical
point. Its value can be obtained from a mapping to the random random walk problem, see Sec.\ref{RW_RTFIC}
and given by the positive root of the equation:
\begin{equation}
\left[\left({J \over h}\right)^{1/z}\right]
_{\rm av}=1\;.
\label{z_I}
\end{equation}
Similarly, there is a Griffiths-phase in the ferromagnetic region, too. Here, the first energy state is
exponentially degenerate, and the time-scale is set by the second gap and follows a relation
like in Eq.(\ref{scales_gr}). To obtain the value of $z$ one should interchange $h$ and $J$ in
Eq.(\ref{z_I}), which also follows from duality\cite{kogut79}.

We note that these exact results illustrate the general features of infinite and strong disorder fixed
points as summarized in Sec.\ref{features}. In the following we turn to systematically use the strong disorder
RG method for the model.

\subsection{RG rules}

\label{RTFIC_3}

The decimation rules for the model have been announced in Sec.\ref{RG_RTFIC} to illustrate the
general features of
the Ma-Dasgupta RG method. Here we present a detailed derivation and show the physical picture behind these
rules. As already mentioned the strongest term in the Hamiltonian in Eq.(\ref{hamilton_I}) can be either a
bond or a transverse field and different decimation procedure has to be adopted in the two cases.

\subsubsection*{Strong bond decimation}

\label{RTFIC_RG_B}

Here we consider a pair of spins, which are connected by the strongest bond of the system, the strength of which is denoted by $J_2=\Omega$, where $\Omega$ is the energy scale in the system.
By definition $J_2$ is larger than any transverse field, thus typically $J_2 \gg h_2, h_3$, where $h_2$ and $h_3$ are the transverse fields acting on the spins of the two-site cluster. Under these circumstances the two spins
are strongly correlated so they flip in a longitudinal magnetic field coherently. In a good approximation the
two spins can be considered as a composite spin with double momentum in an effective transverse field,
$\tilde{h}_{23}$, which is calculated perturbatively. The first two energy levels of the spectrum of the
two-spin cluster with free boundary conditions is given by
$E_0 = -\sqrt{J_2^2+(h_2+h_3)^2}$ and $E_1 = - \sqrt{J_2^2+(h_2-h_3)^2}$, respectively, which are separated
by a large gap of $\approx 2J_2$ from the two other levels. With a small error we drop the two
highest energy states and keep the two lowest, which are identified as the two states of a composite Ising spin. The renormalized value of the transverse field follows from the relation:
\begin{equation}
\tilde{h}_{23}=\frac{E_1-E_0}{2} \approx \frac{h_2 h_3}{J_2} \;,
\label{hdecimation}
\end{equation}
where the second equation holds for $J_2 \gg h_2, h_3$.

\subsubsection*{Strong transverse field decimation}

\label{RTFIC_RG_F}

In this case the strongest term is a transverse field, say $h_2=\Omega$, acting on site $i=2$. The
connecting couplings, $J_2$ and $J_3$ are typically much smaller, $h_2 \gg J_2, J_3$. In a small external
longitudinal field this spin has a negligible response, consequently in a good approximation
it can be considered as "dead", as far as susceptibility and magnetic correlations are concerned.
Therefore this spin is decimated out and a new coupling, $\tilde{J}_{23}$, is generated between the
remaining spins, the strength of which is calculated perturbatively. Most easily the result is obtained
through the duality properties of the RTFIC\cite{kogut79},
which amounts to interchange couplings and transverse fields, $h_i \leftrightarrow J_i$ in
Eq.(\ref{hdecimation}):
\begin{equation}
\tilde{J}_{23}\approx \frac{J_2 J_3}{h_2} \;.
\label{Jdecimation}
\end{equation}

\subsection{RG flow}

\label{RTFIC_32}

Under the repeated use of the decimation transformations
in Eqs.(\ref{hdecimation}) and (\ref{Jdecimation}) the energy scale, $\Omega$, is gradually lowered and at the
same time the distribution of the transverse fields, $P_0(h,\Omega)$, and that of the couplings, $R_0(J,\Omega)$,
are subject of variation. We are interested in the renormalization of these distribution functions, in particular
we want to obtain their scaling behavior around the fixed point of the transformation, which is located at $\Omega^*=0$.

\subsubsection{Renormalization of the distribution functions}

\label{RTFIC31}

Let us denote the energy scale in the initial model by $\Omega_{\rm in}$, which during renormalization is
decreased to a value of $\Omega < \Omega_{\rm in}$. Further infinitesimal decrease of the energy scale as
$\Omega \to \Omega - {\rm d} \Omega$ amounts
to eliminate a fraction of ${\rm d} \Omega [ P_0(\Omega,\Omega)+R_0(\Omega,\Omega)]$
spins, during which the coupling distribution changes as:
\begin{eqnarray}
R_0(J,\Omega-{\rm d} \Omega)=\left\{R_0(J,\Omega)+{\rm d} \Omega P(\Omega,\Omega)
\int_0^{\Omega} {\rm d} J_1 \int_0^{\Omega} {\rm d} J_3 R_0(J_1,\Omega)R_0(J_3,\Omega)
\times \right.
\nonumber\\
\left. \left[ \delta\left(J-\frac{J_1 J_3}{\Omega}\right)-\delta(J-J_1)-\delta(J-J_3)\right]
\right\}\left\{1-{\rm d} \Omega [ P_0(\Omega,\Omega)+R_0(\Omega,\Omega)] \right\}^{-1}
\;.
\label{Jdistr}
\end{eqnarray}
Here in the r.h.s. the three delta functions represent the generated one new coupling and the
decimated two old couplings during one RG step and the second factor ensures normalization. A similar
equation is obtained for the distribution of the transverse fields, just from duality one should
make the interchange $h \leftrightarrow J$ and $P \leftrightarrow R$.

Now expanding $R_0(J, \Omega-{\rm d}\Omega)$ one arrives to the integral-differential
equation:
\begin{eqnarray}
\frac{{\rm d} R_0}{{\rm d} \Omega}=R_0(J,\Omega)\left[P_0(\Omega,\Omega)-R_0(\Omega,\Omega)\right]
\nonumber\\
-P_0(\Omega,\Omega) \int_{J}^\Omega {\rm d} J' R_0(J',\Omega)
R_0(\frac{J}{J'}\Omega ,\Omega) \frac{\Omega}{J'}
\;,
\label{Rdiff}
\end{eqnarray}
and similarly for the distribution $P_0(h,\Omega)$:
\begin{eqnarray}
\frac{{\rm d} P_0}{{\rm d} \Omega}=P_0(h,\Omega)\left[R_0(\Omega,\Omega)-P_0(\Omega,\Omega)\right]
\nonumber\\
-R_0(\Omega,\Omega) \int_{h}^\Omega {\rm d} h' P_0(h',\Omega)
P_0(\frac{h}{h'}\Omega ,\Omega) \frac{\Omega}{h'}
\;.
\label{Pdiff}
\end{eqnarray}
The two integral-differential equations in Eqs.(\ref{Pdiff}) and (\ref{Rdiff})
have to be supplemented by the initial conditions,
represented by the distributions $P_{\rm in}(h)=P_0(h,\Omega_{\rm in})$ and
$R_{\rm in}(J)=R_0(J,\Omega_{\rm in})$.

\subsubsection{Fixed-point solution}

\label{RTFIC32}

A special solution to the problem in Eqs.(\ref{Pdiff}) and (\ref{Rdiff}) is given by the functions:
\begin{eqnarray}
P_0(h,\Omega)&=&\frac{p_0(\Omega)}{\Omega}\left(\frac{\Omega}{h}\right)^{1-p_0(\Omega)}
\label{Psol}\\
R_0(J,\Omega)&=&\frac{r_0(\Omega)}{\Omega}\left(\frac{\Omega}{J}\right)^{1-r_0(\Omega)}
\label{Rsol}
\;,
\end{eqnarray}
thus they depend only on the values of the distributions at their edges, at
$P_0(\Omega,\Omega)=p_0/\Omega$ and at $R_0(\Omega,\Omega)=r_0/\Omega$.
It is argued in Refs.\cite{danielrtfic,danielantiferro} and has been checked trough the numerical solution of
Eqs.(\ref{Rdiff}) and (\ref{Pdiff}) that
this special solution represents the true solution of the problem at the
fixed point, i.e. as $\Omega \to 0$. Later we also show how the parameters
of the special solution can be related with the initial distributions,
$P_{\rm in}(h)$ and $R_{\rm in}(J)$.
Putting Eqs.(\ref{Psol}) and (\ref{Rsol}) into Eqs.(\ref{Pdiff}) and (\ref{Rdiff}) one
obtains ordinary differential equations for $p_0$ and $r_0$, what can be written as:
\begin{equation}
\frac{{\rm d} y_0}{{\rm d} \Gamma} + y_0^2=\Delta^2\;.
\label{dif3}
\end{equation}
in terms of $y_0=p_0-\Delta=r_0+\Delta$. Here $\Gamma=-\ln \Omega$ is the log-energy variable and
$\Delta$ is the asymmetry parameter, which is related to the relative strengths of the couplings and
the transverse fields and can be calculated from
the initial distributions. At the critical point, $\Delta=0$, since the couplings and the
transverse-fields have identical distributions. On the other hand for $\Delta>0$ ($\Delta<0$) we are
in the disordered (ordered) phase.

{\it Critical point solution}

At the critical point, $\Delta=0$, the solution to Eq.(\ref{dif3}) is given by:
\begin{equation}
y_0=p_0=r_0=\frac{1}{\Gamma-\Gamma_0}=\frac{1}{\ln(\Omega_0/\Omega)},
\quad \delta=\Delta=0\;,
\label{sol0}
\end{equation}
where $\Gamma_0=-\ln \Omega_0$ is a reference (log)energy scale. It is instructive
to consider the distribution of the reduced log-coupling variable $\eta=-(\ln \Omega
-\ln h)/\ln \Omega=-(\ln \Omega  -\ln J)/\ln \Omega \ge 0$, which is given from
Eqs.(\ref{Psol}) and (\ref{sol0}) as
\begin{equation}
\rho(\eta){\rm d} \eta = \exp(-\eta){\rm d} \eta \;.
\label{soleta}
\end{equation}
This solution has been obtained by Fisher\cite{danielrtfic} and corresponds to the distribution
of barriers in the Sinai model in Eq.(\ref{ptfixeeta}).

{\it Off-critical solution}

The solution to Eq.(\ref{dif3}) {\it  in the off-critical region}, $\Delta \ne 0$, is
given by:
%


\begin{equation}
y_0=\frac{\Delta \overline{y}_0+\Delta^2 {\rm th}\left[\Delta(\Gamma-\Gamma_0)\right]}
 {\Delta + \overline{y}_0 {\rm th}\left[\Delta(\Gamma-\Gamma_0)\right]}
=|\Delta| \left( 1 + 2 \frac{\overline{y_0}-\Delta}{\overline{y}_0+\Delta}
\left({\Omega / \Omega_0} \right)^{2 \Delta}+\dots \right)\;,
\label{ysol}
\end{equation}
where the solution goes through the point $y_0=\overline{y}_0$ at the reference
(log)energy cut-off, $\Gamma_0$. The second equation in Eq.(\ref{ysol}) is the approximate
form of the solution close to the line of fixed points, where in terms of the original
energy-scale variable $\Omega/\Omega_0 \ll 1$.

It is shown in Ref.\cite{igloi02} that there is a conserved quantity, say $\kappa$,
along the RG trajectory
provided $[(J/h)^{2 \kappa}]_{\rm av}=1$. Evaluating the average with the fixed point solution
in Eqs.(\ref{Psol}) and (\ref{Rsol}) we obtain:
\begin{equation}
\left[\left(\frac{J^2}{h^2}\right)^{\Delta}\right]_{\rm av}=1
\;.
\label{zeq}
\end{equation}
Consequently the asymmetry parameter, $\Delta$, does not change during the RG transformation
and related to the initial distributions. Close to the critical point $\Delta=\delta+O(\delta^2)$,
therefore it is called the non-linear quantum control parameter of the RTFIC.

\subsubsection{Asymptotic exactness of the results}
\label{asymp}
Next we show that the RG equations in Eqs.(\ref{hdecimation}) and (\ref{Jdecimation}) become
asymptotically exact as the line of fixed points is approached, i.e., as $\Omega/\Omega_0 \to 0$.

First we consider the disordered Griffiths phase, $\Delta>0$, but the
reasoning holds for $\Delta<0$ from duality and can be applied at the critical point, too. For
 $\Delta>0$ the ratio of
decimated bonds, $\Delta n_J$, and decimated transverse fields, $\Delta n_h$,
goes to zero as $\Delta n_J/\Delta n_h=R_0(\Omega,\Omega)/P_0(\Omega,\Omega)r_0/p_0 \sim \Omega^{2 \Delta}$, thus close to the fixed point almost exclusively
transverse fields are decimated out.
Then the probability, $Pr(\alpha)$, that the value of a coupling, $J$, being neighbor
to a decimated transverse field is $\Omega>J>\alpha \Omega$ with $0<\alpha<1$ is
given by
\begin{equation}
Pr(\alpha) \simeq \int_{\alpha\Omega}^{\Omega} R_0(J,\Omega) {\rm d} J =1-\alpha^{r_0}\approx
r_0 \ln(1/\alpha)\;,
\label{Pralpha}
\end{equation}
which goes to zero during iteration, since according to Eq.(\ref{ysol})
$r_0=R_0(\Omega,\Omega)\Omega \to 0$. At the critical point, where couplings and transverse fields
are decimated with the same rate Eq.(\ref{Pralpha}) is still valid, and $Pr(\alpha)$ goes to zero
as $r_0=1/(\Gamma-\Gamma_0)$. Consequently the RG transformation becomes
asymptotically exact and the dynamical singularities, which are characterized by the
parameter $\Delta$ are also exact.

As far as static quantities or spatial correlations are concerned the asymptotic exactness of the
RG method is valid only at the critical point, which is controlled by an infinite disorder
fixed point. The origin of this is the existence of a diverging correlation length
and the infinitely broad distributions of both the couplings and the transverse fields.
On the other hand in the disordered Griffiths phase, which is controlled by a line of strong disorder
fixed points the correlation length is finite and the distribution of the transverse fields have a
finite width. Therefore static quantities calculated by the RG are only exact in the vicinity of the
critical point. This phenomena is analogous to that of the biased Sinai walk in Sec.\ref{bSDFP},
which is discussed in detail in Section \ref{chapsinaibiais}.

\subsubsection{Relation between energy- and length-scale}

\label{RTFIC_34}

The length-scale of the problem, $L_{\Omega}$, which is the typical distance between remaining spins is
related to the fraction of non-decimated spins,
$n_{\Omega}$, as $L_{\Omega} \sim n_{\Omega}^{-1}$. When the energy
scale is decreased by an amount of ${\rm d}\Omega$ a fraction of spins.
${\rm d} n_{\Omega}= n_{\Omega}[P_0(\Omega,\Omega)+R_0(\Omega,\Omega)]$, is decimated
out, so that we obtain the differential equation:
\begin{equation}
\frac{{\rm d} n_{\Omega}}{{\rm d} \Omega}= n_{\Omega}[P_0(\Omega,\Omega)+R_0(\Omega,\Omega)].
\;,
\label{dnomega}
\end{equation}
This can be integrated by
using the solution in Eq.(\ref{ysol}) as:
\begin{equation}
n_{\Omega}=\left\{{\rm ch}\left[\Delta \ln\frac{\Omega_0}{\Omega}\right]+
\frac{\overline{y}_0}{\Delta} {\rm sh}\left[\Delta \ln\frac{\Omega_0}{\Omega}\right]
\right\}^{-2}
\;.
\label{nomega}
\end{equation}
At the critical point: $\Delta \to 0$ and
$\Gamma=-\ln \Omega \to \infty$, with however $\Delta \times \Gamma \to 0$ one obtains:
\begin{equation}
L_{\Omega}\sim \frac{1}{n_{\Omega}} \sim \left[ \ln \frac{\Omega_0}{\Omega}\right]^2,
\quad \Delta=0
\;,
\label{lomega0}
\end{equation}
which is just the relation in Eq.(\ref{scales}) as found by Fisher\cite{danielrtfic}.

In the Griffiths phases, $|\Delta|>0$, one obtains in Eq.(\ref{nomega}),
$n_{\Omega} \sim \Omega^{2|\Delta|}$,
in the limit $\Omega \to 0$. Consequently the relation between typical distance
between remaining spins, $L_{\Omega} \sim 1/n_{\Omega}$, and the energy scale is given by:
\begin{equation}
L_{\Omega}\simeq L_{\Omega_0} (\Delta+y_0)^2 \left( \frac{2}{\Delta} \right)^2
\left( \frac{\Omega_0}{\Omega}\right)^{2|\Delta|}
\sim \Omega^{-2 |\Delta|}
\;.
\label{lomega1}
\end{equation}
Thus $\Delta$ is simply related to the dynamical exponent, $z$,
\begin{equation}
z=\frac{1}{2 |\Delta|}\;,
\label{zeq2}
\end{equation}
which is in accordance with the result in Eq.(\ref{z_I}), which is obtained through a mapping to the
random RW.

\subsection{ RG detailed results}
\label{RTFIC_det}
In the following we complete the RG analyses
by calculating auxiliary variables (lengths and magnetic moments), dynamical and thermodynamic quantities.

\subsubsection{Renormalization of lengths and magnetic moments}

During renormalization sites and bonds are decimated out and the distance between the remaining
spins as well as the size (moment) of the spin can be expressed with the original variables.
To keep track of this process we assign a length to each bond, $l^b_i$, (connecting sites $i$ and $i+1$) and a moment, $\mu_i$ and a length, $l^s_i$. to each spin. In the initial model $l^b_i=l^s_i=1/2$ and $\mu_i=1$.
During a strong bond decimation, see Sec. \ref{RTFIC_RG_F}, we have the RG rules for the auxiliary variables:
\begin{equation}
\tilde{l}_{23}^b=l_2^s+l_2^b+l_{3}^s,
\quad \tilde{\mu}_{23}=\mu_2+\mu_{3} \;.
\label{hdecimation_2}
\end{equation}
which completes the relation in Eq.(\ref{hdecimation}). Similarly, decimating out a strong transverse field,
see Sec. \ref{RTFIC_RG_B}, we obtain for the lengths:
\begin{equation}
\tilde{l}^s_{23}=l_{1}^b+l_2^s+l_{2}^b\;.
\label{Jdecimation_2}
\end{equation}
what should be consider together with Eq.(\ref{Jdecimation}). Note that lengths for the Sinai model is
introduced in Eq.(\ref{rulelonguor}) with analogous definition and renormalization of magnetic
moments is announced in Eq.(\ref{rulertficm}).

 The joint distribution functions are given by:
$P(h,l^s,\mu;\Omega)$,
and $R_l(J,l^b;\Omega) $, from which we obtain through integration the reduced distribution functions:  $P_0(h,\Omega)$, $P_l(h,l^s,\Omega)$, $P_{\mu}(h,\mu,\Omega)$ and $R_{0}(J,\Omega)$. Renormalization
of $P_0(h,\Omega)$ and $R_{0}(J,\Omega)$ have already been presented in Sec. \ref{RTFIC_32}.

Solution of the other reduced distribution functions can be obtained in the following
steps\cite{danielrtfic,igloi02},
what we illustrate in the example of the moment distribution.

$\bullet$ In the first step, one should write down the RG equation for the distribution functions, in this way to generalize
the relation in Eqs.(\ref{Rdiff}) and (\ref{Pdiff}). For the moment distribution we have:
\begin{eqnarray}
& &\frac{{\rm d} P_{\mu}(h,\mu,\Omega)}{{\rm d} \Omega}=P_{\mu}(h,\mu,\Omega)
\left[R_0(\Omega,\Omega)-P_0(\Omega,\Omega)\right]-
\nonumber\\
& &R_0(\Omega,\Omega)\int_{h}^\Omega {\rm d} h' \frac{\Omega}{h'} \int_0^{\mu} {\rm d} \mu'
P_{\mu}(h',\mu',\Omega)P(\frac{h}{h'}\Omega,\mu-\mu',\Omega)
\;.
\nonumber\\
\label{h-mu-diff}
\end{eqnarray}

$\bullet$ In the second step the solution is searched in the form of a Laplace transform, c.f.:
\begin{equation}
\tilde{P}_{\mu}(h,s,\Omega)=\int_0^{\infty} {\rm e}^{-\mu s} P_{\mu}(h,\mu,\Omega) {\rm d} \mu
\;.
\label{lapl-tr-s}
\end{equation}

$\bullet$ The third step is to observe that the different $s$-components of the transformed function are
separated, which makes possible to obtain the solution in the fixed point.

$\bullet$ In the fourth step we generalize the solution found in Eq.(\ref{Psol}) for the component, $s=0$,
and search the solution for general $s$ in the form:
\begin{equation}
\tilde{P}_{\mu}(h,s,\Omega)=\frac{\pi_{\mu}(s,\Omega)}{\Omega}
\left(\frac{\Omega}{h}\right)^{1-p_{\mu}(s,\Omega)}
\;.
\label{P-s-sol}
\end{equation}

$\bullet$ In the fifth step we write ordinary differential equations for the functions, $p_{\mu}(s,\Omega)$ and $\pi_{\mu}(s,\Omega)$, which appear in Eq.(\ref{P-s-sol}).

$\bullet$ The solution, in the sixth step, is found in linear order of $s$: $p_{\mu}(s,\Omega)=p_0(\Omega)+s\tilde{p}_1(\Omega)$ and $\pi_{\mu}(s,\Omega)=p_0(\Omega)+s\tilde{\pi}_1(\Omega)$,
where $\tilde{p}_1$ satisfies the differential equation:
\begin{equation}
(y_0^2-\Delta^2)\frac{{\rm d}^2 \tilde{p}_1}{{\rm d} y_0^2}=\tilde{p}_1\;,
\label{p1-y}
\end{equation}
where $y_0=y_0(\Omega)$ is given in Eq.(\ref{ysol}).

$\bullet$ Finally, in the seventh step the average cluster moment is obtained through the relation:
\begin{equation}
\overline{\mu}=\frac{\tilde{p}_1-\tilde{\pi}_1}{p_0}-\frac{\int_{\overline{y}_0}^{y_0} {\rm d} y_0^{'} \tilde{p}_1(y_0^{'})/(y_0^{'}-\Delta)}{y_0+\Delta}\;.
\label{mu-y}
\end{equation}
{\it  At the critical point} with $\Delta=0$ the solution of Eq.(\ref{p1-y}) is $\tilde{p}_1=y_0^{-\tau}$, with $\tau=(\sqrt{5}-1)/2$, thus the average cluster moment is given by
\begin{equation}
\overline{\mu}={\rm const}~y_0^{-(1+\tau)} = \overline{\mu}_0 \left[ \ln
\left(\frac{\Omega_0}{\Omega}\right)\right]^{\phi},~~
\phi=\frac{1}{\tau}=\frac{1+\sqrt{5}}{2}\;.
\label{Phi}
\end{equation}
This result, which is announced in Eq.(\ref{golden}) has been first obtained by Fisher\cite{danielrtfic}.

{In the Griffiths phases} with $\Delta \ne 0$ the solution of the differential equation in Eq.(\ref{p1-y})
in terms of the variable $y=y_0/\Delta$ can be expressed by the hypergeometric function\cite{AS}, $F(a,b;c;z)$, as
\begin{equation}
\tilde{p}_1= |\Delta|^{-\tau} y^{-\tau}F\left(\frac{\tau}{2},
\frac{1}{2}+\frac{\tau}{2};\frac{3}{2}+\tau;
\frac{1}{y^2}\right)=|\Delta|^{-\tau} f_1(y)\;.
\label{pysol}
\end{equation}
We obtain similarly:%
\begin{eqnarray}
\tilde{\pi}_1&=&-|\Delta|^{-\tau}(y-1) y^{-(\tau+1)}F\left(\frac{\tau}{2}+1,
\frac{1}{2}+\frac{\tau}{2};\frac{3}{2}+\tau;
\frac{1}{y^2}\right)
\nonumber\\
&=&|\Delta|^{-\tau} \phi_1(y)\;,
\label{piysol}
\end{eqnarray}
so that the average cluster moment is given by:
\begin{equation}
\overline{\mu}={\rm const}|\Delta|^{-\tau-1} \frac{f(y)}{y+1}\;
\label{mu_o}
\end{equation}
where $f(y)=f_1(y)-\phi_1(y)$. Here one should differentiate between the paramagnetic
($\Delta>0,~y>0$) and the ferromagnetic ($\Delta>0,~y>0$) phases. In the former case the
average cluster moment grows slowly $\sim \ln|\Omega/\Omega_0|$, as $\Omega/\Omega_0 \to 0$,
whereas in the ferromagnetic phase, where $y \to 1^{-}$ in the fixed point, thus
$\overline{\mu}$ is divergent, as $\overline{\mu}(\Omega) \sim \Omega^{-2 |\Delta|}$.

Before going to deal with the average magnetization defined by: $m=\overline{\mu}/\overline{l}_s$, we
quote results about the average lengths, $\overline{l}_s$ and $\overline{l}_b$, which are calculated from the distribution functions, $P_l(h,l^s,\Omega)$ and $R_l(J,l^b;\Omega)$, in the steps outlined above. The
average lengths are given by\cite{igloi02}:
\begin{equation}
\overline{l}_s=\overline{l}_s(\Omega_0)\frac{\overline{y}_0^2-\Delta^2}
{{y}_0^2-\Delta^2},\quad \overline{l}_b=\overline{l}_b(\Omega_0)\frac{\overline{y}_0^2-\Delta^2}
{{y}_0^2-\Delta^2}\;,
\label{l_s_av}
\end{equation}
thus at the line of fixed points, $\Omega \to 0$, we have
$\overline{l}_s \sim \overline{l}_b \sim L_{\Omega}$.
Consequently the interpretation of the dynamical exponent, $z$, in Sec.\ref{RTFIC_34} in Eq.(\ref{lomega1}) is
justified also with the average lengths-scales.

The correlation length, $\xi$, in the paramagnetic phase is measured by the size of non-decimated, i.e. correlated spins in a cluster. This quantity stays constant as the energy scale is lowered and close
to the critical point it is given by:
\begin{equation}
\xi \sim \Delta^{-2} \sim \delta^{-2}\;.
\label{RTFIC_nu}
\end{equation}
Thus the correlation length critical exponent is $\nu=2$,
which is Fisher's result\cite{danielrtfic}.

With the results of the average cluster moment in Eq.(\ref{mu_o}) and the average lengths in Eq.(\ref{l_s_av})
we can write for the magnetization:
\begin{equation}
m=m_0 \frac{(1-y) f(y)}{(1-\overline{y}) f(\overline{y})}\;,
\label{magn-f}
\end{equation}
where $m_0$ is the average magnetization at $\Omega=\Omega_0$  and $\overline{y}$ denotes the
value of the variable $y$ at the same energy-scale. In the ferromagnetic phase
taking $\Omega/\Omega_0 \to 0$ we obtain close to the critical point: $(1-\overline{y})^{-1} \sim |\Delta|$
and $f(\overline{y}) \sim |\Delta|^{\tau}$ so that:
\begin{equation}
m={\rm const} |\Delta|^{1-\tau}={\rm const} |\delta|^{1-\tau}\;.
\label{magn-d}
\end{equation}
From Eq.(\ref{magn-d}) one can read the critical exponent of the average magnetization as:
\begin{equation}
\beta=1-\tau=2-\phi\;,
\label{betan}
\end{equation}
which has been first derived by Fisher\cite{danielrtfic}.

\subsubsection{Scaling of thermodynamic quantities}

\label{SC_term}

In this Section we show the scaling form
of singular thermodynamic quantities as a function of a small, but finite temperature,
$T>0$, or magnetic field, $H>0$.

To treat the effect of a small finite temperature in the RG scheme one should first notice
that the thermal energy sets in an energy scale, $\Omega_T \sim T$, and the RG decimation
should be stopped as $\Omega$ is lowered to $\Omega_T$.
At that energy scale a fraction of spin clusters, $n_{\Omega_T}$, in Eq.(\ref{nomega})
is not decimated out and these spins are loosely coupled comparing with the temperature, $T$.
Consequently the entropy per spin, $s$, is given as the contribution of non-interacting spin
clusters:
\begin{equation}
s \simeq  n_{\Omega_T} \ln 2 \;,
\label{entropy}
\end{equation}
whereas the specific heat can be obtained through derivation: 
$c_V=T \frac{\partial s}{\partial T}$.
From Eqs.(\ref{entropy}) and (\ref{nomega}) we obtain for the singular behavior:
\begin{equation}
s(T) \sim c_V(T) \sim T^{1/z}
\label{entropy_d}
\end{equation}
with $1/z=2 |\Delta|$, which is valid both in the ordered and in the disordered Griffiths phases.

Next, we consider the effect of a small longitudinal field, $H>0$, at zero temperature.
During renormalization the local longitudinal field, $H_l$, at site $l$ is transformed as
\begin{equation}
\tilde{H}_l=H \mu_l\;,
\end{equation}
so that the energy-scale related to the longitudinal field is given by
$\Omega_H=H \overline{\mu}(\Omega)$. As $\Omega$ is lowered to $\Omega_H$, i.e. when
the energy scale satisfies the equation
\begin{equation}
\Omega_H=H \overline{\mu}(\Omega_H)\;,
\end{equation}
the RG procedure is stopped and the remaining spin clusters are practically uncoupled.
Then the average magnetization and the average susceptibility satisfy the equations:
\begin{equation}
m(H)=m(\Omega=\Omega_H),\quad \chi=\frac{\partial m}{\partial H}\;.
\end{equation}
In the {\it  disordered Griffiths phase}, 
where $\overline{\mu}(\Omega_H)$ has a $\Omega_H$ independent
limiting value, we have $\Omega_H \sim H$, consequently from Eq.(\ref{magn-f}) the
singular behavior is given by
\begin{equation}
m(H) \sim \left(\frac{H}{H_D}\right)^{1/z},\quad \Delta>0\;.
\label{m_H}
\end{equation}

Similarly one obtains for the scaling of the susceptibility in the disordered Griffiths phase:
\begin{equation}
\chi(H) \sim \left(\frac{H}{H_D}\right)^{-1+1/z},\quad
\chi(T) \sim T^{-1+1/z},\quad
\Delta>0\;,
\label{chi_T}
\end{equation}
where the temperature dependence follows from the scaling relation, $\Omega_H \sim \Omega_T$.

In the {\it  ordered Griffiths phase}, where $\overline{\mu}(\Omega_H) \sim \Omega_H^{-1/z}$,
as given below Eq.(\ref{mu_o}) we have
$\Omega_H \sim H^{1/(1+1/z)}$. Putting this result into Eq.(\ref{magn-f}) and using the
asymptotic expansion for the hypergeometric functions\cite{AS} in Eqs.(\ref{pysol}) and
(\ref{piysol}) we obtain for
the leading field dependence of the magnetization:
\begin{equation}
m(H)-m(0) \sim \left(\frac{H}{H_D}\right)^{1/(1+z)} \ln\left(\frac{H}{H_D}\right),
\quad \Delta<0\;,
\end{equation}
and similarly for the susceptibility:
\begin{equation}
\chi(H) \sim \left(\frac{H}{H_D}\right)^{-z/(1+z)} \ln\left(\frac{H}{H_D}\right),
\quad \Delta<0\;.
\end{equation}
Note that in the ordered Griffiths phase the singularity exponent is different from that in
the disordered Griffiths phase and there is a logarithmic correction term. The temperature
dependence of the susceptibility, which follows from the relation $\Omega_H \sim \Omega_T$,
is given by:
\begin{equation}
\chi(T) \sim T^{-1+1/z} \ln T,\quad
\Delta<0\;.
\end{equation}
We can conclude that all the singularities of different physical
quantities, both in the (strongly) ordered and disordered Griffiths phases can
be expressed by the non-linear quantum control parameter, $\Delta$, and thus with the dynamical exponent,
$z$.

\subsubsection{Renormalization of dynamical correlations}

The autocorrelation functions in imaginary time can be obtained by scaling
considerations\cite{rieger97,igloi98,igloi99}, both at the critical point, see Sec..\ref{SC_st}, and in the
Griffiths phase, see Sec.\ref{SC_gr}. In the framework of the strong disorder RG dynamical
correlations have been calculated in Ref.\cite{rsrgdyna,eigenhuse}.

The basic quantity used in the renormalization is the local dynamical susceptibility:
\begin{equation}
\chi^{\alpha \alpha}_{jj}(\omega)=\sum_k | \langle k | \sigma_j^{\alpha}|0\rangle |^2
\delta(\omega-\epsilon_k)\;,
\label{chi_om}
\end{equation}
where the sum runs over the excited states, $|k \rangle$, with excitation energy, $\epsilon_k$, and
$\alpha=x,z$. $\chi^{\alpha \alpha}_{jj}(\omega)$ is related through a Laplace transform to the
imaginary time local autocorrelation function,
$G^{\alpha \alpha}_{jj}(t)=[\langle \sigma_j^{\alpha}(t)\sigma_j^{\alpha}(0)\rangle]_{\rm av}$.
The low-frequency limit of the local susceptibilities are related to the long time asymptotic of the
autocorrelation function. The $\alpha=x$ component is the magnetization, or spin autocorrelation function
and for this we shall omit the superscript in the following. The $\alpha=z$ component is generally
called the local-energy autocorrelation function, for which we use the notation $G_{jj}^e(t)$.
As usual in the calculation we consider the average quantities, which are different at the bulk, in this case
the site-index, $j$, is omitted, or at a boundary site with $j=1$.

\paragraph{ Average magnetization autocorrelation function}
\label{Gt_RG}

We start with the average magnetization autocorrelation function in the bulk and in this example we
illustrate the method of the strong disorder RG. During renormalization the energy scale is gradually
lowered, bonds and sites are removed, new bonds and composite spins of moment, $\mu$, are generated. As
a rule of renormalization the matrix-elements in Eq.(\ref{chi_om}) have the same value, both in terms of
the original and in the transformed states, which are denoted by $|\tilde{k} \rangle$. In the renormalized model there is a broad distribution
of variables and the existing spins are very loosely coupled to each other, thus can be considered as free.
Performing the decimation up to the energy-scale: $\Omega=\omega/2$, only spin clusters with a transverse
field, $\tilde{h}=\Omega$, contribute to the average spin susceptibility.
In this system the matrix-element, $\langle \tilde{k} | \sigma_j^x|\tilde{0}\rangle$, can be easily calculated: it is one, if $j$ belongs to a non-decimated cluster and zero otherwise. The fraction of spin-clusters
at this energy is given from Eq.(\ref{dnomega}) as $n_{\Omega} P_0(\Omega,\Omega)= n_{\Omega} p_0(\Omega)/\Omega$,
each of which contribute by an average moment, $\overline{\mu}(\Omega)$, thus we obtain:
\begin{equation}
[\chi]_{\rm av}(\omega) \sim \frac{n_{\omega} p_0(\omega) \overline{\mu}(\omega)}{\omega} \;,
\label{chi_om1}
\end{equation}
where we wrote $\omega$ instead of $\omega/2$, which makes no difference in the asymptotic expressions.

{\it  At the critical point} using the solution in Eqs.(\ref{sol0}), (\ref{lomega0}) and (\ref{Phi}) we obtain:
\begin{equation}
[\chi]_{\rm av}(\omega) \sim \frac{1}{\omega|\ln \omega|^{3-\phi}} \;,
\label{chi_om2}
\end{equation}
for $\omega \ll \Omega_0$. From this the average magnetization autocorrelation function is given for
$t \gg \Omega_0^{-1}$ as:
\begin{equation}
G(t) \sim |\ln t|^{2-\phi} \;,
\label{G_t}
\end{equation}
which is just the scaling result in Eq.(\ref{lauto2}), since $x_m/\psi=2 x_m=2-\phi$.

In the disordered phase, $\delta>0$, with the results in Eqs.(\ref{ysol}), (\ref{lomega1}) and (\ref{mu_o})
we obtain:
\begin{equation}
[\chi]_{\rm av}(\omega) \sim |\delta|^{4-\phi}\frac{|\ln \omega|}{\omega^{1-1/z}} \;,
\label{chi_om3}
\end{equation}
and
\begin{equation}
G(t) \sim |\delta|^{4-\phi} \frac{|\ln t|}{t^{1/z}} \;.
\label{G_t3}
\end{equation}
This corresponds to the scaling result in Sec.\ref{SC_gr} and the origin of the logarithmic correction
will be discussed, too.

In the ordered phase, $\delta<0$, the value of $p_0$ in Eq.(\ref{ysol}) is different of that
in the disordered phase, which leads to a different functional form of the dynamical susceptibility:
\begin{equation}
[\chi]_{\rm av}(\omega) \sim |\delta|^{4-\phi}\frac{|\ln \omega|}{\omega^{1-2/z}} \;,
\label{chi_om4}
\end{equation}
and the average magnetization autocorrelation function:
\begin{equation}
G(t) \sim |\delta|^{4-\phi} \frac{|\ln t|}{t^{2/z}} \;.
\label{G_t4}
\end{equation}
Thus the decay of the magnetization autocorrelation function in the ordered phase involves an exponent $2/z$.
The origin of this is that in the ordered phase excitations, $|k \rangle$, which contribute to $\chi^{xx}_{jj}(\omega)$ are rare disordered regions, flipping the spin at site $j$ back and forth in a
time-interval $\sim t \sim \omega^{-1}$. This type of excitation represents the second gap in the
spectrum and involves the scaling exponent, $z/2$, see in Ref.\cite{igloi99}.

The local-energy susceptibility and autocorrelation function can be calculated in a similar way, leading to the results\cite{rsrgdyna,eigenhuse} at the critical point:
\begin{equation}
[\chi^e]_{\rm av} \sim \frac{\omega}{|\ln \omega|^{4-\phi}} \;,
\label{chi_om_e}
\end{equation}
\begin{equation}
G^e(t) \sim \frac{1}{t^2|\ln t|^{4-\phi}} \;,
\label{G_te}
\end{equation}
and in the ordered and disordered Griffiths phases:
\begin{equation}
[\chi^e]_{\rm av}(\omega) \sim |\delta|^{5-\phi}\omega^{1+2/z} |\ln \omega|\;,
\label{chi_om_e1}
\end{equation}
\begin{equation}
G^e(t) \sim |\delta|^{5-\phi} \frac{|\ln t|}{t^{2+2/z}} \;.
\label{G_te1}
\end{equation}
We note that from scaling considerations and numerical results in Ref.\cite{igloi99} the decay
in the Griffiths phase is predicted in a different form as $G(t) \sim  t^{-2-1/z}$.

Autocorrelations at a boundary site involve the local scaling dimensions, therefore
they generally differ from their bulk counterparts. The surface magnetization autocorrelation at
the critical point is given by:
\begin{equation}
G_1(t) \sim \frac{1}{|\ln t|} \;,
\label{G_t5}
\end{equation}
and in the ordered and disordered Griffiths phase:
\begin{equation}
G^e_1(t) \sim |\delta|^{2} \frac{1}{t^{1/z}} \;.
\label{G_t6}
\end{equation}
For the surface energy autocorrelations one obtains at the critical point:
\begin{equation}
G^e_1(t) \sim \frac{1}{t^2|\ln t|^{3}} \;,
\label{G_te2}
\end{equation}
in the disordered phase
\begin{equation}
G^e_1(t) \sim  \frac{|\delta|^{3}}{t^{2+2/z}} \;,
\label{G_te3}
\end{equation}
and in the ordered phase:
\begin{equation}
G^e_1(t) \sim  \frac{|\delta|^{3}}{t^{2+1/z}} \;.
\label{G_te4}
\end{equation}

\subsection{Finite-size systems : RG results and numerical studies}

\subsubsection{RG results for finite-size scaling properties}

Remarkably, the strong disorder RG method does not only give results in
the thermodynamic limit, but it yields finite-size properties, as well.
We refer the reader to the paper \cite{rgfinitesize}, where
the strong disorder RG method is used to obtain the statistical properties
of the end-to-end spin-spin correlation
\begin{eqnarray}
C(L) \equiv < \sigma_1^z \sigma_L^z >
\label{defcl}
\end{eqnarray}
and of the gap, i.e. the energy difference between the two lowest levels
\begin{eqnarray}
\Delta(L) \equiv E_1-E_0
\label{defgapl}
\end{eqnarray}
over the ensemble of random chains of length $L$.

These results for finite-size systems are important, since
they have been tested in details against numerical studies
\cite{rgfinitesize,dharyoung}.
Moreover, the asymptotic distribution of end-spin magnetization obtained by the RG method \cite{rgfinitesize} is in full agreement
with the direct analytical results \cite{microcano},
obtained from the exact formula (\ref{peschel}).

Finally, the strong disorder RG approach has been
further extended in \cite{refael} to compute
the end-to-end energy-density correlations.

\subsubsection{Ensemble dependence of the results}

In the study of disordered systems, it is usual to consider
that the random variables defining
the disorder in a given sample are independent :
following \cite{igloi98,dharyoung},
this procedure will be called here the ``canonical ensemble"
 (this procedure is also called the ``grand-canonical
ensemble" in \cite{paz1,aharonyharris,domany,paz2}). 
However, it has been argued in \cite{paz1}
that it is much more interesting in some cases
to consider the so called ``microcanonical ensemble"
as in \cite{igloi98,dharyoung} ( this
procedure is called the ``canonical
ensemble" in \cite{paz1,aharonyharris,domany,paz2}) : 
in the microcanonical ensemble, 
there exists a global constraint on 
the random variables defining the disorder in a given sample
of $N$ sites. On one hand, it has been strongly argued in \cite{paz1,paz2} that the microcanonical
ensemble should be preferred to the canonical ensemble,
because the latter introduces an ``extra noise" that may hid the ``intrinsic''
properties of the system.
On the other hand, if one divides
the system of size $L$ into two halfs of size $L/2$,
each half will present fluctuations of order $\sqrt L$ 
in both ensembles :
in the canonical ensemble, these two halfs are independent,
whereas in the microcanonical ensemble,
the two halfs are completely correlated, i.e. they have 
 exactly opposite fluctuations. From this point of view, 
the microcanonical constraint can thus appear to be quite artificial
or even biased. Of course, it seems a priori natural to expect that these two ensembles 
should be equivalent in the thermodynamic limit, 
as was shown in \cite{aharonyharris} for the case of a random classical ferromagnet.
However, it is clear that their finite-size properties can be
very different. But since the finite-size scaling theory
 of phase transitions relates the thermodynamic exponents
to finite-size effects obtained in numerical simulations,
the discussion about these two ensembles actually leads
to the general problem of the finite-size scaling theory
for disordered systems \cite{paz1,domany,paz2}.

The finite-size properties of the RTFIC have actually been studied \cite{igloi98,dharyoung,microcano}
in the canonical ensemble
and in the microcanonical ensemble, in which case there is the global constraint on the disorder variables:
\begin{eqnarray}
\sum_{i=1}^L ( \ln J_i - \ln h_i) =0
\label{microcriticality}
\end{eqnarray}
(With one fixed and one free boundary conditions so that there are 
exactly the same number $L$ of random bonds and random fields).
The microcanonical constraint (\ref{microcriticality})
has been chosen to impose, in some sense,
the criticality condition $[\ln h]_{av} = [\ln J]_{av}$
on finite samples, whereas in the canonical ensemble,
the l.h.s. of (\ref{microcriticality}) presents fluctuations of order $\sqrt L$ around its mean value, zero.

The probability distributions of the appropriate rescaled variables
have been analytically computed for the surface magnetization
\cite{dharyoung,microcano}, as well as for the gap and the end-to-end
correlation via the real space RG method \cite{microcano} :
these distributions are different at criticality in the two ensembles, in particular in their asymptotic behaviors. As a consequence, the size dependences of the averaged observables, are found to be quite different in the two ensembles : these differences can be explained in terms of the rare events that dominate a given averaged observable, and whose measure can be very sensitive to the microcanonical constraint.

\subsubsection{Numerical studies based on free-fermions}

Numerical studies of the RTFIC are generally based on the free-fermionic representation, as described in
Sec.\ref{free_fer}. In this way relatively large systems up to linear size $L \ge 500$ can be studied.
Numerical results \cite{young96,igloi97,igloi98,igloi99} both at the critical point and in the disordered and
ordered Griffiths phases are in
good agreement with the strong disorder RG results. Dynamical correlations\cite{rieger97,young97,igloi99}
and their distribution functions\cite{kisker98}
as well as thermodynamic quantities\cite{young97,igloi98} have been also investigated.

\subsection{Relation with Dirac-type equations with random mass}

In a related work the RTFIC and the random XY spin chain is
investigated in such a way that the low-energy behavior near the
critical point is described by a Dirac-type equation with a random
mass for which an exact analytic treatment is
possible\cite{McKenzie96}. Results obtained for the dynamical critical
exponent, the specific heat, and transverse susceptibility agree with
results of the strong disorder RG. The approach of random Dirac fermions
is used also in related problems of quasi-one-dimensional systems\cite{brouwer03,
titov01,brouwer00,brouwer98}


\section{Random quantum chains with discrete symmetry}

\label{PCAT}

\subsection{Quantum Potts and clock models}

The Ising model has been generalized for $q$-state spin variables
distinct states: $|s_l\rangle=|1\rangle,|2\rangle,\dots,|q\rangle$ with two different forms of
the interaction energy\cite{wu}. For the {\it  Potts-model} in 1d it is given by:
\begin{equation}
U_{Potts}=-\sum_l J_l \delta(s_l,s_{l+1})\;,
\label{Upotts}
\end{equation}
whereas for the {\it  clock model} it is given in the form of a scalar
product
\begin{equation}
U_{clock}=-\sum_l J_l \cos \left[\frac{2 \pi}{q}(s_l-s_{l+1})\right]\;.
\label{Uclock}
\end{equation}
The two models are equivalent for $q=2$ (Ising model) and for $q=3$, but they are
different for higher values of $q$.

The quantum version of the models is obtained analogously to the Ising model (see Sec.\ref{M_W}),
by introducing appropriate transverse fields, the corresponding term in the Hamiltonian is
denoted by $K$. $K$ is defined in such a way that the transfer matrix of the classical
2d model, $T$, and the 1d quantum Hamiltonian as $H=U+K$ commute at the transition point. In
this way the critical properties of the 2d classical and 1d quantum models are isomorph.
Explicit way of construction of $H$ and thus $K$ is made in the extreme anisotropic
limit\cite{kogut79} of the 2d system, as described for the Ising model in Sec.\ref{M_W}.

The transverse-field term in the Potts model is given by\cite{solyom81}:
\begin{equation}
K_{Potts}=-\sum_l \frac{h_l}{q} \sum_{k=1}^{q-1} M_l^k\;,
\label{Kpotts}
\end{equation}
where $M_l | s_l \rangle=| s_l+1, {\rm mod}~q \rangle$. For the clock model one obtains\cite{QCM}:
\begin{equation}
K_{clock}=-\sum_l \frac{h_l}{2} (M_l+M_l^{-1})\;.
\label{Kclock}
\end{equation}
Both models are invariant under the duality transformation, $J_l \leftrightarrow h_l$. From this
follows that the non-random models with $J_l=J$ and $h_l=h$ have a self-duality point located at $h=J$.
In the non-random Potts model there is one phase transition point, which is just the self-duality
point, which separates the disordered and the ordered phases. The phase transition is second order
for $q \le 4$ and first order for $q>4$\cite{baxter}. In the non-random
clock model there is also one phase transition point for $q \le 4$. For $q>4$, however, there are three
phases in the system, the ordered and disordered phases are separated by an extended critical region,
which is the precursor of the low-temperature phase of the $XY$-model, obtained in the limit $q \to \infty$.

\subsection{Ashkin-Teller model}

The quantum Ashkin-Teller model is defined by the Hamiltonian\cite{kohmoto}:
\begin{eqnarray}
H_{\rm AT} &=& -\sum_l J_l(\sigma_l^z \sigma_{l+1}^z + 
\tau_l^z \tau_{l+1}^z) -\sum_l h_l (\sigma_l^x+\tau_l^x)
\nonumber \\
&-&\epsilon \sum_l (J_l \sigma_l^z \sigma_{l+1}^z \tau_l^z \tau_{l+1}^z
+h_l \sigma_l^x \tau_l^x)\;,
\label{HAT}
\end{eqnarray}
in terms of two sets of Pauli matrices, $\sigma_l^{x,z}, \tau_l^{x,z}$. Here
the couplings, $J_l$, and the transverse fields, $h_l$, are
independent random variables, while the coupling between
the two Ising models, $\epsilon$, is disorder independent.
The Hamiltonian in Eq.~(\ref{HAT}) is invariant under the duality transformation $J_l \leftrightarrow
h_l$.

The non-random model is critical along the self-duality line, $J=h$, for $-1<\epsilon \le 1$
with the critical exponents $x_{\rm AT}=1/8$, $x_{\rm AT}^s=\rm{arccos}(-\epsilon)/\pi$, whereas
$\nu_{\rm AT}=2 x_{\rm AT}^s/(4 x_{\rm AT}^s -1)$ for $-1/\sqrt{2}<\epsilon< -1/2$, and, in
the {\it  critical fan}\cite{kohmoto} for $-1<\epsilon<-1/\sqrt{2}$, $\nu_{\rm AT}$ is
formally infinity.

\subsection{Decimation equations}

In the random models, in which the couplings, $J_l$, and the transverse fields, $h_l$, are independent
and identically distributed random variables, one can perform the strong disorder RG transformation. The
renormalization equations, which are obtained by decimating a strong coupling, $J_2=\Omega$, or a strong
transverse field, $h_2=\Omega$, can be obtained analogously to the RTFIC. Repeating the steps described
in Sec.\ref{RTFIC_3} we arrive for all the three models to the same form of equations\cite{senthil,
carlon-clock-at}:
\begin{equation}
\tilde{h}_{23}=\kappa\frac{h_2 h_{3}}{J_2},\quad
\tilde{J}_{23}=\kappa \frac{J_2 J_{3}}{h_2} \;,
\label{deciCM}
\end{equation}
where the first (second) equation refers to the elimination of a strong
bond (field). The value of the prefactor, $\kappa$, which for the RTFIC is $\kappa=1$, for the different
models are the following:
\begin{equation}
 \kappa_{Potts}=\frac{2}{q},\quad \kappa_{Clock}=\frac{1+\delta_{2,q}}{1-\cos(2\pi/q)},\quad
 \kappa_{AT}=\frac{1}{1+\epsilon}\;.
\label{kappas}
\end{equation}
For the Potts and clock models $q$ does not renormalize under the transformation, whereas for
the Ashkin-Teller model we have $\tilde{\epsilon}=\epsilon^2(1+\epsilon)/2$.
The control-parameter of the random models, $\delta$, can be defined
in the same form as for the RTFIC in Eq.(\ref{delta_I}), and the
self-dual point of the models is at $\delta=0$.
We note that the
decimation equations for the random contact process (directed percolation) in Eq.(\ref{RG_CP})
are in the same form as in Eq.(\ref{deciCM}) with $\kappa=\sqrt{2}$, this model, however, does
not have the property of self duality.

\subsection{Disorder induced cross-over effects}

\label{PCAT1}

The renormalization equations in Eq.(\ref{deciCM}) have different characteristics for $\kappa \le 1$ and
for $\kappa > 1$, respectively. In the former case, including the RTFIC, the generated new
terms in the Hamiltonian are smaller than the decimated ones, thus during renormalization the
energy scale is monotonously decreasing. In this case the fixed point of the problem is expected
to be strongly attractive, i.e. the critical behavior is the same for any weak disorder. This type
of universality has been demonstrated for the RTFIC \cite{danielrtfic}. On the contrary for $\kappa > 1$ some generated
new terms of the Hamiltonian are larger than the decimated one, thus the decrease of the energy scale
is non-monotonic. In this case the behavior of the system for sufficiently weak disorder (when the
non-monotonic steps in the energy renormalization are frequent) and for sufficiently strong disorder
(when the non-monotonic steps are rare and their fraction is vanishing as the renormalization goes on)
are expected to be different. The strong disorder RG is valid in this second region, what we will discuss
in the following. The weak disorder region will be considered afterwards.

{\it Strong disorder regime}

Here we assume that after the starting decimation steps, when the energy scale can behave non-monotonically,
we arrive to the stable, attractive part of the RG trajectory, which is controlled by an infinite disorder
fixed point. The RG equations of the distribution function are very similar to that of the RTFIC, as
described in Sec.\ref{RTFIC31}. Using the same notations as for the RTFIC the relation in Eq.(\ref{Rdiff})
is modified to:
\begin{eqnarray}
\frac{{\rm d} R_0}{{\rm d} \Omega}=R_0(J,\Omega)\left[P_0(\Omega,\Omega)-R_0(\Omega,\Omega)\right]
\nonumber\\
-P_0(\Omega,\Omega) \int_{J/\kappa}^\Omega {\rm d} J' R_0(J',\Omega)
R_0(\frac{J}{J'\kappa}\Omega ,\Omega) \frac{\Omega}{J'\kappa}
\;,
\label{Rdiff_k}
\end{eqnarray}
and similarly for the distribution $P_0(h,\Omega)$. Solution of these equations at the fixed point are still
in the form of Eqs.(\ref{Psol}) and (\ref{Rsol}), however for a finite $\delta$ one can not proceed
via Eq.(\ref{dif3}). Thus outside the critical point (
$\delta \ne 0$) one can not obtain an asymptotically exact analytical solution,
for $\kappa \ne 1$. At the
critical point, however, the solution in Eq.(\ref{soleta}) stays valid. This is due to the fact that the
scaling variable in the solution in Eq.(\ref{soleta}), $\eta$, is modified by an additive term, $\ln \kappa/
\ln \Omega$, which goes to zero at the fixed point.

As a consequence for strong enough disorder the critical behavior of all these models (together with the random contact process in Sec.\ref{RCP}) is controlled by the infinite disorder fixed point of the RTFIC, thus they
have the same critical exponents, scaling functions, etc. This universality has been numerically checked
on different models\cite{carlon99,carlon-clock-at}.

In the ordered and disordered Griffiths phases, in which no analytical solution can be found a weak universality
hypotheses is suggested\cite{igloi01}. Two (self-dual) models which are described by the same
decimation rules, thus they have the
same value of $\kappa$, have the same value of the dynamical exponent for the same form of disorder. This
conjecture is based on scaling theory of rare events and has been numerically checked\cite{igloi01}
on the examples of
the $q=4$ state Potts model and the dimerized AF Heisenberg chain (see Sec.\ref{XYZ_dec}, for both $\kappa=1/2$.

{\it Weak and intermediate disorder regimes}

For weak disorder the first question one can pose is the relevance or irrelevance
of the perturbation caused by weak disorder at the fixed point of the non-random model. In this respect one
generally invokes the Harris criterion \cite{harris,chayes} according to which weak disorder is
irrelevant if:
\begin{equation}
\nu_0 > 2/d\;,
\label{harris}
\end{equation}
where $\nu_0$ is the correlation length critical exponent of the pure system.
In the models we consider here having $\kappa<1$ the correlation length exponent
satisfies
the relevance criterion, $\nu_0 < 2/d$. Thus the strong
disorder fixed point is probably strongly attractive as for the RTFIC.

The behavior is different of systems with $\kappa>1$, which corresponds to the Potts-model with $q < 2$, such as
the classical 2d percolation ($q=1$) with correlated (layered) disorder, or to the clock-model with $q>4$,
or the Ashkin-Teller model for $\epsilon<0$. (To this group belongs also the random contact process as
described in Sec.\ref{RCP}.) Now, depending on the sign of $\nu_0 - 2/d$ different scenarios could happen.

\begin{figure}
\centerline{ \includegraphics[width=0.58\linewidth]{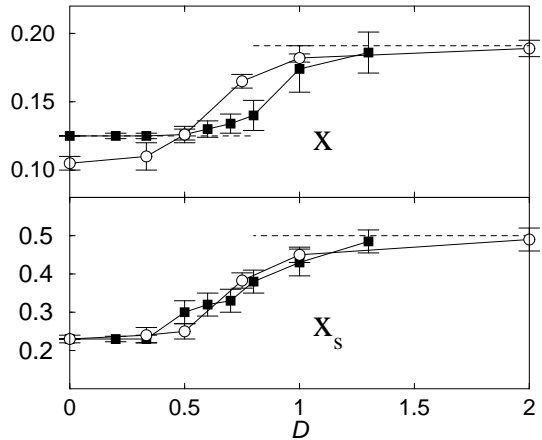}}
\caption{\label{cat}Scaling exponents of the average bulk and surface magnetization as 
a function of the strength of disorder for the $q=5$ clock model
($\circ$) and for the Ashkin-Teller model with $\epsilon=-0.75$
($\blacksquare$).
The limiting values of 
the exponents, corresponding to that of the infinite disorder fixed point are 
denoted by dashed lines.\cite{carlon-clock-at}}
\end{figure}

For irrelevant weak disorder, $\nu_0 > 2/d$, which is realized in the clock model for $q>4$ and in
the Ashkin-Teller model with $-1<\epsilon <-1/2$ the critical behavior is expected to be the same as in the
pure system if the strength of disorder, $D$, is smaller than a finite limiting value, $D_0$, $0<D<D_0$.
The strength of disorder is characterized by the variance of the initial probability distributions
of the couplings and the transverse fields, and directly given by the parameter, $D$, of a power-law
distribution, $P(\lambda)=D^{-1} \lambda^{-1+1/D}$. see in Eq.(\ref{distr_D}). Increasing the
strength of disorder over $D_0$ the
critical behavior is controlled by a
new fixed point. According to numerical investigations\cite{carlon-clock-at} there is a line of
conventional random fixed
points (see Appendix \ref{SC_conv}) in the range of disorder: $D_0<D<D_{\infty}$. In this region, which is called
the intermediate disorder regime the magnetization critical exponents, such as $x_m$ and $x_m^s$, are
disorder dependent and also the disorder induced dynamical exponent, $z'$, is a continuous function of $D$.
(The true dynamical exponent is given by ${\rm max}(z_0,z')$, where $z_0=1$ is the dynamical exponent
due to pure quantum fluctuations.)
On the other hand the energy density, according to numerical studies\cite{carlon-clock-at},
is a marginal operator so that the
correlation length critical exponent is $\nu=2$, which is just the borderline case of the Harris-criterion
in Eq.(\ref{harris}). As the upper critical value of the disorder, $D_{\infty}$, is approached the dynamical
exponent starts to diverge, $1/z \to 0$, and at the same time the magnetization exponents reach their value
in the infinite disorder fixed point. For $D>D_{\infty}$ we arrive at the attractive region of the strong
disorder fixed point.

For relevant weak disorder (but with $\kappa>1$), which is realized for 2d percolation with correlated
disorder, or in the Ashkin-Teller model with $-1/2<\epsilon<0$, first one enters for $0<D<D_{\infty}$
into the intermediate disorder regime and for $D>D_{\infty}$ into the strong disorder regime. This type
of scenario is checked numerically for the percolation problem\cite{igloiperco}, see Sec.\ref{PERC}.
Note, however,
that the contact process, which has anisotropic scaling behavior in the non-random case, in the
intermediate disorder regime show such a scaling behavior which could be of infinite disorder type
with varying exponents, see Sec.\ref{RCP4}.

\begin{figure}
\centerline{\includegraphics[width=0.58\linewidth]{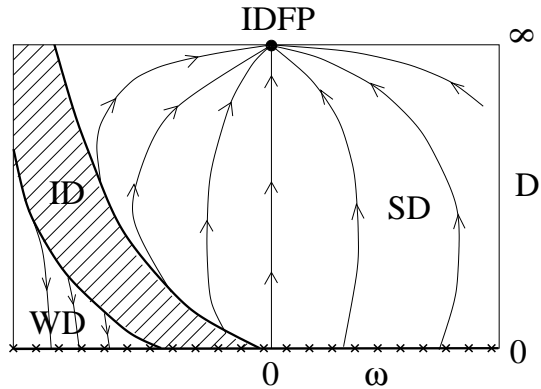}}
\caption{\label{cross-over} Schematic RG phase diagram of a quantum spin system having a
critical line parameterized by $\omega$, in the presence of disorder of
strength, $D$.\cite{carlon-clock-at}}
\end{figure}

A schematic phase diagram of the disorder induced cross-over effects can be found in Fig.\ref{cross-over}.
Here the
model dependent parameter, $\omega$, is given by: $\omega=\epsilon$ for the Ashkin-Teller model,
$\omega=4-q$ for the clock model and $\omega=q/2-1$ for the quantum Potts model.  For
$\omega \ge 0$ the system is in the strong disorder (SD) phase, where
the infinite disorder fixed point (IDFP) is strongly attractive. In the other part of the phase diagram,
for $\omega < 0$, there are two more phases: the weak disorder (WD) and the
intermediate disorder (ID) regions. Here the SD part of the phase diagram is
attracted by the IDFP, while the critical behavior in the WD regime is
governed by the fixed points of the pure system, which are located at
$D=0$. In the ID regime, where there is a competition between quantum
fluctuations and disorder effects, dynamical scaling is anisotropic,
$1<z<\infty$, and the static and dynamical critical exponents are disorder
dependent. At the boundaries of the ID region there are $1/z=0$ and $1/z=1$,
respectively, while at $D=0$, i.e. in the pure system limit they are
at $\omega=0$ and $\omega=\omega_0$. In the latter case for the pure model
with a parameter $\omega_0$ the Harris criterion in Eq.(\ref{harris}) is
saturated, i.e. $\nu_0(\omega_0)=2/d$.

We note that disorder induced cross-over effects are also present in higher spin random
AF Heisenberg chains. The scenario in these systems
is, however, somewhat different, see in Sec. \ref{S=1} and \ref{S>1}.

\section{Random $S=1/2$ AF Heisenberg chains}
\label{S=1/2}

\subsection{Models}

\label{heisenberg3}

The Heisenberg model has been introduced to describe the basic features of interacting localized
magnetic moments. Most generally, with random couplings and anisotropy, the model is described
by the Hamiltonian:
\begin{equation}
H_H=\sum_l\left( J_l^x S_l^x S_{l+1}^x+J_l^y S_l^y S_{l+1}^y+J_l^z S_l^z S_{l+1}^z \right)  \;,
\label{HH}
\end{equation}
in terms of the spin-$1/2$ operators, $S_l^{x,y,z}$, at site $l$.  For terminology, the models for different anisotropies
are respectively called :

 $XYZ$-model : $J_l^x \ne J_l^y \ne J_l^z$ 

 $XXZ$-model : $J_l^x = J_l^y \equiv J^{\perp} \ne J_l^z$ 

 $XXX$-model : $J_l^x = J_l^y = J_l^z$ 

  $XY$-model : $J_l^x \ne J_l^y $ and $ J_l^z =0$ for any $l$

  $XX$-model : $J_l^x = J_l^y $ and $ J_l^z =0$ for any $l$

Generally we consider the situation when the sign of the $J_l^x$ couplings is the same, and the same is true for
the $y$ and $z$ components, as well. Then rotating spins around the $z$-axis one can always take
$J_l^x>0$ and $J_l^y>0$, so that the sign of $J_l^z$ characterizes the model: for $J^z_l>0$ it is antiferromagnetic and for $J^z_l<0$ it is ferromagnetic.

The pure system with $J_l^x=J^x$,$J_l^y=J^y$ and $J_l^z=J^z$ has a reach phase
diagram\cite{baxter72} with four ordered phases ($z$-ferromagnet, $z$-antiferromagnet, $x$-antiferromagnet and $y$-antiferromagnet) which are separated by phase boundaries in which there is quasi-long-range order (QLRO).
For example for the $XXZ$ chain with $J^x=J^y=J^{\perp}$ we have
$G^{\alpha}(r)=\langle S_l^{\alpha}S_{l+r}^{\alpha}\rangle= \sim (-1)^r r^{-\eta_{\alpha}}$,
$(\alpha=x,z)$, where
the decay exponents are coupling dependent and given by\cite{luther75}:
\begin{equation}
\eta_x=\frac{1}{\eta_z}=1-\frac{1}{\pi}{\rm arccos \Delta}\;.
\label{etaXXZ}
\end{equation}
with $1 \ge \Delta=J^z/J^{\perp} \ge -1$.

{\it Dimerization}

The isotropic antiferromagnetic chains, i.e. the $XXX$ and $XX$ models are at the phase boundary and they exhibit QLRO. They can be driven out of their critical state, which is gapless, by introducing alternating
bond strengths for even $(J_e)$ and odd $(J_o)$ bonds. The corresponding control parameter is the
dimerization, which is defined by:
\begin{equation}
d=\frac{J_e-J_o}{J_e+J_o} \;.
\label{dimer}
\end{equation}
(For the $XXZ$-chain one can work similarly by keeping the same $J^{\perp}/J^z$ on even and odd bonds.)
In the dimerized phase we define the {\it  string correlation function}:
\begin{equation}
O^z(r)=-4 \langle S_l^z \exp\left[ i \pi \left( S_{l+1}^z+S_{l+2}^z+\dots+S_{l+r-1}^z \right)\right] S_{l+r}^z \rangle\;.
\label{string}
\end{equation}
which can be written for spin-$1/2$ operators and for $r$ odd as:
\begin{equation}
O^z(r)=2^{r+1} \langle S_l^z S_{l+1}^z \dots S_{l+r}^z \rangle\;.
\label{string1/2}
\end{equation}
Here we used the identity: $S^z=\exp\left(i \pi S^z \right)/2i$, which follows immediately by considering
its action on the full basis states $|\uparrow \rangle$, $|\downarrow \rangle$. The expression in
Eq.(\ref{string1/2}) can be simplified further in terms of dual spin operators\cite{igloixy}, 
$\tilde{S}^x_{l+1/2}=S_1^z S_2^z\dots S_{l+1}^z$, and one
obtains:
\begin{equation}
O^z(r)=4 \langle \tilde{S}_{l+1/2}^x \tilde{S}_{l+r+1/2}^x \rangle\;,
\label{string1/2t}
\end{equation}
thus in an obvious notation $O^z(r)=4 \tilde{G}^x(r)$. In a dimerized state
$O^z(r)$ is different for $l$ even $(O^z_e(r))$ and $l$ odd $(O^z_o(r))$, and the order is measured as:
\begin{equation}
O^z_d(r)= O^z_o(r)-O^z_e(r)\;,
\label{stringd}
\end{equation}
and one should take the limiting value: $\lim_{r \to \infty} O^z_d(r)$.

\subsection{RG rules}

\label{XYZ_dec}

The renormalization method of Ma, Dasgupta and Hu\cite{madasgupta} is originally developed for the
random antiferromagnetic $S=1/2$ chain. Here we consider the general $XYZ$-chain, in which a unit of four
spins contains the strongest bond of the chain, $J_2=\Omega$, between the two central ones.

In the limit, $J_1/J_2 = J_3/J_2 = 0$, the ground state for $J^z_2 \ge 0$ is a singlet having the energy:
$E_0=-(J_2^x+J_2^y+J_2^z/4$, which is shifted by $\Delta E_0^{\uparrow \uparrow} \approx
-(J_1^z-J_3^z)^2/8(J_2^x+J_2^y)$ if the $J_1$ and $J_2$ couplings are switched on and the two
edge spins are fixed into parallel, $(\uparrow \uparrow)$,
states. For antiparallel states, $(\uparrow \downarrow)$, one obtains
$\Delta E_0^{\uparrow \downarrow} \approx -(J_1^z+J_3^z)^2/8(J_2^x+J_2^y)$, which is the
consequence of the transformation $J_3^z \to -J_3^z$. Decimating out the strongest bond, $J_2$, in the
renormalized Hamiltonian the $z$-component of the generated new coupling between
sites $1$ and $4$ is given as:
\begin{equation}
\tilde{J}_{14}^z=2\left(E_0^{\uparrow \uparrow}-E_0^{\uparrow \downarrow}\right)=\frac{J_1^z J_3^z} {J_2^x + J_2^y} \;.
\label{Jz}
\end{equation}
Similarly, by rotating the spins as $S^x \to S^y$, $S^y \to S^z$ and $S^z \to S^x$, etc. we get
for the other components of the renormalized coupling as:
\begin{equation}
\tilde{J}_{14}^x=\frac{J_1^x J_3^x} {J_2^y + J_2^z},\quad
\tilde{J}_{14}^y=\frac{J_1^y J_3^y} {J_2^z + J_2^x}\;.
\label{Jxy}
\end{equation}
In the special case of the $XXX$-model with $J_l^x=J_l^y=J_l^z=J_l$ we have
\begin{equation}
\tilde{J}=\frac{J_1 J_3} {2 J_2},\quad XXX-{\rm model}\;,
\label{JXXX}
\end{equation}
whereas for the $XX$-model with $J_l^x=J_l^y=J_l$ and $J_l^z=0$ we obtain:
\begin{equation}
\tilde{J}=\frac{J_1 J_3}{J_2},\quad \tilde{J}^z=0,
\quad XX-{\rm model}\;.
\label{JXX}
\end{equation}
Note, that both for the $XXX$ and the $XX$ models the generated new coupling is smaller than the decimated
bond, therefore the energy-scale, $\Omega$, gradually decreases during the renormalization process.

\subsection{ RG detailed results}

\label{S=1/2_det}

\begin{figure}
\centerline{\includegraphics[width=0.58\linewidth]{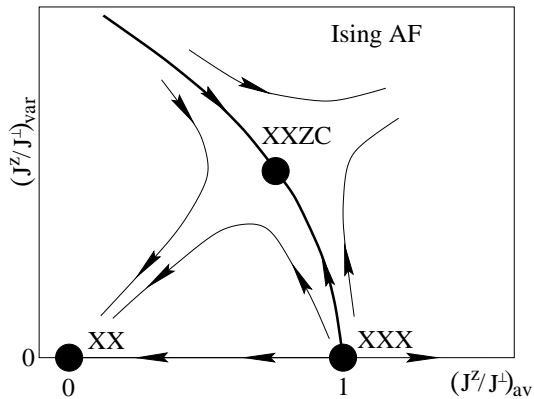}}
\caption{\label{XXZ} Schematic RG phase diagram of the random $XXZ$ chain as obtained
in Ref.\cite{danielantiferro}.}
\end{figure}

The phase diagram of the random $XXZ$ chain, as studied in detail in \cite{danielantiferro}, depends on
the relative strength of the couplings, $J^{\perp}$ and $J^z$. If $J^{\perp}$
dominates the ground state of the system is composed of singlets (random singlet (RS) phase). On the
other hand if $J^z$ dominates in the ground state of the system there is Ising antiferromagnetic (IAF)
order, but due to randomness the gap is vanishing, thus the system is in an IAF-ordered Griffiths phase.
The schematic RG phase diagram as obtained by Fisher \cite{danielantiferro} is shown in Fig.\ref{XXZ}.

It contains three RS fixed points, among which the RG equations have been solved for the
random $XX$ and $XXX$ models\cite{danielantiferro}.

\subsubsection{Renormalization of the random $XX$ chain}

The decimation equation in Eq.(\ref{JXX}) are in a simple additive form
\begin{equation}
\tilde{\xi}=\xi_1+\xi_3\;,
\label{Jxi}
\end{equation}
in terms of the logarithmic variables: $\xi=\ln(\Omega/J),~ 0<\xi<\infty$. Here we have used that $\xi_2=0$, since  $J_2=\Omega$ being
the strongest coupling of the chain. We are looking for the probability
distribution $\rho(\xi,\Gamma)$ of the variable $\xi$ at a log-energy
scale $\Gamma=-\ln \Omega$. By decimating out bonds we change the scale as
$\Gamma \to \Gamma+{\rm d}\Gamma$, which amounts of a change as
$\xi \to \xi-{\rm d}\Gamma$, thus
\begin{equation}
\left.\rho(\xi,\Gamma)\right|_{\Gamma = \Gamma+{\rm d}\Gamma}
= \rho(\xi,\Gamma) + \frac{\partial \rho}{\partial \Gamma}{\rm d}\Gamma-
\frac{\partial \rho}{\partial \xi}{\rm d}\Gamma  \;.
\label{rho1}
\end{equation}
This variation can also be written into the form of a rate equation:
the distribution function $\rho(\xi,\Gamma)$ is reduced due
to the eliminated couplings $\xi_1$ and $\xi_3$, but is increased
due to the generated one with $\xi_1+\xi_3$:
\begin{eqnarray}
\left.\rho(\xi,\Gamma)\right|_{\Gamma = \Gamma+{\rm d}\Gamma} = \left\{\rho(\xi,\Gamma) +{\rm d}\Gamma\rho(0,\Gamma)
\int_0^{\infty} {\rm d}\xi_1\int_0^{\infty} {\rm d}\xi_3
\rho(\xi_1,\Gamma)\rho(\xi_3,\Gamma)\right.
\nonumber\\
\left.\times\left[\delta(\xi-\xi_1-\xi_3)-\delta(\xi-\xi_1)-\delta(\xi-\xi_3)\right]\right\}
\left[1-2 {\rm d}\Gamma\rho(0,\Gamma) \right]^{-1} \;.
\label{rho2}
\end{eqnarray}
Here the last factor in the r.h.s. is to ensure normalization since
the total number of bonds is reduced by a fraction of
$2 {\rm d}\Gamma\rho(0,\Gamma)$ during renormalization. Comparing Eq.(\ref{rho1}) with
Eq.(\ref{rho2}) and changing the notation as $\xi_1 \to \xi_-$ and $\xi_3 \to \xi_+$ we
arrive to the integral-differential equation:
\begin{equation}
\frac{\partial \rho(\xi,\Gamma)}{\partial \Gamma}
= \frac{\partial \rho(\xi,\Gamma)}{\partial \xi}+
\rho(0,\Gamma)\int_0^{\infty} {\rm d}\xi_-\int_0^{\infty} {\rm d}\xi_+
\rho(\xi_-,\Gamma)\rho(\xi_+,\Gamma)\delta(\xi-\xi_-\xi_+) \;.
\label{rho3}
\end{equation}
The solution of Eq.(\ref{rho3}) is tried in the form:
\begin{equation}
\rho(\xi,\Gamma)=\frac{1}{\Gamma} Q(\xi/\Gamma,\Gamma)\;,
\label{rho4}
\end{equation}
which leads to the equation in terms of the variable $\eta=\xi/\Gamma$ as:
\begin{equation}
\Gamma \frac{\partial Q(\eta,\Gamma)}{\partial \Gamma}
= \frac{\partial Q(\eta,\Gamma)}{\partial \eta}(1+\eta)+Q(\eta,\Gamma)+
Q(0,\Gamma)\int_0^{\eta} {\rm d}\eta' Q(\eta',\Gamma)Q(\eta-\eta',\Gamma)\;,
\label{rho5}
\end{equation}
We are looking for the fixed point solution of this equation, $Q^*(\eta)$,
which does not depend on $\Gamma$, thus the l.h.s. of Eq.(\ref{rho5}) is
zero. After Fisher\cite{danielantiferro}, for non-singular initial distributions the fixed point
solution is characterized by the initial value $Q^*(0)=1$, when it is
given as:
\begin{equation}
Q^*(\eta)=\Theta(\eta) e^{-\eta}\;,
\label{Q*}
\end{equation}
where $\Theta(\eta)$ is the Heaviside step-function: $\Theta(\eta)=0$, for $\eta<0$ and $1$
for $\eta\ge0$. Note that the fixed point
distribution of the couplings is identical to that of the RTFIC in Eq.(\ref{soleta}), which is due to
an exact mapping between the two models, as described in Sec.\ref{XY_RTFIC}.

Transforming the solution in Eq.(\ref{Q*}) in terms of the original variables $J$ and $\Omega$
we obtain:
\begin{equation}
P(J,\Omega)=\frac{1}{\Omega \Gamma} \left(\frac{\Omega}{J}\right)^{1-1/\Gamma}\Theta(\Omega-J),
\quad \Gamma=-\ln \Omega\;,
\label{P*}
\end{equation}
which becomes singular at the fixed point, as $\Gamma \to \infty$.
Due to this singularity the decimation transformation in Eq.(\ref{JXX}) becomes exact at the fixed
point. This can be shown by calculating the probability that one of the neighboring
couplings, beside the strongest bond with $\tilde{J}=\Omega$, has a value of $J>\alpha\Omega$,
with $\alpha<1$:
\begin{equation}
P(\alpha) \simeq \int_{\alpha\Omega}^{\Omega} P(J,\Omega) {\rm d} J=\frac{1}{\Gamma}
\int_{\alpha}^1 x^{-1+1/\Gamma}{\rm d} x=1-\alpha^{1/\Gamma} \approx
\frac{1}{\Gamma} \ln(1/\alpha)\;,
\label{Palpha}
\end{equation}
which indeed goes to zero as the iteration proceeds, since $\Gamma \to \infty$. Consequently
the RG transformation becomes asymptotically exact in a similar way as shown for the RTFIC in
Sec.\ref{asymp}.

\subsubsection{Renormalization of the random $XXX$ chain}

The decimation transformations of the $XX$ and the $XXX$ models in Eqs.(\ref{JXXX}) and (\ref{JXX}),
respectively differ by a constant factor of 2. As a consequence the derivation of the
probability distribution in the previous Section is changed at the following points:

i) the decimation transformation in Eq.(\ref{Jxi}) is extended by a term $\ln 2$ at the r.h.s.,

ii) the argument of the first delta-function on the r.h.s. of Eq.(\ref{rho2})
is extended by a term $-\ln 2$,

iii) the integral in the r.h.s. of Eq.(\ref{rho5}) is modified as
\begin{equation}
\int_0^{\eta-\Delta} {\rm d}\eta' Q(\eta',\Gamma)Q(\eta-\Delta-\eta',\Gamma)\;,
\label{correct}
\end{equation}
where $\Delta=\ln 2/\Gamma$.

As the renormalization proceeds $\Gamma \to \infty$ and so $\Delta\to 0$, thus the
RG equation in Eq.(\ref{rho5}) becomes identical to that of the $XX$ model and
consequently the limiting distributions of the two models are the same. Note that the
reasoning used here is analogous to that presented in Sec.\ref{PCAT1} for quantum spin models
with discrete symmetry.

\subsubsection{Properties of the random-singlet phase}

The ground state of the random $XX$ (and $XXX$) models by construction of the strong
disorder RG method is composed of singlet pairs. The two spins
of a pair can be arbitrarily remote and the effective interaction between them is rapidly
decreasing with the distance. In this random-singlet phase
relation between energy-scale, $\Omega=e^{-\Gamma}$, and
length-scale, $L_{\Gamma} \sim 1/n_{\Gamma}$, where $n_{\Gamma}$ is the fraction of
non-decimated spins, can be obtained along the lines of Sec.\ref{RTFIC_34}. As
$\Gamma$ is increased by an
amount of ${\rm d}\Gamma$ (i.e. $\Omega \to \Omega(1-{\rm d}\Gamma)$) a fraction
of spins $2 {\rm d}\Gamma \rho(0,\Gamma)$ is decimated out, thus
${\rm d} n_{\Gamma}=-2 {\rm d}\Gamma \rho(0,\Gamma) n_{\Gamma}$. Now using
Eq.(\ref{rho4}) and with the fixed-point solution in Eq.(\ref{Q*}) we arrive to
the differential equation:
\begin{equation}
\frac{{\rm d} n_{\Gamma}}{{\rm d}\Gamma}=-2\frac{Q^*(0)}{\Gamma} n_{\Gamma} \;,
\label{dngamma}
\end{equation}
with the solution, $n_{\Gamma}=\Gamma^{-2Q^*(0)}=\Gamma^{-2}$,
at the fixed point. Consequently the typical distance between remaining spins is
\begin{equation}
L_{\Gamma} \sim \frac{1}{n_{\Gamma}} \sim \Gamma^2 \sim \left[\ln
\frac{\Omega_0}{\Omega}\right]^2 \;,
\label{lomega}
\end{equation}
where $\Omega_0$ is some reference energy cut-off. This is the usual form of dynamical
scaling at an infinite disorder fixed point, see in Sec.\ref{SC_st}.

The {\it  low-temperature susceptibility} can be estimated by studying
the response of the system on external fields for different ratios of the
thermal energy, $\sim T$, and the energy scale, $\Omega$.
In the low-temperature case, $\Omega \gg T$, the strongly coupled singlet pairs are
very weakly excited by thermal fluctuations. In the opposite limit,
$\Omega \ll T$, the remaining non-decimated spins are very weakly coupled,
since $\tilde{J} \ll T$, and therefore they are essentially free and contribute
to a Curie susceptibility, which goes as $\sim 1/T$. Then one should stop the
renormalization at $\Omega=T$, when the remaining spins with a density of,
$n_{\Gamma_T} \sim [\ln(\Omega_0/T)]^2$, all contribute by a Curie susceptibility
leading to the result:
\begin{equation}
\chi_{\perp} \sim \chi_z \sim \frac{n_{\Gamma_T}}{T} \sim \frac{1}
{T\left[\ln \frac{\Omega_0}{T}\right]^2} \;.
\label{suscXX}
\end{equation}
Note that the transverse and longitudinal susceptibilities have the same singular behavior
where the Curie-type susceptibility is modified by logarithmic corrections. Since these
corrections are very strong they usually lead in measurements effective, temperature
dependent critical exponents.

The {\it  average pair correlation function} between two spins at a distance, $r \sim L$,
is dominated by those spins which have not been decimated out at a length-scale, $L$. The
decimated spins already form singlets and correlation between two spins which belong
to different singlets is negligible. The probability to have a free spin at this length-scale
is $n_{\Gamma_L} \sim 1/L$, for two spins it is $\left( n_{\Gamma_L} \right)^2 \sim 1/L^2$.
Then, under further decimation, there is a finite probability that these two spins form
a singlet, thus will have a correlation $C(r)=O(1)$. Averaging the correlations over the
spin-pairs with a mutual distance, $r$, we obtain:
\begin{equation}
\left[C(r)\right]_{\rm av} \sim \frac{(-1)^r}{r^2} \;.
\label{corrXX}
\end{equation}
If we consider two randomly chosen spins at a distance $r$ they typically belong to different
singlet pairs and the correlations between them, the {\it  typical correlations} are very weak.
If during decimation the length-scale is $L=r$, then the two spins becomes nearest neighbors with
an effective coupling, $\tilde{J}_L \sim \Omega_L$, which measures the size of correlations.
Thus we have
\begin{equation}
-\ln C^{typ}(r) \sim \ln \Omega_L \sim \Gamma_L^{-1} \sim \frac{1}{L^{1/2}} \sim \frac{1}{r^{1/2}}\;,
\label{typXX}
\end{equation}
which is completely different from its average value in Eq.(\ref{corrXX}). Thus the correlation
function in the RS phase is non-self-averaging.

\subsubsection{Properties of the random dimer phase}

Here we consider the random $XX$ and $XXX$ chains with enforced dimerization, see Sec. \ref{heisenberg3},
when couplings at odd, $J_o$, and even, $J_e$, sites are taken from different distributions. For the
random model the control parameter is defined as:
\begin{equation}
\delta={[\ln J_o]_{av}-[\ln J_e]_{av} \over \rm{var}[\ln J_o]+\rm{var}[\ln J_e]}\;,
\label{delta_d}
\end{equation}
thus for $\delta>0$ ($\delta<0$) the odd (even) bonds are stronger in average. Outside the quantum
critical point we are in the random dimer phase, which is a Griffiths phase of the system\cite{hyman-dimer}.

For the random $XX$ chain the distribution of the odd and even couplings at the fixed point can be
exactly obtained\cite{igloixy} through the mapping to the RTFIC in Sec.\ref{XY_RTFIC}. With this help results in
Sec. \ref{RTFIC32} can be translated to the random dimer phase, too. In particular the dynamical exponent, $z$,
is given from Eq.(\ref{z_I}) as
\begin{equation}
\left[\left(\frac{J_o}{J_e}\right)^{1/z}\right]_{\rm av}=1
\;.
\label{zHeq}
\end{equation}
which is related to $\delta$, as $z \approx 1/2 |\delta|$ in linear order.

For the random $XXX$ model the dynamical exponent in the random dimer phase is not known exactly. Here
there is a conjecture\cite{igloi01} that dynamical exponents for the random dimerized $XXX$ chain
and for the random
$q=4$ state quantum Potts models with the same disorder distribution are identical. This is based on the
same form of the decimation equations see in Sec. \ref{PCAT1} and on scaling theory, which has been
checked by numerical calculations\cite{igloi01}.

Scaling of the thermodynamic quantities can be obtained along the lines as described for the RTFIC in
Sec.\ref{SC_term}. In this respect the random dimer phase of the random AF Heisenberg chain corresponds to
the disordered Griffiths phase of the RTFIC. For example the same reasoning as for Eq.(\ref{chi_T}) leads to
the low-temperature susceptibility:
\begin{equation}
\chi(T) \sim T^{-1+1/z}\;,
\label{chi_T1}
\end{equation}
whereas the singularity of the specific heat from Eq.(\ref{entropy_d}) follows as:
\begin{equation}
c_V(T) \sim T^{1/z}\;.
\label{Cv_T}
\end{equation}

\subsubsection{Renormalization of dynamical correlations}

The average autocorrelation functions both in the random singlet and in the random dimer phases are
obtained by the strong disorder RG method\cite{rsrgdyna,eigenhuse} in the same way as for the RTFIC in Sec.\ref{Gt_RG}.
One calculates the local dynamical susceptibilities,
\begin{equation}
\chi^{\alpha \alpha}_{jj}(\omega)=\sum_k | \langle k | S_j^{\alpha}|0\rangle |^2
\delta(\omega-\epsilon_k)\;,
\label{chi_om5}
\end{equation}
with $\alpha=x,z$. The low-frequency behavior of $\chi^{\alpha \alpha}_{jj}(\omega)$ is related to the long-time
limit of the corresponding autocorrelation function:
\begin{equation}
G^{\alpha \alpha}_{jj}(t)=\langle  S_j^{\alpha}(t) S_j^{\alpha}(0) \rangle \;.
\label{G_t_S}
\end{equation}
Repeating the steps of the calculation for the RTFIC in Sec.\ref{Gt_RG} we obtain for the local dynamical susceptibilities:
\begin{equation}
[\chi]_{\rm av}(\omega) \sim \frac{n_{\omega} (p_0^o(\omega) + p_0^e(\omega))}{\omega} \;,
\label{chi_omS}
\end{equation}
which is valid in leading order both for $\alpha=x$ and $\alpha=z$. Here $p_0^o(\omega)$ and $p_0^e(\omega)$
(``o'' and ``e'' for odd and even bonds, respectively) are parameters of the fixed-point solution of
the RG equations, which corresponds to $p_0(\omega)$ and $r_0(\omega)$, respectively, in
the solution of the RTFIC in Sec.\ref{RTFIC32}. see the mapping in Sec.\ref{XY_RTFIC}.

In RS phase with $p_0^o(\omega)=p_0^e(\omega)$ we obtain:
\begin{equation}
[\chi]_{\rm av}(\omega) \sim \frac{1}{\omega|\ln \omega|^{3}} \;,
\label{chi_omS1}
\end{equation}
and
\begin{equation}
G(t) \sim |\ln t|^{2} \;.
\label{G_tS1}
\end{equation}
This corresponds to the scaling result in Eq.(\ref{lauto2}) with $x_m/\psi=2 x_m=2$, where we have
used Eq.(\ref{corrXX}) for the average correlation function.

In the random dimer phase the results are:
\begin{equation}
[\chi]_{\rm av}(\omega) \sim \frac{|\delta|^{3}}{\omega^{1-1/z}} \;,
\label{chi_omS2}
\end{equation}
and
\begin{equation}
G(t) \sim \frac{|\delta|^{3}}{t^{1/z}} \;,
\label{G_tS2}
\end{equation}
in agreement with the scaling results in Sec.\ref{SC_gr}. Dynamical correlations at the boundary spin
can be calculated similarly, see Ref.\cite{rsrgdyna,eigenhuse}.

\paragraph{Another investigations and numerical studies}

The random $XX$ and $XY$ chains can be transformed into free fermion models, from which some exact results
have been obtained\cite{igloixy}. Identifying the rare events the scaling behavior of the bulk and boundary order parameter has been obtained from which the asymptotic decay of the bulk and the end-to-end correlations follows.

Numerical study of the $XX$ and $XY$ models based on the free fermion mapping has given support to the validity
of the strong disorder RG results\cite{henelius98,igloixy}. On the other hand numerical investigation of the random $XXZ$-chains, which
has been done by the DMRG method and thus restricted to comparatively smaller sizes, has found some discrepancies
with the strong disorder RG results\cite{hamacher}. These are debated in\cite{laflorencie03} and attributed to the presence of possible logarithmic corrections\cite{igloixy} or to cross-over effects\cite{laflorencie03}. Scaling of the spin stiffness in random
spin-1/2 chains has been studied recently\cite{laflorencie04}.

The scaling behavior of typical autocorrelations and their distribution function have been studied in
 Ref.\cite{igloixy}.
In the random singlet phase typical autocorrelations decay algebraically, $G^{typ}(t) \sim t^{-\gamma}$, with a
varying parameter, $\gamma$. The appropriate scaling combination is thus $\gamma=-\ln G(t)/\ln t$, which has
a small $\gamma$ behavior as $P(\gamma)=A+B\gamma+O(\gamma^2)$. On the other hand in the random singlet
phase typical autocorrelations are in a stretched exponential form: $G^{typ}(t) \sim \exp(-\gamma' t^{1/(z+1)}$,
thus the appropriate scaling combination is $\gamma'=-\ln G(t)/t^{1/(z+1)}$. The distribution function
$P'(\gamma')$ is analyzed in Ref.\cite{igloixy}, by making use of similar results about the RTFIC\cite{kisker98}.


\section{Random $S=1$ AF Heisenberg chain}

\label{S=1}

\subsection{The antiferromagnetic chain $S =1$ without disorder}

 { \bf Differences between half-integer and integer spin chains}

Whereas the pure antiferromagnetic chain  $S = 1/2 $ presents power-law correlations and excitations without gap, the antiferromagnetic chain  $S = 1$ is characterized by 
exponential correlations and a gap for excitations.  A simple way to understand these differences
 which more generally exist between half-integer and integer spin chains is related to the
`` Valence-Bond-Solid "(VBS) wavefunction, which allows to highlight a long range order
for a topological order parameter which is non-local in terms of the spins. 

 { \bf The VBS wavefunction} 

\begin{figure}[b ] \centerline{\includegraphics[height=3cm]{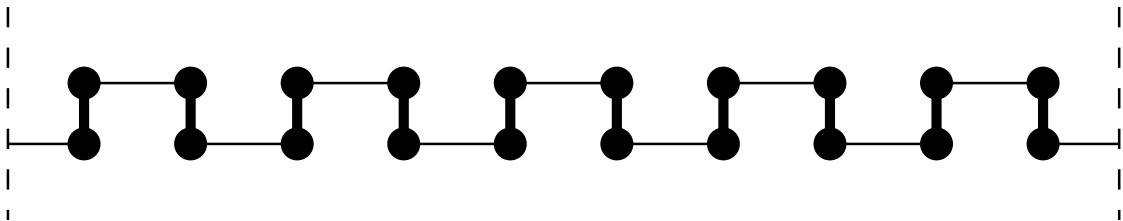}} \caption{\it  
VBS wavefunction:  each spin  $S=1$ is represented by the symmetrization of 2 spins 
$S=1/2$ (black balls connected by a fat vertical bond);  the horizontal lines which connect two spins 1/2 represent the singulets between constitutive spins 
$S=1/2$.} \label{figdefvbs} 
\end{figure} 

If one represents each spin $S =1$ like the symmetrization of two elementary spins  $S=1/2$, the VBS wavefunction \cite{AKLT} is the state where there exists a singlet on each bond between two constitutive spins $1/2$ (see Figure \ref{figdefvbs}).  This wave function is the exact ground state of the Hamiltonian \cite{AKLT} 
\begin{eqnarray} 
H_{AKLT} = \sum_i P_2(\vec S_i + \vec S_{i+1}) = \sum_i \left[ \frac{1}{2} \vec S_i \vec S_{i+1} + \frac{1}{6} (\vec S_i \vec S_{i+1})^2 + \frac{1}{3} \right ] 
\end{eqnarray} 
where  $P_2$  is the projector on the subspace $s=2$. 

 { \bf The String Order Parameter} 

The so called `String Order Parameter' \cite{SOP} which allows to characterize the topological order
 of the VBS wave function is non local in the spins
 \begin{eqnarray} 
t_{ij} = - \langle \psi_o \big \vert S_i^z \exp\left[i \pi \sum_{i<k<j} S_k^z \right ] S_{j}^z\big\vert \psi_0\rangle \label{tij} 
\end{eqnarray}
(Note that this definition corresponds to Eq.(\ref{string}), which was analyzed for the $S=1/2$ model in
Sec.\ref{S=1/2}.)

For the pure VBS state on a ring of $N$ spins, one obtains $t_{ij}=4/9+O(3^{-N})$. 
 For the ground state of the pure antiferromagnetic $S=1$ chain, this order parameter also 
converges towards a finite value, which indicates a relationship with the VBS wave function.  

\subsection{Construction of appropriate RG rules}

\subsubsection*{ Renormalization of an antiferromagnetic bond}

 The Hamiltonian of an antiferromagnetic bond between two spins $S =1$ 
\begin{eqnarray} 
h_0 = J_1 \vec S_1.  \vec S_2={J_1 \over 2} \left[\left(\vec S_1+ \vec S_{2} \right)^2 -\vec S_1^2-\vec S_{2}^2 \right ] = { J_1 \over 2} \left[\left(\vec S_1+ \vec S_{2} \right)^2 - 4\right ]
 \end{eqnarray} 
has three energy levels indexed by the value $s=0,1,2$ of the total spin 
\begin{eqnarray} 
e_{s}={J_1 \over 2} \left[s(s+1)-4 \right ] 
\end{eqnarray} 
the singlet $e_0=-2J_1$, the triplet $e_1=-J_1$ and the quintuplet $e_2=J_1$. 
 If one naively generalizes the Ma-Dasgupta rule for the $S =1/2$ case by projecting onto the singlet, the new effective coupling is 
\begin{eqnarray} 
J_0' ={4 \over 3} { { J_0 J_2} \over { J_1}} 
\label{rules1}
\end{eqnarray} 
Since the coefficient ${4 \over 3}$ is larger than $1$, this rule is not automatically consistent, and the procedure can be justified only if one starts from a rather broad disorder.

To avoid this problem, Hyman and Yang \cite{HY} have proposed an effective $S=1/2$ model,
containing both Ferromagnetic and Antiferromagnetic bonds
to mimic the physics of the weak disorder spin-1 chain.
This effective model 
has the advantage of being consistent and soluble via real-space RG.

Since the mapping between the spin-1 chain and the Hyman-Yang effective model
is heuristic and not quantitative, 
various propositions have been made to define consistent RG rules for arbitrary disorder directly on the spin-1 chain,
in particular :

$\bullet$ Monthus, Golinelli and Jolicoeur \cite{spin1} have proposed the following principle : instead of projecting onto the lowest level of $h_0$, 
the correct generalization of the Ma-Dasgupta principle
consists in {\it  projecting out the highest level}.  Thus, for the Hamiltonian $h_0$, one eliminates the quintuplet, but one should keep the singlet and the triplet, by replacing the two spins $S =1$ by two spins $S =1/2$.  This partial decimation 
of course enlarges the initial space of random chains, but it is 
nevertheless possible to define a closed RG procedure with 4 types of bonds.

$ \bullet $ Saguia, Boechat and Continento \cite{saguia} have more recently
proposed a procedure where the rule (\ref{rules1}) is applied only if 
$max(J_0,J_2) < (3/4) \Omega$, whereas otherwise the 3 spins coupled by $max(J_0,J_2)$
and $\Omega$ are replaced by a new spin with two new couplings.

In the following, we will describe the RG procedure with 4 types of bonds \cite{spin1},
the numerical study of the RG flow, and we will finally describe how
the numerical estimates of the critical exponents
are in full agreement with the analytic critical exponents
that can be computed \cite{spin1} for the Hyman-Yang effective model.
Finally we will describe the controversy between various direct numerical studies.

\subsubsection*{  RG procedure with 4 types of bonds} 

This RG procedure is defined for the enlarged set of chains made of spins of size  $S=1/2$ 
and $S  =1$, in which the couplings $\{J_i\}$ are either ferromagnetic (F) or antiferromagnetic (AF), with the following constraint:  for any segment $\{i, j\}$, the classical magnetization has to satisfy  $\vert m_{i, j} \vert \leq 1$.  This condition for two neighbors $j=i+1$ shows that there are 4 types of possible bonds: 

 1)  F Bond between two S=1/2 spins, 

2) AF Bond between two S=1/2 spins, 

3) AF Bond between a S=1 spin and a S=1/2 spin, 

4) AF Bond between two S=1 spins.  

The four corresponding rules of renormalization are as follows
\cite{spin1} :

\bigskip
\bigskip
\hbox to 450pt {(1)\hfill
\vbox{\hsize=10pt \centerline{$s_0$}\par \centerline{$\bullet$}} 
\kern -7pt \raise 2pt
\vtop{\hsize=45pt \centerline{\hrulefill} \par \centerline{$J_0$}} 
\kern -7pt
\vbox{\hsize=10pt \centerline{$s_1={1\over 2}$}\par 
\centerline{$\bullet$}} \kern -7pt \raise 2pt
\vtop{\hsize=45pt \centerline{\hrulefill} \par \centerline{$J_1<0$}} 
\kern -7pt
\vbox{\hsize=10pt \centerline{$s_2={1\over 2}$}\par 
\centerline{$\bullet$}} \kern -7pt \raise 2pt
\vtop{\hsize=45pt \centerline{\hrulefill} \par \centerline{$J_2$}} 
\kern -7pt
\vbox{\hsize=10pt \centerline{$s_3$}\par \centerline{$\bullet$}} 
\hfill$\longrightarrow$\hfill
\vbox{\hsize=10pt \centerline{$s_0$}\par \centerline{$\bullet$}} 
\kern -7pt \raise 2pt
\vtop{\hsize=67pt \centerline{\hrulefill} \par 
\centerline{$J^\prime_0={J_0\over 2}$}} \kern -7pt
\vbox{\hsize=10pt \centerline{$s^\prime_1=1$}\par 
\centerline{$\bullet$}} \kern -7pt \raise 2pt
\vtop{\hsize=67pt \centerline{\hrulefill} \par 
\centerline{$J^\prime_1={J_2\over 2}$}} \kern -7pt
\vbox{\hsize=10pt \centerline{$s_3$}\par \centerline{$\bullet$}} 
\hfill}
\bigskip
\bigskip
\hbox to 450pt {(2)\hfill
\vbox{\hsize=10pt \centerline{$s_0$}\par \centerline{$\bullet$}} 
\kern -7pt \raise 2pt
\vtop{\hsize=45pt \centerline{\hrulefill} \par \centerline{$J_0$}} 
\kern -7pt
\vbox{\hsize=10pt \centerline{$s_1={1\over 2}$}\par 
\centerline{$\bullet$}} \kern -7pt \raise 2pt
\vtop{\hsize=45pt \centerline{\hrulefill} \par \centerline{$J_1>0$}} 
\kern -7pt
\vbox{\hsize=10pt \centerline{$s_2={1\over 2}$}\par 
\centerline{$\bullet$}} \kern -7pt \raise 2pt
\vtop{\hsize=45pt \centerline{\hrulefill} \par \centerline{$J_2$}} 
\kern -7pt
\vbox{\hsize=10pt \centerline{$s_3$}\par \centerline{$\bullet$}} 
\hfill$\longrightarrow$\hfill
\vbox{\hsize=10pt \centerline{$s_0$}\par \centerline{$\bullet$}} 
\kern -7pt \raise 2pt
\vtop{\hsize=135pt \centerline{\hrulefill} \par 
\centerline{$J^\prime_0={J_0\,J_2\over 2\,J_1}$}} \kern -7pt
\vbox{\hsize=10pt \centerline{$s_3$}\par \centerline{$\bullet$}} 
\hfill}
\bigskip
\bigskip
\hbox to 450pt {(3)\hfill
\vbox{\hsize=10pt \centerline{$s_0$}\par \centerline{$\bullet$}} 
\kern -7pt \raise 2pt
\vtop{\hsize=45pt \centerline{\hrulefill} \par \centerline{$J_0$}} 
\kern -7pt
\vbox{\hsize=10pt \centerline{$s_1=1$}\par \centerline{$\bullet$}} 
\kern -7pt \raise 2pt
\vtop{\hsize=45pt \centerline{\hrulefill} \par \centerline{$J_1>0$}} 
\kern -7pt
\vbox{\hsize=10pt \centerline{$s_2={1\over 2}$}\par 
\centerline{$\bullet$}} \kern -7pt \raise 2pt
\vtop{\hsize=45pt \centerline{\hrulefill} \par \centerline{$J_2$}} 
\kern -7pt
\vbox{\hsize=10pt \centerline{$s_3$}\par \centerline{$\bullet$}} 
\hfill$\longrightarrow$\hfill
\vbox{\hsize=10pt \centerline{$s_0$}\par \centerline{$\bullet$}} 
\kern -7pt \raise 2pt
\vtop{\hsize=45pt \centerline{\hrulefill} \par 
\centerline{$J^\prime_0={4\,J_0\over 3}$}} \kern -7pt
\vbox{\hsize=10pt \centerline{$s^\prime_1={1\over 2}$}\par 
\centerline{$\bullet$}} \kern -7pt \raise 2pt
\vtop{\hsize=90pt \centerline{\hrulefill} \par 
\centerline{$J^\prime_1=-{J_2\over 3}$}} \kern -7pt
\vbox{\hsize=10pt \centerline{$s_3$}\par \centerline{$\bullet$}} 
\hfill}
\bigskip
\bigskip

\hbox to 450pt {(4)\hfill
\vbox{\hsize=10pt \centerline{$s_0$}\par \centerline{$\bullet$}} 
\kern -7pt \raise 2pt
\vtop{\hsize=45pt \centerline{\hrulefill} \par \centerline{$J_0$}} 
\kern -7pt
\vbox{\hsize=10pt \centerline{$s_1=1$}\par \centerline{$\bullet$}} 
\kern -7pt \raise 2pt
\vtop{\hsize=45pt \centerline{\hrulefill} \par \centerline{$J_1>0$}} 
\kern -7pt
\vbox{\hsize=10pt \centerline{$s_2=1$}\par \centerline{$\bullet$}} 
\kern -7pt \raise 2pt
\vtop{\hsize=45pt \centerline{\hrulefill} \par \centerline{$J_2$}} 
\kern -7pt
\vbox{\hsize=10pt \centerline{$s_3$}\par \centerline{$\bullet$}} 
\hfill$\longrightarrow$\hfill
\vbox{\hsize=10pt \centerline{$s_0$}\par \centerline{$\bullet$}} 
\kern -7pt \raise 2pt
\vtop{\hsize=45pt \centerline{\hrulefill} \par 
\centerline{$J^\prime_0=J_0$}} \kern -7pt
\vbox{\hsize=10pt \centerline{$s^\prime_1={1\over 2}$}\par 
\centerline{$\bullet$}} \kern -7pt \raise 2pt
\vtop{\hsize=45pt \centerline{\hrulefill} \par 
\centerline{$J^\prime_1=J_1$}} \kern -7pt\vbox{\hsize=10pt 
\centerline{$s^\prime_2={1\over 2}$}\par \centerline{$\bullet$}} 
\kern -7pt \raise 2pt
\vtop{\hsize=45pt \centerline{\hrulefill} \par 
\centerline{$J^\prime_2=J_2$}} \kern -7pt
\vbox{\hsize=10pt \centerline{$s_3$}\par \centerline{$\bullet$}} 
\hfill}
\bigskip
\bigskip

\subsubsection*{  Interpretation of the renormalization in terms of VBS clusters}

\begin{figure}[ht ] \centerline{\includegraphics[height=3cm]{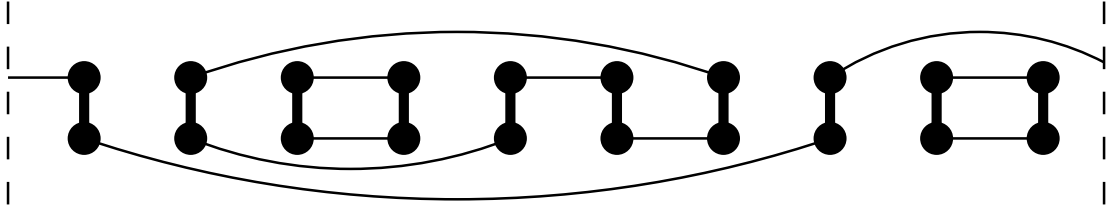}} \caption{\it  The ground state obtained by renormalization, in terms of VBS cluster :  each spin $S =1$ is represented by the symmetrization of 2 spins 1/2 (black balls connected by a fat vertical bond);  the lines which connect two spins 1/2 represent singulets.  Initial spins 1 are thus gathered in clusters:  here, for example, there are two clusters of size 2, a cluster of size 4, and a larger cluster which exceeds the limits of the figure.
\cite{spin1}} \label{figamasvbs} \end{figure}

 If one represents each spin initial  $S=1$ like the symmetrization of two spins $S =1/2$, 
the rules 2, 3, and 4 for AF bonds can be interpret as the formation of a singlet between two  constitutive spins $S =1/2$.  Rule 1 for a F bond between two $S =1/2$ spins corresponds to their symmetrization.  At the end of the RG procedure, when there is no free spin anymore, the chain is broken into a set of disjointed clusters which have a VBS structure  (see Figure \ref{figamasvbs}). 

For the disordered chain, the string order parameter (\ref{tij}) takes the value $t_{ij}=4/9$ if the two sites belong to the same cluster and $t_{ij}=0$ if not.  For a chain of size N, the space average $\sum_{i, j} t_{ij}/N^2$ of the order parameter $t_{ij}$ is proportional to the probability T 
that two spins belong to the same cluster 
\begin{eqnarray} 
T \equiv \sum_c { n_c^2\over N^2}={9\over 4} { 1\over N^2} \sum_{i, J} t_{ij} +O(1/N) 
\end{eqnarray} 
where $n_c$ is the number of spins in a cluster C, which is not directly related to its space extension.  This order parameter T can be non-zero in the thermodynamic limit only if there exists a VBS cluster containing a finite fraction of the spins of the chain. 

 \subsection{ Numerical study of the RG procedure} 

The numerical study of the RG procedure with 4 types of bonds on the basis 
has been made for cyclic chains of N spins (for instance $N=2^{22} \sim 4.10^6$), 
with initial couplings $J_i$ uniformly distributed in the interval $[1,1+d]$. 
 The parameter $d$ thus represents the width of the initial disorder. 
 For each size, the results are averaged over a certain number of samples (typically 100). 
The flow of the following quantities according as a function of the RG scale $\Gamma$
has been computed : 

 (i) the number $N(\Gamma)$ of effective spins $S =1/2$ and $S =1$ still present at the scale $\Gamma$

 (ii) the proportion $ \{ N_{(S=1)} (\Gamma)/N (\Gamma)\}$ of spins $S =1$ among the effective spins still present.

(iii) the proportions $\rho_i(\Gamma)=\{N_i(\Gamma)/N(\Gamma)\}$ of the bonds of the type $i=1,2,3,4$ 

(iv) the probability distributions $P_i(J, \Gamma)$  of the remaining effective couplings $J$ for the four types of bonds $i=1,2,3,4$. 

\begin{figure}[t ] \centerline{\includegraphics[height=5cm]{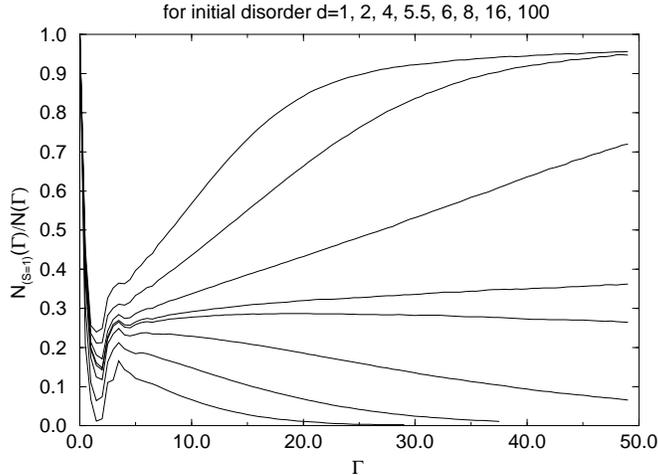}} \caption{\it  Proportion of  $S =1$ spins among the effective spins at scale $\Gamma$ for various values of the initial disorder $d=1,2,3,4,5.5,6,8,16,100$:  this proportion converges towards $0$ for weak initial disorder, and towards $1$ for strong initial disorder.  At the point $d \simeq d_c =5.75(5)$ the proportion remains stationary around the intermediate critical value $0.315(5)$ . \cite{spin1}} \label{figs1global} \end{figure}

 The results of the RG procedure present a qualitative 
change for a certain critical value 
 $d_c \simeq 5.75(5)$ of the initial disorder.
 Figure \ref{figs1global} represents the flow of the proportion $ { { N_{(S=1)} (\Gamma)} \over { N (\Gamma)}} $ of spins $S =1$ for various values of the initial disorder:  there are two attracting values, namely $0$ for weak initial disorder and $1$ for strong initial disorder.

\subsubsection*{  Results for strong initial disorder} 

\begin{figure}[p ] \centerline{\includegraphics[height=6cm]{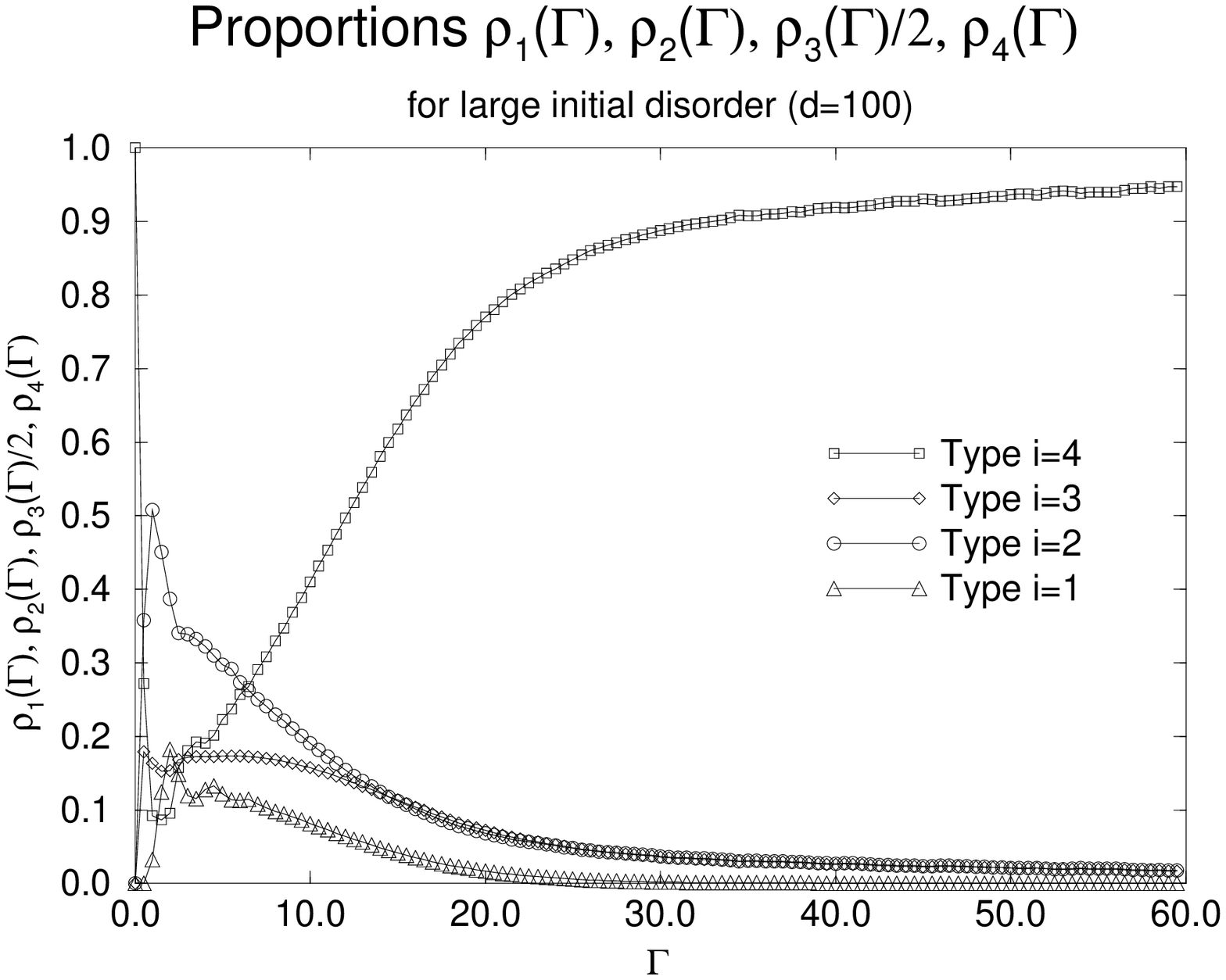}} \caption{\it  The proportions $\rho_i(\Gamma)$ of the 4 types of bonds $i=1,2,3,4$ as functions of the 
RG scale $\Gamma$ for a very strong initial disorder $d=100$.
\cite{spin1}} \label{figrho100} \end{figure}

In the phase of strong disorder $d > d_c$, the number $N(\Gamma)$ of effective spins decreases as in the `` Random Singlet Phase "of the chain  $S=1/2$ chain: 
 \begin{eqnarray} 
N(\Gamma) \oppropto_{\Gamma \to \infty} { 1 \over \Gamma^2}.  
\label{ngafort} 
\end{eqnarray} 
and the proportions $\rho_i(\Gamma)$ of the four types of bonds converge towards 
the following asymptotic regime  (Fig \ref{figrho100}) 
\begin{eqnarray} 
\rho_1(\Gamma)\sim 0 \qquad \ \rho_2(\Gamma) \sim \epsilon(\Gamma) \qquad \ \rho_3(\Gamma) \sim 2 \epsilon(\Gamma) \qquad \ \rho_4(\Gamma) \sim 1-3\epsilon(\Gamma) 
\label{rhofort} 
\end{eqnarray} 
where $\epsilon(\Gamma)$ slowly converges towards $0$ as $\Gamma \to \infty$. 
The chain contains almost everywhere bonds of the type 4, with some defects of the type (bond of the type 3, bond of the type 2, bond of the type 3) which come from the temporary partial decimation of the bonds of the type $4$.  For strong initial disorder, the renormalization thus converges towards the Random Singlet Phase.

\subsubsection*{ Results for weak initial disorder} 

\begin{figure}[p ] \centerline{\includegraphics[height=6cm]{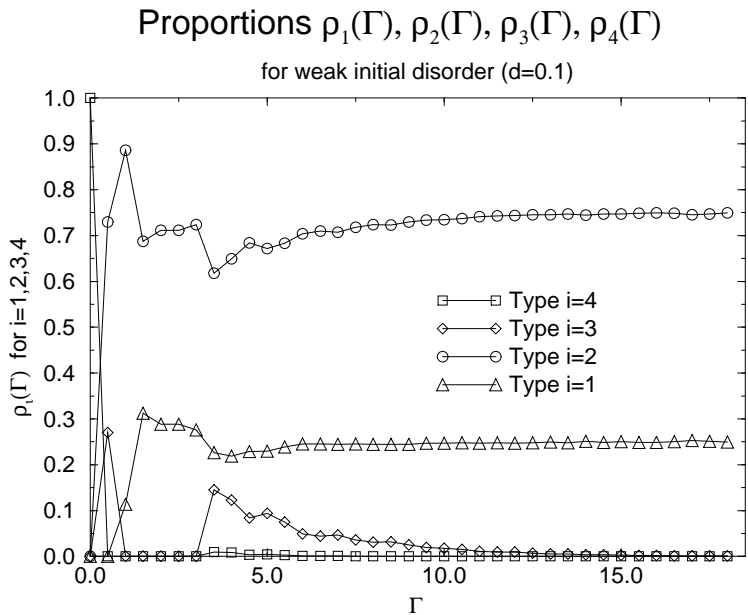}} \caption{\it  The proportions $\rho_i(\Gamma)$ of the 4 types of bonds $i=1,2,3,4$ as functions of the scale $\Gamma$ for a weak initial disorder $d=0.1$ \cite{spin1}} \label{figrho01} \end{figure}

In the weak disorder phase $d<d_c$, the number $N(\Gamma)$ of effective spins decreases exponentially 
\begin{eqnarray} \label{ngafaible}
 N(\Gamma) \oppropto_{\Gamma \to \infty} e^{-\alpha(d) \Gamma} 
\end{eqnarray} 
with a coefficient $\alpha(d)$ which decreases towards $0$ as $d \to d_c^-$.  The proportions $\rho_i(\Gamma)$ of the four types of bonds converge towards 
the asymptotic regime (Figure \ref{figrho01}) 
\begin{eqnarray} 
\rho_1(\Gamma)\simeq 0.25 \qquad \ \rho_2(\Gamma) \simeq 0.75 \qquad \ \rho_3(\Gamma) \simeq 0 \qquad \ \rho_4(\Gamma)  \simeq 0
\label{rhofaible} 
\end{eqnarray} 

\subsubsection*{  Results at the critical point} 

\begin{figure}[ht ] \centerline{\includegraphics[height=8cm]{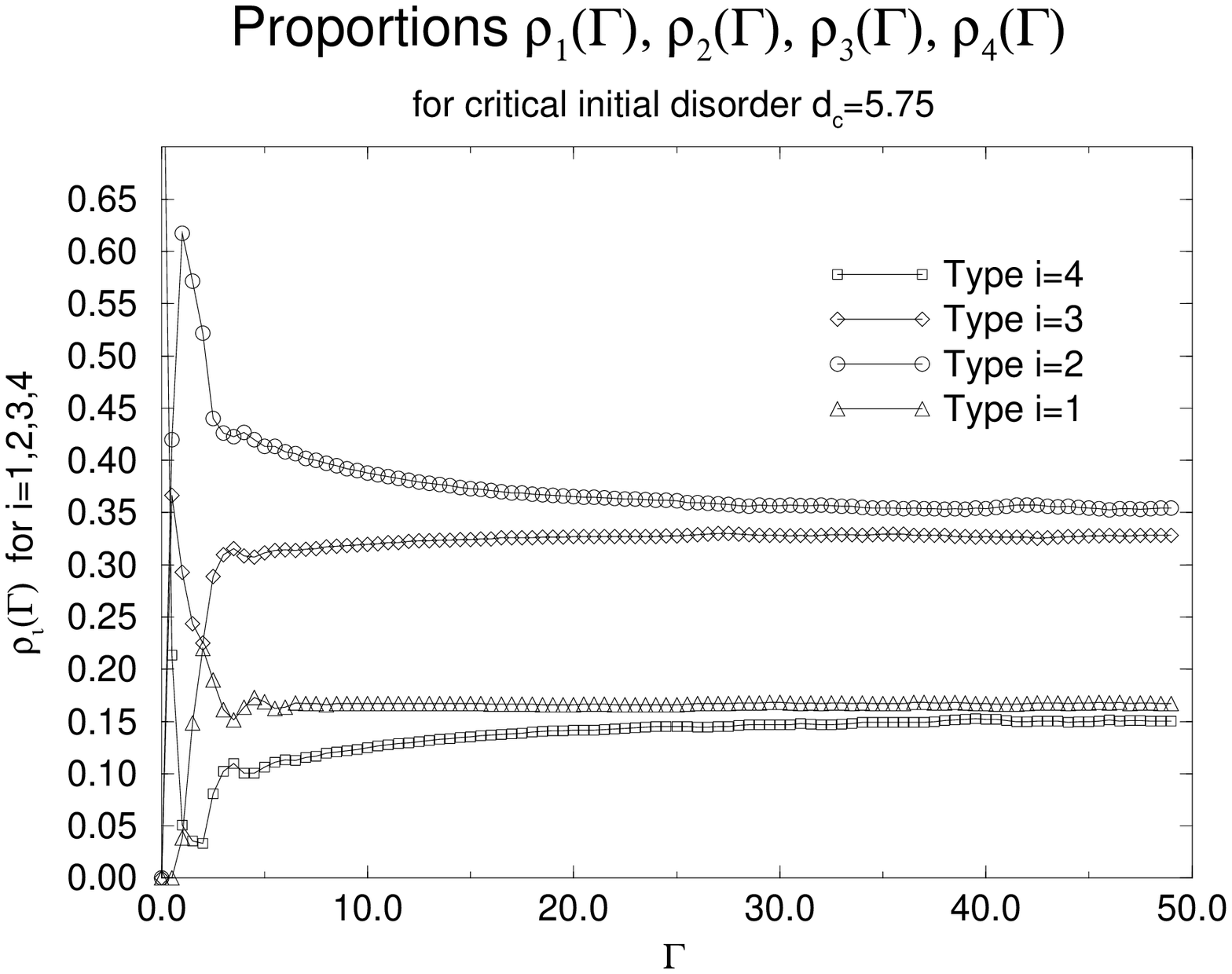}} \caption{\it  The proportions $\rho_i(\Gamma)$ of the 4 types of bonds $i=1,2,3,4$ as functions of the RG
scale $\Gamma$ for a critical initial disorder $d_c=5.75$.
\cite{spin1}} \label{figrhocrit}
 \end{figure}

 At the critical point $d=d_c$, the number of effective spins decrease algebraically  as \begin{eqnarray} 
N(\Gamma) \oppropto_{\Gamma \to \infty} { 1 \over \Gamma^3}.  \label{ngacriti} \end{eqnarray} 
and the proportions $\rho_i(\gamma)$ of the four types of bond converge towards the asymptotic regime (Figure \ref{figrhocrit}) 
\begin{eqnarray} \rho_1(\Gamma)\sim 0.17 \ , \qquad \ \rho_2(\Gamma) \sim 0.35 \ , \qquad \ \rho_3(\Gamma) \sim 0.33 \ , \qquad \ \rho_4(\Gamma) \sim 0.15 \ . 
 \label{rhocnume} 
\end{eqnarray}

\subsubsection*{  Numerical study of the percolation transition from the VBS clusters} 

From the point of view of VBS clusters, the quantum phase transition corresponds to a percolation transition :  in the strong disorder phase, there are only finite clusters, whereas in the weak disorder phase, there exists a macroscopic cluster which contains a finite fraction of the spins. 

 { \it  String Order Parameter} 

\begin{figure}[ht ] \centerline{\includegraphics[height=5cm]{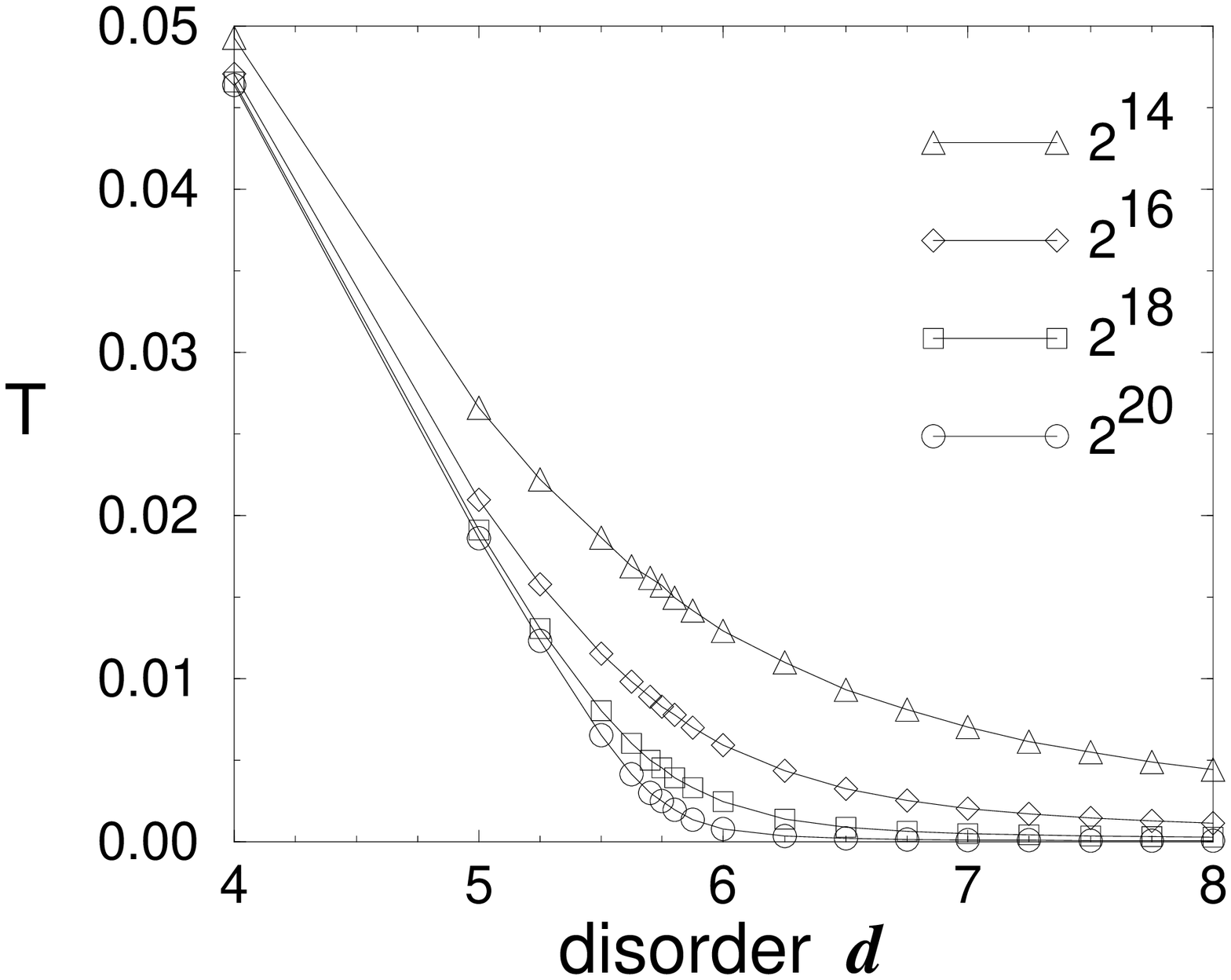}} \caption{\it  The string order parameter as a function of the initial disorder d for chains of sizes $N=2^{14}-2^{20}$.  The critical point is located at $d_c = 5.76(2)$
\cite{spin1}} \label{figordret} \end{figure}

Let $\beta$ be the exponent describing the vanishing of the fraction $n_1/N$ of spins in the macroscopic cluster at the transition.
The string order parameter then vanishes as $T \sim (d_c - d)^{2\beta}$ for $d<d_c$.  The finite size scaling study of Figure (\ref{figordret}) leads to the numerical
estimate 
\begin{eqnarray} 
2\beta = 1.0(1).  \label{nbeta} 
\end{eqnarray} 

{ \it  Susceptibility} 

\begin{figure}[ht ] \centerline{\includegraphics[height=5cm]{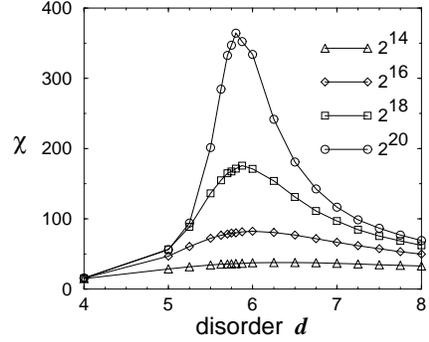}} \caption{\it  The susceptibility $\chi$ as a function of the initial disorder d for chains of sizes $N=2^{14}-2^{20}$.
\cite{spin1}} \label{figchi} \end{figure}

The Figure (\ref{figchi}) represents the average size of the finite clusters, which plays the role of a susceptibility 
\begin{eqnarray} 
\chi \equiv \sum_{c > 1} { n_c^2 \over N}. 
 \end{eqnarray} (c=1 is the largest cluster). 
  The critical exponent $\gamma$
controlling the divergence $\chi \sim|d_c - d|^{-\gamma}$
has been measured to be 
\begin{eqnarray} 
\gamma = 1.2(1).  \label{numegamma} 
\end{eqnarray} 

{ \it  Distribution of the cluster sizes at the critical point} 

\begin{figure}[ht ] \centerline{\includegraphics[height=5cm]{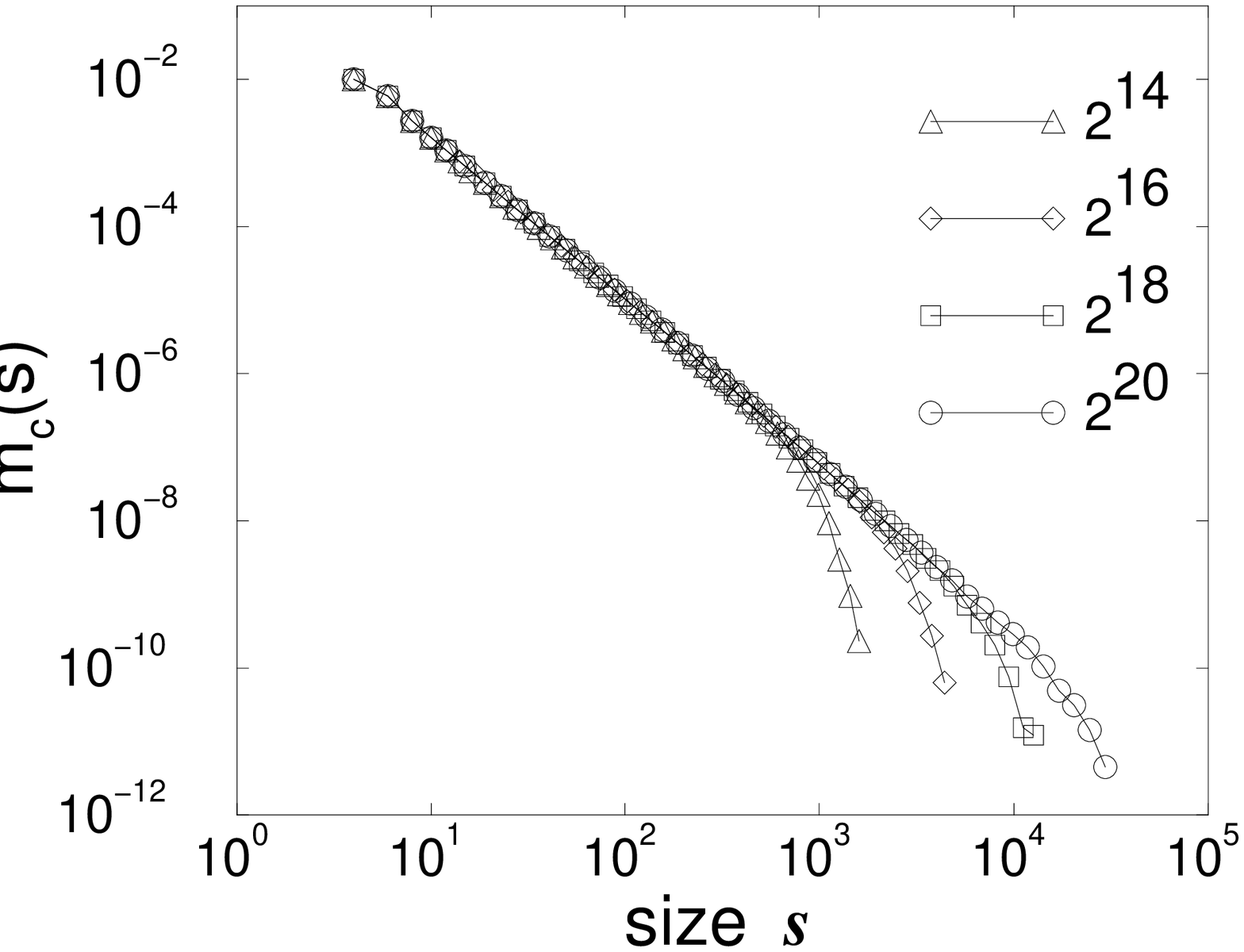}} \caption{\it  Distribution $m_c (s)$ of the size $s$ of the clusters at the critical point:  the measure of the slope leads to the estimate $\tau  = 2.2(1)$
\cite{spin1}} \label{figdistricriti} \end{figure}

At the critical point, the distribution $m_c(s)$ of the size $s$ of the clusters 
presents an algebraic decay (Figure \ref{figdistricriti})
 \begin{eqnarray} 
m_c(s) \sim \ { 1 \over { s^{\tau}}} \label{scaling}
 \end{eqnarray} 
with the exponent 
\begin{eqnarray} 
\tau = 2.2(1).  \label{numetau} 
\end{eqnarray} 

\subsection{ Exact critical exponents via a soluble effective model}

\label{eff_mod}

The effective model introduced by Hyman and Yang \cite{HY}
to describe the physics of the random $S=1$ AF chain
is a {\it random dimerized
spin-{1/2} chain} containing both Ferromagnetic and Antiferromagnetic bonds.
More precisely, this model is defined as follows :
all even bonds are
AF, whereas the odd bonds are either F or  AF. 
The physical motivation for this last point
is that in finite open pure $S=1$ chains, 
 the low energy physics corresponds to two effective $S=1/2$ spins
at the two ends coupled via a weak coupling $J_{eff}(L) \sim (-1)^{L+1} e^{-c L}$
that can be either Ferro or Antiferro.

The renormalization
procedure is then defined as follows \cite{HY} :
 the energy scale $\Omega$ is given by the strongest AF
bond of the system, so that the odd bonds separate into two groups :
{\it group A} contains F bonds weaker than $\Omega$ and all AF bonds, while
{\it group B} contains all F bonds stronger than $\Omega$. In this
effective model, perfect VBS order means singlets
over all even bonds of the initial chain. Indeed, at the stable
fixed
point of the Haldane phase found in \cite{HY} by solving the
renormalization flow equations, all
odd
bonds are much weaker than even bonds so that only singlets over
even
bonds are generated by decimation. At the critical point
between the Haldane phase and the strong disorder phase,
the length scale is
$l \sim \Gamma^3$, in agreement with the behavior (\ref{ngacriti})
found in the 4-bond model.

To compute how the VBS order parameter vanishes at the transition,
one needs 
to introduce an {\it auxiliary variable}
$\mu$ for each odd bond still surviving at scale $\Omega$. The
variable $\mu$ is by definition the number of singlets already made
over even bonds of the initial chain that are contained in this odd
bond. This $\mu$ evolves as follows~:
when an odd bond of variable ($\mu$) is decimated, a
finite cluster of size $(\mu+1)$ is terminated; when an even bond
surrounded by two B-odd bonds of variables $\mu_1$ and
$\mu_2$ respectively, a finite cluster of size $(\mu_1+\mu_2+2)$
is terminated; when an even bond surrounded by two A-odd bonds,
or surrounded by one A-odd bond and one B-odd bond,
with respective variables $\mu_1$ and $\mu_2$, the new odd
bond generated by this decimation inherits the variable
($\mu=\mu_1+\mu_2+1$). At the fixed point describing the transition of the effective
model,  one finds that $\mu$ scales as~:
\begin{equation}
\mu \propto \Gamma^{\varphi}
\qquad \hbox{with} \ \ \varphi=\sqrt 5 .
\end{equation}
This has to be
compared with the scaling $l \propto \Gamma^3$ of the auxiliary
variable $l$ that counts the number of initial bonds in a surviving
bond at scale $\Gamma$.
Deviations from the critical point are driven by a relevant
perturbation\cite{HY} that scales as $\Gamma^{\lambda_+}$ with
$\lambda_+=({\sqrt{13} -1})/2$.
As a consequence, the exact exponent for the string topological order
parameter (\ref{nbeta}) is  \cite{spin1}
\begin{equation}
2\beta ={ {2(3-\varphi)} \over
{\lambda_+}} = { {4(3-\sqrt 5)} \over {\sqrt{13} -1}} =
1.17278...,
\end{equation}
 the
exponent of the percolation susceptibility is  
\begin{equation}
\gamma={{(2\varphi-3)} \over {\lambda_+}} ={ {2(2\sqrt 5-3)} \over
{\sqrt{13} -1}}=1.13000...,
\end{equation}
and the
exponent $\tau$ of the scaling form (\ref{scaling}) for the
distribution of cluster sizes :
\begin{equation}
\tau=1+{3 \over
\varphi} =1+{3 \over \sqrt{5}} =2.34164...,
\end{equation}
 which are in very good agreement with the numerical estimates (\ref{nbeta}), (\ref{numegamma}), (\ref{numetau}) obtained in the 4-bond RG analysis.
It is thus believed that the soluble effective Hyman-Yang model  
correctly describes the disorder-induced phase transition of
the random AF spin-1 chain.

 \subsection{Direct numerical studies}  

We now have to discuss results of various direct
numerical studies on the existence of the various phases 
of the random $S=1$ chain, which are controversial. Generally, one uses here a square-type
distribution of the  initial couplings, which is uniform in the interval $[1-W/2,1+W/2]$.
The correspondence with the flat distribution on the interval $[1,1+d]$
used for the previous numerical RG approach can be obtained from the matching for $J_{max}/J_{min}$
leading to
\begin{eqnarray} 
W=\frac{2 d}{ 1+d}
\end{eqnarray}
The limit of strongest disorder of this form, $d \to \infty$ corresponds to $W \to 2$,
whereas the critical initial disorder $d_c \sim 5.75$ found in the numerical RG study
of the four-bond model corresponds to $W_c \sim 1.48$. Note that the power-law distribution,
$P(J)=D^{-1} J^{-1+1/D}$ as given in Eq.(\ref{distr_D}) for $D=1$ corresponds to $W=2$
and represents more strong disorder for $D>1$.

$\bullet$ on one hand, the following direct numerical studies have concluded
that there was no random singlet phase even for the limiting disorder $W \to 2$ :

(i) Nishiyama via exact diagonalization for sizes  $N \leq 14$  \cite{nishiyama1},

(ii) Nishiyama via quantum Monte-Carlo method for sizes $N \leq 32$ \cite{nishiyama2},

(iii) Hida via Density-Matrix-RG for sizes $N \leq 42$ \cite{hida}

$\bullet$  on the other hand, more recent studies have found the random singlet phase
 for $W_c<2$ :

(iv) Todo {\it  et al} via quantum Monte-Carlo method with the continuous-time loop algorithm
for sizes  $N \leq 256$  \cite{todo} 

(v) Bergkvist {\it  et al} via stochastic series expansion quantum Monte-Carlo 
for sizes  $N \leq 64$  \cite{bergkvist}

$\bullet$ there were new claims of the theoretical side :

(vi) in their Comment \cite{commentonhida}
on the work of Hida \cite{hida}, Yang and Hyman argue that the random singlet
phase appears only for a power-law distribution with stronger disorder, $D>D_c \sim 1.5$.

(vii) Sagia {\it  et al.} \cite{saguia} claim from the numerical study of their
alternative Ma-Dasgupta RG procedure for sizes $N \leq 9000$
 that the random singlet
phase appears for flat initial distributions only at the point $W_c=2$.

We believe that the discrepancies in the numerical results are due to  
strong finite-size effects :
the initial flat distribution for the couplings is very far from
the RG asymptotic power-law distributions for the remaining effective couplings
at low energy.
As a consequence, there is a long transient regime to converge
towards the asymptotic regime, and this is why the RG for the four-bond
model was studied numerically on very large chains 
to obtain satisfying results. 

Also one has to note that the transition point between the gapless Haldane and the RS phases
is a tricritical point, if dimerization is included\cite{damles1}. Therefore the influence of
the critical RS fixed point results in strong cross-over effects. On the side of the numerical
RG studies there are a number of initial approximative decimation steps until the system
approaches sufficiently close the correct asymptotic RG trajectory. Due to these initial
steps the position of the transition in the physical model can be somewhat shifted.

To summarize the basic features of the disorder induced phase transitions in the random
$S=1$ AF Heisenberg chain are verified by direct numerical studies, but still further work
is necessary to locate the precise position of the tricritical point, as well as
to verify the values of the (tri)critical exponents.

 \subsection{Generalizations : dimerization, dynamics}

The effects of enforced dimerization on random $S=1$ chains
have also been studied via real space RG in \cite{damles1},
and numerically in \cite{arakawa} : 
in the phase diagram in the plane (dimerization $\delta$,randomness $R$),
the critical point at $(\delta=0,R_c)$ now becomes multicritical.
This kind of multicritical point has been further discussed in \cite{damlegenechain}.

Finally let us mention that dynamical properties of the $S=1$ chain
(as well as other random spin chains)
have also been studied via strong disorder RG in \cite{rsrgdyna,eigenhuse}.


\section{Other 1D quantum models}

Besides the random AF $S=1/2$ and $S=1$ Heisenberg models there are another problems of interacting
Heisenberg spins in one (and quasi-one) dimension which have been intensively studied. Here we review recent developments obtained for higher spin, $S \ge 3/2$, AF chains, for $S=1/2$
chains with mixed ferro- and antiferromagnetic couplings and for random spin ladders. These theoretical investigations are often initiated by experimental work.

\subsection{Higher spin AF Heisenberg chains}

\label{S>1}

The spin-$S$
random antiferromagnetic Heisenberg chain is defined by the Hamiltonian:
\begin{equation}
H = \sum_i J_{i} \vec{S}_i \cdot \vec{S}_{i+1}\;
\label{Hamilton_S}
\end{equation}
where the $J_i > 0$ are quenched random variables. As before we introduce enforced dimerization of strength,
$\delta$, so that the couplings are in the form:
\begin{equation}
J_i=J(1+\delta (-1)^i)\exp(D \eta_i)\;
\label{J_S}
\end{equation}
where $\eta_i$ are random numbers of mean zero and variance unity. Thus the strength of disorder is measured
by $D$, see Eq.(\ref{distr_D}). The properties of this model for $S=1/2$ and for $S=1$ have already been
presented in Sec.\ref{S=1/2}
and \ref{S=1}, respectively. Here we consider higher values of $S\ge 3/2$.

{\it Non-random models}

As already discussed at the beginning of Sec.\ref{S=1} the non-random models with $\delta=0$ have
different low-energy properties for half-integer and integer values of the spin, respectively\cite
{haldane83}. Half-integer
spin chains, as the $S=1/2$ model, have a gapless spectrum and quasi long range order.  It is
believed that they all belong to the same (bulk) universality class
independently on $S$ \cite{s32}. This was explicitly verified numerically
for the $S=3/2$ chain \cite{Hall96}, which was found to have the same
bulk decay exponent as for the $S=1/2$ chain. In the presence of dimerization a gap opens
in the spectrum, which behaves for small $\delta$ as
\begin{equation}
\epsilon_1(\delta) \sim |\delta|^{\nu}\;.
\label{e_1}
\end{equation}
Here the gap (or correlation-length) exponent for the $S=1/2$ model is given from a bosonization
study\cite{cross79} as $\nu=2/3$, for recent numerical work see Ref.\cite{papenbrock03}.

Integer spin chains are instead gapped and have a hidden topological order \cite{SOP}, as
defined for the $S=1$ model in Eq.(\ref{tij}). The topological order stays even for a small finite dimerization.
If, however, $|\delta|$ exceeds a limiting value, $\delta$, there is a quantum phase transition in the system and for $|\delta|>\delta$ there is dimer order in the ground state.

{\it Effect of disorder}

Weak disorder is expected to have different effect of the two types of chains. The Haldane gap for integer spin
chains is robust against weak disorder, whereas the behavior of half-integer spin chains can be predicted by the
Harris relevance-irrelevance criterion in Eq.(\ref{harris}). Since the value of $\nu=2/3$ seems to be universal
for all half-integer chains weak disorder is predicted to be a relevant perturbation. This is indeed the case
for the $S=1/2$ chain, however, for the $S=3/2$ chain numerical investigations show\cite{carlon04} that
weak disorder is
irrelevant and the properties of the quasi-long-range order are the same as in the non-random system. We shall
come back later to discuss this point.

For stronger disorder one can use the strong disorder RG method. If in the decimation procedure two $S$ spins
with the strongest bond, $J_2=\Omega$, are replaced by a singlet, the new coupling generated between the
remaining sites is given by:
\begin{equation}
\tilde{J}=\frac{2}{3}S(S+1) \frac{{J}_1 {J}_3} {{J}_2} \;.
\label{JS}
\end{equation}
Thus for $S > 1$ we encounter the same problem as for $S=1$: some of the generated couplings are larger than the decimated one, therefore this RG works only for strong enough disorder. The larger the spin the larger the
disorder needed for the procedure. In order to obtain the behavior of the system
for weaker disorder the strategy used for the random $S=1$ chain in Sec-\ref{S=1} is generalized in
Ref.\cite{spin3/2}. In this case each spin-$S$ is represented by the maximally symmetrized multiplet
of $2S$ identical $S=1/2$ spins. Renormalizing a strong term, $J_2 \vec{S}_1 \cdot \vec{S}_{2}$, is equivalent
to eliminate the highest spin combination in the spectrum of the two site cluster, thus replacing
$S_1 \to S_1-1/2$ and $S_2 \to S_2-1/2$. In other words one $S=1/2$ singlet bond is eliminated. In this way,
during renormalization we obtain an effective model in which at each site there are spin
$S$, $S-1/2$,...,$1/2$ degrees of freedom with effective
couplings, which can be ferro- or antiferromagnetic. However during renormalization no spin larger than $S$
can be generated. The ground state at an energy scale, $\Omega$, is represented by singlets (valence bonds)
formed between spin $S=1/2$'s.

If the disorder is very strong all the $2S$ valence bonds between the original spins are formed, thus the
random singlet phase is given in terms of $S$ spins and called as $RS_S$ phase. For somewhat weaker disorder
the valence bonds form two different structures: i) there is a valence bond solid which involves two spin $S=1/2$
at each site, and ii) there is a random singlet phase of effective $S-1$-spins. The low-energy excitations are
given by this $RS_{S-1}$ phase. Continuing this reasoning by decreasing disorder, $D$, there is a sequence
of $RS_S$, $RS_{S-1}$, $RS_{S-2}$, ... phases, and there are multicritical points which separate these strong
disorder phases. Within the different RS phases the singular behavior follows the same rules as in the
traditional $RS_{1/2}$ phase, as described in Sec.\ref{S=1/2}. At the multicritical points, however, there
are new exponents, which are calculated in Ref.\cite{damlegenechain}. For example the exponent,
$\psi$, describing the
relation between the size- and the log-time-scale in Eq.(\ref{logscale}) is given by: $\psi=1/N$, where
$N$ is the number of Griffiths phases in the $D-\delta$ phase diagram, which meet at the multicritical point.
At the principal multicritical point, separating the $RS_S$ and $RS_{S-1}$ phases we have $N=2S+1$.
Similarly, the correlation length exponent in the milticritical point is given by: $\nu=(1+(4N+1)^{1/2})/2$.
Note, that this latter result formally corresponds to Eq.(\ref{aux}) with $b=N-1$. For the $S=1$ model
these results are previously calculated, see in Sec.\ref{eff_mod}.

Among the higher-spin AF Heisenberg chains, the random $S=3/2$ model is investigated in more detail. The phase
diagram and the singular properties of the system is studied by a numerical implementation of the strong
disorder RG method\cite{spin3/2}, as outlined above. The two random singlet phases ($RS_{3/2}$ and $RS_{1/2}$)
are identified and the properties of the multicritical point are numerically calculated. These
are generally in good agreement with the analytical results in Ref.\cite{damlegenechain},
although there are larger deviations for the correlation length exponent, $\nu$.

Another type of decimation scheme in the strong disorder RG method is used by
Saguia {\it  et al.} \cite{saguia03}. In their work the renormalization flow indicates
that the weak disorder is an irrelevant
perturbation of the system. This is in agreement with the numerical, DMRG results\cite{carlon04},
as mentioned above.
The possible origin of the irrelevance of weak disorder, as argued in Ref\cite{carlon04}, is due to localized
edge states in the $S=3/2$ (and higher half-integer spin) chains. The correlation length associated
to surface excitations
is shown to diverge with a larger exponent, $\nu' \approx 2$. This type of excitation could be relevant
in the present of bond disorder, thus, according to the Harris criterion in Eq.(\ref{harris})
it is a (marginally) irrelevant perturbation.

\begin{figure}[t]
\centerline{\includegraphics[width=6.2cm]{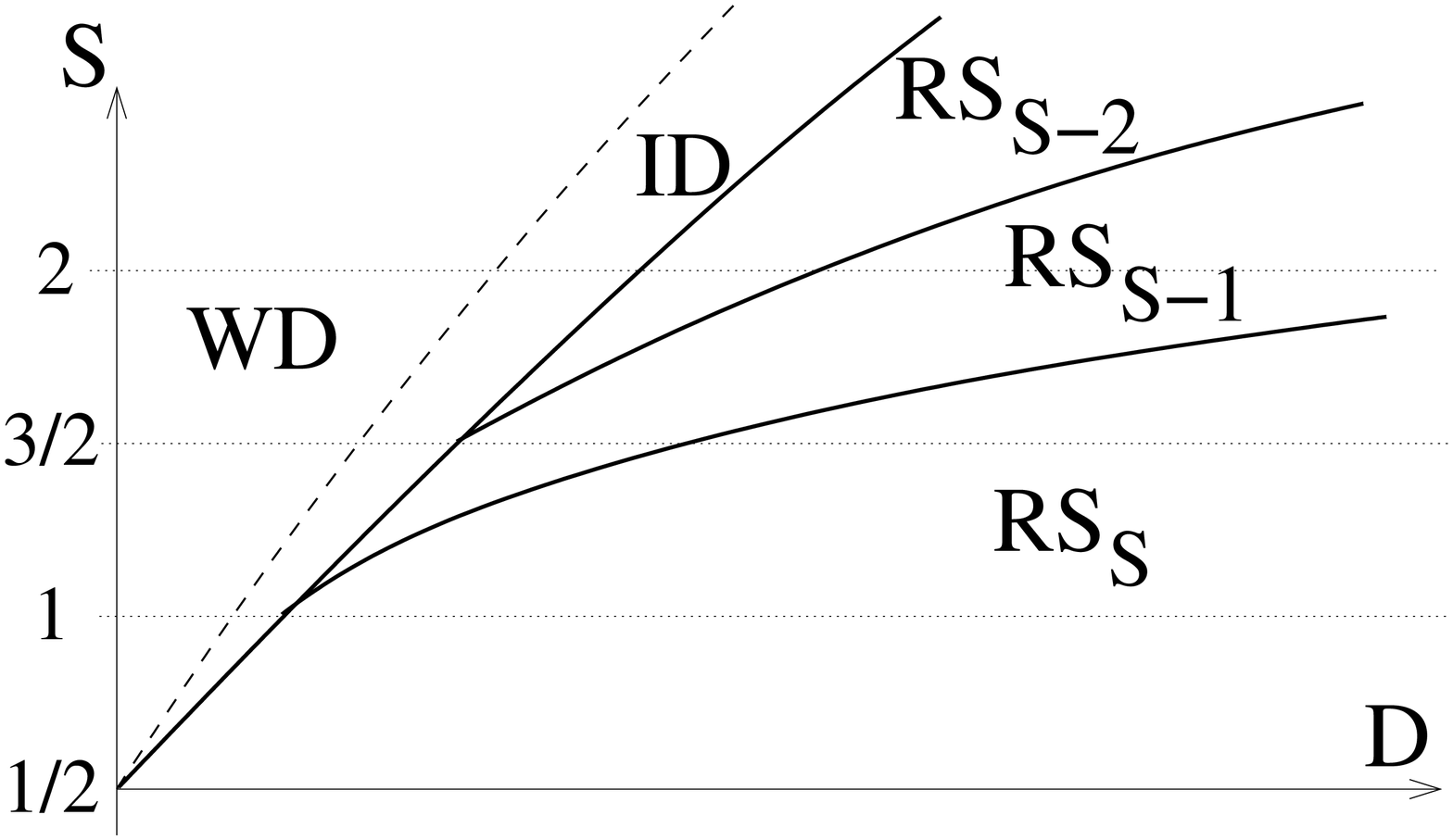}}
\caption{
Schematic phase diagram of the spin $S$ AFH chain as function of the strength of
disorder $D$. The RS$_S$ denotes a random singlet phase where the
relevant degrees of freedom are effective $S$-spins.  The WD and ID are
the region at weak and intermediate disorder.
}
\label{RS_S}
\end{figure}

We close this part by presenting a possible phase diagram\cite{damlegenechain,carlon04} for higher $S$
in Fig. \ref{RS_S}. The transition lines between the different random singlet phases for the time
being cannot be supported by numerical results.  The weak disorder (WD) region
may be separated from the RS phase(s) by an intermediate disorder (ID)
region where exponents vary continuously with $D$, as observed in
other models\cite{carlon-clock-at} in Sec.\ref{PCAT}. For the gapped (integer S) case this would
correspond to a region of Griffiths singularities.

\subsection{Heisenberg chain with random ferro- and antiferromagnetic couplings}

\label{LSFP}

Heisenberg spin chain with randomly mixed ferromagnetic and antiferromagnetic couplings can be
realized in the system $Sr_3 Cu Pt_{1-x} Ir_x O_6$. Here the pure compounds, $Sr_3 Cu Pt O_6$ and
$Sr_3 Cu Ir O_6$, are antiferromagnetic and ferromagnetic spin chains, respectively, the mixed
compound contains randomly both types of couplings\cite{wilkinson91}. Examples of another experimental
realizations can be found, c.f. in\cite{westerberg,melinladders}.

{\it Generalized RG rules}

For a theoretical investigation of the low-energy and low-temperature properties of the system the
strong disorder RG method by Ma and Dasgupta have to be generalized. If, during renormalization, the
strongest term in the Hamiltonian in Eq.(\ref{Hamilton_S}) is, $J_{2} \vec{S}_2 \cdot \vec{S}_{3}$, with
a ferromagnetic coupling, $J_2<0$, then the two spins form a spin cluster of effective size, $\tilde{S}=1$.
Repeating the RG procedure the system transforms into a chain of spins of arbitrary size with mixed
antiferromagnetic and ferromagnetic couplings. One term of the renormalized Hamiltonian is given in
the form: $J \vec{S}_L \cdot \vec{S}_{R}$, where $\vec{S}_L$ ($\vec{S}_{R}$) denotes the effective
spin variable at the left (right) of the two-site cluster. During renormalization the two-site cluster
with a very strong bond is projected into its ground state multiplet, thus replaced by a single effective
spin, $S$, with $S=|S_L \pm S_R|$, where the $+(-)$ sign refers to ferromagnetic (antiferromagnetic)
coupling.
The energy-scale of the two-site cluster is measured by the energy gap, $\Delta$, defined by:
$\Delta=|J|(S_L+S_R)$ for $J<0$ and $\Delta=J(|S_L-S_R|+1)$ for $J>0$, respectively.

During renormalization spin clusters with $\Delta=\Omega$ are transformed by two basic decimation
processes. If $S_L=S_R$ and the coupling is antiferromagnetic one performs standard singlet formation, when the effective coupling generated between the remaining sites is given in Eq.(\ref{JS}).
If $S \ne 0$ there is effective spin (cluster)
formation, when the interaction to the neighboring site of spin $S_L$ and coupling $J_L$ is
renormalized as $\tilde{J}_L=J_L c_L$ with\cite{westerberg,melin00}:
\begin{equation}
c_L= \frac{S(S+1)+S_L(S_L+1)-S_R(S_R+1)}{2 S(S+1)}\;.
\label{J_cl}
\end{equation}

{\it Properties of the fixed point}

As renormalization goes on the energy scale, $\Omega$, is lowered and the number of non-decimated sites, $n_{\Omega}$ is decreased. At the same time the length-scale, $L_{\Omega} \sim 1/n_{\Omega}$ and
the size of the effective spin, $S_{eff} \approx S_L,S_R$ is increased. This latter follows the asymptotic relation:
\begin{equation}
S_{eff} \sim L_{\Omega}^{\zeta}\;,
\label{S_N}
\end{equation}
where $\zeta$ is called the spin moment exponent.
The following random walk argument \cite{westerberg} gives $\zeta=1/2$:
The total moment of a typical cluster of size $L$ can be expressed as
$S_{\rm eff}=|\sum_{1=1}^L \pm S_i|$, where neighboring spins with ferromagnetic
(antiferromagnetic) couplings enter the sum with the same (different) sign. If the
position of the two types of bonds are uncorrelated and if their
distribution is symmetrical, one has $S_{\rm eff}\propto L^{1/2}$,
i.e.\ Eq.(\ref{S_N}) with $\zeta=1/2$.

A non-trivial relation constitutes the connection between the energy
scale $\Omega$ and the size of the effective spin:
\begin{equation}
S_{\rm eff} \sim \Omega^{-\kappa}\;.
\label{S_Omega}
\end{equation}
For not singular initial disorder in numerical calculation the exponent is
found independent of the disorder distribution, as $\kappa=0.22(1)$\cite{westerberg}.
Comparing Eq.(\ref{S_N}) with
Eq.(\ref{S_Omega}), the relation between the length scale
and the energy scale is:
\begin{equation}
\Omega \sim L^{-z},\quad z=\frac{ \zeta}{\kappa}=\frac{1}{2 \kappa}\;,
\label{L_Omega}
\end{equation}
where $z$ is the dynamical exponent.

The average spatial correlations function, $C(r)$, is studied by numerical application of the
RG procedure and by DMRG calculations\cite{hikihara99} leading to a very slow, probably logarithmic dependence:
\begin{equation}
C(r) \sim \frac{1}{\ln(r/r_0)}\;.
\label{C_rFA}
\end{equation}
Thus we can conclude that the low-energy fixed point of the $S=1/2$ Heisenberg chain with
mixed ferromagnetic and antiferromagnetic couplings has special characteristics: there is
a large spin formation, the dynamical exponent is finite and the average correlation function
is logarithmically slow. This type of fixed point is generally called a {\it  large spin fixed point}.

{\it Thermodynamic quantities}

From the strong disorder RG calculation one can obtain the singularities of the thermodynamic
quantities similarly
as for the RTFIC in Sec.\ref{SC_term}. At finite, but small temperature, $T$, the renormalization
stops at the energy scale, $\Omega=T$, when the existing spin clusters of size, $S_{eff}$, are practically
independent, since couplings between them are extremely weak. The entropy per site, ${ S}/N$,
is simply the contribution of non-interacting clusters:
\begin{equation}
\frac{{S}}{N} \sim \left.\frac{\ln(2 S_{eff}+1)}{n_{\Omega}}\right|_{\Omega=T}\sim T^{1/z}|\ln T|\;.
\label{S_LSFP}
\end{equation}
and similarly for the specific heat:
\begin{equation}
\frac{{C}}{N}  \sim T^{1/z}|\ln T|\;.
\label{CV_LSFP}
\end{equation}
The magnetic susceptibility is given by the Curie-type contribution of large, independent effective spins:
\begin{equation}
\frac{\chi}{N} \sim \left.\frac{ \left[ S_{eff} \right]^2}T {n_{\Omega}}\right|_{\Omega=T} \sim \frac{1}{T}\;.
\label{CHI_LSFP}
\end{equation}
In a small, finite magnetic field, $h$, the energy scale is set by the Zeeman-energy,
$E_{ZM}\sim  h S_{eff} \sim \Omega$. Thus the RG stops at $\Omega \sim h^{1/(1+1/2z)}$,
and the existing spin clusters align parallel with the magnetic field. Consequently the magnetization
per site, $m$, is given by
\begin{equation}
m \sim \left.\frac{ S_{eff}}{n_{\Omega}}\right|_{\Omega=E_{ZM}} \sim h^{1/(1+2z)}\;.
\label{M_LSFP}
\end{equation}

\subsection{Disordered spin ladders}

\label{Sec:ladder}

Spin ladders are quasi-one-dimensional Heisenberg systems, in which two or more spin chains are coupled together
by interchain bonds. Experimentally they have been realized in different compounds, for a review,
see\cite{dagotto96}. According to theoretical investigations spin ladders with
even number of legs have a gapped spectrum, whereas the spectrum of
odd-leg ladders is gapless\cite{scalapino92}. For two-leg ladders, which
are analogous objects to $S=1$ spin chains, the ground state structure
can be related to nearest-neighbor valence bonds and a topological hidden
order parameter, similar to that in Eq.(\ref{tij}) can be
defined \cite{kim00}.

More recently, ladder models with competing interactions,
such as with staggered
dimerization\cite{delgado} and with rung and diagonal couplings\cite{kim00},
have been introduced and studied. In these models, depending on the relative
strength of the couplings, there are several gapped phases with
different topological
order, which are separated by first- or second-order phase transition lines.

Disorder in a spin ladder material is realized in Sr(Cu$_{1-x}$Zn$_x$)$_2$ O$_3$,
which is a two-leg ladder, and can be doped
by Zn, a non magnetic ion\cite{azuma}. The specific heat and spin
susceptibility experiments indicate that the doped system
is gapless even with low doping concentrations.
We note that the experimentally found phase diagram of this compound,
as well as other
quantities, such as staggered susceptibility have been obtained by quantum
Monte Carlo simulations\cite{miyazaki}.

{\it Spin ladder models}

\begin{figure}
\centerline{\includegraphics[width=0.58\linewidth]{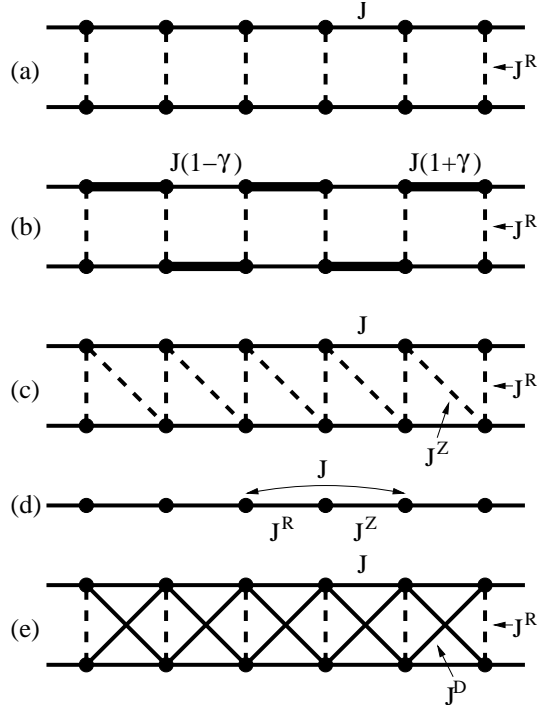}}
\caption{\label{ladder} Spin-ladder models: a) conventional
two-leg ladder, $H=H_1+H_2+H_R$, (b) with
staggered dimerization in the chain couplings, (c) zig-zag ladder, $H=H_1+H_2+H_R+H_Z$,  (d) its
representation as a chain with first and second neighbor couplings, (e) full ladder
with rung and diagonal couplings, $H=H_1+H_2+H_R+H_D$.}
\end{figure}

To model a two-leg ladder with different types of interactions one starts with the
Hamiltonians of the $\tau=1,2$ spin chains:
\begin{equation}
H_{\tau}=\sum_{l=1}^L J_{l,\tau} {\bf S}_{l,\tau} {\bf S}_{l+1,\tau}\;,
\label{Htau}
\end{equation}
in which dimerization is introduced in the couplings as:
\begin{equation}
J_{l,\tau}=J\left[1+\gamma (-1)^{l+n(\tau)}\right],\quad 0\le \gamma < 1\;,
\label{Jdimer}
\end{equation}
with $n(\tau)=0,1$. The interchain interactions are 
\begin{itemize}
\item rung couplings:
\begin{equation}
H_{R}=\sum_{l=1}^L J_{l}^R {\bf S}_{l,1} {\bf S}_{l,2}\;,
\label{HR}
\end{equation}

\item one type of diagonal coupling, generating the zig-zag ladder:
\begin{equation}
H_{Z}=\sum_{l=1}^L J_{l}^Z {\bf S}_{l,2} {\bf S}_{l+1,1}\;.
\label{HZ}
\end{equation}
\item two types of diagonal couplings:
\begin{equation}
H_{D}=\sum_{l=1}^L J_{l}^D ({\bf S}_{l,1} {\bf S}_{l+1,2} + {\bf S}_{l,2} {\bf S}_{l+1,1})\;.
\label{HD}
\end{equation}
\end{itemize}
The different type of ladder models, which can be obtained from these ingredients are shown in
Fig.\ref{ladder}.

{\it Strong disorder RG rules}

\begin{figure}
\centerline{\includegraphics[width=0.58\linewidth]{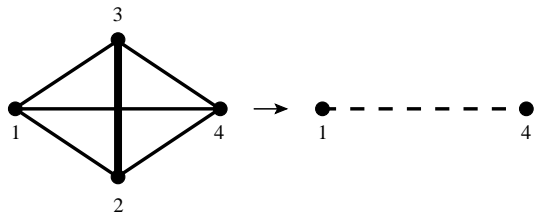}}
\caption{\label{singl} Singlet formation and decimation in the ladder geometry.}
\end{figure}

With a ladder geometry, spins are more interconnected than in a chain, which leads to a
modification of the decimation procedure used for a single chain in Sec.\ref{heisenberg3}.
As shown in Fig.~\ref{singl}
both spins of a strongly coupled pair, say $(2,3)$, are generally connected to
the nearest neighbor spins, denoted by $1$ and $4$. After decimating out the singlet pair the
new, effective coupling between $1$ and $4$ is of the form:
\begin{equation}
{\tilde J}^{eff}_{14}=\kappa \frac{(J_{12}-J_{13})(J_{43}-J_{42})}{\Omega}.\quad
\kappa(S=1/2)=1/2\;,
\label{deci_sing}
\end{equation}
With the rule in Eq.(\ref{deci_sing}) ferromagnetic couplings are also generated. Repeating
the reasoning used for the random Heisenberg chain with mixed ferromagnetic and antiferromagnetic
bonds in Sec.\ref{LSFP}, the ladder Hamiltonian will renormalize into a set of
effective spin clusters having different moments and connected by both
antiferromagnetic and ferromagnetic bonds. The renormalization rules are similar to that in
Sec.\ref{LSFP} and consists of i) singlet formation and ii) cluster formation. In the latter
case the renormalization rule in Eq.(\ref{J_cl}) has to be modified, since the two-site cluster
with spins $S_L$ and $S_R$ has generally two couplings, denoted by $J_L$ and $J_R$, to a
neighboring spin, $S_1$. During renormalization the term in the Hamiltonian: $J_L \vec{S}_1 \cdot \vec{S}_{L}
+ J_R \vec{S}_1 \cdot \vec{S}_{R}$ is transformed to $\tilde{J_1} \vec{S}_1 \cdot \vec{S}$ with:
$\tilde{J_1}=c_L J_L + c_R J_R$. Here $c_L$ is given in Eq.(\ref{J_cl}), whereas $c_R$ can be obtained
from Eq.(\ref{J_cl}) by interchanging $S_L$ and $S_R$.

{\it Numerical renormalization methods}

Due to the ladder topology and the complicated renormalization rules the RG equations
can not be treated analytically and one resorts to numerical implementations of the
renormalization procedure, during which one keeps information about the spin moments,
$S_{eff}$, the gaps of a two-site cluster, $\Delta$, and the energy scale, $\Omega=max\{ \Delta\}$.
The numerical RG calculations are generally made in two different forms.

$\bullet$ In the infinite lattice method a few very large ("infinite") samples are decimated
and the critical exponents are deduced from the scaling form of the distribution function:
\begin{equation}
P(\Delta,S_{eff},\Omega)=\Omega^{\omega} \tilde{P}(\Delta/\Omega,S_{eff} \Omega^{\zeta/z})\;,
\label{P_LSFP}
\end{equation}
which is characterized by the gap exponent, $\omega$, by the spin moment or cluster
exponent, $\zeta$, and
by the dynamical exponent, $z$.

$\bullet$ In finite lattice method one starts with a finite system of $L$ sites (and with periodic boundary conditions) and perform the decimation procedure until the last remaining spin or singlet pair. The
distribution of the gap in the last step, $P_L(\Delta)$ obeys the scaling relation:
\begin{equation}
P_L(\Delta)=L^z \tilde{P}(L^z \Delta)
\sim L^{z(1+\omega)} \Delta^{\omega}\;.
\label{PDelta}
\end{equation}
from which $\omega$ and $z$ can be deduced. In this case the cluster exponent is obtained from the size
dependence of the average effective spin, $S_{eff} \sim L^{\zeta}$, which corresponds to Eq.(\ref{S_N}).

If the low-energy excitations are localized there is a simple relation between the
dynamical exponent, $z$, and the gap exponent, $\omega$ \cite{thill95,igloi98,igloi99}. In this case
the gap distribution should be proportional to the volume of the system,
$P_L(\Delta) \sim L^d$, and for a ladder the dimension is $d=1$. From Eq.(\ref{PDelta}) then follows:
\begin{equation}
z=\frac{d}{1+\omega}\;,
\label{z_omega}
\end{equation}
In the Griffiths phase with localized rare events (\ref{z_omega}) is expected to hold.

The infinite disorder fixed point is signaled by a diverging $z$,
or more precisely the $P_L(\Delta) {\rm d} \Delta$
distributions have strong $L$ dependence, so that the appropriate scaling combination is
\begin{equation}
\ln\left( L^{\psi} P_L(\Delta)\right) \simeq
f\left(L^{-\psi} \ln \Delta\right)\;.
\label{IRFP}
\end{equation}

{\it Renormalization of spin ladders}

Results of numerical renormalization group method can be summarized as follows.

$\bullet$ Random conventional ladders ((a) in Fig.\ref{ladder}) are in the random rung singlet phase,
which is controlled by a Griffiths-type fixed point having a dynamical exponent which depends on
the strength of the initial disorder, $D$\cite{melinladders,yusuf02}.

$\bullet$ Random ladders with staggered dimerization ((b) in Fig.\ref{ladder}) have two different
phases: random rung singlet and random dimer, both are controlled by Griffiths-type fixed points with
finite value of the dynamical exponent\cite{melinladders}.
At the phase transition point the transition is controlled by an infinite disorder fixed point with
an exponent, $\psi=1/2$.

$\bullet$ In random zig-zag ladders ((c) and (d) in Fig.\ref{ladder}) there is a very narrow region
for small second neighbor interaction with infinite disorder characteristics\cite{yusuf,hoyos04}. For larger
second neighbor interaction the fixed point is a large spin fixed point with varying dynamical
exponent\cite{melinladders}.

$\bullet$ In random $J_1$-$J_2$ ladders ((e) in Fig.\ref{ladder})there are two topologically distinct
phases with Griffiths-type singular behavior. In
between there is a quantum phase transition with infinite disorder scaling properties\cite{melinladders}.

$\bullet$ Random conventional ladders with site-dilution or doped with non-magnetic impurities are
found to have a large spin fixed point\cite{yusuf03a}.

For related studies of disordered spin ladders we mention Refs.\cite{arlego04,giamarchi,gogolin,hoyos04}.

\subsection{Other investigations in 1d}

\subsubsection*{ Dissipative random quantum spin chains}

The RTFIC coupled to an Ohmic bath of quantum harmonic oscillators is studied by Monte-Carlo
simulations using the two-dimensional classical counterpart of the coupled system\cite{leticia04}
(see Sec.\ref{M_W}). For small, finite chains this problem constitutes a generalization of the
Caldeira-Leggett localization transition\cite{caldeira}. In the thermodynamic limit the coupling to
the dissipative bath is found to enhance the extent of the ordered phase. However, from the
numerical data it was not possible to distinguish between infinite disorder and strong disorder
scaling at the transition point.

\subsubsection*{AF Heisenberg chains with alternating bonds}

$S=1/2$ antiferromagnetic Heisenberg chains with alternating
bonds and quenched disorder is introduced in Ref\cite{hida03a} as a theoretical model of the
compound\cite{ajiro03} ${\rm Cu Cl}_{2x}{\rm Br}_{2(1-x)}(\gamma-{\rm pic})_2$.
The low-energy properties of the system as a function of the concentration, $x$,
and the type of correlation of disorder is studied in Ref.\cite{lin04} by the numerical implementation of
the strong disorder RG method. For perfect correlation of disorder
the system is in the random dimer (Griffiths) phase having a
concentration dependent dynamical exponent. On the contrary for weak or vanishing
disorder correlations the system is in the
random singlet phase.
These results are compared with the experimentally measured
low-temperature susceptibility\cite{ajiro03} of
${\rm Cu Cl}_{2x}{\rm Br}_{2(1-x)}(\gamma-{\rm pic})_2$.

\subsubsection*{Random AF $S=1$ chains with quadratic and biquadratic interactions}

Spin-1 chains with random nearest neighbor couplings that are
rotationally invariant, but include both Heisenberg and biquadratic
exchange, are studied in Ref.\cite{yang98}, see also Ref.\cite{boechat96}.
Under renormalization also ferromagnetic couplings are generated, which
leads to formation of large magnetic moments.

\subsubsection*{Weakly coupled AF spin chains}

Weakly coupled $S=1/2$ spin chains with random antiferromagnetic bonds are studied in
Ref.\cite{yang03,yang05}. Here the strong disorder RG is used to treat intrachain coupling
and combined it with mean-field/random-phase approximation treatment of interchain coupling to obtain the
phase diagram and collective modes of such systems.

\subsubsection*{Random isotropic AF SU(N) spin chains}

Random isotropic antiferromagnetic SU(N) spin chains are studied by
the strong disorder RG\cite{hoyos04a} and an infinite disorder fixed
point with $\psi=1/N$ is found. The mean correlation function involves
the exponent, $\eta=4/N$, compare with Eq.(\ref{corrXX}).

\subsubsection*{Kondo necklace model}

Griffiths phases in the strongly disordered Kondo necklace model is studied in Ref.\cite{rappoport03} by the
numerical strong disorder RG method and coupling dependent dynamical exponents are found.

\subsubsection*{Dirty superconductors}

The strong disorder RG approach is used to study Griffiths effects and quantum critical point in dirty
superconductors without spin-rotation in\-variance\cite{motrunich01}. In the one-dimensional problem the critical
point is found to belong to the same infinite disorder universality class as the particle-hole symmetric
Anderson localization. Note that in quasi-one-dimensional systems similar results can be obtained
by the use of random Dirac fermions\cite{brouwer00,titov01}.

\subsubsection*{Bosons in 1D}

One-dimensional disordered bosons with large commensurate filling, which is described by a random
$O(2)$ rotor model is studied by the strong disorder RG method\cite{bosonrefael}. A strong disorder
fixed point is found to control the phase transition between an incompressible-Mott glass and
a superfluid phase. The phase transition is in the Kosterlitz-Thouless universality class with $z=1$.


\section{Quantum models in $d>1$ dimensions}

The infinite disorder fixed point scenario is first observed in one-dimensional systems, in which case several
independent exact and numerical results support the validity of the scaling picture in Sec.\ref{SC_st}.
It is an interesting and important question if infinite disorder
fixed points do exist also in higher dimensional systems.
This question can be decided only by numerical investigations, which have been done first for the
random transverse-field Ising model by Motrunich { \it  et al}\cite{motrunich} and later for the Heisenberg model
by Lin {\it  et al} \cite{numedlin}. These results indicate that systems with discrete and continuous symmetry
have different types of random fixed points in $d>1$, in contrast with the similarities observed in $d=1$.
Systems with discrete symmetry, such as the random transverse-field Ising model, have
an infinite disorder fixed point, at least for strong enough disorder\cite{motrunich}.
On the other hand models with a continuous symmetry, c.f. the random Heisenberg model,
have a low-energy fixed point which is either a
Griffiths-type fixed point or a large spin fixed point\cite{numedlin}.
In these systems no infinite disorder fixed point is
observed, even at a quantum critical point. In the following we review the known numerical results.

\subsection{Random transverse-field Ising model in 2d}

\label{RTIM2d}

In the strong disorder RG method the elementary decimation steps are the same as in one-dimension as
described in Sec.\ref{RTFIC_RG_B}. The main difference, however, is that during
renormalization, the topology of the lattice is not preserved and the renormalized
lattice contains bonds between remote sites, too. In this lattice the decimation of a strong transverse
field, $h_i=\Omega$, will generate a new coupling between nearest neighbors, $J'_{jk}=J_{ji} J_{ik}/h_i$,
as in Eq.(\ref{Jdecimation}). Since between sites $j$ and $k$ there could be already an interaction, $J_{jk}$,
after renormalization the new coupling is given by:
\begin{equation}
\tilde{J}_{jk}= {\rm max}(J_{jk},J'_{jk})\;.
\label{J_2d}
\end{equation}
Here to use the maximum is justified in the infinite disorder fixed point since the couplings are of
very different magnitude.

Similarly, after a strong bond, $J_{ij}=\Omega$, decimation a new spin cluster of moment,
$\tilde{\mu}_{i}=\mu_i + \mu_j$ is created in a transverse field, $\tilde{h}_i=h_i h_j/J_{ij}$, see
Eq.(\ref{hdecimation}). The interaction of the spin cluster to a remaining spin, $k$, is given by
the maximum rule:
\begin{equation}
\tilde{J}_{ik}= {\rm max}(J_{ik},J_{jk})\;.
\label{h_2d}
\end{equation}
Numerical analysis of the RG trajectories shows a disordered phase, when the ratio of
average log-fields and average log-couplings exceeds a critical value,
$[\ln h]_{\rm av}/[\ln J]_{\rm av}=\rho>\rho_c$, and an ordered phase in the opposite limit,
$\rho<\rho_c$. The control parameter is then defined by $\delta=\rho-\rho_c$.

\subsubsection{Scaling at the critical point}

At the critical point infinite disorder scaling (see Sec.\ref{SC_st}) is observed\cite{motrunich}, which is manifested
by a logarithmic relation between the energy (time) and the
length-scale, $L \sim \ln(\Omega_0/\Omega)^{1/\psi}$, as in Eq.(\ref{logscale}).
Also the effective cluster moment follows the relation, $\mu \sim \ln(\Omega_0/\Omega)^{\phi}$,
as in Eq.(\ref{Phi}). At the critical point the largest spin
cluster is a fractal, with a fractal dimension, $d_f=\phi \psi$, thus random quantum criticality has
a geometrical, percolative interpretation, see also in Sec.\ref{S=1}.
The scaling dimension of the
magnetization, $x_m$, in Eq.(\ref{scaling_rhom}) is given by:
\begin{equation}
x_m= d-d_f=d-\phi \psi\;.
\label{x_m_d_f}
\end{equation}
Finally, infinite disorder scaling theory involves the correlation length exponent, $\nu$. These critical
exponents have been numerically calculated by the infinite system algorithm\cite{motrunich} and independently
by the finite system algorithm\cite{lin00,karevski01}, using different type of initial disorder distributions.
These results are summarized in Table\ref{TAB2d}.

\begin{table}
\caption{Critical exponents of the two-dimensional random transverse-field Ising model obtained
by numerical strong disorder RG calculations: $^{(i}$ Ref.\cite{motrunich},
$^{(ii}$ Ref.\cite{lin00}, $^{(iii}$ Ref.\cite{karevski01}.}
\label{TAB2d}
\centerline{
\begin{tabular}{cccc}
\hline\noalign{\smallskip}
 $\psi$ & $\phi$ & $\nu$ & $x_m$   \\
\hline\noalign{\smallskip}
 $0.42^{(i}$ & $2.5^{(i}$ & $1.072^{(i}$ & $1.0^{(i}$   \\
 $0.5^{(ii}$ & $2.0^{(ii}$ & & $0.94^{(ii}$  \\
 $0.6^{(iii}$ & $1.7^{(iii}$ & $1.25^{(iii}$ & $0.97^{(iii}$  \\
\noalign{\smallskip}\hline
\end{tabular}
}
\end{table}

We note that results by quantum Monte Carlo simulations about the 2d system in Ref.\cite{pich98}
are: $\psi \simeq .4$ and $x_m \simeq 1.$, which are consistent with that obtained by
strong disorder renormalization.

One important consequence of the fact that the critical point of the random transverse-field Ising model
in 2d is controlled by an infinite disorder fixed point is that here frustration does not matter. In any
elementary plaquette the couplings and transverse fields have different magnitudes and the renormalization
is not affected if the plaquette is frustrated or not. Therefore the
quantum Ising spin-glass (i.e. in which there are positive and negative couplings)
belongs to the universality class of the random (ferromagnetic) transverse-field  Ising model,
at least for strong enough disorder. In this respect the quantum
Monte Carlo results in Refs.\cite{rieger94,guo94}, which predict a conventional random fixed point,
could be due to the
fact that the original disorder is not sufficiently strong. (In the generic phase diagram of disorder induced cross-over
effects in Fig.\ref{RCP1} it could be in the ID region.)

\subsubsection{Disordered phase}

\label{RTIM2d_GR}

In the disordered phase, $\delta>0$, during renormalization the transverse fields are more frequently
decimated. However, until the spin clusters reach a characteristic size,
$\xi \sim \ln \Omega_{\xi}^{1/\psi} \sim \delta^{-\nu} $, also some couplings are decimated.
The characteristic size of clusters, $\xi$, corresponds to the correlation length associated
to the average spin correlation function, which is dominated by spin pairs located in the same cluster.
For further decrease of the energy scale, $\Omega < \Omega_{\xi}$,
almost exclusively transverse fields are decimated and the typical distance between
existing spin clusters: $L_{\Omega} \sim \Omega^{-1/z}$ is divergent as the fixed point,
$\Omega^*=0$ is approached. (For the similar quantity in 1d see Eq.(\ref{lomega1})). A third length-scale
can be defined through the typical correlation function, $C_{typ}(r)$, which is measured between spins
which are in different clusters. We have asymptotically the relation:
\begin{equation}
\ln C_{typ}(r) \sim r/\xi_{typ}, \quad \xi_{typ} \sim \xi^{1-\psi} \sim \delta^{-\nu(1-\psi)}\;,
\label{C_typ}
\end{equation}
where the value of the typical correlation length, $\xi_{typ}$, follows from the following argument. Let us
consider typical correlations between two sites of distance, $r \sim \xi$. Performing the decimation up to
the log-energy-scale, $\ln \Omega_{\xi} \sim \xi^{\psi}$, the system renormalizes into isolated single
spin clusters with a typical distance $L_{\Omega_{\xi}}$ among them. The typical correlation function at
this distance is given by:
$C_{typ}(r=\xi) \sim n_{\Omega_{\xi}}^2 \sim L_{\Omega_{\xi}}^{-2d}$. Now making use the relation between
$\ln \Omega_{\xi}$ and $\xi$ we arrive to the result in Eq.(\ref{C_typ}).

The value of the dynamical exponent, $z$, close to the critical point can be obtained from the consideration,
that up to the log-energy scale, $-\ln \Omega_{\xi} \sim \xi^{\psi}(\delta)$ the RG trajectory is close
to the critical trajectory and the distribution function of the couplings and the transverse fields
can be approximated by that at the critical point. From this follows, that $z(\delta) \sim \ln \Omega_{\xi}$
and therefore:
\begin{equation}
z \sim \xi^{\psi} \sim \delta^{-\nu \psi}\;.
\label{z_delta}
\end{equation}
Singularities in the disordered Griffiths phase can be analyzed similarly as for the 1d case in
Sec.\ref{SC_term}.
The only difference, that in d-dimensions the density, $n_{\Omega}$, is related to the length-scale as,
$n_{\Omega} \sim L_{\Omega}^{-d}$. Consequently in the singularities in Eqs.(\ref{entropy_d}), (\ref{m_H})
and (\ref{chi_T}) $z$ should be replaced by $z/d$. In phenomenological scaling theory in Sec.\ref{SC_gr} in
the scaling relation in Eq.(\ref{auto_g1})we have the scaling factor in the r.h.s. as $b^{-d}$.

\subsubsection{Ordered phase}
\label{Gr_ord}
Properties of the ordered phase, $\delta<0$, are qualitatively different in $d>1$ and in $d=1$, due to
different topology in the two cases. In the RG procedure as the energy scale is lowered the typical
cluster size reaches the correlation length, $\xi \sim |\delta|^{-\nu}$, at $\Omega_{\xi}$, with
$\ln \Omega_{\xi} \sim \xi^{\psi}$. At this point the existing renormalized
sites becomes nearest neighbors, almost exclusively couplings are decimated and a giant,
infinite cluster is formed. The formation of the infinite
cluster, as argued in Ref.\cite{motrunich}, is a classical percolation process, which should be
observed also in finite temperature ordering as the temperature is lowered below $T_{\xi} \sim \Omega_{\xi}$.
At $T=T_c$, however, at which point the classical $d$-dimensional random bond Ising model has a phase
transition the RG-approach breaks down, since close enough to $T_c$ the quantum fluctuations are irrelevant.
In this respect as the temperature is lowered there is a double cross-over effect in the critical behavior.

Next we turn to analyze the properties of quantum Griffiths effects, when the energy and temperature scales
are much lower then $T_c$. As in 1d, singularities in the dynamical quantities are due to rare domains, however,
in $d>1$ these events are much more rare and as a consequence singularities are weaker, than in 1d.
To see this we consider
the system at the percolation point $\Omega=\Omega_{\xi}$ and see for a large sub-cluster composed of
$n$ effective spins, which has an effective field of $\tilde{h} \sim \Omega_{\xi}^{n}$.
During further renormalization this sub-cluster will generally either be decimated or
be connected to the infinite cluster, unless it is sufficiently isolated from the giant cluster.
For an efficient isolation the droplet should be at a linear distance of
$l \sim \ln \tilde{h} \sim n \xi^{\psi}$ from the infinite cluster, which has a
very low probability of $\exp(-c n^d)$. Note, that the probability of existence of such droplet in the disordered
phase, when there is no infinite cluster, is $\exp(-c n)$. As a consequence the low-energy tail of the excitation
energies, which are proportional to $\tilde{h}$, is given by:
\begin{equation}
P(|\ln \tilde{h}|) \sim \exp\left( - \tilde{c} |\ln \tilde{h}|^d \right) \;,
\label{P_ln_h}
\end{equation}
which is less singular than the power-low tail in the disordered phase or in the ordered phase in 1d. Therefore
the autocorrelation function assumes an enhanced power-low form:
\begin{equation}
G(t) \sim \exp(-A |\ln t|^{d})\;,
\label{G_d}
\end{equation}
and also the thermodynamic quantities have weaker singularities.


\subsection{Random Heisenberg models}

Random antiferromagnetic Heisenberg (and $XY$) models in one dimension constitute the simplest realization
of infinite disorder scaling, as described in Sec.\ref{S=1/2}. In the presence of chain-chain interaction between
two or more chains, i.e. for random spin ladders infinite disorder scaling is restricted to quantum critical
points, otherwise the low energy behavior is controlled by conventional Griffiths or by large spin fixed points,
see Sec.\ref{Sec:ladder}. For two- and three-dimensional systems, in which infinite number of chains are coupled
together the strong disorder RG approach has been first applied numerically by Bhatt and Lee\cite{bhattlee}. Later
investigations\cite{motrunich} and scaling considerations\cite{igloixy} indicate that the low energy fixed point of $d \ge 2$
random Heisenberg antiferromagnets is not an infinite disorder fixed point. A comprehensive numerical analysis of
the problem, in which non-frustrated and frustrated lattices, as well as models with competing interactions are
considered can be found in Ref.\cite{numedlin}.

{\it Numerical renormalization rules}

During renormalization higher dimensional random Heisenberg models transform in the same way, as
random spin ladders, as described in Sec.\ref{Sec:ladder}. Thus clusters with various value of the spin are formed
and the couplings between
them are antiferromagnetic or ferromagnetic. The decimation rules are described in Sec.\ref{Sec:ladder}
and if more than one coupling is present between two sites the maximum rule in Eq.(\ref{h_2d}) is applied.
In the actual calculation the finite lattice method of Sec.\ref{Sec:ladder} is used, and one has monitored the
distribution of the gaps as well as the size of typical effective spins, $S_{eff}$.

\subsubsection{Numerical RG results}

{\it Non-frustrated models}

Random antiferromagnetic Heisenberg models on the square and on the simple cubic lattices exhibit conventional
Griffiths-type behavior, see Sec.\ref{SC_gr}. There is no large spin formation and the dynamical exponent, $z$.
is finite. In the square lattice, in which case qualitative estimate was possible, the gap exponent and
the dynamical exponent are found, $\omega=0.7$ and $z=1.2$, respectively, which  satisfy the relation in
Eq.(\ref{z_omega}).

{\it Frustrated models}

Frustration in the Heisenberg model can be of different origin. i) In the case of spin glass models
there are random antiferromagnetic and ferromagnetic couplings. ii) Frustration of geometrical origin
is found in random antiferromagnetic models on the triangular and kagom\'e lattices. iii) Finally, frustration
can be a result of competition between first-, $J_1$, and second-neighbor, $J_2$, couplings. In two-dimensional
problems all frustrated models have the same, so called spin glass (SG) fixed point. This is a large spin
fixed point with the special properties:
\begin{equation}
\omega_{SG}=0,\quad z_{SG}=2, \quad \zeta_{SG}=1/2,\quad d=2\;.
\label{SG_2}
\end{equation}
Note that in 2d relation in Eq.(\ref{z_omega}) is satisfied, thus the excitations seem to be localized.
Also the quantum phase transitions present in the non-random $J_1$-$J_2$ models are washed out by the
disorder.

In three-dimensional frustrated problems, such as the spin glass and the $J_1$-$J_2$ models the
gap exponent is found, $\omega \approx 0$, which is consistent with the 2d results in Eq.(\ref{SG_2}).
The dynamical exponent is practically universal, $z \approx 3/2$, but the relation in Eq.(\ref{z_omega}) is
not valid, thus the excitations are not localized. The moment exponent, $\zeta$, is disorder and parameter dependent and generally larger than $1/2$, found in one- and two-dimensions.

\subsubsection{Related numerical studies}

Quantum Monte Carlo studies of the Heisenberg antiferromagnet on a diluted square
lattice show that N\'eel-type long-range-order disappears at the classical percolation
point\cite{kato00}. While in earlier investigations a novel,
$S$-dependent critical behavior was found\cite{kato00}, recent studies
identify the transition as an $S$-independent classical percolation
transition with the well known exponents\cite{sandvik02}.
The Heisenberg model on the square lattice with random antiferromagnetic bonds is studied
by large scale quantum Monte Carlo simulations, as well as by L\'anczos exact diadonalization
and modified spin-wave theory\cite{laflorencie05a}. In agreement with the numerical strong
disorder RG results no infinite disorder fixed point is observed, but rather some Griffiths
singularities (for sufficiently strong disorder) with a disorder-dependent
dynamical exponent. Moreover, the antiferromagnetic order parameter is found surprisingly very
robust against bond randomness and a "coexistence" between an antiferromagnetically ordered
backbone with some localized region with easily flipable spins
(contributing to a divergent uniform susceptibility) is observed.
Another work
studied the $\pm J$ Heisenberg (quantum) spin glass and found that for
a concentration of ferromagnetic bonds $p>p_c\approx0.11$ the N\'eel-type long-range-order in
the ground state vanishes and is replaced by a so-called spin glass
phase\cite{nonomura95}. Within the spin-glass phase, the average ground
state spin, $S_{\rm tot}$, scales as $S_{\rm tot} \sim \sqrt{N}$, and
the gap as $\Delta E \sim 1/N$, where $N$ is the number of
spins.\cite{oitmaa01} This is in accordance with the strong disorder
RG results in Eq.(\ref{SG_2}).

A 2d bilayer Heisenberg antiferromagnet
with random dimer dilution is studied by quantum Monte Carlo simulations in Ref.\cite{sandvik02a,vajk02}
and its 3d classical counterpart in
Refs.\cite{sknepnek04}. In this system as the ratio of inter-layer and intra-layer couplings is
varied a quantum phase transition takes place, which is governed by a conventional
random fixed point. This problem with random antiferromagnetic couplings is studied
by a numerical implementation of the strong disorder RG method\cite{lin05}. For strong enough disorder
the antiferromagnetic order and thus the phase transition is found to be destroyed so that the
system is in the quantum Griffiths phase. A somewhat related problem, a single layer
Heisenberg antiferromagnet with staggered dimers and dimer dilution has been studied
recently by quantum Monte Carlo simulations\cite{sandvik05}. Surprising critical properties are
found, among others at the persolation point there is
a whole critical line with varying exponents, including a dynamic
exponent that appears to diverge continuously in one limit. 

Ground state and finite temperature properties of a system of coupled
frustrated and/or dimerized spin-1/2 chains modeling e.g. the
CuGeO$_3$ compound are studied in Ref\cite{laflorencie05}. This system
is mapped into a low-energy effective model, which describes a two-dimensional system of
effective spin-1/2 local moments interacting by spatially anisotropic
long range spin exchange interactions. By a
strong disorder RG analysis large spin formation is observed.

\subsection{Other  problems}

Here we list some higher dimensional problems in which (a variant of) the strong disorder
RG method has been successfully used.

\paragraph*{Doped spin-Peierls model}

The problem of antiferromagnetism in a two-dimensional Heisenberg model of doped spin-Peierls system
is studied by a numerical application of the strong disorder RG method. The low-energy fixed point of
the problem is of finite randomness type\cite{fabrizio97,melin00}.

\paragraph*{Random tight-binding models}

The problem of particle-hole symmetric localization in two dimensions is studied in terms of a bipartite
hopping Hamiltonian with random hopping rates by the strong disorder RG method\cite{motrunich02}. The low-energy
fixed point of this model is infinite disorder type, the energy- and length-scales are expected to related
as: $|\ln \Omega| \sim |\ln L|^{x}$, where the conjectured values: $x=2$ in Ref.\cite{gade93} or $x=3/2$ in
Ref.\cite{motrunich02} (see also Ref.\cite{mudry03}). Note, that this singularity formally corresponds to a divergent $z$ or to a vanishing
$\psi$ in Eq.(\ref{logscale}). Introducing two types of hopping amplitudes with a ratio of $e^{\delta}$
in the brick-wall (honeycomb) lattice the system is found delocalized for $\delta<\delta_c$ and localized for $\delta>\delta_c$.

\paragraph{Random Heisenberg and tight-binding models on fractal lattices}

Numerical implementation of the strong disorder RG method is used to study the
low-energy fixed points of random Heisenberg and tight-binding models on different
types of fractal lattices\cite{melin05}.
For the Heisenberg model new types of infinite disorder and strong disorder fixed points are found. 
For the tight-binding model an orbital magnetic field is added and both diagonal and off-diagonal disorder
is considered. For the latter model, besides the gap spectra also the fraction of frozen sites,
the correlation function, the diamagnetic response and the two-terminal current is studied.
The magnetoresistive effects are found qualitatively
different for the bipartite and non bipartite lattices..


\section{Variations : Correlations, disorder Broadness, etc...}

In the random systems we considered till now the random variables (couplings, transverse fields)
are independent and identically distributed, furthermore their
distribution is not too broad, generally we assume that the second moment exists.
In some problems, however, these assumptions are not satisfied. Disorder is often correlated or
broadly distributed and the strength of disorder can be spatially inhomogeneous. Finally,
a somewhat related problem when the variables follow non-random,
but aperiodic or quasiperiodic sequences. Here we shortly review these developments.

\subsection{Correlated disorder}

Here we consider the effect of (isotropic) spatial
correlations in the disorder that can be modeled with a disorder
correlator $G_d({\bf r})$:
\begin{equation}
[\delta({\bf r})\delta({\bf r'})]_{\rm av}=G_d({\bf r}-{\bf r'})\;.
\label{corr}
\end{equation}
For uncorrelated disorder $G_d({\bf r})$ is a
delta-function. Regarding the recent experiments on $f$-electron systems
there is evidence \cite{neto} that the spatial
correlations in the metallic compound U$_{1-x}$Th$_x$Pd$_3$ decay like
$G_d({\bf r}) \sim r^{-3}$. 

The Harris criterion for correlated disorder
\cite{weinrib} shows that any disorder correlator that
falls off faster than $r^{-2/\nu}$ (i.e.\ $G({\bf r})\sim{
O}(r^{-\rho})$ with $\rho>2/\nu$, where $\nu$ is the correlation
length exponent for uncorrelated disorder) does not change the
universality class of a model with uncorrelated disorder.
On the other hand for
\begin{equation}
G({\bf r})\sim r^{-\rho}\quad{\rm with}\quad0<\rho\le 2/\nu\quad(\le d)\;,
\end{equation}
where the last inequality holds generally for disordered system
with uncorrelated disorder \cite{chayes}, the
disorder correlations are relevant, the
critical exponents become different from the uncorrelated case and the
critical point constitutes a new universality class.

Detailed study of the RTFIC with correlated disorder\cite{rieger99a} has lead to the following
exact results for relevant correlations, i.e. with a decay of $0 < \rho < 1$.
The strength of singularities at the infinite disorder fixed are enhanced and the
critical exponents are $\rho$ dependent:
\begin{equation}
\psi(\rho)=1-\rho/2, \quad x_s(\rho)=\rho/2, \quad \nu(\rho)=2/\rho.
\end{equation}
The bulk magnetization exponent, $x_m(\rho)$, according to numerical studies is a decreasing function
of $\rho$, which behaves for small $\rho$ as  $x_m(\rho) \approx \rho/2$.

In the disordered phase the strength of Griffiths singularities are also enhanced, this means that
the dynamical exponent, $z(\delta,\rho)$, is larger than for uncorrelated disorder, as given in
Eq.(\ref{z_I}). Close to the transition point $z$ is given by: $z(\delta,\rho)\propto\delta^{1-2/\rho}$,
what should be compared with $z(\delta)=(2\delta)^{-1}$, for uncorrelated disorder, see Eq.(\ref{zeq2}).

In higher dimensions one still has the result $\nu=2/\rho$
for $\rho<2/\nu_{\rm unc}$, according to a general argument given
in \cite{weinrib}. Moreover, $\psi$ increases with increasing disorder
correlations, since its value is connected to the geometric
compactness of strongly coupled clusters. Thus, the dynamical exponent
$z \sim \delta^{-\nu\psi}$ grows, again enhancing the Griffiths singularities.

\subsection{Broad disorder distribution}

Here we consider the effect of broad disorder distributions of parameters  on the critical properties of
random quantum magnets, similar investigations for random random walks have been reviewed in\cite{jpbreview}.
Keeping in
mind that in the infinite disorder fixed point of the RTFIC (and also for the 2d model) the logarithm of
the couplings and the transverse fields follows a smooth probability distribution, see Eq.(\ref{sol0}) the
appropriate parameterization is:
\begin{equation}
J_{ij}= \Lambda^{\Theta_{ij}}\;.
\label{parametr}
\end{equation}
whereas $h_i=h_0$. The exponents, $\Theta_{ij}$ are independent
random variables, which are taken from a broad distribution, $\pi(\Theta)$, such
that for large arguments they decrease
as, $\pi(\Theta) \sim |\Theta|^{-1-\alpha}$ (``L\'evy flight''). The L\'evy index, $ \alpha >1$, and
the $\kappa$-th moment of the distribution exists for $\kappa<\alpha$.

The random transverse-field Ising model in one- and two-dimensions is studied by numerical
strong disorder RG method, and in 1d several exact results are also obtained\cite{karevski01}.
The broadness of the disorder distribution is found relevant, if the L\'evy index is lowered below a critical
value, $\alpha_c$. In the region of $1<\alpha<\alpha_c$ the critical points of the systems are
governed by infinite disorder fixed points, in which the critical exponents
are continuous functions of $\alpha$ and for $\alpha>\alpha_c$ they are the same as in
the model with normal (i.e. non-broad) disorder.

In 1d, for the RTFIC  $\alpha_c=2$, in close analogy with random walks, where the central limit
theorem is valid for $\alpha>2$. This analogy is due to the  exact mapping, which is shown in
Sec.\ref{RW_RTFIC}. For the RTFIC several critical exponents are exactly calculated for $\alpha<2$:
\begin{equation}
\psi(\alpha)=1/\alpha, \quad x_s(\alpha)=1/2, \quad \nu(\alpha)=\alpha/(\alpha-1).
\end{equation}
In 2d numerical results indicate that $\alpha_c \approx 4.5$, thus in the region of
$2<\alpha<\alpha_c$ the broadness of disorder is relevant for the random transverse-field Ising model,
whereas it is irrelevant for the random random walk.

In the off-critical region the strength of Griffiths singularities are also enhanced for $1<\alpha<\alpha_c$.
In 1d the low-energy excitations have a typical size-dependence:
\begin{equation}
\ln \epsilon(L) \sim L^{1/(1+\alpha)},\quad \delta>0\;.
\label{eps+}
\end{equation}
Thus the dynamical exponent is formally infinite in the whole Griffiths region.

\subsection{Inhomogeneous disorder}

Here we consider the critical and off-critical properties at the boundary of
the RTFIC when the distribution
of the couplings,  and/or transverse fields, at a distance $l$ from the
surface, deviates from its uniform bulk value by terms of order
$l^{-\kappa}$:
\begin{equation}
\pi_l(J)-\pi(J) \simeq A l^{-\kappa}, \quad {\rm and/or} \quad
\rho_l(h)-\rho(h) \sim A l^{-\kappa}\;.
\label{inhm}
\end{equation}
Exact results are obtained\cite{karevski99} using the
correspondence between the surface magnetization of the RTFIC and the
surviving probability of a random walk with time-dependent absorbing
boundary conditions, see Sec.\ref{m_s}.
For slow enough decay, $\kappa<1/2$, the
inhomogeneity is relevant: Either the surface stays ordered at the bulk
critical point or the average surface magnetization displays an essential
singularity, depending on the sign of $A$. In the marginal situation,
$\kappa=1/2$, the average surface magnetization decays as a power law
with a continuously varying, $A$-dependent, critical exponent which is
analytically known. The surface critical behavior of the model is summarized
in Table \ref{TABinhm}.

\begin{table}
\caption{Summary of the surface critical properties of the inhomogeneous
RTFIC.}
\centerline{
\begin{tabular}{cccc}
&$\ln t_r$&$[m_s(\delta)]_{\rm av}$
&$[G_s(t)]_{\rm av}$\\
\ms
\hline
\ms
$\kappa >1/2$&$\sim\xi^{1/2}$
&$\sim\vert\delta\vert$
&$\sim(\ln t)^{-1}$\\
\hs
$\kappa=1/2$&$\sim\xi^{1/2}$
&$\sim\vert\delta\vert^{\beta_s(A)}$
&$\sim(\ln t)^{-\beta_s(A)}$\\
\hs
$\kappa<1/2$&$\sim\xi^{1-\kappa}$
&---
&---\\
\ms
$A>0$&---
&$\sim\vert A\vert^{1\over1-2\kappa},\,\delta=0$&$\sim$ const\\
\ms
$A<0$&---&$\sim\exp[-{\rm const}\,
\vert\delta\vert^{-(1-2\kappa)/\kappa}]$ &$\sim\exp[-{\rm
const}\,(\ln t)^{1-2\kappa\over1-\kappa}]$\\ \ms
 \end{tabular}
}
 \label{TABinhm}
 \end{table}

In the off-critical region the properties of the Griffiths singularities are not
affected by the inhomogeneity. A somewhat different form of the inhomogeneity in the RTFIC
is studied in Ref.\cite{turban99}.

\subsection{Aperiodic systems}

Quasiperiodic, or more generally aperiodic sequences have several similarities with random
systems. In both cases i) the systems are inhomogeneous, ii) the perturbations have non-periodic character,
i.e. there is no finite length-scale associated with the spatial modulation of the couplings
and iii) the fluctuations in the energy generally grow with the size of the system
(see Eq.(\ref{fluctuate})). To clarify the
possible similarities and differences in the critical behavior of random and aperiodic systems
several investigations have been performed, both for the diffusion process\cite{igloi99a} and in more
details for quantum spin systems\cite{luck94}. Recently, the analysis is extended by the use of the
strong disorder RG method\cite{hida04,vieira04}.

An aperiodic or quasiperiodic sequence is generated through substitutional rules\cite{queffelec87}.
For example the
Fibonacci sequence is built on two letters, ${\bf A}$ and ${\bf B}$, following the rules:
${\bf A} \to {\bf AB}$ and ${\bf B} \to{\bf A}$. Putting a transverse-field Ising spin chain in
this lattice (see Eq.(\ref{hamilton_I}) the coupling at bond $i$ is $J_i=J_A$ ($J_i=J_B$)
if the letter at this position is  ${\bf A}$  (${\bf B}$), whereas the transverse fields are
chosen constant, $h_i=h_0$. Relevance or irrelevance of aperiodic perturbations are related
to the value of the wandering exponent, $\omega$, defined through the fluctuations of the
couplings\cite{dumont90}:
\begin{equation}
\Delta(L)=\sum_{i=1}^L\left( J_i-[J]_{\rm av} \right) \sim L^{\omega}\;,
\label{fluctuate}
\end{equation}
for large size, $L$. Note, that for the Fibonacci sequence, $\omega=-1$, thus the fluctuations
are bounded, whereas for a random system, $\omega=1/2$. According to an extension of the Harris
criterion due to Luck\cite{luckaper} the aperiodic perturbation is irrelevant for:
\begin{equation}
\nu_0 > 1/(1-\omega)\;,
\label{luck}
\end{equation}
what should be compared with the Harris criterion in Eq.(\ref{harris}).

For the transverse-field Ising chain with uniform couplings $\nu_0=1$, thus the border-line value
is $\omega=0$. Therefore with bounded fluctuations, $\omega <0$, the critical behavior of the aperiodic
chain is the same as that of the homogeneous one, whereas for $\omega  \ge 0$, there is a new type of
fixed point of the system. Calculations on specific sequences are in accordance
with the above criterion\cite{turban94}.
For marginal perturbations, $\omega=0$, coupling dependent critical behavior is found with a varying
dynamical exponent, $z$. These calculations are based on an exact RG method, which is proposed in
Ref.\cite{igloiturban96}, applied in Ref.\cite{igloi97a} and generalized in Ref.\cite{hermisson97}.
Similar calculations are made
for the aperiodic $XY$ model\cite{hermisson90} and for the diffusion process\cite{igloi99a}.
This aperiodic RG method
is equivalent to the  strong disorder RG method in Sec.\ref{RTFIC_RG_B} and \ref{RTFIC_RG_F} in the
limit of $J_A \gg J_B$, or vice versa.

Aperiodic sequences with unbounded fluctuations, $\omega >0$, are in close analogy with random systems.
In particular the Rudin-Shapiro sequence\cite{queffelec87}, which is built on four letters,  ${\bf A,~B,~C}$ and ${\bf D}$ with
the substitutional rule:
\begin{equation}
{\bf A} \to {\bf AB}~~,~~{\bf B} \to {\bf AC}~~,~~{\bf C} \to {\bf DB}~~,~~
{\bf D} \to {\bf DC}\;,
\label{rs}
\end{equation}
have the same wandering exponent, $\omega=1/2$, as the random system. A comparison of the singular behavior
of the two systems is performed in Ref.\cite{igloi98a}, here we summarize the main findings.

In an aperiodic system the energy- , $\Delta E$, and length-scale ,$L$, are related as:
\begin{equation}
|\ln \Delta E| \sim L^{\omega}\;,
\label{Delta_L}
\end{equation}
and the energy has a logarithmically broad distribution, as for random chains, see Eq.(\ref{epscrit}).
Typical and average values of physical quantities at the critical point are very different and the average
is dominated by rare realizations, as observed at an infinite disorder fixed point, see Sec.\ref{SC_st}.
For example the surface magnetization in the Rudin-Shapiro chain is typically\cite{igloi94},
$m_s^{typ}(L) \sim \exp(-{\rm const} \times \sqrt{L})$, whereas the average, which is obtained by
averaging over all chains, when the starting
$l=0,1,\dots$ letters from the sequence are omitted, has a power low dependence,
$[m_s(L)]_{\rm av} \sim L^{-x_m^s}$, with $x_m^s=1/2$. Another singularities at the critical point are
different for the Rudin-Shapiro and the random model. For example the correlation length exponent is
$\nu=4/3$, what should be compared with $\nu=2$ for the random chain in Eq.(\ref{xi_av}).

In the off-critical region aperiodic and random chains behave very differently. In an aperiodic chain
the low-energy excitations are bounded\cite{igloi98a}, therefore there are no Griffiths-singularities in this case.


\part{  RG STUDY OF CLASSICAL MODELS}


\section{ Sinai walk : Random walk in Brownian potential}

\label{chapsinai}

\subsection{Model}

The so called `Sinai model', describing a random walk in a random Brownian potential,
has interested both the mathematicians since the works
of Solomon \cite{solomon}, Kesten { \it  et al..}  \cite{kestenetal},Sinai \cite{sinai}..., and the physicists since the works of Alexander { \it  et al.}\cite{alexander}, Derrida-Pomeau \cite{derridapomeau}... 
 There have been many developments in the two communities for the last thirty years: we refer the reader to the recent review \cite{zhanshi} for the mathematical side and to the reviews \cite{haus, havlin, jpbreview}
for the various physicists's approaches.  For the physicists, the interest of the Sinai model is double. On one hand, the Sinai walk represents a simple dynamical model containing quenched disorder, in which many properties that exist in more complex systems
can be studied exactly.
On the other hand, the Sinai model naturally appears in various contexts, for instance in the dynamics of a domain wall in the random field Ising chain (see section \ref{chaprfim}) or in the unzipping of DNA
 in the presence of an external force \cite{lubenskinelson}.  

The continuous version of the Sinai model corresponds to the 
Langevin equation \cite{jpbreview} 
\begin{eqnarray} 
\frac{dx}{dt} = - U' (x(t)) + \eta(t) 
\label{langevin} 
\end{eqnarray}
where $\eta(t)$ represents the usual thermal noise 
\begin{eqnarray} 
< \eta(t) \eta(t ') > = 2 T \delta(t-t ') 
\end{eqnarray} 
and where $U(x)$ is the quenched Brownian potential  
\begin{eqnarray} 
\overline { (U(x)-U(y))^2} = 2 \sigma \vert x-y \vert 
\label{defsigma} 
\end{eqnarray} 
More generally, in the whole review, thermal averages of observables are denoted by
 $<f>$, whereas disorder averages are denoted by $\overline{f}$.

In the discrete Sinai model on the 1D lattice, the particle which is on site $i$ has a probability
$\omega_i \equiv w_{i,i+1}$ of jumping to the right and a probability $(1-\omega_i) \equiv w_{i,i-1}$ of jumping to the left.  The $\omega_i$ are independent random variables in $]0,1[$.  The random walk is recurrent only if $\overline{ \ln \omega_i} = \overline{ \ln (1-\omega_i)}$, which corresponds to the absence of bias of the random
potential $U(x)$ (\ref{defsigma}) of the continuous version. 

The time evolution of $P_i(t)$,
the probability for the particle to be on site $i$ at time $t$, is governed by
the Master equation:
\begin{equation}
{d P_i\over d t}=w_{i-1,i} P_{i-1}-(w_{i,i-1}+w_{i,i+1})P_i
+w_{i+1,i} P_{i+1}\,,
\label{mastersinai}
\end{equation}
from the solution of which one can obtain different physical quantities. Here we present the drift
velocity, $v_d$, and the persistence, $P_{pr}$, which is useful to define
the phase diagram and an appropriate order parameter of the model.

\subsubsection{Drift velocity and dynamics}

For a given sample of length, $L$, i.e. for a given set of the forward ($w_{i,i+1}$) and the backward
($w_{i+1,i}$) transition rates the drift velocity has been calculated by Derrida\cite{derridapomeau}, as:
\begin{equation}
v_d=\frac{L}{\sum_{i=1}^{L} r_i} \left[1-\prod_{i=1}^L \frac{w_{i,i+1}}{w_{i+1,i}} \right]\;,
\label{v_d}
\end{equation}
where the $r_i$ stands for:
\begin{equation}
r_i=\frac{1}{w_{i+1,i}} \left[1+\sum_{n=1}^{L-1}\prod_{j=1}^n \frac{w_{i+j-1,i+j}}{w_{i+j+1,i+j}}\right]\;.
\label{v_d1}
\end{equation}
As for the surface magnetization (\ref{peschel}) in the random transverse field Ising chain,
we recognize again Kesten random variables, whose properties
are discussed in more details in Appendix \ref{appkesten}.

In the thermodynamic limit, $L \to \infty$, and for asymmetric transition rates, $w_{i,i+1} \ne w_{i+1,i}$, one
can define a control parameter:
\begin{equation}
\delta={[\ln w_\leftarrow]_{\rm av}-[\ln w_\rightarrow]_{\rm av}
\over \rm{var}[\ln w_\leftarrow]+\rm{var}[\ln w_\rightarrow]}\;,
\label{deltarw}
\end{equation}
where $w_\rightarrow$ ($w_\leftarrow$) stands for transition
probabilities to the right (left), i.e.\ $w_{i,i+1}$ ($w_{i,i-1}$). For the biased Sinai walk, considered
in Section \ref{chapsinaibiais}, $\delta>0$ or $\delta<0$ and the particle moves to the right or to the left, respectively. The critical situation, $\delta=0$, is called the Sinai walk, when
ultra-slaw diffusion takes place\cite{sinai}:
\begin{equation}
\overline{ <x^2>} \sim (\ln t)^4\;.
\label{sinai}
\end{equation}

\subsubsection{Persistence and order}
\label{persistence}
For a random walk persistence, $P_{pr}(t,L)$, is the probability that the walker has not crossed the starting
position at $i=1$ until time, $t$\cite{redner}. In a finite system one has an absorbing wall at $i=L+1$ and the probability:
$p_{pr}(L)=\lim_{t \to \infty} P_{pr}(t,L)$ plays an analogous role to the order parameter in magnetic systems.
For a given sample it is given by the expression\cite{igloi98c}:
\begin{equation}
p_{\rm pr}(L)
=\left(1+\sum_{i=1}^L\prod_{j=1}^i
\frac{w_{j,j-1}}{w_{j,j+1}}\right)^{-1}\;,
\label{persl}
\end{equation}
and in the thermodynamic limit $[p_{\rm pr}]_{\rm av}>0$, for $\delta>0$, whereas $[p_{\rm pr}]_{\rm av}=0$,for $\delta \le 0$, which explains analogy with the order parameter.

At the critical point, $\delta=0$, the average persistence is vanishing, its size dependence, however, can
be obtained by simple arguments analogous to those used in Sec.\ref{m_s} for the surface magnetization of the
RTFIC. In a {\it  typical sample} the
largest term of the sum in Eq.(\ref{persl}) grows as $\sim \exp( {\rm cst} L^{1/2})$, which follows from the central limit theorem. Consequently,
\begin{equation}
p_{\rm pr}^{typ}(L) \sim \exp( -{\rm cst} L^{1/2}),
\label{persl_typ}
\end{equation}
which goes to zero very fast with $L$. The {\it  average behavior} is dominated
by {\it  rare events}, in which the persistence is of $O(1)$. It is easy
to see that in a rare event:  i) non of the $L$ products in Eq.(\ref{persl}) are larger than $O(1)$, and ii)
a typical product goes to zero fast enough with $L$. In such a sample the transition rates are
distributed in such a way,
that the quantity: $\varepsilon_j=\ln \frac{w_{j,j+1}}{w_{j,j-1}}$, satisfies the relation:
$\sum_{j=1}^i \varepsilon_j \le 0$, $i=1,2,\dots,L$. This problem is equivalent to a random walk having a
length, $\varepsilon_j$, at step $j$, and which never crosses the origin,
thus has a surviving character. The fraction of the rare events is just the survival probability of an
$L$-step random walk:
\begin{equation}
P^{rare}(L) \sim P_{\rm surv}(L) \sim L^{-1/2}.
\label{p_rare}
\end{equation}
At this point it is evident that the average of the persistence is dominated by the rare events and the contribution of the typical samples is totally negligible. We obtain finally:
\begin{equation}
p_{\rm pr}^{av}(L) \sim P^{rare}(L) \sim L^{-\theta},
\label{persl_av}
\end{equation}
thus the persistence exponent is $\theta=1/2$. (See also the RG derivation in Eq.(\ref{singl_pers})).

The singular behavior of the random random walk has many similarities with that of the RTFIC, which is due to an
exact mapping, as described in Appendix \ref{RW_RTFIC}.

\subsection{RG rules for the Sinai model : Physical motivations} 

The aim of this Section is to justify in details the physical origin
of the RG rules that we have given in Eqs (\ref{rulesinai1},\ref{rulesinai2})
as an example of a strong RG rule.

As we have seen before in this review,
the strong disorder RG for quantum spin models corresponds to
 an elimination of degrees of freedom in the Hamiltonian:  it is thus rather close to usual renormalizations, the only difference being that the decimation is done in an iterative way on the extremal coupling, instead of being done in a homogeneous way on the whole chain at each stage.  On the contrary, in the 
strong disorder RG for statistical physics models, the way of thinking deviates much more from the usual procedures. Indeed, the starting point is 
usually not an exact or approximate integration on the degrees of freedom in the microscopic model, i.e. on the partition function for static problems or on the master equation for  dynamic problems.  The starting point is rather a heuristic physical argument, which allows to identify the degrees of freedom which will be important at large scale.  One then defines the renormalization directly on these important
degrees of freedom, and one obtains in the end, in favorable cases, 
results that become exact in the asymptotic limit where the RG procedure is applied a large number of times.  As this mixture between heuristic arguments 
at the beginning and exact results at the end can appear disconcerting at first
sight, and even often meets a certain incomprehension, it seems useful to analyze in details the various stages of the reasoning on the case of the Sinai model.

 \label{exemplesinai}

\subsubsection*{ Identification of the degrees of freedom which will be
important at large scale} 

There exists for a long time a simple qualitative argument \cite{jpbreview} to predict the typical displacement $x \sim (\ln t)^2$ in the Sinai model, instead of the usual behavior  $x \sim \sqrt{t}$ of the pure diffusion. It can be summarized as follows:  
the time $t(x)$ necessary to reach the point $x>0$ will be dominated by the Arrh\'enius factor of $e^{\beta B_x}$ associated to the largest barrier $B_x$
which should be passed by thermal activation to go from the starting point $x=0$ to the point $x$ (This approximation by the Arrhenius factor amounts 
to apply a saddle-point method on the exact expression for the first passage
time).  In a Brownian potential, the typical behavior of the barrier $B_x \sim \sqrt{x}$ leads to an Arrhenius time $t \sim e^{\beta \sqrt{x}} $, which indeed corresponds to the scaling $x \sim (\ln t)^2$ after inversion. 

This heuristic argument suggests that the degrees of freedom which will be important
at large scale are the large barriers which exist in the random potential.  More precisely, at a given time $t$, the particle will not have been able to cross by thermal activation any barrier larger than the scale $ (T \ln t)$. 

 \subsubsection*{ Definition of the RG rules directly on the important degrees of freedom of the disorder} 

The RG procedure is defined directly on the barriers of the random potential.
It simply consists in the iterative elimination of the smallest barrier.  This smallest barrier remaining in the system defines the RG scale $\Gamma$. The renormalized landscape at scale $\Gamma$ only contains barriers larger than $\Gamma$, all the smaller barriers having already been eliminated.  As the scale $\Gamma$ increases, the probability distribution of barriers $F$
 in the landscape at scale $\Gamma$ converges towards a scaling form
 $\theta(F \geq \Gamma) P^* \left(\frac{F-\Gamma}{\Gamma} \right)$, where 
the stationary distribution $P^*$  that characterizes the infinite disorder
fixed point has been given in Eq (\ref{ptfixeeta}).

\subsubsection*{ Correspondence with the initial model} 

To each time $t$ of the initial model, one associates the renormalized landscape at  scale $\Gamma=T \ln t$.  One defines an effective dynamics without thermal fluctuations, in which the particle is at time $t$ 
near the minimum of the renormalized valley at scale
 $\Gamma=T \ln t$ that contains the initial position at $t=0$.
More generally, the various observables of the initial model
can be associated to the various properties, either static or dynamic, of the renormalized landscape.

 \subsubsection*{ Check of the consistency in the asymptotic limit and study
of first corrections}

The probability that the particle is not in the renormalized valley of the 
effective dynamics, is of order $1/(\ln t)$ and thus tends towards zero in the limit of infinite time. This shows the consistency of the RG procedure and its asymptotic exactness.  One can then study the thermal fluctuations around the effective dynamics, by considering on the one hand the probability distribution inside the renormalized valley, and on the other hand the rare events of order $1/(\ln t)$ where the particle is not in the renormalized valley of effective dynamics.

  \subsubsection*{ Discussion} 

This example shows very well how this way of reasoning allows to obtain a very detailed description of the asymptotic dynamics at large time.  
It also shows the interest of this inhomogeneous RG  
to fully adapt to the local extrema of disorder on various scales, compared to the usual renormalizations based on identical cells at each stage.  In a certain sense, the way of reasoning we have just described uses to the maximum the qualitative ideas contained in the concept of renormalization, like the irrelevance of the details of the microscopic model on the behaviors at large scale, and the
convergence towards simpler theories representing universality classes, 
{ \it  before} the definition of a quantitative procedure.  As a consequence,
one should not criticize the strong disorder RG approaches 
 for taking as starting point some qualitative physical arguments, because it is precisely from there that all their effectiveness comes! 
 Indeed, for many statistical physics models that we will consider, whereas a usual renormalization on the disordered microscopic model would have no hope to be closed and to lead to exact results, the strong disorder approach allows to obtain a closed renormalization directly on the degrees of freedom which are really important at large scale, and to obtain in the end
asymptotic exact results.

 \subsection{ Notions of effective dynamics and localization} 

In
 the strong disorder RG approach of
the Sinai model, the essential idea is to decompose the process $x_{U, \eta}(t)$, representing the position of the random walk generated by the thermal noise $\eta(t)$ in the random potential Brownian $U(x)$, into a sum of two terms \begin{eqnarray} 
x_{\{U, \eta\}}(t) = m_{\{U\}}(t) + y_{\{U, \eta\}}(t) 
\end{eqnarray} 

$ \bullet $  the process $m_{\{U\}}(t)$ called the ``effective dynamics" 
depends only on the disorder but not on the thermal noise :
 it represents the most probable position of the particle at the moment $t$.
It simply corresponds to the best local minimum of the random potential $U(x)$ that the particle has been able to reach at time $t$.  As the escape over a potential barrier $F$ requires an Arrh\'enius time of order $t_F=\tau_0 e^{\beta F}$, one can study in detail this effective dynamics by using a strong disorder RG  which consists in the iterative decimation of the smallest barriers remaining in the system.  One then associates to time $t$ the renormalized landscape in which only barriers larger than the RG scale $\Gamma=T \ln t$
have been kept.   The position $m_{\{U\}}(t) \sim \Gamma^2=(T \ln t)^2$ then corresponds at the bottom of the renormalized valley at scale $\Gamma=T \ln t$ which contains the initial condition at $t=0$.

$ \bullet $ the process $y_{\{U, \eta\}}(t)$ represents the thermal fluctuation 
with respect to effective dynamics.  In the limit of infinite time, it 
remains a finite random variable. This very strong result is the Golosov localization
phenomenon \cite{golosovlocali} :  all the particles which diffuse in the same sample starting from the same starting point with different thermal noises $\eta$ are asymptotically concentrated in the same renormalized valley of minimum $m_{\{U\}}(t)$.  More precisely, if one considers the first corrections at large time, the probability that a particle is not in the valley corresponding to effective dynamics $m_{\{U\}}(t)$ is of order $1/(\ln t)$, in which case the particle is at a distance of order $(\ln t)^2$ from $m_{U}(t)$.  These events are thus rare (their probability tends towards zero to large time) but they nevertheless dominate  certain observables, such as the thermal width $\overline{\Delta x^2(t)} \sim \overline { < y^2(t) >} \sim (\ln t)^3$ which diverges.  

\begin{figure}

\centerline{\includegraphics[height=8cm]{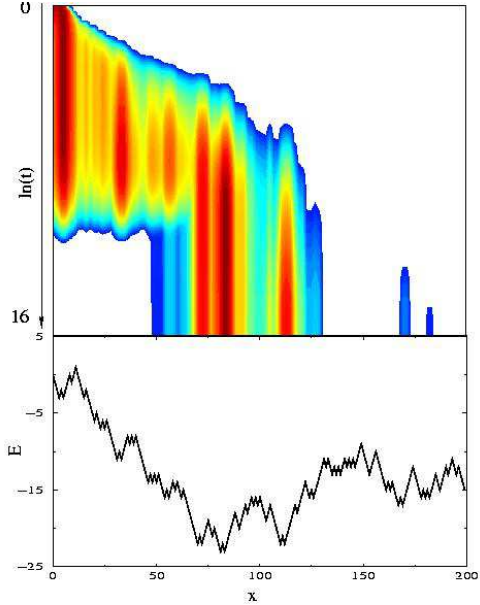}} 
\caption{\it Example of diffusion front in a given sample,
obtained by J. Chave and E. Guitter \cite{chave} ``Evolution with time (in logarithmic scale) of the distribution $P(x,t)$ in a given energy landscape (drawn below). 
The evolution runs over $10^7$ iterations. The intensity in the grey scale is proportional to $(-\ln P(x,t))$,
i.e. darker regions correspond to higher values of $P(x,t)$."} 
\label{figchave}
\end{figure}

 \subsection{ Properties of the effective dynamics} 

\subsubsection { Diffusion front}

As a consequence of the Golosov localization, the distribution of the rescaled 
variable $X=\frac{x_{\{U, \eta\}}}{ (T \ln t)^2}$, with respect to the thermal noise $\eta$ in a given sample is asymptotically a Dirac delta distribution  $\delta(X-M)$ where $M=\frac{m_{\{U\}}(t)}{ (T \ln t)^2}$ is the rescaled variable of the effective dynamics.  To compute the averaged diffusion front over the samples (or 
equivalently over the initial conditions), it is then enough to study the distribution of $M$ over the samples \cite{rgsinaishort,rgsinailong}. This leads to the Kesten law \cite{kestenlaw,golosovdemi,golosovhitting}
\begin{eqnarray} 
P(X) = LT^{-1}_{p \to \vert X \vert} \left[ \frac{1}{p} \left(1-\frac{1}{\cosh \sqrt p} \right) \right ] = \frac{4}{\pi} \sum_{n=0}^{+\infty} \frac{(-1)^n}{2n+1} e^{ - \frac{\pi^2}{4} (2n+1) \vert X \vert} 
\end{eqnarray} 
which is an exact result of mathematicians \cite{kestenlaw, golosovhitting}.  This example explicitly shows how the strong disorder RG 
allows to obtain asymptotic exact results, and gives confidence in the new results of the method concerning finer properties. 

\subsubsection { Energy distribution} 

Similarly, the rescaled variable for the energy 
\begin{eqnarray} 
w=\frac{U(x(0))-U(x(t))}{(T \ln t)} \simeq \frac{U(m(0))-U(m(t))}{(T \ln t)} 
\end{eqnarray} 
is entirely given at large time by the effective dynamics.  
The reduced variable $w$ has the following limit law as $t \to \infty$:  \begin{eqnarray}
 {  D}(w) = \theta(w < 1) \left(4- 2 w-4 e^{-w} \right) + \theta(w \geq 1) \left(2 e - 4 \right) e^{-w} 
 \end{eqnarray} 
This law is continuous, like its derivative at $w=1$, but the second derivative 
 is discontinuous at $w=1$, which can seem surprising at first sight.
  Indeed, for any finite time, the energy distribution is analytic, and it is only in the limit of infinite time that the discontinuity appears for the rescaled variable.  It is interesting to note that in the recent mathematical work \cite{hu} over the return time to the origin after time $t$,
another not-analytical asymptotic distribution for a rescaled variable also
appears. The joint limit distribution of the position $X = \frac{x(t)-x(0)}{(T \ln t)^2}$ and the energy $ w=\frac{U(x(0))-U(x(t))}{(T \ln t)}$
may also be computed in Laplace transform \cite{energysinai}
\begin{eqnarray} 
 && \int_0^{+\infty} dX e^{-sX} { P}(X,w>1)   =  \frac{ \sinh {\sqrt s}}{ \sqrt s}
\left( e^{ {\sqrt s} \coth {\sqrt s}}-  2 \cosh {\sqrt s}   
\right)  e^{- w {\sqrt s} \coth {\sqrt s}} \nonumber \\
&& \int_0^{+\infty} dX e^{-sX} { P}(X,w<1) 
 =   \frac{ \sinh {\sqrt s} (2-w)}{ {\sqrt s}}
- \frac{ \sinh 2 {\sqrt s}}{ {\sqrt s}} e^{- w {\sqrt s} \coth {\sqrt s}}
 \end{eqnarray}

\subsubsection{ Aging properties}

The two-time diffusion front $\overline{P(x, t;  x', \vert 0,0)}$ 
presents an aging regime in $({{\ln t}/{\ln t'}})$.  In the rescaled variables  $X={(x/\ln^2t)}$ and $X'={(x'/\ln^2t)}$, the diffusion front is again determined by the effective dynamics.  The RG procedure allows to compute the joint law of the positions $\{m(t), m(t_w)\}$ at two successive times $t \geq t_w$ 
\cite{rgsinailong}.  In particular, this two time diffusion front presents a 
Dirac delta function $\delta(X-X')$, which means that the particle can be trapped in a valley from which it cannot escape between $t'$ and $t$.  The weight $D(t, t_w)$ of this delta function thus represents the probability of having $m(t)=m(t_w)$
\begin{eqnarray} 
D(t, t_w) = \frac{1}{3} \left(\frac{\ln t_w}{\ln t}\right)^2 
\left(5-2 e^{ 1 - \left(\frac{\ln t}{\ln t_w} \right)} \right) 
\label{dttw} 
\end{eqnarray}

\subsubsection{ Statistics of returns to the origin} 

The strong disorder RG also yields that the distribution of the sequence $\Gamma_1=T \ln t_1$, $\Gamma_2=T \ln t_2$... of the successive return times to the origin of effective dynamics $m(t)$ has a simple structure \cite{rgsinailong}:  it is a Markovian multiplicative process defined by the recurrence $\Gamma_{k+1}=\alpha_k \Gamma_k$, in which the coefficients $\{\alpha_i\}$ are independent random variables, distributed
with the law 
\begin{eqnarray} 
\rho(\alpha) = \frac{1}{\sqrt{5}} \left(\frac{1}{\alpha^{1+\lambda _-}}- \frac{1}{\alpha^{1+\lambda_+}}\right) \ \hbox{with} \ \lambda_{\pm} = \frac{3 \pm \sqrt 5}{2} 
\end{eqnarray} 
Here are two important consequences:  

$\bullet$ the total number $R(t)$ of returns to the origin during $[0, t]$ behaves as \begin{eqnarray}
 R(t) \sim \frac{1}{3} \ln (T \ln t) 
\end{eqnarray} 
whereas the total number $S (t)$ of jumps during $[0, t]$ behaves as
 \begin{eqnarray} 
S(t) \sim \frac{4}{3} \ln (T \ln t) 
\end{eqnarray} 
(but here there are correlations between times of jumps.)  

$\bullet$ the probability that $m(\tau)>0$ for $\tau \in ]0, t]$ 
involves an irrational persistence exponent \cite{rgsinailong}
\begin{eqnarray} 
\Pi(t) \sim \left[\frac{1}{(T \ln t)^2}\right]^{\overline{ \theta}} \ \ \hbox{with} \ \ { \overline{\theta}=\frac{3-\sqrt{5}}{4}=0.19...} 
 \nonumber
  \end{eqnarray}
 whereas the probability that a given walker $x(t)$ does cross the origin  during $]0, t]$ has the simpler persistence exponent
\begin{eqnarray}
 \Pi_1(t) \sim \left[\frac{1}{(T \ln t)^2}\right]^{ \theta} \ \ \hbox{with} \ \ { \theta=\frac{1}{2}}
\label{singl_pers} 
\end{eqnarray}

The number of returns to the origin for the effective dynamics
translate for quantum spin chain models
into the number of decimations above a given point,
a quantity which has been recently used to characterize
the entanglement properties of these random quantum
spin chains \cite{refaelentanglement}.

\subsection{ Localization properties}
 
\subsubsection{ Distribution of the thermal packet} 

The asymptotic distribution of the relative position $y = x(t)-m(t)$ with respect
 to the effective dynamics $m(t)$ 
corresponds to the Boltzmann distribution in an infinite Brownian valley \begin{eqnarray}
 P(y) = \left< \frac{e^{ - \beta U_1( \vert y \vert))}} { \int_0^{\infty} dx e^{ - \beta U_1(x)} + \int_0^{\infty} dx e^{-\beta U_2(x)}} \right>_{\{U_1, U_2 \}} 
\end{eqnarray} 
where the average is over two Brownian trajectories ${\{U_1, U_2\}}$ 
forming an infinite valley.  This formulation is equivalent to the Golosov
theorem \cite{golosovlocali}.  The law can be explicitly computed in Laplace transform in terms of Bessel functions \cite{locgolosov}, and
presents in particular the algebraic decay 
\begin{eqnarray} 
P(y) \opsim_{y \to \infty} \frac{1}{y^{3/2}} 
\label{loi3/2} 
\end{eqnarray} 
This can be understood as follows:  whereas  
 the Brownian potential $U(y)$ yields a typical decay  of order $e^{ - \beta \sqrt{\sigma y}} $ for the Boltzmann factor, there are rare configurations which return close to $U \sim 0$ at a long distance $y$, with a probability of order $1/(y^{3/2})$.  The correlation of two 
independent particles in the same sample computed in \cite{locgolosov}
\begin{eqnarray} 
C(l) = \lim_{t \to \infty} 2 \int_{-\infty}^{+\infty} dx \overline{ \left[ P(x, t|x_0,0) P(x+l, t|x_0,0) \right ]} 
\end{eqnarray} 
presents the same algebraic decay in $1/l^{3/2}$.  

\subsubsection{Localization parameters} 

The localization parameters, which measure the average probabilities to find $k$ particles at the same point at infinite time \cite{locgolosov}
\begin{eqnarray} 
Y_k = \lim_{t \to \infty} \int_{-\infty}^{+\infty} dx \overline{ \left[ P(x, t|x_0,0) \right]^k} = \frac{\Gamma^3(k)}{\Gamma(2 k)} (\sigma \beta^2)^{k-1} \end{eqnarray} 
are dominated for large $k$ by the very narrow valleys having a small 
partition function.  

\subsubsection{ Comparison with equilibrium in a Brownian potential}

These various observables which characterize the asymptotic statistics of the thermal packet in the Sinai diffusion actually coincide with their static analogs defined as the thermodynamic limit of the Boltzmann
distribution in a Brownian potential on a finite interval \cite{locgolosov}.  This convergence towards equilibrium of the thermal packet (whereas the effective dynamics remains forever out of equilibrium) is not true any more as soon as one adds a bias (cf section \ref{chapsinaibiais})

\subsubsection{ Thermal width and rare events}

 The algebraic decay (\ref{loi3/2}) for the limit law of the relative position $y$ implies that the second moment $\overline{<y^2>}$ diverges at infinite time.  To obtain its leading behavior at large time, it is in fact necessary to take into account the following rare events \cite{rgsinailong}:  (a) a renormalized valley can have two minima which are almost degenerated in energy;  (b) two neighboring barriers can be almost degenerated;  (c) 
a barrier can be near the decimation threshold $(\Gamma+\epsilon)$.
 These rare events appear with a weak probability of order $1/\Gamma$, but they give rise to a splitting of the thermal packet into two sub-packets, separated by a long distance of order $\Gamma^2$.  As a consequence, these 
rare events dominate the thermal width \cite{rgsinailong}
\begin{eqnarray} 
\overline{<x^2(t)>-<x(t)>^2} \oppropto_{t \to \infty} \frac{T}{\Gamma} (\Gamma^2)^2 = T (T \ln t)^3
 \end{eqnarray} 
 This behavior of the thermal width has been measured numerically \cite{chave}.
More generally, all divergent moments of order $k>1/2$ have for leading behavior
\begin{eqnarray} 
\overline{ \vert x(t)-<x(t) > \vert^k} \opsimeq_{t \to \infty} c_k T (T \ln t)^{2k-1} 
\end{eqnarray} 
where the constants $c_k$ can be calculated from the statistical properties of the rare events (a,b,c) described above \cite{rgsinailong}.

\subsection{ Relations with the general theory of slow dynamics and  metastable states} 

\subsubsection{ Metastable States} 

In the usual qualitative description of slow dynamics, 
 the idea of metastable states plays an important role. 
 As the true metastable states only exist in the mean-field approximation
or in the zero temperature limit, if one wishes to use this concept for systems in finite dimension at finite temperature, it is necessary to consider metastable states of finite lifetime \cite{biroli}, by 
decomposing the dynamics into two parts. There are on the one hand fast degrees of freedom, which convergence quickly towards a local quasi-equilibrium : they correspond to the `` metastable states ". On the other hand, there is a slow
out-of-equilibrium dynamics which corresponds to the evolution of the metastable states.  In this language, the strong disorder RG description 
of the Sinai random walk can be reformulated as follows: 

 $ \bullet $ the metastable states at time $t$ are the valleys of the 
renormalized landscape at scale $\Gamma = T \ln t$:  indeed, the walkers who were at $t=0$ inside this valley have not been able to escape from this valley
before time $t$.  

$ \bullet $ In each renormalized valley, there is a quasi-equilibrium described by a Boltzmann distribution inside the valley.  

$\bullet$ the slow dynamics corresponds to the evolution of the 
renormalized landscape with the scale $\Gamma = T \ln t$:  some metastable states disappear and are absorbed by a neighbor. 

 {\bf One-time observables and Edwards Conjecture} 

As the ` Edwards Conjecture', which proposes to calculate dynamical quantities via a flat average over all metastable states, has given rise to many recent works \cite{barratetc, desmedt}, it is interesting to reconsider from this point of view the strong disorder RG.  In the RG approach, all one-time observables are indeed computed via averages over the renormalized valleys (which are the metastable states), but with a measure which is not flat,
but is proportional to the length of the valleys.
Indeed, for a uniform initial condition, the length of a renormalized
valley represents the size of the attraction basin of the valley.

\subsubsection { Decomposition of the diffusion front over metastable states} 

If one wishes to describe at the same time the effective dynamics of the 
renormalized valleys and the Boltzmann equilibrium in each renormalized valley, one can write the diffusion front in a sample as
 \begin{eqnarray} 
P(x t|x_0 0) \simeq \sum_{V_\Gamma} \frac{1}{Z_{V_\Gamma}} e^{ - \beta U(x)} \theta_{V_\Gamma}(x) \theta_{V_\Gamma}(x_0)
 \label{valleysum} 
\end{eqnarray} 
where the sum is over all renormalized valleys $V_\Gamma$ existing at scale $\Gamma=T \ln t$.  The notation $\theta_{V}(x)$ indicates the 
characteristic function of the valley $V$, i.e. $\theta_{V}(x)=1$ if $x$ belongs to the valley and $\theta_{V}(x)=0$ if not.  Finally $Z_{V} = \int_{V} dx e^{ - \beta U(x)}$ represents the partition function of the valley $V$.  This expression of the diffusion front (\ref{valleysum}) exactly corresponds to the general construction in the presence of metastable states (see \cite{biroli, kurchanfermion} and references therein), in which the evolution operator 
 $e^{-t H_{FP}} $ is replaced by a projector on the states $(i)$ of energy $E_i < 1/t $ 
\begin{eqnarray} 
e^{-t H_{FP}} \sim \sum_{i} \vert P_i > < Q_i \vert 
\end{eqnarray} 
whose interpretation is clear:  `` everything fast has happened and everything slow has not taken place " \cite{kurchanfermion}.  For the Sinai model, the explicit expressions are as follows:  the right eigenvectors 
\begin{eqnarray} 
P_i(x) = \frac{ e^{-\beta U(x)}} { \int_{V_{\Gamma}^{(i)}} dx' e^{-\beta U(x')}} \theta (x \in V_{\Gamma}^{(i)}) 
\label{pi} 
\end{eqnarray} 
are positive, normalized, and their supports do not overlap, whereas the 
 left eigenvectors 
\begin{eqnarray} 
Q_i(x) = \theta (x \in V_{\Gamma}^{(i)})
 \end{eqnarray} 
are simply equal to 1 on the support of their corresponding right
eigenvector, and zero elsewhere.  In the Sinai model, one can in fact go beyond this one-time description by considering the dynamics of the metastable states to obtain information on the spectral properties of the Fokker-Planck operator.

\subsection{ Eigenfunctions of the Fokker-Planck operator} 

\subsubsection{ Fokker-Planck eigenfunctions associated to the effective dynamics} 

It is interesting to see how the one-time expression (\ref{valleysum}) 
changes upon the decimation of a renormalized valley, i.e. 
when a metastable state disappears.  The decomposition of the evolution
operator on the eigenvalues $E_n \geq 0 $ and the right $\Phi^r_n$ and  left $\Phi^l_n$ eigenfunctions of the Fokker-Planck operator $H_{FP}$ reads \begin{eqnarray} 
{ P}(x t|x_0 0) = < X \vert e^{-t H_{FP}} \vert x_0 > = \sum_{n} e^{ - { E_n} t} { \Phi}^R_n(x) { \Phi}^L_n(x_0) 
\label{relationschro} 
\end{eqnarray} 
Apart from the ground state $n=0$ of zero energy $E_0=0$ which corresponds to the Boltzmann equilibrium on the full sample 
\begin{eqnarray} 
\Phi_0^L(x) && = 1/\sqrt{Z_{tot}} \\ \Phi_0^R(x) && = e^{ - U(x)/T}/\sqrt{Z_{tot}}
 \end{eqnarray} 
the comparison with equation (\ref{valleysum}) leads to the following identifications for the excited states $n \geq 1$:  the energies $ { E_n}$ are determined by the RG scales $\Gamma_n=T \ln t_n = - T \ln { E_n}$ 
 corresponding to barrier decimations.  When a decimation takes place, two valleys $V_1$ and $V_2$ merge into one new renormalized valley $V'$, and the associated  eigenfunctions read \cite{locgolosov}
 \begin{eqnarray} 
&& { \Phi}_n^L(x) = \sqrt{\frac{Z_{V_1} Z_{V_2}}{Z_{V_1} + Z_{V_2}}} (\frac{1}{Z_{V_1}} \theta_{V_1}(x) - \frac{1}{Z_{V_2}} \theta_{V_2}(x)) \\ && { \Phi}_n^R(x) = e^{ - U(x)/T} { \Phi}_n^L(x) 
\label{eigen} 
\end{eqnarray} 
One can check that these eigenfunctions satisfy all the necessary properties of orthonormalization 
\begin{eqnarray} 
\int dx { \Phi}_n^L(x) { \Phi}_m^R(x) = \delta_{n, m} 
\end{eqnarray} 
and of the normalization of the thermal packet \begin{eqnarray} 
\int dx { \Phi}_n^R(x) = 0 
\end{eqnarray} 
Beyond the Sinai model, the structure (\ref{eigen}) in terms of partial partition functions seems more generally to describe the eigenfunctions of the Fokker-Planck operator for slow dynamics in which metastable states disappear in a hierarchical way.

\subsubsection{Spatial Structure of eigenfunctions:  2 peaks and 3 length scales}

The eigenfunction (\ref{eigen}) presents two peaks which correspond to the minima of the valleys $V_1$ and $V_2$. Each of the two peaks has a finite width, which represents the characteristic length associated to
the Boltzmann weight around the valley minimum.  The distance between the two peaks is of order $l(E)  \sim \Gamma^2 \sim (\ln E)^2$.  Far from the minima, but inside the renormalized valley $ r \leq \Gamma^2$, the decay
of the eigenfunction $\Phi_n^r(x)$ is controlled by the Boltzmann weight
$e^{-\beta U(r)}$ of behavior typical $e^{ - c \sqrt{r}} $. In particular at the edge of the valley $r \sim \Gamma^2$, the typical amplitude is of order
 $e^{ - c' \Gamma} $. Beyond the two valleys concerned, the simple approximation (\ref{eigen}) with theta functions becomes insufficient.  To estimate the decay of the eigenfunction at a distance $r \geq \Gamma^2$, one has to consider \cite{eigenhuse} that the two points are separated by a number of 
renormalized valleys of order $\frac{r}{\Gamma^2}$ and that 
the overlap between two neighboring valleys is not zero, but of order
 $e^{ - c '' \Gamma} $. A perturbation theory then leads to an exponential decay
of order $e^{ - C '' \frac{r}{\Gamma}} $, which indeed corresponds to 
the localization length $\lambda(E) \sim \Gamma \sim (- T \ln E)$ 
that has been computed for the associated Schr\"odinger problem via the exact Dyson-Schmidt method \cite{jpbreview}.  In conclusion, the properties of the eigenfunctions involve three length scales which coexist:  

$ \bullet$ the finite scale $l \sim 1$ which characterizes the width of a peak, and which is connected to the Golosov localization of the thermal packet.

  $ \bullet$ the scale $ l(E) \sim (\ln E)^2 $ which represents the distance between the two peaks and which is connected to the total distance reached
by the random walk at time $t \sim 1/E$.  

$ \bullet$ the scale $ \lambda(E) \sim (- \ln E) $ which characterizes the asymptotic exponential decay of the eigenfunction, and which corresponds to the localization length of the associated Schr\"odinger problem.
  \label{twolengtheigen}

{\bf Discussion}

 The Sinai model is thus a perfect `infinite disorder fixed point'.  The RG procedure gives a very complete picture of the asymptotic dynamics : it allows to obtain many explicit exact results on the effective dynamics of a particle, on the aging properties, on the localization
properties of the thermal packet, and on the rare events which control the thermal width.  
In addition, the description of the Sinai model in terms of renormalized valleys is an explicit example of metastable states of finite life time, and allows to understand the spatial structure of the Fokker-Planck eigenfunctions. 

 Let us mention to finish that various results obtained via the strong disorder
RG have now been confirmed by mathematicians, in particular 
the results concerning the weight of the singular part of the two time diffusion front \cite{dembo} and the statistics of the returns at the origin of effective dynamics \cite{cheliotis}. 


\section{Biased Sinai walk and associated directed trap model}

\label{chapsinaibiais}

\subsection{Models}

\subsubsection{Biased Sinai walk}

 The introduction of a constant force $F_0$ into the Langevin equation (\ref{langevin}) of the Sinai model is very natural.  This biased
model has even more interested mathematicians and physicists for a long time, because it presents a series of dynamic phase transitions \cite{kestenetal, derridapomeau, jpbreview} in terms of the dimensionless parameter $\mu=F_0 T/\sigma$.  In particular, there exists an anomalous diffusion phase
for $0<\mu<1$, which is characterized by the asymptotic behavior 
 \begin{eqnarray} 
\overline{ < x(t) >} \opsimeq_{t \to \infty} t^{\mu} 
\end{eqnarray} 
whereas for $\mu>1$, the velocity becomes finite:  $\overline{ < x(t) >} \sim V(\mu) t$ with $V(\mu)=F_0 (1-1/\mu)$.  
In the anomalous diffusion phase, the exact diffusion front is given in
terms of L\'evy stable distributions \cite{kestenetal,jpbreview,hushiyor}

The discrete Sinai model corresponds to the Master equation (\ref{mastersinai})
with asymmetric rates. The dynamical exponent $\mu$ is then defined by 
the positive root of the equation\cite{derridapomeau}:
\begin{equation}
\left[\left({w_{\rightarrow} \over w_{\leftarrow}}\right)^{\mu}\right]_{\rm av}=1\;.
\end{equation}
The anomalous diffusion phase $0<\mu<1$ is analogous to the Griffiths phase of a random magnet.

\subsubsection{Directed trap model}

It has been proposed for a long time \cite{feigelman, jpbreview,jpbreview2} 
that the biased Sinai model should be asymptotically equivalent 
to a directed trap model defined by the master equation
 \begin{eqnarray} 
\frac{d P_t(n)}{dt} = - \frac{P_t(n)}{\tau_n}+\frac{P_t(n-1)}{\tau_{n-1}} 
\label{masterdirected} 
\end{eqnarray} 
in which the $\tau_n$ are independent random variables 
distributed with the algebraic law
 \begin{eqnarray} 
q(\tau) \opsimeq_{\tau \to \infty} \frac{\mu}{ \tau^{1+\mu}} 
\label{lawtrap} 
\end{eqnarray} 
The anomalous diffusion phase $0<\mu<1$ then corresponds to
the case where the averaged trapping time is infinite. 
For a given trap $\tau$, the distribution of the escape time $t$ 
is exponential
\begin{eqnarray}
f_{\tau}(t) = \frac{1}{\tau} e^{- \frac{t}{\tau}}  
\end{eqnarray}
which yields after averaging over $\tau$  (\ref{lawtrap})
\begin{eqnarray}
\overline{ f_{\tau}}(t)  = \int_0^{+\infty} d\tau q(\tau) f_{\tau}(t)
= \int_0^{+\infty} \frac{dv}{v} q \left( \frac{t}{v} \right) e^{-v} 
\opsimeq_{t \to \infty}  \frac{ \mu \Gamma(1+\mu)}{t^{1+\mu}}
\label{meanf}
\end{eqnarray}
For a given sample $(\tau_0,\tau_1,...)$, the probability $P_t(n)$ 
for the particle to be in the trap $n$ at time $t$ reads
\begin{eqnarray}
P_t(n) = \int \prod_{i=0}^{+\infty} dt_i f_{\tau_i} (t_i) \theta(t_0+t_1...+t_{n-1}<t< t_0+t_1+...+t_{n} ) 
\end{eqnarray}
and it is thus directly related to the sum of a large number $n$
of independent variables $t_i$ distributed with
the law (\ref{meanf}) presenting an algebraic decay.
The diffusion front can be thus expressed in terms of L\'evy stable laws \cite{jpbreview}.

The directed character of this trap model allows to obtain many exact results,
since the particle visit sites only once in a fixed order, from left to right.
In particular, the thermal width has been exactly computed in \cite{aslangul}
\begin{eqnarray}
\overline { < \Delta n^2(t) >}  \equiv  
 \overline { \sum_{n=0}^{+\infty} n^2 P_t(n) -
[ \sum_{n=0}^{+\infty} n P_t(n) ]^2}
= \frac{1}{\Gamma (2 \mu)} \left( \frac{ \sin \pi \mu}{\pi \mu} \right)^3 I(\mu) t^{2 \mu} 
\label{aslanguleq}
\end{eqnarray}
where the integral $I(\mu)$ \cite{aslangul}
can be rewritten after a change of variables as
\begin{eqnarray}
I(\mu) 
&& =  \int_0^{1} dz \frac{ (1+z) z^{\mu} (1-z)^{2 \mu}}
 {  z^{2 \mu +2} + 2 \cos \pi \mu z^{\mu+1} +1}
\label{integralmu}
\end{eqnarray}
The result (\ref{aslanguleq}) shows that the the thermal
packet is spread over a length of order $t^{\mu}$.

\begin{figure}[ht ] \centerline{\includegraphics[height=6cm]{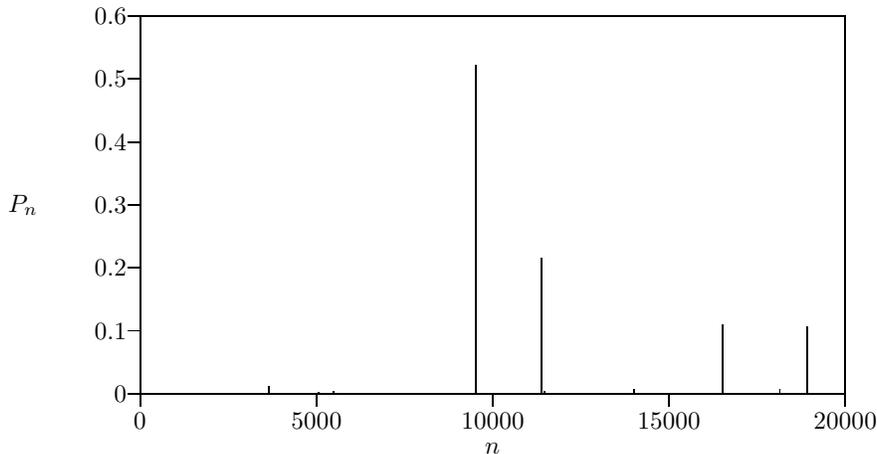}} \caption{\it Figure from Compte and Bouchaud \cite{compte} with its caption :
``Distribution of probability after a time $t=7 \times 10^{10}
$ for a particular sample of disorder in our 1D
directed random walk model with $\mu = 0.4$. The simulation was done
with $1.000$ particles in a lattice of $20.000$ sites.
One sees that this probability distribution is made of several sharp
peaks that gather a finite fraction of the particles. The
position of these peaks is scattered on a region of space of width
$t^\mu$."} \label{figcompte} 
\end{figure}

On the other hand, the infinite-time limit
of the localization parameter
for $k=2$ has been exactly in \cite{compte} :
their result (24)  
may be rewritten after a deformation of the contour
in the complex plane as
\begin{eqnarray}
Y_2 (\mu) && \equiv \lim_{t \to \infty} 
\sum_{n=0}^{+\infty} \overline{ [P_t(n)]^2} = \int_{-\pi}^{+ \pi} \frac{ d\theta}{2 \pi}
\  \frac{e^{i \theta \mu}-e^{i \theta}}{ 1- e^{i \theta (\mu+1)}}
\label{y2exact}
\end{eqnarray}
This expression shows that $Y_2$ is finite in the full phase $0 \leq \mu<1$
and vanishes in the limit $\mu=1$. 
How can this property coexist with the result \cite{aslangul}
for the thermal width ? The numerical
simulations of \cite{compte}
show that for a single sample at fixed $t$, the probability distribution 
$P_t(n)$ is made out of a few sharp peaks that have a finite weight
but that are at a distance of order $t^{\mu}$ (see Figure \ref{figcompte}). This explains why
at the same time, there is a finite probability to find two particles
at the same site even at infinite time, even if the thermal width 
diverges as $t^{2 \mu}$ at large time.

 \subsection{ Principle of the generalized RG} 

\begin{figure}[ht ] \centerline{\includegraphics[height=6cm]{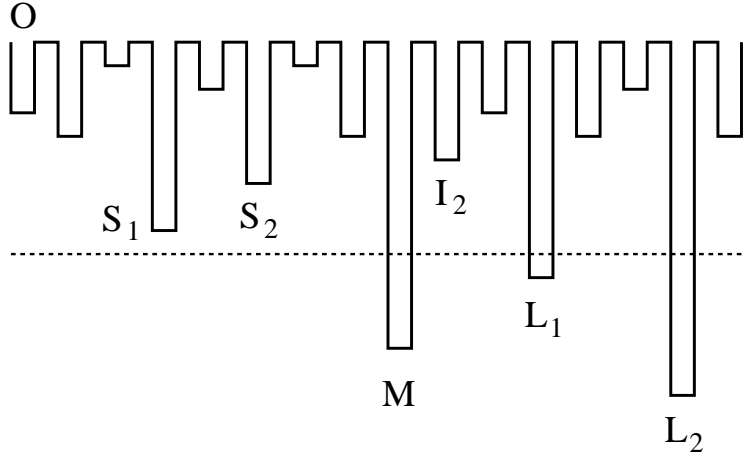}} \caption{\it  Hierarchical structure of the important traps for a particle starting at the origin.
The dashed line separates the ``small" traps (that have a trapping time
smaller than $t$) from the ``big" traps (that have a trapping time bigger 
than $t$). The first big trap called $M$ is occupied with
a weight of order $O(\mu^0)$. The next big trap $L_1$ and the biggest 
small trap $S_1$ before $M$ are occupied with weights of order $O(\mu)$.
The third big trap $L_2$, the biggest small trap $I_2$ between $M$ and $L_1$,
and the second biggest small trap $S_2$ before $M$ are occupied
with weights of order $O(\mu^2)$.} \cite{trapdirected} \label{figpieges} 
\end{figure}

The strong disorder RG presented in the previous section \ref{chapsinai} for the symmetric Sinai model can be extended to the biased case, but the
obtained results are exact in the limit of infinite time $t \to \infty$ only 
if the bias is very small $\mu \to 0$ \cite{rgsinailong}:
for instance, the RG yields an exponential diffusion front for the rescaled
variable $X = \frac{x(t)}{t^{\mu}}$ that coincide with the exact result involving
a L\'evy distribution \cite{kestenetal, jpbreview} only in the limit $\mu \to 0$.  The reason why the effective dynamics is not exact any more when $\mu$ is finite is that the distribution of the barriers $F_-$ against the bias (\ref{soluppdelta})  converges towards an exponential distribution of finite width proportional to $1/(2\delta) = T/\mu$.  This shows that the localization of the full thermal packet in a single renormalized valley at large time, which is valid in the limit $\mu \to 0$, is not exact any more for finite $\mu$.  
It is thus necessary to generalize the strong disorder RG approach to include the spreading of the thermal packet into several renormalized valleys
\cite{trapdirected}. Let us first describe this generalized procedure
for the directed trap model (\ref{masterdirected}) : in a given sample,
the diffusion front is a sum of delta functions with the hierarchic structure
explained on the figure (\ref{figpieges}). 

\subsection{ Results for the associated directed trap model} 

From this description of the diffusion front in each sample, one can compute
exact series expansion in $\mu$ for all observables.  
In particular, the explicit computations up to order order $\mu^2$
\cite{trapdirected} of the diffusion front for the rescaled variable $X=\frac{x}{t^{\mu}}$, of the thermal width 
\begin{eqnarray} 
\lim_{t \to \infty} \frac{\overline { < \Delta x^2 (t) >}} { t^{2 \mu}} = \mu (2 \ln 2) + \mu^2 [ - \frac{\pi^2}{6} + 2 \ln 2 (\ln2 -2+2 \gamma_E) ] +O(\mu^3) \label{widthexact} \end{eqnarray} 
and of the localization parameter
\begin{eqnarray} Y_2 (\mu) = 1 - \mu (2 \ln 2) + \mu^2 (4 \ln 2 \frac{\pi^2}{6}) +O(\mu^3) 
\label{y2exactexpansion} 
\end{eqnarray} 
coincide with the series expansions of the corresponding
results, for the diffusion front
 \cite{kestenetal, jpbreview}, for the thermal width given in Eq \ref{aslanguleq} and for the localization parameter 
 given in Eq \ref{y2exact}.  These comparisons with exact results obtained independently shows that the generalized RG procedure is exact order by order in $\mu$.  More generally, to compute observables at order $\mu^n$, it is enough to consider that the diffusion front is spread over $(1+n)$ traps and to average over the samples with the appropriate measure \cite{trapdirected}.

This approach allows to understand how the anomalous diffusion phase $0<\mu<1$
presents at the same time a diverging thermal width as $t^{2\mu}$ (\ref{widthexact}) 
together with a finite probability $Y_2(\mu)$ (\ref{y2exactexpansion}) of
finding two particles in the same trap at large time.  
The qualitative structure of the diffusion front in a given sample 
is in full agreement with the numerical simulations of A. Compte and J.P. Bouchaud (see Fig \ref{figcompte}).

\subsection{ Quantitative mapping between the biased Sinai model and the
directed trap model} 

\begin{figure}[ht ] \centerline{\fig{6cm}{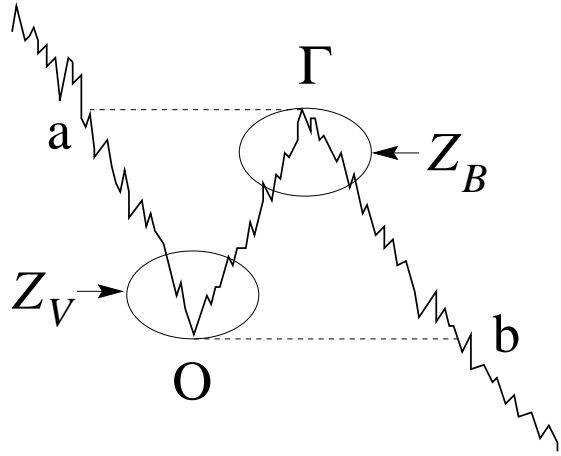}} \caption{\it 
Computation of the escape time from a renormalized valley
of barrier $\Gamma$ : the first-passage-time at $b$
for a particle starting at $0$
is dominated by the Arrhenius factor $e^{\beta \Gamma}$, and the prefactor
is the product of two partition functions :
$Z_V$ represents the partition function
of the bottom of the valley and $Z_B$ represents the partition function
of the inverse potential $(-V)$ near the top of the barrier $\Gamma$.
\cite{trapdirected}}  \label{figescapeti} \end{figure}

The saddle point analysis of the mean thermal exit time from a given renormalized 
valley of barrier $\Gamma$ shows that this time is exponentially distributed
as in the trap model, with trapping time given by
 \begin{eqnarray} 
\tau \opsimeq_{\Gamma \to \infty} \beta Z_B Z_V e^{\beta \Gamma} 
\label{theta1V}
 \end{eqnarray} 
which mostly depend on the barrier $\Gamma$ via the usual Arrh\'enius factor $e^{\beta \Gamma}$, but also on some details of the valley via the prefactor
that involves two partition functions functions $Z_V$ and $Z_B$ 
of Brownian valleys (see Figure \ref{figescapeti}).

The probability distribution of the trapping time over the renormalized valleys at scale $\Gamma$ can be computed from the probability distribution of the barriers and prefactors \cite{trapdirected} :  the result takes the same form as in the directed
trap model 
\begin{eqnarray} 
q_t(\tau) = \theta(t<\tau) \frac{\mu}{\tau} \left(\frac{t}{\tau} \right)^{\mu} \label{qttau} 
\end{eqnarray} 
with the following quantitative prescription
for the renormalization scale $\Gamma$ as a function of time
 \begin{eqnarray} 
\Gamma(t) = T \ln \left[ t \sigma^2 \beta^3 \left(\Gamma^{2} (1+\mu) \right)^{\frac{1}{\mu}} \right ] 
\label{gammat} 
\end{eqnarray} 
The corresponding length scale 
\begin{eqnarray} b(t) = \frac{ \Gamma^2 (\mu)}{\sigma \beta^2} \left[ t \sigma^2 \beta^3 \right]^{\mu} 
\label{bt} 
\end{eqnarray} 
is then in full agreement with the constant that has been recently computed
by mathematicians for the diffusion front \cite{hushiyor}.

In conclusion, the generalized strong disorder RG approach
allows to establish that the biased Sinai model and the directed trap model
are asymptotically equivalent from the point of view of their 
renormalized descriptions, up to a length scale that can be also exactly computed.

  \subsection{ Results for the biased Sinai model} 

\begin{figure}[ht ] \centerline{\fig{6cm}{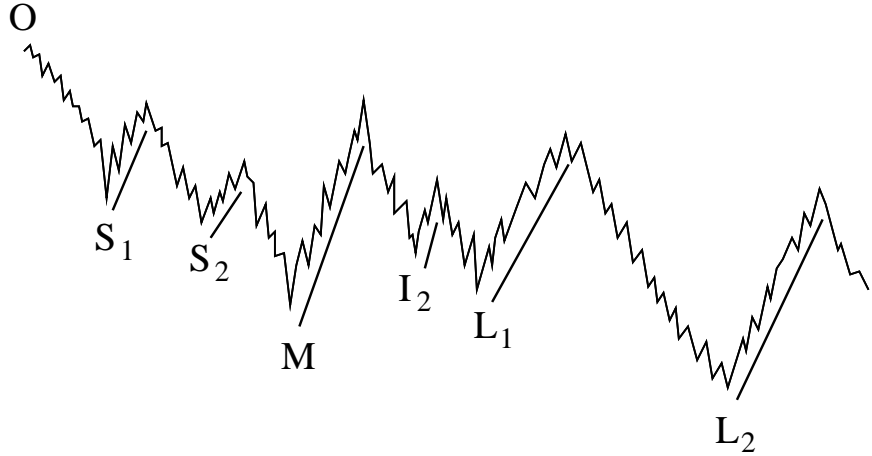}} \caption{\it  
Hierarchical structure of the important valleys for a particle starting at the origin.
The barriers against the bias that are emphasized by the straight lines
correspond to the depths of the trap model represented on Figure \ref{figpieges}.  The bottom $M$ of the renormalized valley that contains the origin
at scale $\Gamma$ is occupied with
a weight of order $O(\mu^0)$. The bottom $L_1$ of the next renormalized
valley and the bottom $S_1$ of the biggest 
sub-valley before $M$ are occupied with weights of order $O(\mu)$.
The next-nearest renormalized valley $L_2$, 
the biggest sub-valley $I_2$ between $M$ and $L_1$,
and the second biggest sub-valley $S_2$ before $M$ are occupied
with weights of order $O(\mu^2)$.} \label{figsinaib} \end{figure}

All the results for the directed trap model can be translated for the biased Sinai
model, one simply has to replace traps by renormalized valleys.  
The diffusion front in a given sample has a hierarchic structure 
sketched on the Figure \ref{figsinaib} which is the analog of Figure \ref{figpieges}.  For the quantitative results, we just need to replace the rescaled variable $X=\frac{n}{t^{\mu}}$ of the trap model
by the rescaled variable $X=\frac{x(t)}{b(t)}$ of the biased Sinai model (\ref{bt}).  In particular, the thermal width has for expansion
 \begin{eqnarray} 
\frac{ \overline { < \Delta x^2(t) >}}{t^{2 \mu}} = \frac { \left(\sigma^2 \beta^3 \right)^{2\mu}} { \sigma^2 \beta^4} \left[ \frac{(2 \ln 2)}{\mu^3} + [ - \frac{\pi^2}{6} + 2 \ln 2 (\ln2 -2-2 \gamma_E) ] \frac{1}{\mu^2} +O(\frac{1}{\mu}) \right ] \label{widthsinaidv} 
\end{eqnarray} 
And if one translates the exact result of the directed trap model \cite{aslangul}, one even obtains the thermal width in the whole phase $0<\mu<1$ of 
anomalous diffusion 
\begin{eqnarray} 
\frac{ \overline { < \Delta x^2(t) >}}   {t^{2 \mu}} = 
\frac { \left(\sigma^2 \beta^3 \right)^{2 \mu}}  { \sigma^2 \beta^4} \frac{\Gamma^4 (\mu)}{\Gamma (2 \mu)} \left(\frac{ \sin \pi \mu}{\pi \mu} \right)^3 I(\mu) \label{widthsinai}
 \end{eqnarray} 

In conclusion, the anomalous diffusion phase $x \sim t^{\mu}$ with $0 < \mu<1$ of the biased Sinai model is characterized by a localization on several  renormalized
valleys, whose positions and weights can be described sample by sample.  
The generalized strong disorder RG allows to compute all observables
via series expansions in $\mu$.  From the strong disorder RG method, this 
 shows that the usual procedure which is exact for infinite disorder fixed point, is an approximation which is already very interesting for
finite disorder fixed point :  the usual procedure corresponds to 
the leading term in a systematic expansion.


\section{Symmetric Trap model}

\subsection{ Model}

Trap models propose a very simple mechanism for aging \cite{jpbtraps}.
A particle performs a random walk in a landscape made of traps,
according to the master equation
 \begin{eqnarray} 
\frac{d P_t(n)}{dt} = - \frac{P_t(n)}{\tau_n}+\frac{P_t(n-1)}{2\tau_{n-1}} +\frac{P_t(n+1)}{2\tau_{n+1}} 
\label{masterdirected1} 
\end{eqnarray} 
The trapping times $\tau_n = e^{\beta E_n}$ are defined in terms of random energies
$E_n$  distributed with the following exponential distribution 
 \begin{eqnarray} 
\rho(E) = \theta(E) \frac{1}{T_g} e^{ - \frac{E}{T_g}} 
\label{rhoe} \end{eqnarray} 
This choice of exponential distribution comes from
the statistics of low energy states in
the Random Energy Model \cite{rem}, in the replica theory \cite{replica}, and more generally from the exponential tail of the Gumbel distribution  which represents an important universality class for extreme statistics.  The exponential distribution of energies (\ref{rhoe}) translates for the trapping time $\tau = e^{\beta E}$ into the algebraic law 
\begin{eqnarray} 
q(\tau) = \theta(\tau>1) \frac{\mu}{\tau^{1+\mu}} 
\label{qtau} 
\end{eqnarray}
 with exponent
 \begin{eqnarray} 
\mu = \frac{T}{T_g} \label{defmu} 
\end{eqnarray} 
At low temperature $T<T_g$, the average trapping time 
$\int d \tau \tau q(\tau)$ diverges, and this directly leads to aging effects.
 The aging properties of trap models have been much studied, either in the 
mean field version \cite{jpbtraps, benarousreview, fielding, junier}, or in the 1D version \cite{isopi, bertinjp,bertinthese}, which naturally appears in various physical contexts \cite{alexander, jpbreview, bubbledna}, and which presents two characteristic time scales for aging,
in contrast with the mean field case. The aim of this section is to understand this phenomenon via an appropriate strong disorder RG.  

\subsection{ Principle of the RG procedure} 

We have already described the strong disorder RG for the { \it  directed} trap model
in the previous section \ref{chapsinaibiais}. Here, in the { \it  symmetric} model, each site can be visited several times, which leads to an essential change \cite{trapsymmetric} :  a trap of the renormalized landscape will be characterized by two important times, namely (i) its trapping time $\tau_i$, which represents the typical time of exit towards its immediate neighbors (ii) its 
escape time, which represents the time needed to reach a deeper trap.

\begin{figure}[ht ] \centerline{\includegraphics[height=8cm]{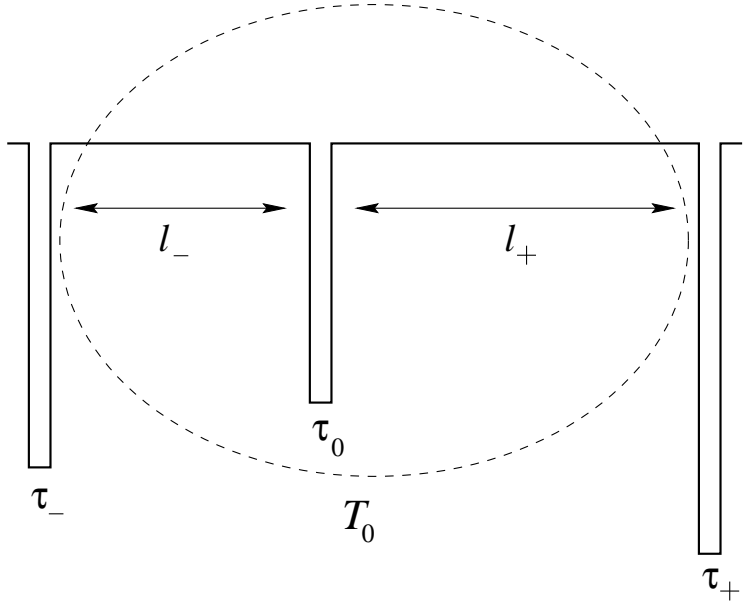}} \caption{\it  Definition of the escape time in the renormalized landscape:  the trap of trapping time $\tau_0$ is surrounded by two traps, the trap $\tau_+$ at distance $l_+$ and the trap $\tau_-$ at the distance $l_-$.  The escape time is the characteristic time necessary to reach $\tau_+$ or $\tau_-$ when one starts from $\tau_0$. \cite{trapsymmetric}} \label{defescape} 
\end{figure} 

The renormalized landscape at scale $R$ is defined as follows:  all the traps $\tau_i<R$ are replaced by a flat landscape, whereas all traps $\tau_i>R$ are kept.  In the renormalized landscape (see Figure \ref{defescape}), when the particle leaves the trap $\tau_0$, it escapes to the right or to the left with probability $1/2$.  If it escapes to the left, it will 
succeed to reach the next trap $\tau_-$ with probability $1/l_-$.  If it escapes to the right, it will succeed to reach the next trap $\tau_+$ with probability $1/l_+$.  If not, it will be re-absorbed by the trap $\tau_0$.  Asymptotically, the number of returns to the trap $\tau_0$ before the escape will thus be large, of order $R^{\mu}$.  One can show that the time spent inside $\tau_0$ during these multiple visits dominates over the time spent in the multiple unfruitful excursions and the final successful excursion \cite{trapsymmetric}.
The final result is that the escape time $t_{esc}$ from the trap $\tau_0$ has an exponential distribution 
\begin{eqnarray} 
P(t_{esc}) \opsimeq_{R \to \infty} \frac{1}{T_0} e^{ - \displaystyle \frac{t_{esc}}{T_0}} \label{pin} 
\end{eqnarray}
 with the characteristic time 
\begin{eqnarray} 
T_0 = \frac{2}{ \frac{1}{l_+} + \frac{1}{l _ -}} \tau_0 
\end{eqnarray} 
To describe the dynamics at time $t$, one has to keep the traps 
whose escape time $t_{esc}>t$ (from which particle have not been able to escape
at time $t$). This leads to the following choice for the RG scale $R(t)$ of the landscape as a function of time 
\begin{eqnarray} 
R(t) \simeq t^{\frac{1}{1+\mu}} 
\end{eqnarray} 
Moreover, within the limit $\mu \to 0$, the following effective dynamics becomes exact:  at time $t$, the particle starting from the origin at $t=0$
will be at time $t$ either on the first renormalized trap $M_+$ at distance $l_+$
on the right or on the first renormalized trap $M_-$ at distance $l_-$  on the left.  The weight of the trap $M_+$ is simply the probability $l_-/(l_+  + l_-)$ of reaching $M_+$ before $M_-$ in a flat landscape
\begin{eqnarray} 
P_{eff}(x, t) \sim \frac{l_+}{l_+ + l _ -} \delta(x+l _ -) + \frac{l_-}{l_+ + l _ -} \delta(x-l_+) 
\label{pefftrap} 
\end{eqnarray}
    This shows that the dynamics always remains out-of-equilibrium :  the weights of the two traps are not given by Boltzmann factors, they do not even depend on the energies of the traps, but only on their distances to the origin.

\subsection{ Results for the symmetric trap model} 

{ \bf Distance between traps in the renormalized landscape}

The distances between successive traps in the renormalized landscape are independent random variables and the reduced variable $\lambda=l/\xi(t)$ is exponentially distributed  \begin{eqnarray}
 P(\lambda)=e^{-\lambda} 
\end{eqnarray} 

with the characteristic length scale at time $t$ 
\begin{eqnarray} 
\xi(t) = \xi_0(\mu) \ t^{\frac{\mu}{1+\mu}} 
\end{eqnarray} 
with the prefactor 
\begin{eqnarray} 
\xi_0(\mu) = 1+O(\mu) 
\end{eqnarray} 

{ \bf Diffusion front} 

In the rescaled variable $X=x/\xi(t)$, the average of the diffusion front (\ref{pefftrap}) 
over the samples reads \cite{trapsymmetric}
\begin{eqnarray} 
g_{\mu}(X) = e^{ - \vert X \vert} \int_0^{+\infty} du e^{-u}  \frac{u}{\vert X \vert +u} +O(\mu) 
\end{eqnarray} 

{ \bf Localization parameters} 

The localization parameters, which represents the averages over the samples of the probabilities to find $k$ independent particles on the same site, are given by \cite{trapsymmetric}
\begin{eqnarray} 
Y_k (\mu) \equiv \lim_{t \to \infty} \overline{ \sum_{n=0}^{+\infty} P^k(n, t\vert 0,0)} = \frac{2}{(k+1)} +O(\mu) 
\label{resyksummary} 
\end{eqnarray} 
This result is in agreement with the numerical simulations of Bertin and Bouchaud \cite{bertinjp} who have obtained $Y_2 \to 2/3$ and $Y_3 \to 1/2$ in the limit $\mu \to 0$.

{ \bf Generating function of thermal cumulants}

 The thermal width reads
 \begin{eqnarray} 
c_2(\mu) \equiv \lim_{t \to \infty} \overline{\frac{<n^2>-<n>^2} { \xi^2(t)}} = 1+O(\mu) \end{eqnarray} 
and more generally, the others thermal cumulants can be derived from the generating function 
\begin{eqnarray} 
Z_{\mu}(s) \equiv \overline{ \ln < e^{-s \frac{n}{\xi(t)}} >} = \int_0^{+\infty} d \lambda e^{ - \lambda} \lambda \left(\frac{s \lambda}{2} \coth \frac{s \lambda}{2}-1 \right) +O(\mu) \end{eqnarray}

{ \bf Two-particle correlation function}

The two-particle correlation function reads
\begin{eqnarray}
C(l,t) && \equiv \overline{  \sum_{n=0}^{+\infty} \sum_{m=0}^{+\infty}
P(n,t\vert 0,0) P(m,t\vert 0,0) \delta_{l,\vert n-m \vert}}  \opsimeq_{t \to \infty} Y_2(\mu) \delta_{l,0}
+ \frac{1}{ \xi(t)} { C}_{\mu} \left( \frac{l}{\xi(t)} \right)
\nonumber \label{correform}
 \end{eqnarray}

The weight of the $\delta$ peak at the origin correspond
as it should to the localization  parameter $Y_2
=2/3 +O(\mu)$ (\ref{resyksummary}),
whereas the second term involves the following scaling function 
\begin{eqnarray}
 { C}_{\mu} (\lambda)= e^{-\lambda} \frac{\lambda}{3}   +O(\mu) 
\label{correlong} 
\end{eqnarray}

{ \bf The two aging correlations}

  The probability $\Pi(t+t_w,t_w)$ of no jump during the time interval
 $[t_w,t_w+t]$ presents a sub-aging scaling form
\begin{eqnarray}
\Pi(t+t_w,t_w) = {\tilde \Pi}_{\mu} \left( g= \frac{t}{ R(t_w)} \right)
= {\tilde \Pi}_{\mu} \left( g= [ {\tilde T}_0(\mu) ]^{\frac{1}{1+\mu}}
 \frac{ t}{ t_w^{\frac{1}{1+\mu}}} \right)
\end{eqnarray}
with
 \begin{eqnarray}
 {\tilde \Pi}_{\mu}^{(0)} (g)
= \int_0^1 dz \mu z^{\mu-1}
 e^{- z  g}
\label{resumepi0}
\end{eqnarray}
In particular, the asymptotic behavior reads
\begin{eqnarray}
\Pi(t+t_w,t_w) \opsimeq_{\frac{ t}{ t_w^{\frac{1}{1+\mu}}} \to + \infty} \left( \frac{t}{ t_w^{\frac{1}{1+\mu}}} \right)^{-\mu}
\left[ \mu +O(\mu^2) \right]
\end{eqnarray}

In contrast, the probability $C(t+t_w,t_w)$ of being at time $(t+t_w)$ 
in the trap where it was at time $t_w$ presents the aging scaling form
\begin{eqnarray}
C(t+t_w,t_w) = {\tilde C}_{\mu} \left( h= \frac{t}{R^{1+\mu}(t_w)} \right)
= {\tilde C}_{\mu} \left( h= {\tilde T}_0(\mu) \frac{t}{t_w} \right)
\end{eqnarray}
with 
\begin{eqnarray}
 {\tilde C}_{\mu}^{(0)} (h)
=  {\tilde C}_{\mu} (h)
= \frac{2 \mu}{ ( 2  h )^{ \mu}}
 \int_0^{\sqrt{2  h}}  dz z^{1+2 \mu}
 K_1^2( z )
\end{eqnarray}

In particular, the asymptotic behavior reads
\begin{eqnarray}
C(t+t_w,t_w) \opsimeq_{ \frac{t}{t_w} \to \infty} 
\left( \frac{t}{t_w} \right)^{-\mu} \left[ \mu +O(\mu^2) \right]
\end{eqnarray}

So these two time correlation present different aging scalings
because they involve the two different time scales, the trapping time and the escape time
of typical traps reached at time $t$.

 \subsection{ Linear and non-linear responses to an external field}

 \label{chapreponsetrap} 

 In the presence of an external field $f$, 
the master equation of the trap model becomes \cite{jpbreview}
 \begin{eqnarray} 
\frac{dP_t^{(f)}(n)}{dt} = P_t^{(f)}(n+1) \frac{e^{ - \beta \frac{f}{2}}}{2 \tau_{n+1}} + P_t^{(f)}(n-1)\frac{e^{+ \beta \frac{f}{2}}} { 2 \tau_{n-1}} - P_t^{(f)}(n) \frac{e^{+ \beta \frac{f}{2}} +e^{ - \beta \frac{f}{2}}}{2 \tau_{n}} 
\label{masterequation} 
\end{eqnarray} 
in which the transition rates satisfy the detailed balance condition 
 \begin{eqnarray} 
e^{-\beta U(n)} W_{ \{n \to n+1 \}}^{(f)} = e^{ - \beta U_{n+1}} W_{ \{n+1 
\to n \}}^{(f)} \label{detailedbalance} 
\end{eqnarray} 
in terms of the total energy $U_n$ containing the trap random energy $(-E_n)$ trap and 
the linear potential energy $(-f n)$ associated the 
external field $f$ 
\begin{eqnarray} 
U_n = - E_n - f n
 \end{eqnarray} 
The response has been recently studied by E. Bertin and J.P. Bouchaud \cite{bertinreponse,bertinthese} with scaling arguments and numerical simulations.  Their main result is as follows:  the Fluctuation-Dissipation theorem for the linear response is valid even in the aging sector, but the response becomes non-linear asymptotically. 
Here we will generalize the previous RG procedure for the unbiased
trap model to take into account the presence of an external field.

 { \bf Explicit Results for $t_w=0$} 

The RG procedure in external field involves a scale length $\xi(t, f)$ which represents the average distance between two traps in the renormalized landscape associated to time $t$
and field $f$.  This scale interpolates between the two behaviors \cite{trapreponse}
\begin{eqnarray} 
&& \xi(t, f) \opsimeq_{t \ll t_{\mu}(f)} t^{\frac{\mu}{1+\mu}} \left[ 1+O(\mu) \right ] \\ && \xi(t, f) \opsimeq_{t \gg t_{\mu}(f)} (\beta f t)^{\mu} \left[ 1+O(\mu) \right ] \label{xifbig} 
\end{eqnarray}

The average of the diffusion front over the samples 
takes the scaling form
 \begin{eqnarray} 
\overline{ P_t(n, f)} = \frac{1}{\xi(t, f)} g_{\mu} \left(X = \frac{n}{\xi(t, f)}, F = \beta f\xi(t, f) \right) 
\end{eqnarray} 
where the scaling function $g_{\mu}$ has the following expression in the limit $\mu \to 0$
\cite{trapreponse}
\begin{eqnarray} 
&& g_0(X, F)  =  e^{ - \vert X \vert}\left[ \theta(X>0) + \theta(X<0) e^{ - F \vert X \vert} \right ] \int_0^{+\infty} d \lambda e^{ - \lambda}
 \frac{1-e^{-F \lambda}} { 1 - e^{-F (\vert X \vert+ \lambda)}} 
\nonumber \\ && = e^{ - \vert X \vert}\left[ \theta(X>0) + \theta(X<0) e^{ - F \vert X \vert} \right ] \sum_{n=0}^{+\infty} \frac{F}{ (1+F N) (1+F+F N)} e^{-n F \vert X \vert} \nonumber 
\end{eqnarray}

The average position takes the scaling form
\begin{eqnarray}
\overline{ <x(t,f)>} \equiv  \sum_{x=-\infty}^{+\infty}
 x  \overline{ P_t(x,f)}
= \xi(t,f) { X}_{\mu} (F= \beta f\xi(t,f) ) 
\end{eqnarray}
with the following scaling function $ { X}_{\mu}$ in the 
 $\mu \to 0$ limit \cite{trapreponse}
\begin{eqnarray}
{ X}_{0}(F)    = \int_{0}^{+\infty} d\lambda \lambda
e^{-  \lambda} \frac{ \frac{ F \lambda}{2} \coth \frac{ F \lambda}{2} -1}{ F} = 1- \frac{1}{F}-\frac{1}{F^3} \psi'' \left( 1+\frac{1}{F} \right)  
\label{calx0f}
\end{eqnarray}

The thermal width reads
 \begin{eqnarray} \overline{ < \Delta x^2(t, f) >} = \xi^2(t, f) \Delta_{\mu}(F=\beta F \xi(t, f)) \label{defthermalwidtf} \end{eqnarray} 
where the scaling function$ { \Delta}_{0}$ is in fact directly connected to the function $ {  X}_{0}$ \cite{trapreponse}
\begin{eqnarray} 
\Delta_0(F) && = \left[ \frac{1}{F} + \frac{d}{d F} \right ] {  X}_0(F) \label{simpledeltafrommean} 
\end{eqnarray}

{ \bf Explicit results for $t_w>0$} 

The state reached at time $t_w$ in the absence of field must be regarded as an initial condition for the dynamics in field with time $(t-t_w)$.  Since at time $t_w$, the particle is typically in a trap $\tau>R(t_w, f=0)$, the effective dynamics starts again only when the decimation procedure becomes again activate, i.e. for $R(t-t_w, f)>R(t_w, f=0)$.  It is thus useful to introduce the parameter 
\begin{eqnarray} 
\alpha(t, t_w, f) \equiv \frac{ \xi(t-t_w, f)} { \xi(t_w, f=0)} = \frac{ R^{\mu}(t-t_w, f)} { R^{\mu}(t_w, f=0)} 
\label{defalpha} 
\end{eqnarray} 

which measures the ratio of the length scales for the two 
renormalized landscapes at $t_w$ and $t$.  In the sector $\alpha<1$, the response is dominated by rare events, whereas in the sector $\alpha(t, t_w, f)>1$, the answer is dominated by the effective dynamics.  The time domain $(t-t_w)$ corresponding to the sector $\alpha>1$ depends on the relative values of $t_w$ and time $t_{\mu}(f)$.  

(i) For $t_w \ll t_{\mu}(f)$, the sector $\alpha>1$ corresponds to the time domain $(t-t_w)>t_w$, and the parameter $\alpha$ behaves according to 
\begin{eqnarray} \alpha(t, t_w, f) && = \left(\frac{ t-t_w} { t_w} \right)^{\frac{\mu}{1+\mu}} \ \ \rm{for} \ \ t_w < t-t_w \ll t_{\mu}(f) \label{ashort} \end{eqnarray}
and
\begin{eqnarray} 
\alpha(t,t_w,f) && =  
\left( \frac{  \frac{\beta f}{2}   (t-t_w)}
{ t_w^{\frac{1}{1+\mu}}} \right)^{\mu}
\ \ \rm{for} \ \   t-t_w \gg   t_{\mu}(f)
\label{along}
\end{eqnarray}

(ii) For $t_w \gg t_{\mu}(f)$, the sector $\alpha>1$ corresponds to the time domain $(t-t_w)>\frac{2}{\beta f} t_w^{\frac{1}{1+\mu}}$, and the parameter $\alpha$ behaves like (\ref{along}) everywhere.  

By studying the statistics of the renormalized landscape at two successive RG scales, one obtains explicit results in all the sector $\alpha>1$ within the limit $\mu \to 0$. 
In terms of the rescaled variables  $Y=\frac{ x(t)-x(t_w)}{\xi(t-t_w, f)}$, $F=\beta F \xi(t-t_w, f)$ and $\alpha = \xi(t-t_w, f)/\xi(t_w, f=0)$, the averaged distribution of $Y$ 
takes the form
 \begin{eqnarray}
 P \left(Y;  F, \alpha \right) = \frac{1}{ \alpha} \delta(Y) + \left[ \theta(Y \geq 0) + e^{-F \vert Y \vert} \theta(Y \leq 0) \right ] G_{ns} \left(\vert Y \vert, F, \alpha \right) \label{reslawy} 
\end{eqnarray} 
The singular part in $\delta$ represents the particles which have not been able to
 escape towards another renormalized trap between $t_w$ and $t$.  The regular part is defined in terms of the function \cite{trapreponse}
\begin{eqnarray} 
G_{ns} \left(Y, F, \alpha \right)  = \frac{e^{-Y}}{2} \int_0^{+\infty} du \frac{1-e^{-F u}} { 1-e^{-F \left(Y+ u\right)}} \left[ (2 - e^{ - (\alpha-1) Y}) e^{ - u} - e^{ - \alpha u} \right ] 
\end{eqnarray}

Complementary results concerning in particular the average position and the thermal width are given in \cite{trapreponse}.

The diffusion front (\ref{reslawy}) presents a very simple asymmetry in $Y \to -Y$ that implies that the Fluctuation-Dissipation theorem for the linear
response is always satisfied. This simple asymmetry of the two-time diffusion front
is actually valid for any $\mu$ as a consequence of a special
dynamical symmetry of the master equation \cite{trapnonlinear} : so 
the validity 
of the Fluctuation-Dissipation theorem does not mean that there is some local
equilibrium, but is a consequence of this special symmetry of the Master Equation
that prevents any FDT violation.


\section{Classical Spin chains : Equilibrium and Coarsening dynamics}

\label{chaprfim}

The thermodynamics of classical disordered spin chains can be studied via 
product of random  $2 \times 2$ transfer matrices.  In particular, the free energy corresponds to the Lyapunov exponent and can be studied via the Dyson-Schmidt method:  the book \cite{luckbook} summarizes the many results obtained within this framework. 
Here we will instead analyze the equilibrium at low temperature in terms of Imry-Ma domains.

\subsection{Models : Random-field Ising chain and spin-glass chain in external field}

The random field Ising chain is defined by the following Hamiltonian for classical spins $S_i=\pm 1$
\begin{eqnarray} {  H} = - J \sum_{i=1}^{N-1} S_i S_{i+1} - \sum_{i=1}^{i=N} h_i S_i 
\label{hamilrfim} 
\end{eqnarray} 
where the fields $\{h_i\}$ are independent random variables, of zero average $ \overline{h_i}=0$ (see \cite{rgrfim} for the non-zero case) and of variance \begin{eqnarray} g \equiv \overline{h_i^2} 
\end{eqnarray} 
According to the Imry-Ma argument \cite{imryma}, the ground state at zero temperature is disordered:  there are $(+)$ and $(-)$ domains which alternate, the typical size of a domain being the Imry-Ma length 
\begin{eqnarray} L_{IM} = \frac{4 J^2}{g} 
\end{eqnarray} 
The Imry-Ma argument is based on the competition between
the energy cost $4J$ of two domain walls, with the typical energy gain $\vert \sum_{i=x}^{x+L} h_i \vert_ { \text{typ}} \sim 2 \sqrt{ g L}$ of a favorable disorder fluctuation on an interval of length $L$.  As a consequence, for $L>L_{IM}$,  it becomes favorable to create domains to benefit from the favorable fluctuations of the random fields.  

 The spin glass chain in external field is defined by the following Hamiltonian for classical spins $\sigma_i=\pm 1$
 \begin{eqnarray} 
 {  H} = -\sum_{i=1}^{N-1} J_i \sigma_i \sigma_{i+1} - \sum_{i=1}^N h \sigma_i 
 \end{eqnarray} 
where the couplings $\{J_i\}$ are independent random variables.  In the particular case where $J_i=\pm J$ with probability $(1/2,1/2)$, there is an equivalence with the random field chain (\ref{hamilrfim}) with $h_i=\pm h$ via a direct gauge transformation.
The physical interpretation of this correspondence between the two models is as follows. In zero field $h=0$, the two ground states of the spin glass 
 correspond to the two ferromagnetic states of the pure Ising chain. 
  In the presence of $h>0$, the domain walls of the random field chain
correspond to the frustrated bonds $J_i \sigma_i \sigma_{i+1} = J S_i S_{i+1} < 0$ of 
the spin glass.  In the following, we will thus describe the results only in the language of the random field chain, because it is immediate to translate them for the spin glass in external field.
For the spin-glass chain in external field with continuous distribution $\rho(J)$ of the couplings $J_i$, the mapping to the ferromagnetic chain
in random fields breaks down, and we refer to the reference \cite{cmtg}
for the statistical properties of the ground state and low energy excitations.

\subsection { RG construction of the ground state}

\label{rfimground}

The Hamiltonian (\ref{hamilrfim}) for the spins corresponds to the following 
Hamiltonian for the domain walls $A_{\alpha}(+\vert-)$ and $B_{\alpha}(-\vert+)$ that appear in the alternate fashion $A_1 B_1 A_2 B_2... $:  
\begin{eqnarray} {  H} = 2 J (N_A + N_B) + \sum_{\alpha=1}^{N_A} V(a_\alpha) - \sum_{\alpha=1}^{N_B} V(b_\alpha) \label{hparois} 
\end{eqnarray} 
in term of the Brownian potential seen by the domain walls
\begin{eqnarray} 
V(x) = - 2 \sum_{i=1}^{x} h_i \label{potsinai} 
\end{eqnarray} 
So each domain wall (A or B) costs an energy $2J$, all
 domain walls of type $A_{\alpha}(+\vert-)$ see the potential $V(x)$ 
whereas all
 domain walls of type $B_{\alpha}(-\vert+)$ see the opposite potential $(-V(x))$.

\begin{figure}[ht] \centerline{\includegraphics[height=3cm]{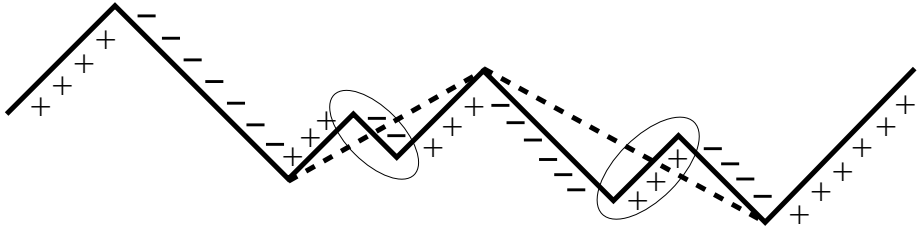}} \caption{\it   Illustration of the RG procedure to construct the ground state in a given sample
(see text for more details)}
 \label{figrgprocedure} \end{figure} 

\begin{figure}[ht] \centerline{\includegraphics[height=5cm]{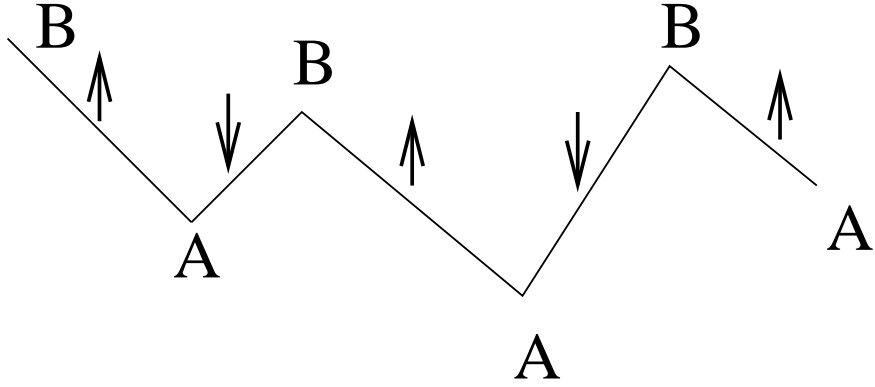}} \caption{\it   Picture of the ground-state : the zig-zag line represents the renormalization
of the random potential $V(x)=2 \sum_{i=1}^x h_i$ seen by the domain walls,
where only extrema separated by $\Delta V> 4 J$ have been kept.
Then all maxima are occupied by a domain wall $B(-\vert+)$, and all
minima are occupied by a domain wall $A(+\vert -)$.
So descending bonds are domains of $+$ spins, whereas ascending bonds are
 domains of $-$ spins. \cite{rgrfim}} \label{figparois} \end{figure}

The ground state may now be constructed via the following RG procedure
(see Fig \ref{figrgprocedure}).
We start from the state where each spin is aligned with its local field
$S_i=sgn(h_i)$, i.e. each local extremum of the Brownian potential
$V(x) =  2 \displaystyle \sum_{i=1}^{x} h_i$ is occupied by a domain wall.
Then we iteratively eliminate the pair of domain walls that are separated by
the smallest potential difference $\Delta V=\Gamma$, as long as it corresponds
to a decrease of the total energy $\Delta E= -4J + \Gamma <0$.
So the RG procedure has to be stopped at scale 
$\Gamma_{eq}=4 J$, the final state giving the structure
of the ground state into Imry-Ma domains, as shown on Fig. \ref{figparois}. 

In the limit where the Imry-Ma length $L_{IM}$ is large, the statistical properties of
the ground state can be obtained from the fixed point associated to the RG rules \cite{rgrfim}.
In particular, the lengths of Imry-Ma domains are independent variables, distributed with the law 
\begin{eqnarray} 
P^*(\lambda) = \pi \sum_{n = -\infty}^{\infty} \left(n+\frac{1}{2}\right) (-1)^n e^{ - \pi^2 \lambda \left(n+\frac{1}{2}\right)^2}
\label{peqlambdarfim} 
\end{eqnarray}
for the rescaled variable $\lambda=\frac{l}{2 L_{IM}}$.  
The two-point correlation function then reads \cite{rgrfim}
 \begin{eqnarray}
  \overline{\langle S_0 S_{x}  \rangle} = \sum_{n=-\infty}^{\infty} \frac{48 + 64 (2n+1)^2\pi^2 g \frac{\vert x \vert}{\Gamma^2}}{(2n+1)^4 \pi^4} e^{-(2n+1)^2\pi^2 2 g \frac{\vert x \vert}{\Gamma^2}} 
\label{spacecorrerfim} 
\end{eqnarray}
with $\Gamma=\Gamma_{eq}=4J$. In particular, the $n=0$ term yields the correlation length
 \begin{eqnarray}
\xi(T=0) = \frac{ 8 J^2}{\pi^2 g}
\end{eqnarray}
in agreement with the exact solution via the Dyson-Schmidt method \cite{luckbook}.

 \subsection{  Low temperature properties from two-level excitations} 

The thermal fluctuations that exist at very low temperature
in disordered systems are often attributed to
the existence of some two-level excitations.
For the random field Ising chain, these thermal excitations
 above the ground-state have been numerically studied
in \cite{alava} via transfer matrix.
In this section, we describe how the strong disorder RG approach allows
to identify precisely the `two-level' excitations 
 the random field Ising chain and to compute their statistical
properties \cite{twolevel}.

The excitations of nearly vanishing energy above the ground state
are rare events of two kinds, which have been called rare events of type (a)
and of type (c)  \cite{rgrfim,twolevel} :

\begin{figure}[ht]
\centerline{\includegraphics[height=6cm]{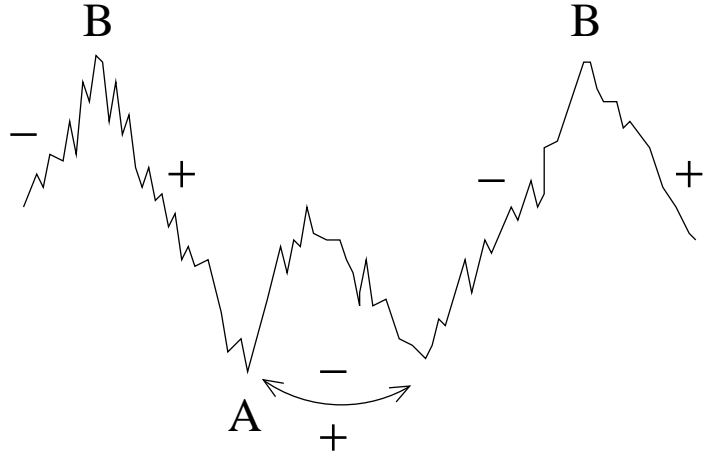}} 
\caption{ \it  Representation of a two-level excitation of type (a) :
a domain wall $A$ of the ground state may have two nearly degenerate optimal positions,
separated by a distance $l$ if $\Delta V = 2 \sum_{i=1}^l h_i \sim 0$. \cite{twolevel}} 
\label{rarea}
\end{figure}

(i) The excitations of type (a) involve a single domain wall
which has two almost degenerate optimal positions (see Figure \ref{rarea}).
These excitations have a length of order $l \sim L_{IM}$,
but concerns only a small fraction of order $1/\Gamma_J$
of the domain walls. Their density reads
\begin{eqnarray} 
\rho_a^{large}(E=0,l) dl = \frac{\pi^2}{ \Gamma_J L_{IM}^2} 
 \sum_{n=1}^{+\infty} n^2 e^{- n^2 \pi^2 \frac{  l}{ 2 L_{IM}}}
\label{rhoal}
\end{eqnarray}

The normalization diverges at small $l$ as a result of the continuum limit,
but is regularized by the lattice (see \cite{twolevel} for more details). 
Indeed, nearly degenerate excitations of small length $l \sim 1$
also exist. They are not described by the universal scaling
regime (\ref{rhoal}), but depend on specific properties
of the initial distribution $P(h_i)$.
For instance, the excitations of length $l=1$ correspond to the domain walls
that have a neighbor with a small random field $  h_i  <T \to 0$.
The density of such excitations
is simply proportional to the density $n = 1/L_{IM}$ 
of domain-walls in the ground-state, and to the weight $P(h_i=0)$ of the initial
distribution at the origin
\begin{eqnarray}
\rho_a(E=0,l=1) = \frac{1}{L_{IM}} 2 P(h_i=0)
\label{small1}
\end{eqnarray}
More generally, the statistics of small excitations $l=2,3,..$
 is governed by the probabilities of returns to the origin
for a constrained sum of $l$ of random variables.

\begin{figure}[ht]
\centerline{\includegraphics[height=6cm]{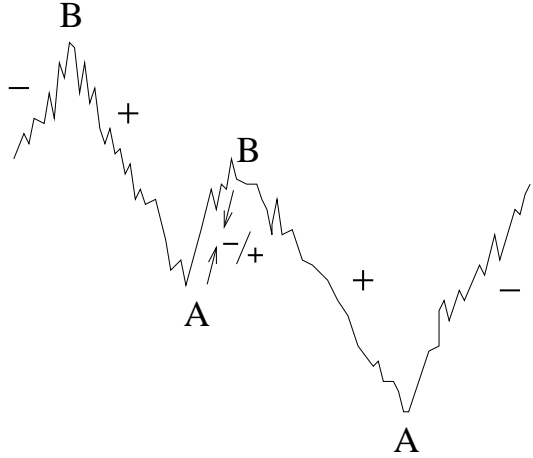}} 
\caption{ \it  Representation of a two-level excitation of type (c) :
a pair (A,B) of neighboring domain walls separated by a distance $l$
may appear or annihilate with almost no energy cost 
if $\Delta V = 2 \sum_{i=1}^l h_i \sim 4 J$. \cite{twolevel}} 
\label{rarec}
\end{figure}

(ii) The excitations of type (c) involve a pair of domain walls which can appear or annihilate with almost no energy cost (see Figure \ref{rarec}).
These excitation have by definition a large length $l \sim L_{IM} \gg 1$,
but they
 concern only a small fraction of order $1/\Gamma_J$
of the domain walls. Their density reads \cite{twolevel}

\begin{eqnarray} 
\rho_c(E=0,l)  = \frac{\pi^2}{ \Gamma_J L_{IM}^2} 
 \sum_{n=1}^{+\infty} (-1)^{n+1} n^2 e^{- n^2 \pi^2 \frac{  l}{ 2 L_{IM}}}
\label{rhocl}
\end{eqnarray}

So there exist on one hand large excitations of length
$l \sim L_{IM} \gg 1$, whose universal probability densities
are given by the explicit expressions (\ref{rhoal},\ref{rhocl}),
and there are on the other hand small excitations of length $l \sim 1,2,..$, whose statistics
depend on the initial random field distribution.

To compare with the existing rigorous results
\cite{theoluck}, we now consider that the initial distribution $P(h_i)$ of the random fields
has the following scaling form
\begin{eqnarray}
P(h_i)= \frac{1}{H} { P} \left(\frac{\vert h_i \vert}{H} \right)
\label{examplescaling}
\end{eqnarray}
where ${ P}(x)$ is a continuous function, such as
for instance the exponential distribution ${ P}(x)=e^{- x}/2$
considered in \cite{theoluck} among other cases.
In the regime where the Imry-Ma length is large,
 the results obtained 
are in agreement (see \cite{twolevel} for more details) 
with the exact results of the Dyson-Schmidt method \cite{theoluck}.
In addition, the RG approach 
shed light on the influence of small/large
excitations on various observables :
indeed, the specific heat 
\begin{eqnarray}
C(T)= T \frac{\pi^2}{6} \int dl \rho(E=0,l) +O(T^2)
\end{eqnarray}
is dominated by the small non-universal excitations of length $l \sim 1$,
that depend on the details of the disorder distribution,
whereas the Edwards-Anderson order parameter
\begin{eqnarray}
 q_{EA}= \overline{ <S_i>^2} = 1- 2 T
 \int_0^{+\infty} dl \ l \rho(E=0,l) +O(T^2)
\end{eqnarray}
and the magnetic susceptibility
\begin{eqnarray}
\overline{\chi} \equiv \frac{N}{T} \left( \overline{ <m^2>-<m>^2} \right)
= 2
\int_0^{+\infty} dl \ l^2 \rho(E=0,l) +O(T)
 \end{eqnarray}
are dominated by the large excitations whose length $l$ is of order of the Imry-Ma length $L_{IM}$, and whose properties are universal with respect
to the initial disorder distribution, since they only depend upon its variance.
 These excitations are rarer
than the small ones, but involve a larger number of spins.

\subsection{  Coarsening Dynamics}

The RG procedure that we have described above to construct the ground state
sample by sample has in fact an interesting dynamical interpretation in
the context of phase ordering kinetics (for a review see \cite{bray1}):
it corresponds to the coarsening dynamics at low temperature
from a random initial
configuration with typical domain size $l \sim 1$ towards the
equilibrium state with typical domain size $l \sim L_{IM}$.
Indeed, the Glauber dynamics involves the transition rate $W(S_j \to - S_j) = \frac{ e^{-\beta \Delta E}} { e^{\beta \Delta E} + e^{-\beta \Delta E}} $ 
with  $ \Delta E=2 J S_j (S_{j-1} + S_{j+1}) + 2 h_j S_j$, i.e. more explicitly
in terms of domain walls 
 \begin{eqnarray} 
&& \Delta E \{ \hbox{ creation of 2 walls} \} = 4 J \pm 2 h_j \nonumber \\ && \Delta E \{ \hbox{ diffusion of a wall} \} = \pm 2 h_j \nonumber \\ && \Delta E \{ \hbox{ annihilation of 2 walls} \} = -4 J \pm 2 h_j \end{eqnarray} 
So in the regime $\{h_i \} \sim T \ll J$ on which we will concentrate
in this section, the coarsening is described by the following  reaction-diffusion model in the Brownian potential 
 \begin{itemize} 
\item the domain walls A diffuse towards the minima 
\item the domain walls B diffuse towards the maxima 
\item Annihilation when meeting $A+B \to \emptyset$ 
\end{itemize} 
One can moreover show \cite{rgrfim} that at large time, all maxima and minima
are occupied by domain walls, as on the Figure (\ref{figparois}), which allows
to derive directly the properties of the reaction-diffusion process at time $t$ from the landscape RG at scale $\Gamma=T \ln t$.

In particular, the lengths of Imry-Ma domains are independent variables 
during the coarsening (this is not true for the pure Ising case \cite{zeitak}), and are distributed with the law (\ref{peqlambdarfim}) for the rescaled
variable $\lambda = \frac{2 g l}{(T \ln t)^2}$. The two spin correlation
also take the same form as in the ground state (\ref{spacecorrerfim}) 
with the RG scale $\Gamma =T \ln t$.

The two-time autocorrelation reads  \cite{rgrfim} 

\begin{eqnarray}  
\overline{ \langle S_i(t) S_i(t') \rangle}  =\frac{4}{3}  \left(\frac {\ln t'}{\ln t} \right)  -\frac{1}{3} \left( \frac {\ln t'} {\ln t} \right)^2  
\end{eqnarray}  
Taking into account the coarsening length scale $L(t) \sim (\ln t)^2$, the non-equilibrium
autocorrelation exponent defined by the asymptotic behavior \cite{huse_auto}
\begin{eqnarray}  
\overline{ \langle S_i(t) S_i(t') \rangle}  \sim  \left(\frac {L( t')}{L( t)} \right)^{\lambda}   
\end{eqnarray}  
has thus for value $\lambda=1/2$ here. 

Various persistence exponents can be also obtained \cite{rgrfim} 

  $\bullet$ The probability that a spin does not flip during the time interval $[0,t]$ decays as $  \left( \frac{1}{(T \ln t)^2}\right)^{\theta} $ with   {$\theta=1$}.     

 $\bullet$ The probability that the thermal average  $<S_i(t)>$ keeps the same sign during $[0,t]$ decays as  $  \left( \frac{1}{(T \ln t)^2} \right)^{\overline \theta} $ with   {$\overline \theta= \frac{3-\sqrt 5}{4}$}.    

$\bullet$ The probability that an initial domain has not disappeared
at time $t$ decays as $  \left( \frac{1}{(T \ln t)^2}\right)^{\psi} $ with   {$\psi= \frac{3-\sqrt 5}{4}$}. 

We now turn to the question of the fluctuation-dissipation violation,
which is quantified via   
the ratio $X$ defined by \cite{cuku}  
\begin{eqnarray}  
T \ R(t,t_w)= X(t,t_w) \  \partial_{t_w} C(t,t_w)  
\end{eqnarray} 
where $C(t,t_w)$ represents the connected thermal correlation
   \begin{eqnarray}  
C(t,t_w)= \sum_x \overline{< S_0(t) S_x(t_w) > -< S_0(t)> <S_x(t_w)>}  \end{eqnarray} 
and where $R(t,t_w)$ represents the linear response to a uniform external field $H$ applied from $t_w$
  \begin{eqnarray}   \overline{< S_0(t)>} = H \int_{t_w}^t du R(t,u)  \end{eqnarray}  
For the coarsening of the random field Ising chain, three
regimes have been found \cite{rgrfim}

(i) quasi-equilibrium of the domain walls in renormalized valleys \begin{eqnarray} 
X(t, t_w) = 1 \qquad \text{for} ~~ 0< \frac{\ln (t-t_w)}{\ln t_w} < 1 \end{eqnarray} 

(ii) a nontrivial ratio X when the effective dynamics of the valleys takes place again 
\begin{eqnarray} 
X(t, t_w) = \frac{t + t_w}{t} \qquad \text{for} ~~ \frac{t-t_w}{t_w} ~~~ \text{fixed} 
\end{eqnarray} 

(iii) a final aging regime
\begin{eqnarray} 
X(t, t_w) = \frac{t}{t_w \ln t_w} (1 + \frac{24}{7} \frac{\ln^2 t_w}{\ln^2 T}) ~ \text{for} ~ \frac{\ln t}{\ln t_w} > 1 
\end{eqnarray} 
In particular,  $X$ grows towards $+\infty$, because the truncated thermal correlations are very weak with respect to the response.

In comparison with mean field models, the ratio $X$ is not a function of the correlation $C(t, t_w)$ alone :  it is because of the two scales $(t, t_w)$ and $(\ln t, \ln t_w)$).  In addition, in mean field the ratio $X$ belongs to the interval $ [ 0,1]$ and is interpreted as an inverse effective temperature $X = 1/T_{eff}$ \cite{cukupeliti}.  Here, the final regime corresponds
to $T_{eff} \to 0$, which can interpreted as a fixed point of zero temperature.
   It is also interesting to compare with what is known on the coarsening in 
pure ferromagnets. At the critical point $T_c$, there exists
a non trivial ratio $X(\frac{t}{t_w})$ which can be identified as an amplitude ratio \cite{golutc}. For instance, in the pure 1D Ising chain at zero temperature, the amplitude ratio \cite{golu1d} is  $X(t, t_w)=\frac{t+t_w}{2t}$:  it decreases from $X(t=t_w, t_w)=1$ to $X(t \to \infty, t_w)=\frac{1}{2}$.

For $T<T_c$, the ratio $X$ vanishes $X=0$ \cite{barratgrowing, berthiergrowing}, which is understood in the following way:  the domain walls respond 
with a factor $O(1)$ but they occupy only a small fraction $1/L(t_w)$ of the volume.
By comparison, for the random field chain, only a small fraction $1/\Gamma_w$ of domain walls respond, but with a very large response, since it corresponds to the flipping of an entire domain with length scale $ \Gamma_w^2$. 

In this section, we have described how the formulation of Glauber
dynamics of the random field Ising chain in terms of a reaction-diffusion process in a Brownian potential for the domain walls allows to
obtain very detailed results as a consequence of the very strong localization of the domain walls by the disorder.
Similarly, more general reaction-diffusion processes in a Brownian potential
have also been studied in details \cite{rgreadiff}.


\section{Localization of a random polymer at an interface}

\subsection{Model}

This section is devoted to the following random polymer
model introduced by Garel, Huse, Leibler and Orland \cite{garelpoly}. An heteropolymer
made of monomers carrying random charges $q_i$ is near an interface located at $z=0$.
The medium $z>0$ is favorable for positive charges $q>0$ whereas the medium $z<0$ 
is favorable for negative charges $q<0$.  More precisely, the 
model is defined by the following partition function
on Random Walks $\{z_i\}$ \cite{garelpoly} 
\begin{eqnarray} 
\label{functional} 
Z_L(\beta;\{q_i \}) = \sum_{ RW \{z_i\}}  \exp \left( \beta \sum_{i=1}^{L} q_i { \rm sgn}(z_i ) \right) \qquad. 
 \end{eqnarray}

At high temperature, Imry-Ma arguments have been proposed \cite{garelpoly} for the symmetric case $\overline{q_i}=0$ as well as for the asymmetric case  $\overline{q_i}>0$:  these arguments are based on energy/entropy balances, unlike the usual Imry-Ma
arguments which are based on an energy/energy balance. 
Since the localization mechanism is different in the symmetric and asymmetric cases,
we will now discuss them separately.

 \subsection{ Symmetric case} 

\subsubsection{  Imry-Ma argument based on typical events}

The Imry-Ma argument for the symmetric case \cite{garelpoly} is as follows. 
 The chain is assumed to be localized around the interface, with a typical loop
length $l$ in each solvent: 

 $\bullet$ the typical energy gain per loop is of order $ \displaystyle \sum_i^{i+l} q_i \sim \sqrt{\sigma l}$, i.e. of order $\sqrt{\sigma / l}$ per site.

$\bullet$ the entropy loss due to the confinement in a band of width $r \sim \sqrt{l}$ around
the interface is of order $1/r^2 \sim 1/l$ per site.

The optimization of the free energy by monomer 
\begin{eqnarray} 
f(l) \sim - \sqrt{ \frac{ \sigma} { l}} +  \frac{T}{l} 
\end{eqnarray} 
with respect to the length $l$ leads to the characteristic length

 \begin{eqnarray} 
l \sim \frac{ T^2} { \sigma} 
  \end{eqnarray} 
and to the following behavior for the free energy
\begin{eqnarray} f(T) \sim -\frac{\sigma}{T}  \label{imrymasym} 
\end{eqnarray}

This argument for the symmetric polymer thus predicts a localization at any temperature,
with the above behaviors at high temperature.

\subsubsection{  Comparison between two strategies} 

To go beyond the above Imry-Ma argument, two extreme strategies have been proposed
for the polymer configurations adapted to a given random sequence :

(a) a sequence-dependent strategy based on a real space RG procedure \cite{heteropolymer},
where only the conformations corresponding
to the optimal loop configuration are taken into account.

(b) a sequence-independent strategy, where all configurations compatible
with a fixed loop distribution $K(l)$ are considered \cite{giacomin04}.

The recent mathematical work \cite{giacomin04} (and references therein) indicates
that the strategy (b) is better for the symmetric case :
this means that  the disorder is not able to dominate over the random walk fluctuations.
On the contrary, the strategy (a) seems better for the asymmetric case. 
We will thus describe the sequence-dependent strategy based on a real space RG procedure \cite{heteropolymer} only for the asymmetric case.

\subsection{ Asymmetric case}

 \subsubsection{ Imry-Ma argument based on rare events} 

\label{imrymarare}

 The Imry-Ma argument proposed for the asymmetric case \cite{garelpoly} is in fact much more subtle
 than the argument for the symmetric case, because the correct description of the loops in the 
minority solvent $(-)$ requires the consideration of the ``rare events" where the sum of $l_-$  random variables $q_i$, of positive average $\overline{q_i}=q_0>0$, turns out to be negative enough to make favorable an excursion in the solvent $(-)$. 
More precisely, the argument is as follows  \cite{garelpoly} :  the polymer is expected to be
in its preferred solvent $(+)$, except when a loop of length $l^-$ in the solvent $(-)$ becomes energetically favorable with a sufficient charge $Q_-=-\sum_{i=j}^{j+l^ -} q_i>0$.  
As the probability of having $\sum_{i=j}^{j+l^ -} q_i=-Q^-$ 
\begin{eqnarray} { \rm Prob}(Q^ -) = \frac{1}{\sqrt{4 \pi \sigma l^ -}} e^{-\frac{(Q^-+q_0 l^-)^2}{4 \sigma l^ -}} 
\end{eqnarray} 
is weak, the typical distance $l^+$  between two such events behaves like the inverse of this probability 
\begin{eqnarray} 
l^+ \sim e^{ \frac{(Q^-+ q_0 l^-)^2}{4 \sigma l^ -}} 
\end{eqnarray} 
This rare event argument gives that the energy gained $Q^-$ in a loop of the solvent $(-)$ behaves as \begin{eqnarray} 
Q^ - \sim \sqrt{4 \sigma l^ - \ln l^+}-q_0 l^ - 
 \end{eqnarray} 

whereas the corresponding entropy loss is of order $ (\ln l^+)$.  The difference in free energy per monomer between this localized state with loops $(l_+, l_-)$ and the delocalized state in 
the preferred solvent $(+)$ is of order 
\begin{eqnarray} 
f(T, l^+, l^-)-f_{deloc}(T) \sim \frac{1}{l^+} \left(- Q^ - + T \ln l^+ \right) \sim \frac{1}{l^+} \left(q_0 l^ - - \sqrt{4 \sigma l^ - \ln l^+} + T \ln l^+ \right) 
  \end{eqnarray} 
The optimization with respect to the length $l^-$  gives the relation
 \begin{eqnarray} 
\label{relationlmlp} 
l^ - \sim \frac{\sigma}{q_0^2} \ln l^+ 
\end{eqnarray}
 and thus finally 
\begin{eqnarray} 
\label{qmlog} 
Q^ - \sim \frac{\sigma}{q_0} \ln l^+ 
 \end{eqnarray} 

This argument thus predicts that both the energy gain $Q^-$ and the entropy cost have the same dependence in $(\ln l^+)$:  the difference in free energy factorizes into 
\begin{eqnarray} 
\label{relationflp} 
f(T, l^+)-f_{deloc}(T) 
\sim (T-T_c) \frac{\ln l_+}{l_+} 
\end{eqnarray} 
This argument predicts a transition at a critical temperature 
\begin{eqnarray} 
T_c \sim \frac{\sigma}{q_0} 
\end{eqnarray} 
between the localized phase and the delocalized phase.
In contrast with the symmetric case, the behaviors of the free energy and length $l_+$ 
in terms of the temperature are not determined by the argument. 

 \subsubsection{ The disorder-dependent strategy based on real space RG}

 The aim of the strong disorder RG is to construct the optimal loop structure
 around the interface for a given disorder realization. 

\begin{figure}[thb ] \null \vglue 1.0cm \centerline{\epsfxsize=10.0 true cm \epsfbox{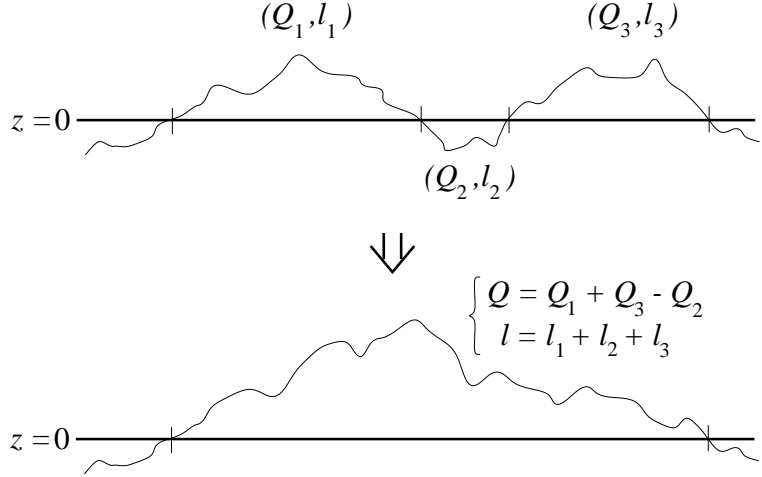}} \vskip 1.0cm \caption{\it  { \label{figRG} Illustration of the real space RG procedure :  the transfer of a loop $(Q_2=\gamma, l_2)$ surrounded by the 
neighboring loops $(Q_1, l_1)$ and $(Q_3, l_3)$ gives rise to a new loop $(Q, l)$ with the RG rules  (\ref{RGrules}). \cite{heteropolymer}}} 
\end{figure}

At zero temperature, each monomer is in its preferred solvent ${\rm sgn}(z_i)={\rm sgn}(q_i)$:
  the polymer is broken into loops $\alpha$ containing $l_{\alpha}$ consecutive monomers of the same sign, and carrying some absolute charge $Q_{\alpha}$.  When the temperature increases, we consider the configurations which can be obtained from the ground state loop structure by 
the iterative transfer in opposite solvent of the loop presenting the smaller absolute charge $Q_{min} \equiv \Gamma$.  When a loop $(Q_2=\Gamma,l_2)$ surrounded by two loops $(Q_1, l_1)$ and $(Q_3, l_3)$
is transferred , one obtains a new large loop whose length $l$ and absolute charge
$Q$ are given by the rules by the rules (see Fig. \ref{figRG}) 
\begin{eqnarray} 
&& Q=q_1+q_3-q_2 \\ 
&& l=l_1+l_2+l_3 \nonumber 
\label{RGrules} 
\end{eqnarray} 

These rules correspond to the RG rules for the Brownian extrema described in Eqs (\ref{rulesinai1},\ref{rulesinai2}) :  
here the corresponding random walk is the sum of the charges $\sum_{0}^i q_j$ up to
the running monomer $i$, and the RG procedure builds the best loop structure with the constraint 
that only loops with absolute charges larger than $\Gamma$ are allowed. To establish the correspondence between the RG scale $\Gamma$ and the temperature, we now have to determine under which conditions the transfer of a loop $(Q_2, l_2)$ in opposite solvent is indeed favorable from the free energy
point of view. The energy cost is
\begin{eqnarray} 
\Delta E^{flip} = 2 Q_2 
\end{eqnarray} 
whereas the entropy gain is 
\begin{eqnarray} 
\label{dsflip} 
\Delta S^{flip} = \ln ({  M} (l_1+l_2+l_3)) - \ln [ {  M} (l_1) {  M} (l_2) {  M} (l_3) ] \end{eqnarray} 
where ${ M} (l)=c^l/l^{3/2}$ represents the number of random walks of $l$ steps going from $z=0$ to $z=0$ in the presence of an absorbing condition at $z=0^-$.
 This leads to the following free energy balance
\begin{eqnarray} 
\Delta F^{flip} = \Delta E^{flip} - T \Delta S^{flip} = 2 Q_2 - T \ln \left(\frac{ {  M} (l_1+l_2+l_3)} { {  M} (l_1) {  M} (l_2) {  M} (l_3)} \right) 
\label{flipcondition} 
\end{eqnarray} 
To obtain the optimal structure at temperature $T$, 
the RG procedure has to be implemented as long as it lowers the free energy ($\Delta F^{flip} < 0$), 
and it should be stopped before the first iteration which would increase the free energy ($\Delta F^{flip} > 0$). 

\subsubsection{ Results of the strong disorder RG} 

The results of the strong disorder RG for the asymmetric case are the following \cite{heteropolymer} :

$\bullet$ in the limit $\sigma \gg q_0$, the delocalization transition takes place at high temperature
when the loops are large, and the critical temperature reads
\begin{eqnarray} 
T_c=\frac{4 \sigma} { 3 q_0} 
\end{eqnarray}

$\bullet $ the free energy presents the following essential singularity 
 \begin{eqnarray} 
f(T) - f(T_c) \opsimeq_{T \to T_c^ -} - 2 q_0 \left(\ln \frac{4 \sigma}{q_0} \right) \exp \left[ - \frac{ \ln \frac{4 \sigma}{q_0}} { \left(1-\frac{T}{T_c} \right)} \right ] 
\end{eqnarray} 
i.e. the transition is of infinite order in this scenario.

$\bullet $  the typical loop length $l^{+}_{  blob}(T)$ in the preferred solvent diverges with an essential singularity at the transition, whereas the typical blob length $l^{-}_{  blob}(T)$  in the unfavorable solvent diverges algebraically 
\begin{eqnarray} 
&& l^{+}_{ blob}(T) \opsimeq_{T \to T_c^ -} \frac{\sigma}{q_0^2} \exp \left[ + \frac{ \ln \frac{4 \sigma}{q_0}} { \left(1-\frac{T}{T_c} \right)} \right ] \\ 
&& l^{-}_{blob}(T) \opsimeq_{T \to T_c^ -} \frac{\sigma}{q_0^2} \frac{ \ln \frac{4 \sigma}{q_0}} { \left(1-\frac{T}{T_c} \right)} 
\end{eqnarray} 

$\bullet$ the scaling variable $\lambda_+=l_+/l^{+}_{ blob}(T)$ for the loops in the preferred solvent is distributed with the exponential distribution $e^{-\lambda_+}$.

$\bullet$ For a large chain of length $L$, the distribution of the delocalization temperature
can be computed \cite{heteropolymer}. In particular the typical value
 presents a logarithmic correction of order $(1/\ln L)$ with respect to the critical temperature of
the thermodynamic limit 
\begin{eqnarray} 
T_{\rm deloc}^{typ} \sim T_c \left(1-\frac{4 \sigma}{q_0^2 \ln L} \right) 
\end{eqnarray} 

In conclusion, the strong disorder RG just described for the asymmetric case 
is a direct extension of the Imry-Ma argument  
based on rare events \cite{garelpoly} given in section \ref{imrymarare}.
It allows to characterize in details the delocalization transition
via the explicit construction of polymer loops in each sample. 
Recent mathematical work \cite{giacomin04} have obtained rigorous bound
for the free-energy from this RG loop construction.


\section{Asymmetric simple exclusion process with disorder}

\label{ASEP3}

The asymmetric simple exclusion process (ASEP) is the paradigmatic example of a model for
non-equilibrium transport. Moreover this model can be related via simple mappings to other
problems of non-equilibrium statistical physics which include e.g. surface growth problems
\cite{hinrichsen,schutzreview}. For homogeneous transition rates the model has a detailed
analytical solution, both for periodic and open boundary conditions.
Disorder in this model can be introduced in two different ways. For particle-wise (pt) disorder\cite{kf,evans}
each particle has specific transition rates, which are the random variables. On the other hand for site-wise (st)disorder\cite{barma} the specific transition rates are assigned to lattice sites, rather than to particles. Our understanding about the random model is more complete for pt disorder, what we are going to consider first in the following.

\subsection{ASEP with particle-wise disorder}

To be concrete we consider here the ASEP on a periodic chain of $N$ sites and with $M$ particles. Particle $i$ may hop to empty neighboring sites with rates $p_i$ to the right and $q_i$ to the left, where $p_i$ and $q_i$ are independent and identically distributed random variables. Representing a configuration by the
number of empty sites $n_i$ in front of the $i$th particle, the
stationary weight of a configuration $n_1,n_2,\dots ,n_M$ is given by
\cite{spitzer,evans}:
\begin{equation}
  f_N(n_1,n_2,\dots,n_M)=\prod_{\mu
=1}^Mg_{\mu}^{n_{\mu}},
\end{equation}
where
\begin{equation}
g_{\mu}=\left[1-\prod_{k=1}^M{q_k\over
p_k}\right]^{-1}\left[\sum_{i=0}^{M-1}{1\over p_{\mu -i}}\prod_{j=\mu
+1-i}^{\mu}{q_j\over p_j}\right]
\label{g}
\end{equation}
provided $p_i> 0$ for all particles. The stationary
velocity is given by:
\begin{equation}
v=\frac{Z_{N-1,M}}{Z_{N,M}},\quad Z_{N,M}=\sum_{n_1,n_2,\dots,n_M} f_N(\{n_{\mu}\})\;.
\label{v}
\end{equation}
where in the summation $\sum_{\mu=1}^M n_{\mu}=N-M$.
In the thermodynamic limit one can define a control parameter:
\begin{equation}
\delta=\frac{[\ln p]_{\rm av} - [\ln q]_{\rm av}}{{\rm var}[\ln
p]+{\rm var}[\ln q]}\;,
\label{delta_A}
\end{equation}
so that for $\delta>0$ ($\delta<0$) the particles move to the right
(left).

Depending on the form of disorder the stationary state can be of three different types.
i) The stationary state has a homogeneous particle density and there is a finite velocity,
$v > 0$. ii) The particle density in the stationary state is not homogeneous: there is a
macroscopic hole before the slowest particle and $v > 0$. iii)  The particle density in the
stationary state is not homogeneous and the stationary velocity is $v=0$. Here we shall
consider the third case, which is realized in the 
partially asymmetric model, which means that i) both $p_i > 0$ and $q_i > 0$, and
ii) there is a finite fraction of particles, for which $p_i > q_i$ and also $q_i > p_i$.
The partially asymmetric model is conveniently studied by the strong disorder RG 
method\cite{igloipartial},
what is going to be described in the following.

\subsubsection{RG rules}

\begin{figure}
\centerline{\includegraphics[width=0.58\linewidth]{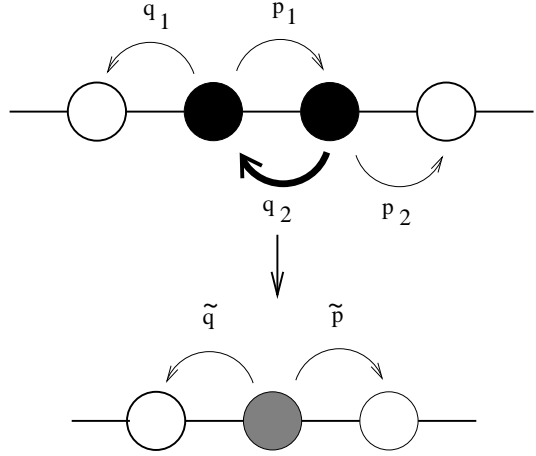}}
\caption{\label{renorm} Renormalization scheme for particle
  clusters. If $q_2$ is the largest hopping rate in a time-scale, $\tau > 1/q_2$,
  the two-particle cluster moves coherently and the composite particle
  is characterized by the effective hopping rates $\tilde q$ and  $\tilde
  p$, respectively, see the text.\cite{igloipartial}}
\end{figure}

In the RG method one sorts the transition rates in descending order
and the largest one sets the energy scale,
$\Omega=max(\{p_i\},\{q_i\})$, which is related to the relevant
time-scale, $\tau=\Omega^{-1}$. During renormalization the largest
hopping rates are successively eliminated, thus the time-scale is
increased. In a sufficiently large time-scale some cluster of
particles moves coherently and form composite particles, which have
new effective transition rates. To illustrate the decimation rules for the
system (see Fig. \ref{renorm}) let us assume
that the largest rate is associated to a left jump, say $\Omega=q_2$.

Typically, $q_2 \gg p_2$, and the same relation is assumed to hold with the
transition rates of the particle to the left: $q_2 \gg q_1,p_1$. In a time-scale,
$\tau>\Omega^{-1}$, the
fastest jump with rate $q_2$ can not be observed and the two particles
$1$ and $2$ form a composite particle. The composite particle has a
left hopping rate $\tilde{q}=q_1$, since a jump of particle $1$ is
almost immediately followed by a jump of particle $2$. The transition
rate of the composite particle to the right, $\tilde{p}$, follows from
the observation that, if the neighboring site to the right of particle
$2$ is empty it spends a small fraction of time:
$r=p_2/(p_2+q_2)\approx p_2/q_2$ on it. A jump of particle $1$ to the
right is possible only this period, thus $\tilde{p}=p_1 r \approx p_1
p_2/q_2$. The renormalization rules can be obtained similarly for a
large $p$ decimation and can be summarized as:
\begin{equation}
\tilde{p}=\frac{p_1 p_2}{\Omega},\quad \Omega=q_2;\quad
\tilde{q}=\frac{q_1 q_2}{\Omega} ,\quad \Omega=p_1\;.
\label{deci_ASEP}
\end{equation}

\subsubsection{Relation with the random $XX$-chain}
\label{ASEP_XX}
The decimation equations in Eq.(\ref{deci_ASEP}) are in a similar form as that of the random $XX$
chain in Eq.(\ref{JXX}). The correspondence properly holds for a chain with $2M$ sites having the exchange couplings: $J_{2i-1}=p_i$ and $J_{2i}=q_i$. Using the results of Sec. \ref{XY_RTFIC} this mapping can be
extended further for the RTFIC, too. Here we list the immediate consequences of these mappings.

Let us consider first the region with $\delta>0$, which corresponds to the Griffiths phase of the
RTFIC. Using the analogy with the solution to the RTFIC in Sec.\ref{RTFIC32} we can state that during
renormalization almost exclusively the left hopping rates are decimated out. After $M$-steps of decimation
we are left with a single particle of mass, $M$, having effective hopping rates, $\tilde{p}$ and $\tilde{p}$,
so that $\tilde{q}/\tilde{p} \to 0$ for large $M$ and $\tilde{p} \sim
M^{-z}$, with a dynamical exponent given from Eqs.~(\ref{zeq}) and (\ref{zeq2}) as:
\begin{equation}
\left[\left({p/q}\right)^{1/z}\right]_{av}=1\;.
\label{z_A}
\end{equation}
This result indicates that in the stationary state of the partially asymmetric process
i) there is a phase separation, with an occupied region, which corresponds to the effective
particle and with a non-occupied region. ii) The stationary velocity $v$ of the system
vanishes in a large ring as:
\begin{equation}
v \sim N^{-z}\;,
\label{z_st}
\end{equation}
provided $M/N=O(1)$. More precisely the accumulated
distance traveled by the particles, $x$, in time, $t$, is given by
\begin{equation}
x \sim t^{1/z}\;.
\label{x_st}
\end{equation}
This relation is very much similar to the behavior of a Brownian particle in a random potential in
the anomalous diffusion regime with the correspondence $\mu=1/z$
( see section \ref{chapsinaibiais}).

 This analogy can
be made even closer by noting that motion of the macroscopically occupied region takes place in such a
way that single and non-interacting holes diffuse into the opposite direction. The
motion of the holes takes place in a position
dependent random potential, the hopping rates of which, $p_i$ and
$q_i$, are generated by the particles. The average distance traveled by a hole is just
the absolute value of the accumulated distance made by the ASEP.

At the critical point, $\delta=0$, left and right hopping rates are decimated
symmetrically and the system scales into an infinite disorder fixed point. After $M$-steps the
remaining effective particle has a symmetric hopping probability:
$\tilde{q} \sim \tilde{p} \sim \exp(- const  M^{1/2})$. Thus the motion
of the system is diffusive and ultra-slow, the appropriate scaling
combination is given by: $(\ln v) M^{-1/2}$. Close to the critical point
the correlation length in the system, $\xi$, which measures the width
of the front, is given by $\xi \sim \delta^{-2}$, compare with Eq.(\ref{RTFIC_nu}) for the RTFIC.

\begin{figure}
\centerline{\includegraphics[width=1.0\linewidth]{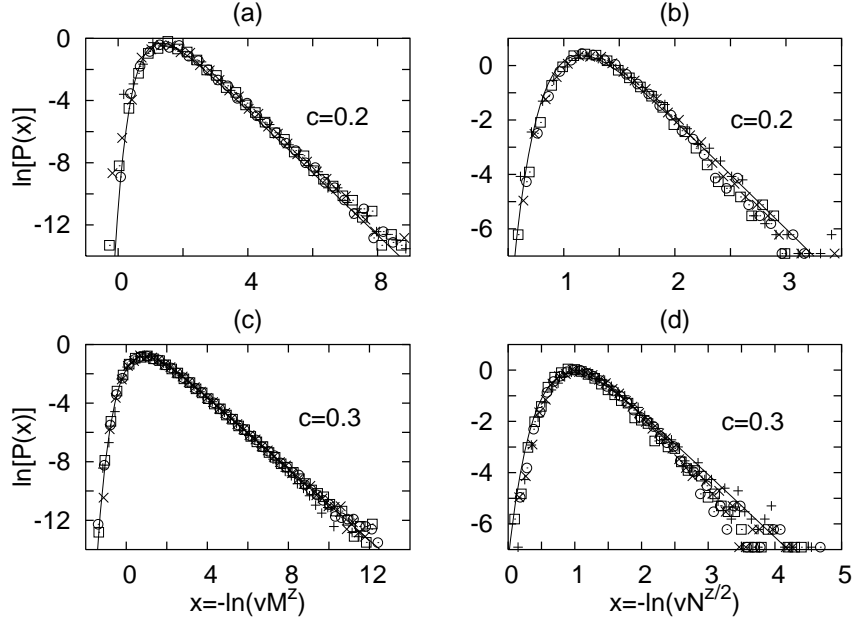}}
\caption{\label{asep2} Scaling of the velocity distribution in the
Griffiths phase at two concentrations ($c=0.2$ and $c=0.3$) for pt
(a,c) and st (b,d) disorder.
Data denoted by symbols +,$\times$,O,$\Box$ correspond to
 $M=64,128,256,512$ for (a,c) and $N=64,128,256,512$ for (b,d), respectively.
The Fr\'echet distribution\cite{galambos} with the dynamical exponent: $z=z_{pt}$,
calculated from Eq.(\ref{z_A}) and
with $z_{st}=z_{pt}/2$,
as given in Eq.(\ref{z_rel}) are indicated by a full line.\cite{igloipartial}}
\end{figure}

Numerical results about the velocity distribution of a periodic system with $M/N=1/2$
is presented in Figs.\ref{asep2} and \ref{asep3}, in the Griffiths phase and at the
critical point, respectively. Here a bimodal distribution is used with $p_iq_i=r$, for all $i$,
and $P(p)=c \delta(p-1)+(1-c) \delta(p-r)$, with $r>1$ and $0<c \le 1/2$. In
this case the control-parameter is $\delta=(1-2c)/[2c(1-c)
\ln r]$ and the dynamical exponent from Eq.~(\ref{z_A}) is $z=\ln
r/\ln(c^{-1}-1)$. 

\subsection{ASEP with site-wise disorder}

\label{ASEP2}

\begin{figure}
\centerline{\includegraphics[width=0.8\linewidth]{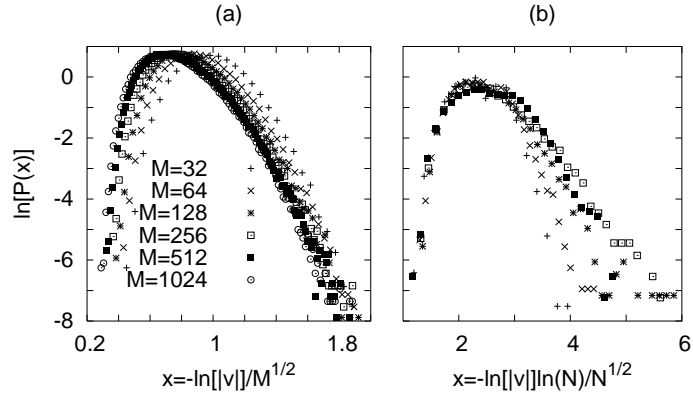}}
\caption{\label{asep3} Scaling plot of the velocity
  distribution at the critical point for pt (a)
  and st (b) disorder. In the latter case
  a logarithmic correction term is included.\cite{igloipartial}}
\end{figure}

For site-wise disorder, i.e. when the hopping rates are assigned to sites rather than
to particles, the RG rules can not be simply generalized. Therefore our knowledge in
this case is mainly based on numerical and scaling results. For st disorder and for the partially
asymmetric process analytical and numerical results show the presence of macroscopic
phase separation on a periodic chain. While for pt disorder the macroscopic occupied domains
move with a stationary velocity given in Eq.~(\ref{z_st}), for st disorder the position of
the domains are fixed in the stationary state. In this case the macroscopic transport is due to
two symmetric processes: i) diffusion of holes in the ordered domains, and ii) diffusion of
particles in the non-occupied domains. It is argued in Ref.\cite{igloipartial} that for st disorder both
diffusive particles and holes should overcome one-one independent large potential barriers, whereas
for pt disorder just the holes should go over a large barrier. As a consequence the probability
of occurrence of rare events with a characteristic time-scale, $\tau$, in the two cases are related as:
$p_{st}(\tau)=p^2_{pt}(\tau)$. Since the distribution of the relaxation times follows a power-low form, as
$p_{\alpha}(\tau) \sim \tau^{-1/z_{\alpha}}$, where $\alpha$ stands for $pt$ or $st$. From the above
argument one  obtains the relation:
\begin{equation}
z_{st}=\frac{z_{pt}}{2}\;,
\label{z_rel}
\end{equation}
which has been checked by numerical simulations, see Figs. \ref{asep2} and \ref{asep3}. In particular
at the critical point with $\delta=0$
also with st disorder the dynamical exponent is formally infinity and the appropriate scaling combination
is $(\ln v) M^{-1/2}$, although in this case it is supplemented by logarithmic corrections.

\section{Reaction-diffusion models with disorder}

\subsection{ Phase transition into an absorbing state}

Stochastic many particle systems with a phase transition into an absorbing
state are of wide interest in physics, chemistry and even biology
\cite{MarDic}. Much recent work has focused on establishing a classification
of possible universality classes for systems having this type
of transition \cite{Dic}. For
models with a scalar order parameter,
absence of conservation laws, and short range
interactions, the critical behavior is conjectured to be that of directed
percolation \cite{Grass}. Well known models with a phase
transition in this universality
class are the contact process \cite{Harris} and the
Ziff-Gulari-Barshad model of catalytic reactions \cite{ZGB}.
When there is a conservation law present,
other universality classes can
appear, the best known of which is the parity conserving class
\cite{CardyTauber}.
Quenched, i.e. time-independent, disorder is an inevitable feature of many real processes and could
play an important role in stochastic particle systems too. As an example, it has been argued that due to the presence of some form of disorder the directed percolation universality class
has not yet been seen in real experiments\cite{Hinexp}, such as in catalytic reactions \cite{ZGB},
in depinning transitions \cite{Barabasi}, and in the flow of granular matter \cite{Sand}
(for a review, see \cite{Hinexp}). In stochastic particle systems, disorder is represented by
position dependent reaction rates and its relevance can be expressed in a $(d+1)$-dimensional
system by a Harris-type criterion \cite{duitser},
\begin{equation}
\nu_{\perp}<2/d\;.
\label{harris_anis}
\end{equation}
Here $\nu_{\perp}$ is the
correlation length exponent in the spatial direction of the pure system, compare with Eq.(\ref{harris})
for isotropic systems. Indeed for directed
percolation at any $d<4$ dimensions the disorder is a relevant perturbation.

In the following we consider the random contact process and study the properties of the new random
fixed point.

\subsection{Random contact process}

\label{RCP}

In the contact process each site of the lattice can be either vacant ($\emptyset$) or occupied by at most one particle ($A$), and thus can be characterized by an Ising-spin variable, $\sigma_i=1$ for $\emptyset$ and $\sigma_i=-1$ for $A$. The state of the system is then given
by the vector ${\bf P}({\bf \sigma},t)$
which gives the probability that the system is in the state ${\bf \sigma}=\{\dots ,\sigma_i, \dots\}$ at time $t$.  A particle can be created at an empty site $i$ with
a rate $p \hat{\lambda_i}/p_0$, where $p$ ($p_0$) is the number of occupied neighbors (the coordination
number of the lattice) and at an occupied site the particle is annihilated with a rate $\mu_i$. The time evolution is governed by a master equation, which
can be written into the form:
\begin{equation}
\frac{{\rm d} {\bf P}({\bf \sigma})}{{\rm d} t}=-H_{CP} {\bf P}({\bf \sigma})\;.
\label{masterRCP}
\end{equation}
Here the generator $H_{CP}$ of the Markov process is given by:
\begin{eqnarray}
H_{CP} = \sum_i
\mu_{i} M_{i}
+ \sum_{\langle ij \rangle} \frac{\hat{\lambda}_{i}}{p_0} (n_{i} Q_{j} + Q_{i} n_{j})
\label{hamilton_CP}
\end{eqnarray}
in terms of the matrices:
\begin{eqnarray*}
M = \left(\begin{array}{rr}
0 & -1\\
0 & 1\end{array}\right),
n = \left(\begin{array}{rr}
0 & 0\\
0 & 1\end{array}\right),
Q = \left(\begin{array}{rr}
1 & 0\\
-1 & 0\end{array}\right)
\end{eqnarray*}
and $\langle ij \rangle$ stands for nearest neighbors. It is well known\cite{schutzreview} that the
steady state probability distribution of a stochastic process coincides with the
ground state of its generator (sometimes also called quantum Hamiltonian of
the stochastic process) while relaxation properties can be determined
from its low lying spectrum.

The average number of particles at site-$i$ and time-$t$ is given by $\langle n_i\rangle(t)$, which evolves to a constant in the stationary state. The order-parameter of the system is given by
$\rho = 1/L \sum_i \langle n_i\rangle$, for surface sites, $i=1$ and $i=L$, we obtain the surface
order-parameter: $\rho_s=[\langle n_1\rangle]_{\rm av}$. In the thermodynamic limit in the active phase
$\rho >0$ and in the inactive phase $\rho =0$.
For non-random couplings the non-equilibrium phase transition, which separates the active and inactive phases, belongs to the universality class of directed percolation\cite{Dic,Grass}. Scaling in this transition point is
governed by an anisotropic fixed point, as described in Sec.\ref{SC_conv}, but here we use the convention $\nu=\nu_{\perp}$. In
one dimension the transition is at $(\mu/\hat{\lambda})_c=0.3032$ and the critical exponents are given
by\cite{hinrichsen}:
$\beta=0.2765$, $\beta_s=0.7337$, $\nu_{\perp}=1.097$ and $z=1.581$.
In the following we often use the variable $\lambda=\hat{\lambda} / p_0$ to characterize the creation
rate. The random contact process and related models are studied in a series of
papers\cite{duitser,weerDick,Janssen, BramDur,Cafiero,Webman} and unconventional critical properties
are observed (logarithmically slow dynamical correlations, disorder dependent Griffiths-like effects, etc.).
Recently, the system is studied by the strong disorder RG method\cite{igloicontact},
the results of which have provided possible explanations of the previous numerical results.
In the following we describe the application of the strong disorder RG for this system.

\subsubsection{RG rules}

In the usual way the transition rates are put in descending order and the largest rate:
$\Omega=max(\{\lambda_i\},\{\mu_i\})$ sets the energy scale in the system.
Here one should use two different ways of decimation depending if the largest transition rate is a branching rate
or it is an annihilation rate.

\bigskip
{\it The largest term is a branching rate: $\Omega=\lambda_2$}

In this case the two-site cluster, $(2,3)$, having
the largest branching rate, $\lambda_2$, spends
most of the time in the configurations $AA$ or $\emptyset\emptyset$ and can be rarely found in one of the
other two configurations, $A \emptyset$ and $\emptyset A$. Consequently for a large time-scale,
$\tau \sim 1/\Omega$, the two sites behave
as an effective cluster with a moment of $\tilde{m}=2$ and with an effective death rate, $\tilde{\mu}_2$,
which can be obtained by the following simple reasoning.
Let us start with the original representation, when the two-site cluster is in the occupied state, $AA$.
In the effective decay process first the particle
at site $(2)$  should decay (with rate $\mu_{2}$), which is then
followed by the decay of the particle at $(3)$. This second
process has a very low probability of $pr(3)=\mu_{3}/(\lambda_{2}+\mu_{3})\approx \mu_{3}/\lambda_{2}$.
Since the same processes can also occur with the role of $(2)$ and $(3)$ interchanged,
we find that $\tilde{\mu}  = pr(3) \mu(2)+pr(2) \mu(3) $, which is given by:
\begin{equation}
\tilde{\mu}=\frac{2 \mu_2 \mu_3}{\lambda_2}\;.
\label{mutilde}
\end{equation}
The renormalization equation
in Eq.(\ref{mutilde}) should be extended by the renormalization of moments
(i.e. the number of original sites in the cluster):
\begin{equation}
\tilde{m}=m_2 + m_3\;,
\label{moments}
\end{equation}
where in the initial situation $m_2=m_3=1$.

\bigskip
{\it The largest term is a death rate: $\Omega=\mu_2$}

In this case the site (2) is almost always empty, $\emptyset$, therefore it does not
contribute to the fractal properties of the $A$ cluster and can be decimated out. The
effective branching rate, $\tilde{\lambda}$ between the remaining sites $(1)$ and $(3)$ can
be obtained from the following reasoning. Let us have the configuration of the three-site cluster in
the original representation as $A\emptyset\emptyset$. The effective branching rate
between sites $(1)$ and $(3)$ is generated
by a virtual process, in which first a particle is created at site $(2)$ (rate
$\lambda_{2}$), and then one at site $3$ (probability $\lambda_{3}/(
\lambda_{3} + \mu_{2})$). Hence, we get for strong disorder the effective branching rate as:
\begin{equation}
\tilde{\lambda}=\frac{\lambda_2 \lambda_3}{\mu_2}\;.
\label{lambdatilde}
\end{equation}
The renormalization equations in Eqs.(\ref{mutilde}) and (\ref{lambdatilde}) can be transformed
into a symmetric form in terms of the variable,
$J= \lambda/\kappa=\hat{\lambda}/(p_0\kappa)$ with $\kappa=\sqrt{2}$ as
\begin{eqnarray}
\tilde{\mu} = \kappa \frac{\mu \mu'}{J},
\quad \tilde{m}=m+m',
\ \ \ \ \ \tilde{J} = \kappa
\frac{J J'}{\mu}\;.
\label{RG_CP}
\end{eqnarray}
We note that the renormalization rules in Eqs.(\ref{mutilde}) and (\ref{lambdatilde}) can be obtained
in the Hamiltonian formalism\cite{igloicontact} with Eq.(\ref{hamilton_CP}) in an analogous way as for
the RTFIC in Sec.\ref{RTFIC_3}.

\subsubsection{Analysis of the RG equations}

\label{RCP_st}

The decimation equations in Eq.(\ref{RG_CP}) are very similar to that
of the RTFIC in Eqs.(\ref{Jdecimation}) and (\ref{hdecimation}), and a
similar prefactor, $\kappa \ne 1$ can be found in the renormalization equations
of random quantum spin chains with discrete symmetry in Sec.\ref{PCAT},
see Eq.(\ref{deciCM}). Using the same arguments as in Sec\ref{PCAT} one
expects that for strong enough initial disorder the RG flow is attracted
by the infinite disorder fixed point of the RTFIC. For weaker disorder, however,
the infinite disorder fixed point is not attractive. This is due to
the prefactor, $\kappa=\sqrt{2}>1$, in Eq.(\ref{RG_CP}). As already argued
in Sec.\ref{PCAT} in this case the variation of the energy-scale during
renormalization is not monotonic and the corresponding fixed point is different
of that observed for strong disorder.

The control parameter of the model, $\delta$, is defined as:
\begin{equation}
\delta=\frac{[\ln \mu]_{\rm av}- [\ln J]_{\rm av}}
{{\rm var}[\ln \mu]+{\rm var}[\ln J]}\;.
\label{contr}
\end{equation}
Its value at the strong disorder fixed point is given by $\delta=0$, which follows
from duality of the RG equations in Eq.(\ref{RG_CP}).

\begin{figure}
\centerline{\includegraphics[width=0.58\linewidth]{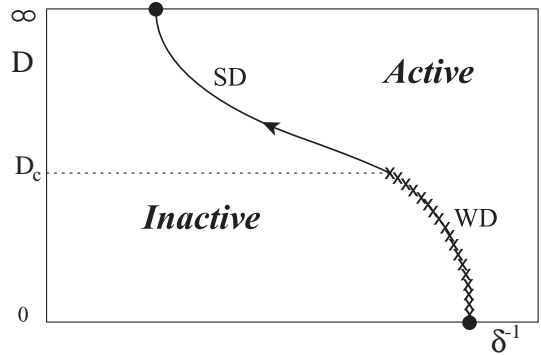}}
\caption{\label{RCP1} Schematic phase diagram of the random contact process as a function
of the control parameter, $\delta$, and the strength of disorder, $D$,
(see text). The inactive and active phases are separated by
a phase transition line, along which the critical exponents are $D$ dependent
for weaker disorder (WD, $ D \leq D_{c}$), or are determined by an infinite disorder fixed point at $D=\infty$,
for strong disorder (SD, $D>D_{c}$).}
\end{figure}

A schematic phase diagram of the model is presented in Fig.\ref{RCP1}  as a function of the control
parameter $\delta$ and the strength of disorder, $D={\rm var}[\ln \mu]+{\rm var}[\ln J]$.
Note that along the phase transition line $\delta \ge 0$.
According to numerical results\cite{igloicontact} and 
analogies with random quantum spin chains in Sec.\ref{PCAT} for strong enough disorder, $D>D_c$,
the critical behavior is controlled by the strong disorder fixed point located at $D \to \infty$.

The critical exponents at the strong disorder fixed point are exactly known
from the analysis of the RTFIC in Sec.\ref{RTFIC_det} and from the mapping between the
two models. In the random contact process the average particle density in the bulk, $\rho$, and at the surface,
$\rho_s$, corresponds to the bulk and surface magnetizations, respectively, of the RTFIC. They
have a singular behavior, $\rho \sim (-\delta)^{\beta}$ and $\rho_s \sim (-\delta)^{\beta_s}$,
$\delta \to 0^-$, respectively. At the strong disorder fixed point: $\beta^{\infty}=(3-\sqrt{5})/2$ and
$\beta_s^{\infty}=1$. The particle-particle correlation function, $[\langle n_i n_j \rangle ]_{\rm av}$ of
the random contact process is analogous to the spin-spin correlation function of the RTFIC, and the
correlation length, $\xi_{\perp} \sim |\delta|^{-\nu_{\perp}}$, involves an exponent, $\nu_{\perp}$, which at
the infinite disorder fixed point is given by: $\nu_{\perp}^{\infty}=2$. Finally, the correlation
length in the parallel direction, $\xi_{\parallel}$, corresponds to the relaxation time of the RTFIC.
At the strong disorder fixed point $\ln \xi_{\parallel} \sim \xi_{\perp}^{\psi}$, with $\psi=1/2$.

\subsubsection{Numerical results and scaling in the weak disorder regime}

\label{RCP4}

\begin{figure}
\centerline{\includegraphics[width=0.58\linewidth]{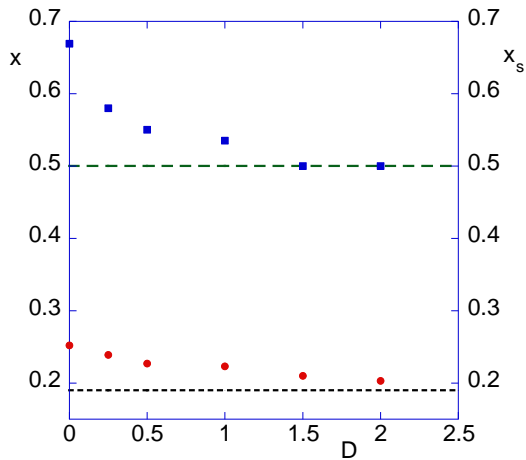}}
\caption{\label{RCP3} Numerical estimates of the exponents $x_m=\beta/\nu_{\perp}$
(circles)
and $x_m^{s}$ (squares). The broken lines indicate the
value at the strong disorder fixed point. The errors are of the same order as the
size of the symbols\cite{igloicontact}.}
\end{figure}

A systematic numerical study of the random contact process with varying strength of disorder is performed in
Ref.\cite{igloicontact}. The results of density matrix renormalization (DMRG) of the Hamiltonian version
in Eq.(\ref{hamilton_CP}) and that of Monte Carlo simulations are consistent with each other.
For weak disorder, $D<D_0$, the order parameter
exponents, $x_m=\beta/\nu_{\perp}$ and $x_m^s=\beta_s/\nu_{\perp}$ are found to be disorder dependent,
which vary continuously between the values of the pure system and that of the infinite disorder
fixed point. This is shown in Fig.\ref{RCP3}. Therefore we can conclude that in the weak disorder regime there
is a line of random fixed points.

For the form of dynamical scaling in this regime two possible scenarios are proposed. It can be either
i) conventional random scaling, see Sec.\ref{SC_conv} or ii) infinite disorder scaling, as in Sec.\ref{SC_st}.
In the former case the dynamical exponent, $z$, is finite for $D<D_0$, but becomes divergent
at $D=D_0$. We remind that this type of behavior is observed in random quantum spin chains, see
Sec. \ref{PCAT} and \ref{S=1}. In the second scenario the dynamical exponent is formally infinity in the
weak disorder regime, too, but the scaling exponent, $\psi$ in Eq.(\ref{logscale}) is disorder dependent,
and approaches $\psi=1/2$ at $D=D_0$. This infinite disorder scaling is noticed in numerical
studies of the 2d random contact process\cite{weerDick}.

For the random contact process in 1d it is difficult to decide between the two scenarios by numerical
studies and there is still no final, definite answer. The reason of this is that for a
finite system one can fit c.f. the autocorrelation function by both types of scaling forms. Conventional
scaling with a large effective $z$ has similar accuracy, as infinite disorder scaling with a small $\psi$.
One should note, however, that the present numerical data can be interpreted with a slight preference
of infinite disorder scaling.

\subsection{Other types of reaction-diffusion models with quenched disorder}

The Reggeon field-theory, which is the Hamiltonian version of directed percolation\cite{QRFT},
has been studied by the strong disorder RG approach\cite{igloicontact}.
The decimation equations are found to be identical to that of the random contact
process in Eq.(\ref{RG_CP}), which is a consequence of an exact mapping\cite{Grass}, which exists between the two
models. Therefore the random Reggeon field-theory has the same critical behavior as the random contact
process. The presence of infinite disorder scaling for the directed percolation in $(1+1)$-dimension is
demonstrated in Ref.\cite{igloicontact}, for extreme strong disorder.

 On the other hand the generalized contact process\cite{Hin}, which for homogeneous transitions rates
belongs to the parity conserving universality class, in the presence of quenched disorder has
conventional random scaling\cite{igloicontact}.


\section{ Classical models in $d \ge 2$}

Quenched randomness could cause strong disorder effects in higher dimensional classical systems, too.
However, not all the systems which have an infinite disorder fixed point in $d=1$ have a similar one
in $d \ge 2$. For example disorder is an irrelevant perturbation
for the random walk in $d \ge 2$, (in $d=2$ there are logarithmic corrections) as has been shown in\cite{fisher84}. On the contrary for reaction-diffusion
models, such as the random contact process infinite disorder scaling is observed\cite{weerDick,igloicontact}
in higher dimensions, too.
$d$-dimensional classical systems with layered randomness, such as the McCoy-Wu model\cite{mccoywu},
which are isomorph with random quantum spin chains, constitute examples of strong disorder scaling, too. 
Finally, one can ask the question, if an isotropic classical spin model can be constructed,
which is isomorphic with the McCoy-Wu model, and thus with the RTFIC and therefore the
critical properties are exactly known. A possible realization of this universality class is the random bond
Potts model in the large-$q$ limit\cite{igloipottsq}. We are going to review these developments in the following.

\subsection{Random contact process in 2d}

\begin{figure}
\centerline{\includegraphics[width=0.58\linewidth]{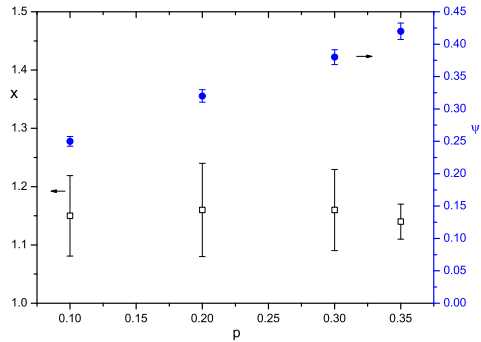}}
\caption{\label{RCP8} Exponents $x_m$(squares) and $\psi$(circles) in the random contact
process in $d=2$ as a function of dilution $p$
assuming logarithmic scaling\cite{igloicontact}. The values are calculated from the data given in
Ref.\cite{weerDick}.}
\end{figure}

Renormalization of the random contact process, as shown in Sec.\ref{RCP}, has many similarities
with the random transverse-field Ising model. The RG rules in Eq.(\ref{RG_CP}) differ only by a prefactor,
$\kappa=\sqrt{2}$ from the similar equations for the random transverse-field Ising model in
Eqs. (\ref{Jdecimation}) and (\ref{hdecimation}) and the topology of the lattice, even in higher
dimensions, are analogous in the two cases. Therefore the conclusion obtained in Sec.\ref{RCP}, which
predicts for strong enough disorder isomorph infinite disorder fixed points for the two problems in
$d=1$, should be valid for $d \ge 2$, too. 
This conjecture has been checked in Ref\cite{igloicontact} in which the numerical data of
Ref.\cite{weerDick} about the strongly random contact process are compared to the critical
parameters of the 2d random transverse-field Ising
model, see Table \ref{TAB2d}. As seen in Fig.\ref{RCP8} for strong dilution
there is a satisfactory agreement. For weaker disorder the random
contact process has an intermediate disorder regime\cite{igloicontact}, the properties of which are
similar to that found in 1d, see in Sec.\ref{RCP4}.

\subsection{Classical systems with layered randomness}

Here we consider $d$-dimensional classical systems in which disorder in the
couplings is strictly correlated in $d'<d$ dimension. For the general problem we mention field-theoretical
investigations\cite{boyanovsky} and mean-field results\cite{bbip98}.
In the following we restrict ourselves to two-dimensional systems with layered randomness
($d=2$ and $d'=1$), the prototype of which is the two-dimensional
Ising model, which has been introduced and partially exactly solved by McCoy and Wu\cite{mccoywu}.
The McCoy-Wu model and the RTFIC are isomorph at their critical points and a
similar relation is true between another classical strip-random models and the corresponding
random quantum spin chains. As an example we mention the $q$-state Potts model the
quantum version of which is studied in Sec.\ref{PCAT}.
The Potts model in the $q \to 1$ limit corresponds to the percolation problem\cite{kasteleyn},
which - in the presence of layered disorder - has been studied in\cite{igloiperco}.

\subsubsection{The McCoy-Wu model and the RTFIC}
\label{M_W}

The McCoy-Wu model is a square lattice Ising model with vertical, $K_1(i)=-J_1(i)/k_B T$,
and horizontal bonds, $K_2(i)=-J_2(i)/k_B T$, the value of both may depend on the position
in the horizontal direction, $i$. The row transfer matrix, ${T}$, which is used to
transfer information in the vertical direction is given by\cite{schultz64}:
\begin{equation}
{ T}=\exp \left[\sum_iK_1^*(i)\sigma_i^z\right]\exp\left[
\sum_iK_2(i)\sigma_i^x\sigma_{i+1}^x\right]\;
\label{Tmatrix}
\end{equation}
in terms of the $\sigma_i^{x,z}$ Pauli matrices and the $K_1^*(i)$ dual couplings ($\tanh K_1^*=\exp
(-2K_1)$). In the extreme anisotropic (or Hamiltonian) limit\cite{kogut79} there are strong vertical and weak
horizontal bonds, so that $K_1^*(i)$ and $K_2(i)$ are
both small and one can combine the exponentials to obtain
\begin{equation}
{ T}=\exp(-\tau H_I)\;.
\label{T_H}
\end{equation}
Here $\tau=K_1^*$ is a reference value, which measures the lattice
spacing in the vertical direction and $H_I$ is the Hamiltonian of the RTFIC as given in
Eq.(\ref{hamilton_I}). This mapping, which works also in higher dimensions, constitutes a relation
between the thermodynamic quantities of a $d$-dimensional classical system with layered
randomness and the ground-state expectation values of a $(d-1)$-dimensional random quantum spin
system. According to the standard relations\cite{kogut79} the equivalent quantities are:

\begin{itemize}
\item
transfer matrix  $\leftrightarrow$ Hamiltonian
\item
free-energy density $\leftrightarrow$ ground-state energy density
\item
correlations in the horizontal direction $\leftrightarrow$ equal-time correlations
\item
correlations in the vertical direction $\leftrightarrow$ imaginary-time autocorrelations
\item
vertical correlation length $(\xi_{\parallel})~~ \leftrightarrow$ relaxation time, inverse energy gap
\item
anisotropy exponent  $\leftrightarrow$ dynamical exponent
\end{itemize}

The McCoy-Wu model is solved analytically for the boundary spin magnetization \cite{mccoywu} and
for the typical correlations in the vertical direction\cite{shankar}. Both agree
with the strong disorder RG results, which gives further credit about the conjecture
that the strong disorder RG calculation leads to asymptotically exact results.

\subsubsection{Percolation in a random environment}

\label{PERC}

In percolation the $i$-th bond (or site) of a regular lattice is occupied with a given
probability, $p_i$, and one is interested in the properties of clusters, in particular
in the vicinity of the percolation transition point\cite{stauffer}. Singularities at the transition
point are insensitive to homogeneous randomness in the values of $p_i$, which follows
from the Harris criterion Eq.(\ref{harris}), since $\nu_0 > 2/d$ in any dimensions.
Layered randomness in two-dimensions, however, is a relevant perturbation, since
$\nu_0=4/3$\cite{stauffer} and the
critical behavior of this system is studied in Ref.\cite{igloiperco}.

\begin{figure}
\centerline{\includegraphics[width=0.58\linewidth]{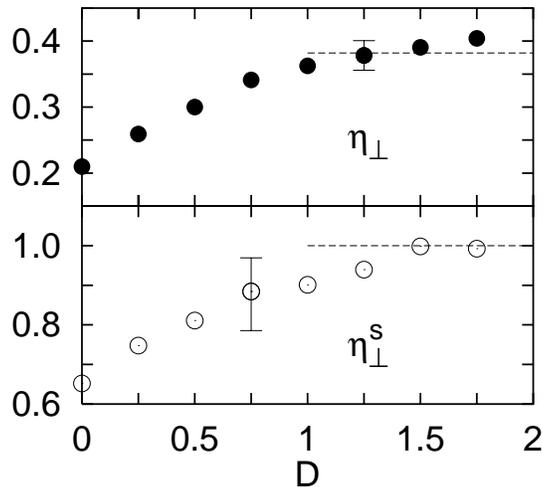}}
\caption{\label{Fig:PERC} Bulk ($\eta_{\perp}=2 x_m$) and surface ($\eta_{\perp}^s=2x_m^s$)
decay exponents versus the strength of disorder for the $2d$ percolation with layered
randomness. Values at the IDFP, as given in Table \ref{TAB1d} are denoted by dashed lines\cite{igloiperco}.}
\end{figure}

In the MC simulation the distribution of the mass of clusters, the anisotropy exponent,
$z$, and the critical bulk and surface correlations in the horizontal direction
are measured for different strength of disorder, $D$. Disorder dependent
critical behavior is found, and the scenario is analogous to that of random
quantum spin chains, as described in Sec.\ref{PCAT}. For weak disorder,
$D < D_{\infty}$,
which corresponds to the {\it  intermediate disorder regime} in Sec.\ref{PCAT}, the critical behavior of the
system is controlled by a line of conventional random fixed points. Here the anisotropy
exponent is finite, $1<z<\infty$, and together with the order-parameter exponents monotonously
increases with the strength of disorder, see Fig.\ref{Fig:PERC}.
In the {\it  strong disorder regime}, $D>D_{\infty}$, the critical behavior of
the system is controlled by the infinite disorder fixed point. Here the anisotropy
exponent is formally infinity and the order-parameter exponents have the same
values as for the RTFIC, see Table \ref{TAB2d}. This means that ordinary percolation
and directed percolation (contact process, see Sec.\ref{RCP}) in the presence of strong layered
randomness have the same fixed point, so that the different type of anisotropy in the
non-random models does not matter.

\subsection{Random-bond Potts model in the large-$q$ limit}

\label{RBPM}

The mapping presented in Sec.\ref{M_W} makes a relation between the RTFIC and an anisotropic
random classical model, the McCoy-Wu model. It is interesting to ask the question if the
RTFIC is isomorph with an {\it  isotropic} random classical model? 
If yes, then the critical singularities of that classical random model should be known exactly from the
properties of the RTFIC. In Ref.\cite{igloipottsq} it is argued that a possible candidate of this r\^ole is
the random ferromagnetic-bond Potts model in the large-$q$ limit, what we will describe in the following.

First we note that disorder has a rounding effect at first-order phase transitions\cite{cardy99}.
In $2d$, according to
rigorous results by Aizenman and Wehr\cite{aizenmanwehr} any amount of quenched disorder will turn a first-order
phase transition into a second-order one. Numerical studies of the $2d$ random-bond Potts
model\cite{picco97,cardy97,chatelain98,olson99,palagyi00} indicate
that the magnetization exponent, $x_m$ is $q$-dependent, and $x_m(q)$ seems to be saturated\cite{jacobsen00}
in the large-$q$ limit. Here we argue that this limiting value is universal and can be obtained from the
strong disorder RG results of the RTFIC.

The partition function of the Potts-model is convenient to express in the random cluster
 representation\cite{kasteleyn}:
\begin{equation}
Z =\sum_{G\subseteq E}q^{c(G)}\prod_{ij\in G}\left[q^{\beta J_{ij}}-1\right]
\label{eq:kasfor}
\end{equation}
where the sum runs over all subset of bonds, $G\subseteq E$ and $c(G)$ stands for the number of connected
components of $G$. The $J_{ij}>0$ are nearest neighbor coupling constants, see Eq.(\ref{Upotts}), and
$\beta=1/(k_B T \ln q)$. In the large-$q$ limit, where $q^{\beta J_{ij}} \gg 1$, the partition function can be written as
\begin{equation}
Z=\sum_{G\subseteq E}q^{\phi(G)},\quad \phi(G)=c(G) + \beta\sum_{ij\in G} J_{ij}\label{eq:kasfor1}
\end{equation}
which is dominated by the largest term, $\phi^*=\max_G \phi(G)$. Thermodynamic quantities are calculated
from the free energy per site, $f=-\beta \phi^*/N$, and the scaling dimension of the magnetization, $x_m$, is
obtained from the fractal dimension of the percolating cluster, $d_f$, as $x_m=d-d_f$. One important
observation that for a given realization of disorder thermal fluctuations are irrelevant and disorder
plays a completely dominant r\^ole, which is a property characteristic for infinite disorder
fixed points, see Sec.\ref{dominance}.

\begin{figure}
 \begin{center}
     \includegraphics[width=2.35in,angle=270]{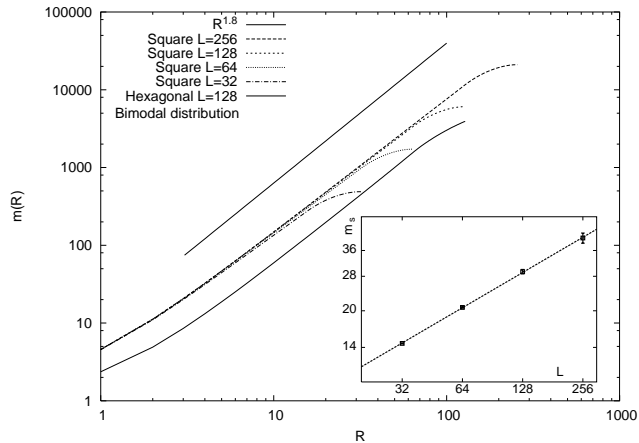}
   \end{center}
\caption{\label{RBPM1} Average mass of the percolating cluster within a region of size $R$ at the
critical point, for the square and the hexagonal lattices. The conjectured asymptotic behavior is
indicated by a straight line: $m(R) \sim R^{d_f}$ with $d_f=(5+\sqrt{5})/4 \approx 1.8$.
Inset: average mass of surface points of the
percolating cluster in the square lattice, $m_s(L) \sim L^{1-x_m^s}$ with the conjectured
value $x_m^s=1/2$.\cite{igloipottsq}}
\end{figure}

In a technical point of view solution of the random bond Potts model in the large-$q$ limit is reduced to
an optimization problem\cite{juhasz01}, which can be solved by a computer algorithm in polynomial
time\cite{angles02}. Numerical results
in 2d \cite{igloipottsq} have given strong support to the conjecture, that the magnetization exponents,
$x_m$ and $x_m^s$, are the same as for the RTFIC in Table \ref{TAB1d}. This is illustrated in
Fig.\ref{RBPM1}. The correlation length exponent is conjectured to be $\nu=1$, which is the half of the
correlation length exponent of the RTFIC.

\begin{figure}
\centerline{\includegraphics[width=0.58\linewidth]{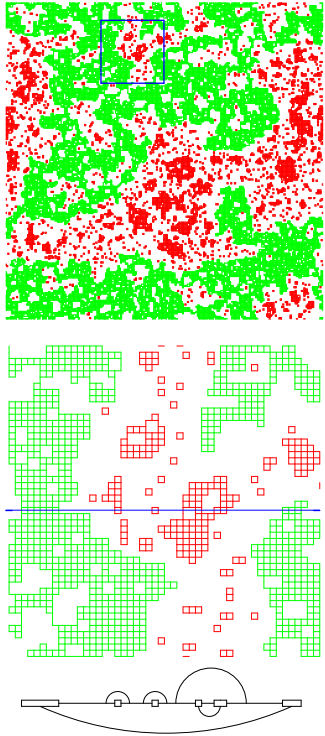}}
\caption{\label{RBPM2}top:  Optimal set for a typical disorder realization of the random bond Potts model:
percolating and finite clusters
are marked with gray (green) and dark (red) lines, respectively. middle:  Enlargement of a square
proportion in the upper-middle part to illustrate
self-similarity. bottom: The connectivity structure of the optimal set along a line, which
consists of six connected units (``spins'') and five open units (``bonds'').
\cite{igloipottsq}}
\end{figure}

To explain this conjectured isomorphism we present two arguments. The first argument concerns the topological
structure of the optimal set, which is show in Fig.\ref{RBPM2} for a typical disorder configuration
at the critical point\cite{kinzel}. The percolating cluster is self-similar, i.e. it is a fractal, and in
a one-dimensional cut it has a connectivity structure, which can be brought in analogy with the
strong disorder RG ground state of the RTFIC. If the connected parts are identified with spin clusters and
the empty parts with (renormalized) bonds to each 1d cut corresponds an RG ground state of the RTFIC,
for which a given set of random couplings and transverse fields can be identified. 
If in the two problems the statistics of the equivalent ground states
is asymptotically similar the singularities of the average quantities are indeed simply related to each
other.

In the second argument we consider the $q$-state Potts model with layered
randomness, when the fractal structure of the dominant graph is the same, as for random
percolation in the strong disorder region, see Sec.\ref{PERC}. Therefore the mass of the giant cluster
scales as: $M \sim L_{\parallel} L_{\perp}^{1-x_m}$. Now let the couplings to be random also in
the vertical direction. In this way
translational symmetry in the vertical direction is broken and in the originally homogeneous (occupied
and non-occupied) strips connected
and disconnected parts will appear. At the critical point, due to duality, these two (creation and destruction)
processes are symmetric, therefore it is plausible to assume that the mass of the largest cluster stays
invariant, i.e. $M \sim L^{d_f} \sim  L^{2-x_m}$, which leads to the conjectured isomorphism.

\newpage

\newpage

\subsubsection{Numerical study in three-dimensions}

In three-dimensions the first-order transition in the non-random model remains unaffected by
weak disorder, however the thermodynamic quantities display essential singularities\cite{mercaldo05}.
Only for strong enough disorder will the transition be soften into a second-order one, in which case the
  ordered phase becomes non-homogeneous at large scales, while the
  non-correlated sites percolate the sample. In the critical regime
  the critical exponents are found universal: $\beta/\nu=0.60(2)$ and
  $\nu=0.73(1)$. The 3d random bond Potts model and the 2d random transverse-field Ising model do not
belong to the same infinite disorder universality class, in contrary to that found in 2d.


\section{Summary and perspectives}

In this review, we have described in details how strong disorder RG methods
allows to study disordered systems  in which space heterogeneities of the disorder dominate at large scale over quantum, thermal or stochastic  fluctuations. 
We have explained how these methods could be applied to different types of random systems.
In all cases, these methods have a very clear physical meaning, because
the RG rules are defined in real space (or in energy space) and use some physical arguments
to construct the degrees of freedom  that are important at large scale:

(i) in quantum models, the RG rules allow to construct clusters of strongly correlated spins,
such as singlets in the $S=1/2$ antiferromagnetic chain,  VBS clusters 
in the $S=1$ antiferromagnetic chain, or ferromagnetic clusters in the Random Transverse Field Ising chain;

(ii) in random walks in random media, the RG rules allow to construct the finite-time metastable
states, and are based on the time needed to cross over a potential barrier or to escape from a trap;

(iii)  in classical random field models, like the random field Ising chain, or the heteropolymer
near an interface, the RG rules can be considered as an extension of the Imry-Ma arguments
based on the energy that can be gained by taking advantage of the fluctuations of the random fields
in an interval.

(iv) in stochastic models of interacting particles, such as the simple exclusion process or
the contact process, the RG rules allow to define effective clusters of strongly correlated particles.

In contrast with usual RG methods which treat space homogeneously, the strong disorder RG procedures 
are inhomogeneous in space to better adapt to the local realizations of the disorder.
Moreover, it has been understood for many years that the probabilistic questions arising in
disordered systems concern extremal statistics  
\cite{derridayk,bouchaudm-extreme}:  
 the equilibrium at low temperature is controlled by the statistics of states of low energy \cite{rem, derridayk, bouchaudm-extreme, carpentier2,deanmaj}, whereas dynamics at large time is controlled by the statistics of large barriers \cite{feigelman,drosselbarrier,balents,vinokuretal}.  From this point of view, the strong disorder RG rules give a new perspective. Indeed, following the Ma-Dasgupta idea, it is always the extreme value of a disorder variable which is decimated iteratively and which generates the RG flow of  the corresponding probability distribution. 
  In dimension $d=1$, it turns out that this special structure makes it possible to obtain explicit solutions most of the time,
whereas in higher dimension $d>1$, the RG rules have to be implemented numerically.  
Whenever these methods can be applied, they give a very detailed description.  Indeed, as the RG procedure is carried out sample by sample, it is possible to study many observables within the same 
framework, their distributions over the samples, and the influence of the rare events/rare regions on averaged quantities.

We now finish with a short list of perspectives that seem to us the most interesting presently 

 \begin{itemize}
 
  \item {\it  New fields of applications for strong disorder RG methods}
 
 It would be of course very interesting to define and study  strong disorder RG rules
 for other disordered systems than the ones that have been considered up to now.
 Indeed, considering a model from the point of view of strong disorder RG
 usually gives completely new insight and new predictions with respect to other methods.
 
 More generally, the strong disorder RG methods show that an essential property of a
 disordered system is the relative importance of disorder inhomogeneities
 with respect quantum, thermal or stochastic  fluctuations. A general classification
 of disordered systems with respect to this ratio `disorder/fluctuations'
 would certainly be very helpful.

\item {\it  Numerical studies of strong disorder RG flows in $d>1$}

The strong disorder RG method has provided several qualitative predictions about the singular
behavior of higher dimensional systems, such as the isomorphism between spin glasses and random
ferromagnets, or the absence of infinite disorder fixed point in Heisenberg antiferromagnets.
Furthermore there are some numerical predictions by the method for the actual values of the
critical exponents, c.f. for the random transverse-field Ising model or for the Heisenberg
model with mixed ferromagnetic and antiferromagnetic interactions. Most of these RG results need
a numerical verification. These future
numerical investigations could also help to explore new problems having disorder dominated singularities.

 \item {\it  Disorder induced cross-over effects}
 
 For several systems considered in this review, such as for instance the $S=1/2$ antiferromagnetic spin chain, or the symmetric Sinai walk, any amount of initial
 disorder drives the system towards an infinite disorder fixed point.
  However for other models, such as for instance the $S >1/2$ antiferromagnetic spin chains,  the quantum clock model or the contact process, the RG rules drive the system towards
 the infinite or strong disorder fixed points only if the initial disorder is larger than a threshold. 
  In these cases, the disorder
induced cross-over effects have not yet been explored in full details.
 A consistent treatment of the different regions of disorder would be of great importance.

  \item {\it  Relations with other methods}

Exact calculations, not using the RG frame-work, are expected to reproduce the  asymptotically
exact results of the strong disorder RG method. For the Sinai model,
some results obtained via strong disorder RG have now be proven by mathematicians \cite{dembo,cheliotis}, and we believe that the mathematical proofs of other strong disorder RG
should be similarly possible.

Another interesting questions concerns the relations of strong disorder RG
with other methods widely used in disordered systems, such as the replica method \cite{replica}, the supersymmetric method \cite{susy}, and the dynamical method
\cite{dynamicreview}  : is it possible to recover some strong disorder RG results
via these other methods which start by averaging over the disorder ?

\end{itemize}

\section*{Ackowledgements}
\addcontentsline{toc}{section}{Ackowledgements}

It is a pleasure to thank our collaborators on strong disorder RG
methods :

Ferenc Igl\'oi wishes to thank J-C. Angl\'es d'Auriac,  E. Carlon, B. Dou\c{c}ot, J. Hooyberghs, 
R. Juh\'asz, D. Karevski,  N. Kawashima, P. Lajk\'o,
Y-C. Lin, R. M\'elin, M.-T. Mercaldo, M. Preissmann, H. Rieger, L. Santen, A. Seb\H o, L. Turban and
C. Vanderzande.

C\'ecile Monthus wishes to thank  
D.S. Fisher, O. Golinelli, Th. Jolicoeur and P. Le Doussal
for their collaborations on strong disorder RG methods, as well as 
G. Biroli and T. Garel for useful comments on the manuscript.

The work of F.I. has been supported by the French-Hungarian cooperation programme
Balaton (Minist\'ere des Affaires Etrang\`eres - OM), by a German-Hungarian
exchange program (DAAD-M\"OB), the Hungarian National Research Fund under  grant No OTKA TO34183, TO37323,
TO48721, MO45596 and M36803. He thanks the SPhT Saclay for hospitality.


\newpage

\part*{APPENDICES}

\addcontentsline{toc}{part}{APPENDICES}

\appendix

\section{Scaling in random systems}

Here we summarize different types of scaling behaviors observed in random systems and
presented in context of specific systems in this review. Our aim is to give here a uniform
description of singularities and list the known results. After setting the general notions we
consider the different scaling types. For a detailed presentation of the theory of scaling in critical
systems we refer to\cite{fisherme,ma1,fisherme1,cardy1} and in quantum systems to \cite{sachdev}.

\subsection{General notions}

\begin{itemize}

\item
Strength of disorder - $D$ 

We consider an interacting many-particle system, in which the interactions $\{ \lambda_{i,j}\}$ are
independent and identically distributed random variables. The strength of disorder, $D$, is related
to the broadness of the distribution. One possibility is to identify $D^2$ with
the variance of the distribution of the logarithmic couplings. In this case a generic distribution
is given by:
\begin{equation}
P(\lambda)=D^{-1} \lambda^{-1+1/D}\;
\label{distr_D}
\end{equation}
with $0 \leq \lambda \leq 1$.
The $\lambda_{i,j}$-s are couplings in spin
models, transition probabilities in a random walk or in reaction-diffusion systems, etc.

\item
Length-scale - $L$, energy-scale - $\Omega$, time-scale - $t_r$ 

The length-scale, $L$, can be most simply defined for a finite system by its linear extent.
The corresponding energy-scale, $\Omega$, is the typical excitation energy in such a finite
system. More generally, in an infinite system for a given type of localized excitation of
typical energy, $\Omega$, the average distance between two such excitations is measured by $L$.
The time-scale associated to such type of excitation is given by $t_r \sim \Omega^{-1}$.

\item
Control parameter - $\delta$

In these systems there are deterministic fluctuations which are generally due to a (disordering) field,
such as the
temperature, $T$, a
set of transverse-fields $\{ h_i \}$, enforced dimerization, bias in a random walk or in a
driven lattice gas, etc. The combined effect of interactions and fields (fluctuations) can be
measured by an average quantity, which is called the control parameter, $\delta$. Generally $\delta$
is defined in such a way that at $\delta=0$ there is a singularity in the system which is
associated to a phase transition. Examples for control parameters of different models
can be found, c.f. in Eqs.(\ref{delta_I}),(\ref{delta_d}),(\ref{deltarw}) and (\ref{delta_A}).

\item
Dynamical variables - $\sigma_i(t)$

These are variables which enter into the definition of the model, c.f. operators in the
Hamiltonian, site occupation variables in random walks or lattice gases, etc. Generally $\sigma_i(t)$
depends on the position, $i$, and time, $t$, and their average value is often related to the order
in the system.

\item
Averaging procedure

For a physical observable, say, $O$, averaging in the presence of quenched disorder is made by
the usual way\cite{brout,binderyoung}. First, for a given realization of disorder one performs thermal averaging,
which is denoted by $\langle  O \rangle$,
and this is followed by an average over disorder, for which we use the
notations $O=\overline{\langle  O \rangle}$ or $O=[\langle  O \rangle]_{\rm av}$.

\item
Equal time correlation function - $C(r)=\overline{\langle \sigma_i(0) \sigma_{i+r}(0) \rangle}$

Its long-distance limit:
\begin{equation}
\lim_{r \to \infty} C(r)=\left\{
\begin{aligned}
m^2, \quad &\delta& <0\\
0, \quad &\delta& >0
\end{aligned}
\right.
\end{equation}
is used to define the order-parameter, $m$, in the system. In the ordered phase, $\delta <0$,
one often uses the connected correlation function: $C^c(r)=C(r)-m^2$.

\item
Correlation lengths - $\xi_{av}$ and $\xi_{typ}$

The correlation length is defined by:
\begin{equation}
\xi^{-1}_{av}=\lim_{r \to \infty} -~\frac{\ln C(r)}{r}\;,
\end{equation}
in the paramagnetic phase, whereas in the ordered phase one uses the connected correlation
function. The typical correlation length, $\xi_{typ}$, is defined by:
\begin{equation}
\xi_{typ}^{-1}=\lim_{r \to \infty} -~\frac{\overline{\ln\ \langle \sigma_i(0) \sigma_{i+r}(0) \rangle}}{r}. \;,
\end{equation}
In systems in which the correlation function is self-averaging $\xi_{av}=\xi_{typ}$, which is
the case c.f. for the $d=2$ random bond Ising model \cite{stinchcombe,lajkoigloi}.

In several systems, however, the correlation function is not self-averaging and therefore
$\xi_{av} \ne \xi_{typ}$. Examples are the $d=1$ Ising model\cite{derr}, the random bond Potts models in
$d=2$ \cite{olson99,palagyi00,parisi} and the RTFIC \cite{danielrtfic}.

\item
Autocorrelation function - $G_i(t)=\overline{\langle \sigma_i(t) \sigma_{i}(0) \rangle}$

Its asymptotic limit is used to define the local order-parameter, $m_i$:
\begin{equation}
\lim_{t \to \infty} G_i(t)=\left\{
\begin{aligned}
m_i^2, \quad &\delta& <0\\
0, \quad &\delta& >0
\end{aligned}
\right.
\end{equation}
For surface sites it is the surface order-parameter, $m_1 \equiv m_s$.
\item
Singular points

Spatial and (or) dynamical correlations are quasi-long-ranged (i.e. they have a power-law asymptotic
decay) at a critical (semi-critical) point.

\begin{itemize}
\item
Critical point

At this point, which is generally located at $\delta=0$, in an infinite system both the physical length scale,
$\xi$, and the time scale, $t_r$, are divergent. In the vicinity of the critical point one has: $\xi \sim
|\delta|^{-\nu}$.

\item
``Semi-critical'' points

In a random system generally there is a line of semi-critical fixed points, in which - due to Griffiths
singularities - the time scale is divergent, although $\xi$ is finite.

\end{itemize}
\item
Relevant scaling fields - $H$, $T$, etc.

At a singular point one generally considers the behavior of different physical quantities as
a function of relevant scaling fields, such as a small ordering
field of strength, $H$, or in a quantum system as a function of a small temperature, $T$.

\item
Scaling transformation

At a singular point, static and (or) dynamical correlations transform covariantly under
a scaling transformation, when lengths are rescaled by a factor, $b>1$, i.e. $L'=L/b$.
For example at the critical point the order-parameter correlation function behaves as\cite{fisherme}:
\begin{equation}
C(r)=b^{-2x_m}C(r/b)\;.
\label{C_b}
\end{equation}
Similarly, static and (or) dynamical densities, such as the local magnetization and (or)
the local susceptibility obey scaling relations\cite{fisherme}.

\end{itemize}

\subsection{Conventional random critical scaling}
\label{SC_conv}
In a conventional random critical point:
\begin{itemize}
\item
dynamical scaling is anisotropic
\item
densities in large scale are homogeneous
\end{itemize}
The correlation length and the relaxation time  are related as:

\begin{equation}
t_r \sim \xi^z\;,
\label{scales_conv}
\end{equation}
with a dynamical exponent, $1 \le z < \infty$. From this follows that in a finite system the
excitation energies are typically:
\begin{equation}
\Omega \sim L^{-z}\;,
\label{scales_conv1}
\end{equation}
and are transformed as $\Omega'=\Omega b^z$.
Distribution of the low-energy excitations, $P(\epsilon)$, ($\epsilon \sim \Omega$) is assumed
to be in the form as given in Eq.(\ref{distr_D}). Under a scaling transformation
the distribution function is expected to satisfy the relation:
\begin{equation}
P(\epsilon) {\rm d} \epsilon =b^{-d} P( \epsilon') {\rm d} \epsilon'
=b^{-d+z/D} P(\epsilon) {\rm d} \epsilon\;,
\label{P_scal}
\end{equation}
provided the low-energy excitations are localized, thus their density is transformed by
a factor of $b^{-d}$, where $d$ is the dimension of the system. The fixed point of Eq.(\ref{P_scal})
is given by:
\begin{equation}
D=z/d\;,
\label{D=z/d}
\end{equation}
thus the strength of disorder is finite and proportional to the dynamical exponent.

The order-parameter (magnetization), $m(\delta,L)$, in a finite system satisfies the scaling relation:
\begin{equation}
m(\delta,L)=b^{-x_m}m(\delta b^{1/\nu},L/b)\;.
\label{scaling_m}
\end{equation}
Taking the scaling parameter $b=\delta^{-\nu}$ we have $m(\delta,L)=\delta^{x_m \nu} \tilde{m}(L \delta^{\nu})$,
where the scaling function in the limiting cases behaves as: $m(y)=const$, if $y \to \infty$
and $m(y) \sim y^{-x_m}$, if $y \to 0$. Thus in the thermodynamic limit, $L \to \infty$,
$m(\delta) \sim \delta^{\beta}$ with $\beta=x_m \nu$. On the other hand at the critical
point, $\delta=0$, the finite-size scaling behavior is: $m(\delta=0,L) \sim L^{-x_m}$. We note that for a
surface spin $x_m$ is replaced by the appropriate surface exponent, $x_m^s$.

The scaling law of the critical correlation function is given in Eq.(\ref{C_b}), which is
generalized outside the critical point:
\begin{equation}
C(\delta,r)=b^{-2 x_m} C(\delta b^{1/\nu},r/b)\;.
\label{scaling_C}
\end{equation}
Here with $b=r$ we obtain $C(\delta,r)=r^{-2 x_m} \tilde{C}(\delta r^{1/\nu})$, where the scaling function 
is $\tilde{C}(0)=const$ and $\tilde{C}(y) \sim \exp(-y^{\nu})$ for $y \to \infty$.

The autocorrelation function transforms similarly to Eq.(\ref{C_b}):
\begin{equation}
G(\delta,t)=b^{-2 x_m} G(\delta b^{1/\nu},t/b^z)\;,
\label{scaling_G}
\end{equation}
and with $b=t^{1/z}$ we obtain $G(\delta,t)=t^{-2 x_m/z} \tilde{G}(\delta t^{1/\nu z})$. Here the scaling
function is $\tilde{G}(0)=const$, whereas $\tilde{G}(y)$ for large $y$ has a decay, which is slower than
$\exp(-y^{\nu z})$. Its proper form is determined by classical Griffiths singularities\cite{griffiths}.

At a conventional random quantum critical point, which takes place at $T=0$, the low-temperature behavior
of singular quantities, such as the local susceptibility, $\chi(T)$, and the specific heat, $C_v(T)$,
can be obtained from the consideration that the temperature sets an energy-scale, $\Omega \sim T$,
which corresponds to a thermal length, $L_T \sim T^{-1/z}$. Taking the rescaling factor, $b=L_T$, we obtain
for small $T$:
\begin{equation}
\chi(T) \sim T^{-\gamma/\nu z},\quad C_v(T) \sim T^{-\alpha/\nu z}\;,
\label{T_conv}
\end{equation}
where $\gamma=(d-2x_m)\nu$ and $\alpha=2-d\nu$ are the standard susceptibility and specific heat
exponents, respectively.

Similarly, for a (longitudinal) ordering field the energy-scale is given by $\Omega_H \sim H m L_H^d$, which
is set a length-scale: $L_H \sim H^{-1/(d+z-x_m)}$. Taking the rescaling factor, $b=L_H$, we obtain
for small $H$:
\begin{equation}
\chi(H) \sim H^{-\gamma/\nu (d+z-x_m)},\quad C_v(H) \sim H^{-\alpha/\nu (d+z-x_m)}\;.
\label{H_conv}
\end{equation}

\subsection{Infinite disorder scaling}
\label{SC_st}

An infinite disorder fixed point has two important properties:
\begin{itemize}
\item
activated scaling
\item
densities are non-homogeneous in large scales, average values are dominated by rare regions.
\end{itemize}
Activated scaling means that the log time-scale is related
to the length-scale:
\begin{equation}
\ln t_r \sim \xi^{\psi}\;,
\label{logscale}
\end{equation}
thus in a finite system the excitation energies are typically:
\begin{equation}
|\ln \Omega| \sim L^{\psi}\;,
\label{scales_conv2}
\end{equation}
and are transformed as $\ln \Omega'=\ln \Omega b^{-\psi}$.

Distribution of the log of the low-energy excitations is assumed
to follow the transformation:
\begin{equation}
P(\ln \epsilon,L)  =b^{-\psi} P(b^{-\psi}\ln  \epsilon,L/b)\;.
\label{P_inf}
\end{equation}
As the length, $L \sim b$, increases the strength of disorder increases,
too, and in the fixed point the disorder becomes infinitely strong. (Note the difference
with Eq.(\ref{P_scal}) for conventional scaling.)

Scaling of a physical observable, $O$, which can be an average density or a correlation function,
is dominated by rare regions (realizations) in which regions (samples) it has a value of
$\langle O \rangle_{rare}=O(1)$,
whereas in typical samples it is (exponentially) small,
$\langle O \rangle_{typ}=O(\exp(-a L^{\omega})$, $\omega=O(1)$.
Explicit examples of rare events can be found in Sec.\ref{m_s} for the
surface magnetization of the RTFIC and in Sec.\ref{persistence} for the persistence of the Sinai walk.
The average value, $\overline{\langle O \rangle}$, has the same scaling properties as the fraction of
rare events or the density of rare regions, $\rho_O$.

Scaling of the average order-parameter, $m$, follows from scaling of the density of locally ordered regions,
$\rho_m$, which is given by:
\begin{equation}
\rho_m(\delta,L)=b^{-x_m}\rho_m(\delta b^{1/\nu},L/b)\;.
\label{scaling_rhom}
\end{equation}
Note, that this relation is in identical mathematical form as for conventional random critical scaling
in Eq.(\ref{scaling_m}), therefore $m$ has the same properties in the two cases, see below Eq.(\ref{scaling_m}).
The same conclusion holds for the spatial correlation function, $C(r)$, too. In this case
a rare event consists of two independent, locally ordered regions which are separated by a distance, $r$.
The density of rare regions of the correlation function, $\rho_C$,
is given by $\rho_C \sim \rho_m^2$ and through $C \sim \rho_C$ we arrive to the scaling law
in Eq.(\ref{scaling_C}) and to the conclusion described below this equation.

In the strong disorder RG one often considers the global order parameter, $\mu=m L^d$, which is
called the cluster moment. This obeys the scaling relation:
\begin{equation}
\mu(\delta,|\ln \Omega|)=b^{d_f}\mu(\delta b^{1/\nu},|\ln \Omega| b^{\psi})\;,
\label{scaling_mu1}
\end{equation}
in which $d_f=d-x_m$ is the fractal dimension of the cluster. Taking the length-scale,
$b=|\ln \Omega|^{1/\psi}$, we have at the critical point:
\begin{equation}
\mu(\delta=0,|\ln \Omega|) \sim |\ln \Omega|^{\phi},\quad \phi=(d-x_m)/\psi\;.
\label{scaling_mu2}
\end{equation}
We note that the typical correlation length at an infinite disorder fixed point is much smaller then the
(true) correlation length and they are related as given in Eq.(\ref{C_typ}): $\xi_{typ} \sim \xi^{1-\psi}_{av}$.

To calculate the average autocorrelation function one should keep in mind that disorder in
the time-direction is strictly correlated. Thus, if there is local order at a given rare region,
say at site, $i$, at time $0$, the
order stays for any later time, $t$. Consequently the density of rare regions of $G$ is given by:
$\rho_G \sim \rho_m$, so that the average autocorrelation function satisfies the scaling relation:
\begin{equation}
G(\delta,\ln t)=b^{-x_m} G(\delta b^{1/\nu},\ln t/b^{\psi})\;.
\label{lauto}
\end{equation}
Now taking $b=(\ln t)^{1/\psi}$ we obtain:
\begin{equation}
G(\delta,\ln t)=(\ln t)^{-x_m/\psi} \tilde{G}(\ln t \delta^{\nu \psi})\;,
\label{lauto1}
\end{equation}
thus the decay of critical autocorrelations is ultra-slow, logarithmic in time. In the disordered
phase the scaling function,
$\tilde{G}(y)$ is expected to behave for large arguments as: $\ln \tilde{G}(y) \sim -y$. Thus the dominant
decay of the average autocorrelation function in the disordered region is in a power-law form:
\begin{equation}
G(\delta,\ln t) \sim (\ln t)^{-x_m/\psi}~t^{-\eta(\delta)}, \quad \delta>0\;,
\label{lauto2}
\end{equation}
which is supplemented by logarithmic corrections. This region is the disordered Griffiths
phase in which the decay exponent, $\eta(\delta)$, depends on the control parameter. For $\delta \to  0$
it goes to zero as $\eta(\delta) \sim \delta^{\nu \psi}$.

The low-temperature singularities at a quantum critical point can be obtained by setting the length-scale
$b=L_T \sim (\log T)^{1/\psi}$. For the specific heat the rare events are low-energy excitations
which are separated by a distance, $L_T$, thus $C_v(T) \sim L_T^{-d}$. For the susceptibility
the rare events bring a Curie-type contribution, thus we obtain in analogy with Eq.(\ref{T_conv}),
$\chi(T) \sim L_T^{d-2x_m}/T$, thus:
\begin{equation}
\chi(T) \sim \frac{\left(\ln T \right)^{(d-2x_m)/\psi}}{T}, \quad C_v(T) \sim \left( \ln T \right)^{-d/\psi}\;.
\label{T_inf}
\end{equation}
For a small ordering field, $H$, the thermal and magnetic energy-scales can be compared: $T \sim H m L_H^d
\sim H L_H^{d-x_m}$, from which we have the relation: $H \sim T (\ln T)^{d-x_m}$. Putting it into
Eq.(\ref{T_inf}) we obtain the small $H$ singularities at $T=0$:
\begin{equation}
\chi(H) \sim \frac{\left(\ln H \right)^{-x_m/\psi}}{H}, \quad C_v(H) \sim \left( \ln H \right)^{-d/\psi}\;.
\label{T_inf1}
\end{equation}
\subsection{Scaling in the Griffiths phases}
\label{SC_gr}

In the Griffiths-phase:
\begin{itemize}
\item
the correlation length is finite, the relaxation time is divergent
\item
dynamical scaling (between $L$ and $t_r$) is anisotropic
\item
densities are non-homogeneous in large scales, average values are dominated by rare regions.
\end{itemize}
The physical origin of the singular behavior of disordered systems outside the critical point is
due to rare regions, in which strong fluctuations of the local couplings prefer the existence of
the thermodynamically non-stable phase locally. Scaling in the ordered and disordered Griffiths
phases is somewhat different, in particular for $d>1$.

\subsubsection{Disordered Griffiths phase}

In the disordered phase a small region is locally ordered if the couplings there are larger then
the average disordering field. To find such a rare region of linear
size, $l_c$, is exponentially small, $p(l_c) \sim \exp(-\alpha l_c^d)$, however these regions are
extremely stable against fluctuations and have a typical relaxation time, $t_r \sim \exp(\sigma l_c^d)$.
Indeed the lowest gap of a finite, ordered system is given by $\epsilon \sim \exp(-\sigma l_c^d)$ and
$t_r \sim \epsilon^{-1}$. Then the distribution of large relaxation times has an algebraic tail:
$p(t_r) \sim t_r^{-d/z-1}$, with $d/z=\alpha/\sigma$, which is supplemented by a logarithmic
correction factor, $(\ln t_r)^{-1+1/d}$.  The leading behavior of the average autocorrelation function:
\begin{equation}
G(t) \sim \int {\rm d} t_r p(t_r) \exp(-t/t_r) \sim t^{-d/z}\;
\label{auto_g}
\end{equation}
is algebraic and the decay exponent is $\delta$ dependent. Comparing with the scaling form in Eq.(\ref{lauto2})
we obtain for small $\delta$:
\begin{equation}
\frac{d}{z} \sim \delta^{\nu \psi}\;,
\end{equation}
see also Eq.(\ref{C_typ}).

From the distribution of the relaxation times one obtains for the distribution of the small gaps:
\begin{equation}
P(\epsilon) \sim \epsilon^{-1+d/z}\;,
\label{P_eps_gr}
\end{equation}
which is in the same form as in Eq.(\ref{distr_D}). Thus in the Griffiths-phase there is finite disorder the
strength of which is given by $D=z/d$. Since the excitations are localized the scaling transformation
of $P(\epsilon)$ is the same as in Eq(\ref{P_scal}) for conventional scaling. Consequently one has the relation:
\begin{equation}
\Omega \sim L^{-z}\;.
\label{scales_gr1}
\end{equation}
The scaling form of the autocorrelation function follows from the observation that the density of rare
regions in the Griffiths phase is just the density of the system which transforms as
\begin{equation}
G(t)=b^{-d}G(t/b^z)\;.
\label{auto_g1}
\end{equation}
Here with $b=t^{1/z}$ we recover the relation in Eq.(\ref{auto_g}).

The low-temperature singularities can be obtained by setting the length-scale, $b=L_T \sim T^{-1/z}$, and
make use of the fact that the density of rare regions is the particle density, both for the specific heat,
$C_v(T) \sim L_T^{-d}$, and for the susceptibility $\chi(T) \sim L_T^{-d}/T$, where in the latter case a
Curie-type contribution per site is taken into account. We thus obtain:
\begin{equation}
\chi(T) \sim T^{-1+d/z},\quad C_v(T) \sim T^{d/z}\;.
\label{T_gr}
\end{equation}
The small $H$ singularities at $T=0$ are obtained by using the length-scale, $b=L_H \sim H^{-1/z}$. Note
that in the disordered phase for $H \sim T$, $L_H \sim L_T$ and we obtain as in Eq.(\ref{T_gr}):
\begin{equation}
\chi(H) \sim H^{-1+d/z},\quad C_v(H) \sim H^{d/z}\;.
\label{H_gr}
\end{equation}
\subsubsection{Ordered Griffiths phase}

In the ordered Griffiths phase the relevant excitations are connected to
such large ordered domains, which are isolated from the macroscopic ordered regions of the system. Therefore
the probability of the existence of such an ordered domain of linear size, $l_c$, is generally smaller, than
the similar expression in the disordered phase. In one dimension one needs $l_c$ weak and consecutive
$l_c$ strong couplings, thus the corresponding probability is $p_o(l_c) \sim \exp(-\alpha 2l_c)$. Since
the typical relaxation time stays the same as in the disordered phase, $t_r \sim \exp(\sigma l_c)$,
the autocorrelation function in Eq.(\ref{auto_g}) involves the decay exponent, $2/z$. This is in agreement
with the strong disorder RG calculation for the RTFIC in Eq.(\ref{G_t4}). 

In $d>1$, as argued in Sec.\ref{Gr_ord} for a successful isolation of a cluster of $l_c^d$ sites one needs a
distance from the ordered domain at least, $l_o \sim l_c^d$. Consequently the probability of the existence
of an isolated large cluster reads as $p(l_c) \sim \exp(-\alpha' l_o^d) \sim  \exp(-\alpha l_c^{d^2})$. Now,
with $t_r \sim \exp(\sigma l_c^d)$ we obtain from Eq.(\ref{auto_g}) an enhanced power-law form:
\begin{equation}
G(t) \sim \exp(-A|\ln t|^d)\;,
\label{auto_g2}
\end{equation}
as already presented in Eq.(\ref{G_d}).

Relation between the length-scale, $L$, and the energy-scale, $\Omega$, can be obtained in the following
way. In a system of linear size, $L$, the typical size of the largest isolated cluster follows from extreme
value statistics and given by: $l_c^{d^2} \sim \ln L$. On the other hand the excitation energy due to
this cluster is $\Omega \sim 1/t_r \sim \exp(-\sigma l_c^d)$, consequently:
\begin{equation}
|\ln \Omega| \sim (\ln L)^{1/d}\;.
\label{lnO_lnL}
\end{equation}
The low-temperature singularities can be obtained as in the disordered phase
however with a length-scale, $L_T \sim \exp\left[ A |\ln T|^d\right]$. In this way we obtain:
\begin{equation}
\chi(T) \sim \frac{1}{T} \exp\left[ - C |\ln T|^d\right],\quad C_v(T) \sim \exp\left[ - C' |\ln T|^d\right] \;,
\label{T_gr_o}
\end{equation}
and similarly for the small $H$ dependence, by the substitution $H \sim T$.

\subsection{Scaling in the large spin phase}

This phase is observed in random Heisenberg models and characterized by 
\begin{itemize}
\item
a size-dependent effective spin
\item
anisotropic dynamical scaling
\item
often non-localized excitations.
\end{itemize}

The typical value of the large spin, $S_{eff}$, grows with the size as:
\begin{equation}
S_{eff} \sim L^{d \zeta}\;,
\label{S_eff_L}
\end{equation}
where $\zeta$ is often equal to (or close to) $1/2$, see Sec.\ref{LSFP}. The energy- and length-scales
are related as in the Griffiths-phase:
\begin{equation}
\Omega \sim L^{-z}\;,
\label{scales_ls}
\end{equation}
so that the relation between spin and energy, $S_{eff} \sim \Omega^{-\kappa}$, involves an exponent,
$\kappa=d\zeta/z$. The gap exponent, $\omega$, defined as $P(\epsilon) \sim \epsilon^{\omega}$,
$\epsilon \to 0$, however, is generally not given by $\omega=-1+d/z$, as in the Griffiths-phase,
see Eq.(\ref{P_eps_gr}), while the excitations are often non-localized.

The low-temperature singularities are given by:
\begin{equation}
\chi(T) \sim \frac{1}{T} ,\quad C_v(T) \sim T^{2\zeta(\omega+1)} |\ln T|\;,
\label{T_ls}
\end{equation}
whereas for a small ordering field the order parameter behaves as:
\begin{equation}
m(H) \sim H^{\zeta(\omega+1)/[1+\zeta(\omega+1)]}\;.
\label{H_ls}
\end{equation}

\section{Mapping between different models}

This Appendix describe the mapping between different models considered in this review.

\subsection{The random RW and the RTFIC}

\label{RW_RTFIC}

Solution of the master equation for the RW in Eq.(\ref{mastersinai}) necessitates the solution of the eigenvalue
problem of the Fokker-Planck operator, $ \underline{\underline{M}}$, the matrix elements of which take the form
$(\underline{\underline{M}})_{i,j}=w_{i,j}$
for $i\ne j$ and $(\underline{\underline{M}})_{i,i}=-\sum_j w_{i,j}$.
All the physical properties of the model can be expressed in terms of
the left and right eigenvectors $\underline{u}_q$ and $\underline{v}_q$,
respectively, and the eigenvalues $\lambda_q$, which are non-positive. The eigenvalue problem in terms of the components of the right eigenvector $v_q(i)$ reads as:
\begin{equation}
w_{i-1,i}\, v_{q}(i-1)-(w_{i,i-1}+w_{i,i+1})\, v_{q}(i)
+w_{i+1,i}\, v_{q}(i+1)=\lambda_q\, v_{q}(i)\;.
\label{master1}
\end{equation}
Here we consider a finite system of size $L$, i.e., we put $w_{0,1}=w_{L+1,L}=0$.
Then we introduce the new variables:
\begin{equation}
v(i)=\alpha_i{\tilde v}(i)\,,\qquad
\alpha_{i+1}=\alpha_{i}\left({w_{i,i+1}\over w_{i+1,i}}\right)^{1/2}\!\!\!=\alpha_{1}\left(\prod_{j=1}^i{w_{j,j+1}\over w_{j+1,j}}\right)^{1/2}\,,
\label{vi}
\end{equation}
in terms of which the eigenvalue problem is transformed into:
\begin{equation}
\left(w_{i-1,i} w_{i,i-1}\right)^{1/2} {\tilde v}_{q}(i-1)
-(w_{i,i-1}+w_{i,i+1}){\tilde v}_{q}(i)
+\left(w_{i+1,i} w_{i,i+1}\right)^{1/2}{\tilde v}_{q}(i+1)=\lambda_q {\tilde
v}_{q}(i)\,,
\label{master2}
\end{equation}
which corresponds to a real symmetric eigenvalue problem $\sum_j
S_{ij}\tilde{v}_{q}(j)=\lambda\tilde{v}_{q}(i)$ with $S_{ij}=S_{ji}$.
Consequently the eigenvalues $\lambda_q$ of the FP operator are real.

The symmetric matrix, $\underline{\underline{M}}$, can be compared with the square of the
matrix, $\underline{\underline{T}}^2$, in Eq.(\ref{trid}), which appears in the eigenvalue problem
of the free-fermion representation of the RTFIC. The two problems are equivalent, if we have the correspondences:
\begin{equation}
\begin{array}{rcl}
J_i &\iff &\left(w_{i+1,i}\right)^{1/2}\\
h_i &\iff &\left(w_{i,i+1}\right)^{1/2}\\
\epsilon_q^2 &\iff & -\lambda_q\,.
\end{array}
\label{corresp}
\end{equation}
Thus there is a {\it  mathematical equivalence} between the random RW
and the RTFIC with the corresponding
random couplings, as described in~(\ref{corresp}). The relation between the values of the low-energy
excitations in Eq.(\ref{corresp}), which leads to the exact result of the dynamical exponent of the RTFIC in
Eq.(\ref{z_I}) can be extended by a relation between the order-parameters:
\begin{equation}
[{m}_s(L)]^2\iff p_{per}(L)\,.
\label{corresppr}
\end{equation}
The correspondence between the two models outlined in this section explains the presence of the same type of strong
disorder fixed points in the two models. This mapping, however, does not establish relations between another fundamental observables, such as the bulk magnetization of the RTFIC (see Ref.\cite{rieger99}).

\subsection{The random $XY$ chain and the RTFIC}
\label{XY_RTFIC}
The decimation equations in Eqs.(\ref{hdecimation}) and (\ref{Jdecimation}) are very similar to the decimation equation for the XX chain in Eq.(\ref{JXX}), in particular if dimerization is considered. This analogy is the consequence of an exact mapping between the $XY$ model and the RTFIC, what we show here for an open $XY$ chain of $L = even$ sites with open boundary conditions. In terms of the spin operators $S_j^{(x,y)}, j=1,2,\dots,L$ we introduce two sets of
Pauli operators $\sigma_l^{(x,z)},\tau_l^{(x,z)}, l=1,2,\dots,L/2$ as:
\begin{eqnarray}
\sigma_i^x&=&\prod_{j=1}^{2i-1} \left( 2S_j^x \right),~~~\sigma_i^z=4 S^y_{2i-1} S^y_{2i}
\nonumber\\
\tau_i^x&=&\prod_{j=1}^{2i-1} \left( 2S_j^y \right),~~~\tau_i^z=4 S^x_{2i-1} S^x_{2i}\;.
\label{oprel}
\end{eqnarray}

The original $XY$ Hamiltonian, $H_{XY}$ with
$L$ spins can be expressed as the sum of two TIM-s with variables $\sigma_l^{x,z}$ and
$\tau_l^{x,z}$ each of which of $L/2$ spins as:
\begin{equation}
H_{XY}=H_{TIM}(\sigma)+H_{TIM}(\tau)\;.
\label{XYTIM}
\end{equation}
where the couplings and the transverse fields are given by:
\begin{eqnarray}
J_l(\sigma)=\frac{1}{4} J^x_{2l},~~~h_l(\sigma)=\frac{1}{4} J^y_{2l-1}
\nonumber\\
J_l(\tau)=\frac{1}{4} J^y_{2l},~~~h_l(\tau)=\frac{1}{4} J^x_{2l-1}\;.
\label{Jhrel}
\end{eqnarray}
Correlations in the two models are related as can be found in Refs.\cite{danielantiferro,igloixy}.

\section{Kesten random variables : exact results versus strong disorder RG}

\label{appkesten}

Since Kesten random variables
naturally appear in various models considered in this review,
we first recall in this Appendix
some important properties of these variables. 
We then explain how the strong disorder RG approach 
corresponds for these variables to a saddle-point approximation 
in each sample, that becomes asymptotically exact
for large samples.

\subsection{ Discrete Kesten random variables}

A Kesten random variable \cite{kestenetal}
has the following specific structure of 
a sum of products of random variables $y_i$
\begin{eqnarray}
Z_L \equiv \sum_{i=1}^L \prod_{j=1}^i y_j
\label{kestenzl}
\end{eqnarray}
This type of random variables appears
in the context of random walks in random media
for various observables \cite{solomon,sinai,derridapomeau}
(see Eq \ref{v_d}), in the surface magnetization 
(\ref{peschel}) of the RTFIC \cite{igloi98,dharyoung},
as well as in the random field Ising chain
via the formulation with $2 \times 2$ random transfer matrices \cite{derridahilhorst,calan}.

It is actually convenient to rewrite the Kesten random variable
(\ref{kestenzl}) with a varying left boundary
\begin{eqnarray}
Z(a,b)=\sum_{i=a}^b \prod_{j=1}^i y_j
\label{kestendiscrete}
\end{eqnarray}
The fundamental property of $Z(a,b)$ is the recurrence equation 
\begin{eqnarray}
Z(a,b)=  y_a \left[ 1+ Z(a+1,b) \right]
\label{rec}
\end{eqnarray}
where the random coefficient $ y_a$ appears multiplicatively :
$Z(a,b)$ is thus a multiplicative stochastic process.

One of the main outcome of the studies on the random variable $Z(a,b)$ 
is that, in the limit 
of infinite length $L=b-a \to \infty$, there exists a limit distribution
$P_{\infty}(Z)$ if 
\begin{eqnarray}
[\ln y ]_{av} <0
\label{lnsnegatif}
\end{eqnarray}
Moreover, the limit distribution then presents
the algebraic tail \cite{kes73,kestenetal,derridahilhorst,calan} 
\begin{eqnarray}
P_{\infty}(Z) \opsim_{Z \to \infty} \frac{1}{Z^{1+\mu}}
\label{zunplusmu}
\end{eqnarray}
where the exponent $\mu$ is defined
as the positive root $\mu>0$ of the equation
\begin{eqnarray}
[ y^{\mu} ]_{av} =1
\label{eqmudiscret}
\end{eqnarray}
In the field of random walks in random media, this exponent $\mu$
is known to govern the anomalous diffusion behavior $x \sim t^{\mu}$
in the domain $0<\mu<1$ \cite{kestenetal,derridapomeau,jpbreview}.
In the context of the RTFIC, the exponent $\mu$ defined by (\ref{eqmudiscret})
has for analog the dynamical exponent $1/z$.

\subsection{ Continuous version of Kesten random variables}

The continuous version of the Kesten random variable (\ref{kestendiscrete})
is the exponential functional \cite{jpbreview,flux}
\begin{eqnarray}
Z[a,b]= \int_a^b dx e^{- \int_a^x dy F(y)}
\label{kestencontinuous}
\end{eqnarray}
where $\{F(x)\}$ is the random process 
corresponding to the random variables $(-\ln y_i)$
in the continuous limit.  The analog of the recurrence equation (\ref{rec})
is the stochastic differential equation
\begin{eqnarray}
\partial_a Z[a,b]= F(a) Z[a,b]-1 
\label{langevenmultipli}
\end{eqnarray}
where the random process $F(x)$ appears multiplicatively,
in contrast with usual Langevin equations where the noise appears additively.
In the limit 
of infinite length $L=b-a \to \infty$, 
the condition (\ref{lnsnegatif}) to have a limit distribution
$P_{\infty}(Z)$ becomes a condition on the mean
value of the process $F(x)$ that should be strictly positive 
\begin{eqnarray}
F_0 \equiv [F(x)]_{av}  >0
\label{meanpositif}
\end{eqnarray}
The exponent $\mu$ (\ref{zunplusmu})
is now determined as the root of the equation (\ref{eqmudiscret}) 
\begin{eqnarray}
[ e^{- \mu \int_0^x dy F(y)} ]_{av}  =1
\label{eqmucontinuous}
\end{eqnarray}
for arbitrary $x$ as long as the process $F(x)$
has no correlation.

It is interesting to note that the exponential functional
(\ref{kestencontinuous}) actually
determines the stationary flux $J_L$ \cite{flux}
that exists in a given Sinai sample $[0,L]$ between two fixed concentration
$c_0$ and $c_N=0$ (i.e. particles are injected via a reservoir at $x=0$
and are removed when they arrive at the other boundary $x=N$) :
it is simply given by the inverse of
the variable $Z_L \equiv Z[0,L]$ 
\begin{eqnarray}
J_L= \frac{c_0}{Z_L}
\label{defflux}
\end{eqnarray}
In some sense, it is the simplest physical observable  
in the Sinai diffusion, as the surface magnetization is
the simplest order parameter in the RTFIC : both
can be expressed in a simple way in terms of the Kesten random variable
of the sample. 

The simplest process for $F(x)$ is of course the case where $F(x)$
is a biased Brownian motion  
\begin{eqnarray}
<F(x) >  && = F_0 \\
<F(x) F(x')>-F_0^2 && = 2 \sigma \delta(x-x')
\label{brown}
\end{eqnarray}
In this case, the exponent $\mu$ 
solution of (\ref{eqmucontinuous}) reads \cite{jpbreview,flux}
\begin{eqnarray}
\mu= \frac{ F_0}{  \sigma}
\label{gaussmu}
\end{eqnarray}
The Brownian process actually corresponds to the fixed point of the real-space renormalization approach and can thus be used to study the universal properties near the critical point.
The probability distribution of 
the random variable $Z_L$ (\ref{kestencontinuous})
in the case where the process $\{F(x)\}$ is a Brownian motion (\ref{brown})
can be determined exactly 
by various methods \cite{flux,yor,yorbook,microcano},
and we refer to these articles for various detailed explicit results.

\subsection{ Meaning of the strong disorder RG}

It turns out that the strong disorder RG for observables
involving Kesten random variables actually
corresponds to the following saddle-point analysis \cite{microcano}.
In the case of a Brownian process (\ref{brown}), the
continuous version of the Kesten process (\ref{kestencontinuous})
can be written as
\begin{eqnarray}
Z_L =  \int_0^L dx e^{- U(x)}
\end{eqnarray}
where the potential $U(x)=  \int_0^{x} dy F(y)$
is a random walk that presents
fluctuations of order $\sqrt{L}$ on the interval $[0,L]$.
As a consequence, it seems natural to evaluate
this integral in the large $L$ limit by the saddle-point method  
\begin{eqnarray}
Z_L \oppropto_{L \to \infty} e^{ E_L} 
\label{saddlezl}
\end{eqnarray}
where $(-E_L)<0$ is defined as the minimum reached by the process $U(x)$ 
on the interval $[0,L]$.
The scaling $E_L \sim \sqrt{L}$
shows that there will exists a limit distribution
for $(\ln Z_L)/\sqrt L $,
and that this limiting distribution is given by
the limit distribution of $E_L/\sqrt{L}$.
We refer the reader to \cite{microcano} for a more detailed discussion
along these lines.

In conclusion, the study of observables that
have a closed expression in terms of the disorder variables,
such as the surface magnetization in the RTFIC,
shed light on the meaning of the strong disorder approach
and of its asymptotic exactness :
for these variables, the strong disorder approach
amounts to perform a saddle-point analysis in each sample.
More generally, for the other observables
one may still consider
that the strong disorder approach gives in some sense a direct access
to an appropriate `saddle-point',
even if there is no closed exact expressions
on which one could perform a usual saddle-point method.

\section{ Extrema of 1D random potentials via RG}

In the Sinai model where a particle diffuses in a 1D Brownian potential,
we have seen that the strong disorder RG
amounts to construct the extrema of the Brownian potential at large scale,
where only barriers bigger than the RG scale $\Gamma$ are kept.
This construction of the extrema at large scale can actually be
defined for arbitrary 1D potentials \cite{rgtoy}.
In particular, the RG procedure may be implemented numerically
for correlated random potentials, which has been done
in particular for logarithmic correlations \cite{castillo}.

On the analytical side, we explain in this Appendix 
what can be said in general for the case of Markovian potentials
and some results that have been obtained for the specific example
of a Brownian potential with quadratic confinement
\cite{rgtoy}.

\subsection{ RG for Markovian potentials}

For ` Markovian ' random potentials $U(x)$ , satisfying a local 
Langevin equation of the form
 \begin{eqnarray} 
\frac{dU(x)}{dx} = F[U(x), X ] + \eta(x) 
\end{eqnarray} 
where $\eta(x)$ is a white noise, the measure of the renormalized landscape 
can be factorized in blocks which satisfy closed RG equations 
\cite{rgtoy}.  In the case of stationary landscapes, where the force $F$ is independent of $x$ 
\begin{eqnarray} 
F[U, X ] = F[U ] = - \frac{d W[U]}{d U} 
\label{defstatio} 
\end{eqnarray} 
(this case generalizes the pure Brownian landscape $F=0$ 
and the biased Brownian landscape $F[U]=F>0$), one obtains explicit solutions for the renormalization equations \cite{rgtoy}.
Other cases may also be explicitly solved, in particular 
the Brownian potential with quadratic confinement
considered in details below, which corresponds to a force independent of $U$ and linear in $x$ 
\begin{eqnarray} 
F[U(x), X ] = F[x ] = \mu X \label{forcetoy} 
\end{eqnarray}
The measure of the renormalized landscape is then expressed in terms
 of Airy functions \cite{rgtoy}.

\subsection{ Results for the Brownian potential with quadratic confinement}

\subsubsection{ Definition and properties of the model}

Let us first briefly explain the physical interests
of the so-called `toy model' 
 \begin{eqnarray} 
U_{toy}(x)=\frac{\mu}{2} x^2 + V(x) 
\label{deftoy} 
\end{eqnarray} containing a deterministic quadratic term and a random 
Brownian term Brownian $V(x)$ 
\begin{eqnarray} \overline{ \left(V(x)-V(y) \right)^2} = 2 \vert x-y \vert \label{corre} 
\end{eqnarray}
This model was introduced by Villain {\it et al.} as a `toy model' for  interfaces in the presence of random field \cite{villain83}.  In addition, within the field of random manifolds of internal dimension 
$D$ living in a random medium of dimension $(N+D)$, the toy model
 is considered as the extreme simplest case $D=0$ and $N=1$ 
\cite{mezardparisi92}.  

As a consequence of the quadratic containment, the absolute minimum of the potential (\ref{deftoy}) is finite and an Imry-Ma
argument can be used to obtain its scaling :
the balance between the elastic energy of order $\mu x^2$ and the random energy of order $\sqrt{x}$ yields the scaling
 \begin{eqnarray} 
x_{min} \sim \mu^{-2/3} 
\label{scaling1} 
\end{eqnarray} 
whereas the usual perturbation methods at all the orders \cite{villain88} or of iteration \cite{villainsemeria} are unable to reproduce this scaling (\ref{scaling1}).  Within the framework of the replica variational method, the result (\ref{scaling1}) requires a replica symmetry breaking \cite{mezardparisi92,engel}.

Another important property of the model 
is its ` statistical tilt symmetry' also present
in other models with random fields. This symmetry implies remarkable identities \cite{identities88} for disorder averaged of thermal cumulants
of the position, that are summarize by  
\begin{eqnarray} 
\overline { \ln < e^{ - \lambda X} >} && = T \frac{\lambda^2}{2 \mu} \label{genlambda2}
 \end{eqnarray} 
This identity on the generating function shows that the second cumulant
is simply 
\begin{eqnarray} 
 \overline { < x^2 > - < x>^2} = \frac{T}{\mu} 
\label{second} 
\end{eqnarray} 
and that the disorder averages of all higher cumulants actually vanish! 
 At low temperature, the result (\ref{second}) implies that the thermal fluctuations are related to the presence of metastable states in rare samples \cite{identities88}.  The renormalization allows in particular to study this phenomenon quantitatively, as explained below.

\subsubsection{ Results on the statistics of the minima} 

 At $T=0$, the particle is at the minimum $x_{min}$ 
of the random potential (\ref{deftoy}).  The final state $\Gamma=\infty$ of the renormalization procedure \cite{rgtoy} allows to find that the distribution of $x_{min}$ over
 the samples is, in agreement with \cite{groeneboom, frachebourgmartin} \begin{eqnarray} 
&& P_{]-\infty, +\infty[}(x) = g(x) g(-x) 
\label{distriminabs} \end{eqnarray} and the auxiliary function \begin{eqnarray} g(x) = \int_{-\infty}^{+\infty} \frac{d\lambda}{2 \pi} \frac{e^{ - i \lambda X}}{a Ai(b i \lambda)} 
\label{defg} 
\end{eqnarray} 
with the notations $a=(\mu/2)^{1/3}$ and $b=1/a^2$.

The probability that a sample presents two almost degenerate minima, 
located at positions $x_1$ and $x_2$,
 with an energy difference $\delta E=\epsilon \to 0$,  
can be written as 
\begin{eqnarray} 
{ \cal D}(\epsilon, x_1, x_2) = \epsilon g(-x_1) d(x_2-x_1)g(x_2) +O(\epsilon^2)
\end{eqnarray}
in terms of the function $g$ (\ref{defg}) and the function
 \begin{eqnarray} 
d(y) = a \int_{-\infty}^{+\infty} \frac{d \lambda}{2 \pi} e^{i \lambda y} \frac{ Ai' (i B \lambda)}{ Ai(i b\lambda)} 
\end{eqnarray}

The probability of having two minima separated by a distance $y>0$ is thus \begin{eqnarray} D(y)  = \int_{-\infty}^{+\infty} dx_1 \lim_{\epsilon \to 0} \left(\frac{ { \cal D}(\epsilon, x_1, x_1+y)}{\epsilon} \right) = B d(y) \int_{-\infty}^{+\infty} \frac{d\lambda}{2 \pi} \frac{e^{-i \lambda y}} { Ai^2(I B \lambda)} 
\label{deffDy}
 \end{eqnarray} 
In particular, the computation of the second moment yields
 \begin{eqnarray} 
\int_0^{+\infty} dy y^2 D(y) = \frac{1}{\mu} \label{y2toy}
 \end{eqnarray} 
The contribution of the samples with two nearly degenerate minima 
to the second thermal cumulant of the position (\ref{second}) can be estimated at first order in temperature as follows : the two minima have for respective Boltzmann weights $ p = \frac{ 1} { 1 + e^{ - \beta \epsilon}} $ and $(1-p)=\frac{ e^{ - \beta \epsilon}} { 1 + e^{ - \beta \epsilon}} $. 
The variable $(x-<x>)$ is thus $(1-p)(x_1-x_2)$ with probability $p$ and $p (x_2-x_1)$ with probability $(1-p)$. The average yields
 \begin{eqnarray} 
&& \overline{ < (x-<x>)^2 >} = \overline{ p(1-p) (x_1-x_2)^2} \\ 
&& = \int_{-\infty}^{+\infty} dx_1 \int_{x_1}^{+\infty} dx_2 \int_{-\infty}^{+\infty} d\epsilon { \cal D}' (\epsilon=0, x_1, x_2) \frac{e^{ - \epsilon/T}}{(1 + e^{ - \epsilon/T})^2} (x_2-x_1)^2 \nonumber \\ 
&& = T \int_0^{+\infty} dy D(y) 
 \end{eqnarray} 
Using (\ref{y2toy}), one then obtains the exact result (\ref{second}).  This shows that the thermal fluctuations at low temperature are entirely 
due to the metastable states which exist in some rare samples. 
 In particular, the susceptibility 
\begin{eqnarray} \chi \equiv \frac{1}{T} \left(< x^2>-<x>^2 \right) \label{suscepti}
 \end{eqnarray}
has a finite average at zero temperature
 \begin{eqnarray}
 \overline{ \chi}  = \frac{1}{\mu} 
\end{eqnarray} 
but only the samples with two nearly degenerate minima actually contribute to this average value, because the typical samples with only one minimum have a susceptibility which vanish at zero temperature. 

Similarly, the even moments of the relative position $(x-<x>)$ behaves in the following way at low temperature 
\begin{eqnarray} 
\overline { < (x-<x>)^{2n} >} && = \frac{T}{n} \int_0^{+\infty} y^{2n} D(y) +O(T^2)
 \label{momentslowT}
 \end{eqnarray}
 in term of the function $D(y)$ defined in (\ref{deffDy}).  The comparison with the identity (\ref{genlambda2}) shows that there are many terms which cancel in the disorder averages of thermal cumulants. 

 \subsubsection{ Results on the statistics of the largest barrier} 

The time $t_{eq}$ necessary to reach equilibrium is directly related to the largest barrier $\Gamma_{max}=t \ln t_{eq}$ existing in the sample.  More precisely, the probability ${\cal P}(t_{eq}<t)$ that the system 
has already reached equilibrium at time $t$, corresponds to the probability that there remains only one renormalized valley at scale $\gamma = T \ln t$:  it is a function of the scaling variable  $\gamma$ \begin{eqnarray} { \cal P}(t_{eq}<t) = \Phi \left(\gamma \equiv \left(\frac{\mu}{2} \right)^{1/3} T \ln T \right) \label{ptequi} 
\end{eqnarray} 
The function $\Phi$ has the following
 explicit expression in terms of Airy functions \cite{rgtoy}
\begin{eqnarray}
\Phi \left(\gamma \right)
&&  = \int_{-\infty}^{+\infty} \frac{d\lambda}{2 \pi}
\frac{1}{ Ai^2(i  \lambda)} e^{- 2 \int_{0}^{+\infty}
 df \tilde{\psi}_{\gamma} (f,\lambda)} 
\\
 \tilde{\psi}_{\gamma}(f,\lambda) && = \frac{  Ai(f+\gamma +  i \lambda)}
{\pi Ai(f +  i \lambda) 
\left[ Ai(f +  i \lambda) Bi(f+\gamma +  i \lambda)
 -Bi(f +  i \lambda) 
Ai(f+\gamma +  i \lambda) \right]}
\nonumber
\end{eqnarray}

This result yields directly by derivation the probability distribution of the scaling variable $\gamma = \left(\frac{\mu}{2} \right)^{1/3} \Gamma_{max}$ 
for the largest barrier $\Gamma_{max}$ existing in the sample 
\begin{eqnarray} P_{max}(\gamma) = \frac{d}{d \gamma} \Phi \left(\gamma \right) \label{distribarrmax} 
\end{eqnarray} 
The asymptotic behaviors of this distribution are as follows 
\begin{eqnarray} 
P_{max}(\gamma) && \opsimeq_{\gamma \to \infty} \frac{9}{4} \sqrt{\frac{\pi}{2}} \gamma^{5/4} e^{ - \frac{3}{2} \gamma^{3/2}} \label{distribarrmaxlarge} \\ P_{max}(\gamma) && \opsimeq_{\gamma \to 0} \frac{6 \zeta(3)}{ \gamma^4} e^{ - 2 \frac{\zeta(3)}{ \gamma^3}} 
\label{distribarrmaxsmall} 
\end{eqnarray} 
where $\zeta(n)$ is the Riemann zeta function.

\section{ Comparison with some growth models without disorder}

\label{growthmodels}

The strong disorder RG rules discussed in Section  \ref{chapRGrules} have actually a close relationship with some growth models, which were introduced in a completely independent way.  These geometrical growth models describe a collection of intervals on the line which evolve by an iterative transformation on the smallest interval remaining
on the line. The various rules that have been considered are
the following : 
 
(i) in the ``cut-in-two model "\cite{yekutieli}, the smallest interval is eliminated and gives one half of its length to each one of these two neighbors.  This model thus introduces correlations between the neighboring intervals and 
has been studied numerically  \cite{yekutieli}. 

 (ii) in the ``paste-all model "\cite{yekutieli}, the smallest interval is eliminated and gives all its length randomly to the one of these two neighbors drawn in an equiprobable way.  This model does not introduce correlations between the neighboring intervals, and the invariant distribution of lengths has been computed \cite{yekutieli}. 

 (iii) in the ``instantaneous collapse model "\cite{rutenberg}, the smallest interval is eliminated with its two neighbors to form one new interval
$l =l_1+l_2+l_3$.  This model actually describes the effective dynamics at large time of the unidimensional scalar field which evolves according to a
Ginzburg-Landau equation at zero temperature \cite{nagai, rutenberg}.
This model thus aroused a great interest as a soluble model of coarsening.  The exact results concern the invariant distribution of the lengths \cite{rutenberg}, the persistence exponent \cite{phi4persist} which characterizes the auxiliary variable $d'=d_1+d_3$, the autocorrelation exponent \cite{phi4autocorre} which characterizes the auxiliary variable $q'=q_1-q_2+q_3$ and finally the 
generalized persistence exponent\cite{majbray} which characterizes the auxiliary variable $m'=m_1+p m_2+ m_3$ with a parameter $p$.  The only technical difference 
 with the strong disorder RG rules presented before is that, in these growth models, it is the length which is the main variable that determines the renormalization, whereas in  disordered models, the length is only an auxiliary variable, the main variable 
that defines the dynamics being a disorder variable.  This explains the analytical differences between the solutions for the fixed points and the exponents in the two types of models. 

 From a physical point of view, this example shows that a dynamics without intrinsic disorder, starting from a random initial condition, can be controlled, in a certain sense, by an `infinite disorder ' fixed point.
More recently, another pure dynamical model starting from a random initial condition
 involving diffusion and annihilation of multi-species
was shown to be described also by some RG of the Ma-Dasgupta type
\cite{hilhorstetal}.

\newpage


\end{document}